\documentclass[fleqn,usenatbib]{mnras}
\usepackage{newtxtext,newtxmath}
\usepackage[T1]{fontenc}

\DeclareRobustCommand{\VAN}[3]{#2}
\let\VANthebibliography\thebibliography
\def\thebibliography{\DeclareRobustCommand{\VAN}[3]{##3}\VANthebibliography}

\usepackage{graphicx}	
\graphicspath{{figures/}{images/}}
\usepackage{amsmath}	
\usepackage{upgreek}
\usepackage[toc,page]{appendix}
\usepackage{longtable}
\usepackage{pifont}
\usepackage{subcaption}
\usepackage{capt-of}
\usepackage{comment}
\usepackage{booktabs}
\usepackage{gensymb} 
\usepackage{siunitx}
\usepackage{pdflscape}

\def\H0{{\rm ~km~s^{-1}~Mpc^{-1}}}

\def\cm3{$\rm ~cm^{-3}$}

\def\la{\mathrel{\hbox{\rlap{\hbox{\lower4pt\hbox{$\sim$}}}{\raise2pt\hbox{$<$}}}}}
\def\ga{\mathrel{\hbox{\rlap{\hbox{\lower4pt\hbox{$\sim$}}}{\raise2pt\hbox{$>$}}}}}

\def\hii{H~{\scriptsize II}$~$}

\title[Higher-resolution follow-up of point-like CORNISH UCHII regions]{The coordinated radio and infrared survey for high-mass star formation -- VI. Higher-resolution follow-up of point-like northern ultra-compact H {\large II} regions}

\author[I. E. Kalcheva et al.]{
Ivayla E. Kalcheva,$^{1}$\thanks{E-mail: ikalcheva@astro.bas.bg}
Melvin G. Hoare$^{2}$
\\
$^{1}$Institute of Astronomy and National Astronomical Observatory, Bulgarian Academy of Sciences, Tsarigradsko Chaussee 72,
BG-1784 Sofia, Bulgaria\\
$^{2}$School of Physics and Astronomy, University of Leeds, Leeds LS2 9JT, UK\\
}

\pubyear{2026}

\begin{document}
\label{firstpage}
\pagerange{\pageref{firstpage}--\pageref{lastpage}}
\maketitle

\begin{abstract}
The CORNISH survey provided the most complete sample of UCHII regions in the northern Galactic plane to date, but at its 1.5\arcsec\ resolution the most compact, and plausibly youngest, members of the sample could not be morphologically classified. We present Karl G. Jansky Very Large Array (JVLA) A-configuration observations at 10~GHz of 69 such CORNISH UCHII regions, reaching an angular resolution of $\sim$0.2--0.3\arcsec\ and a typical rms noise of $\sim$10~$\upmu$Jy~beam$^{-1}$. All but two of the imaged targets are detected; the two non-detections lie in fields that are dynamic-range limited by bright neighbours. Six targets ($\sim$9 per cent) are resolved into pairs of sources, and a further 15 CORNISH UCHII regions detected within the fields reveal complex, multi-component structure. We also report candidate massive young stellar object (MYSO) jet knots and a row of three compact sources coincident with a previously unresolved infrared-dark \hii\ region. The targets have a median deconvolved angular diameter of $\sim$0.9\arcsec\ and physical diameters of 0.004--0.17~pc (median $\sim$0.03~pc), with no correlation between angular size and distance, consistent with a population of intrinsically compact, young regions. Spectral indices between 1.4 and 10~GHz place $\sim$32 per cent of the targets below the thermal free-free limit, possibly indicating non-thermal emission or variability. These data enable morphological classification of nearly the entire CORNISH UCHII sample, which will be presented in a forthcoming paper.
\end{abstract}


\section{Introduction}

Despite their outsized influence on galactic evolution -- through ionising feedback, stellar winds and eventual supernovae that reshape the interstellar medium (ISM) and seed it with heavy elements \citep{yorke:2007} -- massive stars ($M \gtrsim 8~M_{\odot}$) are rare, and the earliest phases of their formation remain poorly constrained observationally, owing to their rapid evolutionary timescales and the heavy dust obscuration of the dense natal cores in which they are embedded \citep{hoare:2007}.

An ultra-compact \hii~(UCHII) region marks a stage at which a newly formed massive star has photoionised a pocket of gas in its immediate surroundings, and represents an intermediate phase between the deeply embedded accretion phase and the more diffuse, evolved \hii~regions visible throughout the Galactic plane. First distinguished as a separate observational class by \citet{churchwell:1989a}, UCHII regions are compact, dense pockets of ionised gas -- typically $\lesssim$0.1~pc across, with emission measures in excess of $10^{7}$~pc~cm$^{-6}$ and electron densities exceeding $10^{4}$~cm$^{-3}$ \citep{churchwell:1989a, churchwell:2002}. Crucially, \citet{churchwell:1989a} also established a morphological classification scheme for UCHII regions -- cometary, shell, core-halo, spherical/unresolved and irregular -- based on higher-resolution radio continuum imaging of a smaller IRAS-selected sample. The relative incidence of these morphological classes provides constraints on the distribution of ionised gas around newly formed massive stars and, potentially, on the physical mechanisms (champagne flows, bow shocks, stellar-wind confinement, or ongoing accretion) shaping this early visible phase \citep{depree:2005}. Revisiting individual well-studied regions such as W49A and Sgr\,B2 at higher angular resolution, \citet{depree:2005} found shell morphologies to be more prevalent than previously recognised and proposed a modified classification scheme -- dropping core-halo, adding a bipolar category, and relabelling shell as shell-like -- underlining that morphological classification is highly sensitive to the resolution at which a sample is imaged.

The CORNISH survey \citep{hoare:2012, purcell:2013} was the first arcsecond-resolution radio continuum survey of the northern Galactic plane ($10\degr<l<65\degr$, $|b|<1\degr$), with a resolution of 1.5\arcsec~at 5~GHz and coverage and resolution matched to those of the mid-infrared GLIMPSE survey \citep{benjamin:2003}. It yielded the largest complete and unbiased sample of UCHII regions to date, comprising 240 visually identified candidates, 239 of which were confirmed as UCHII regions and characterised in \citet[][hereafter Paper~III]{kalcheva:2018}. The survey was subsequently extended to the southern Galactic plane by the CORNISH-South survey \citep{irabor:2023}. This sample has formed the basis for statistically robust studies of the physical properties, spectral indices, distances and luminosities of the UCHII population as a whole. However, at the CORNISH resolution, a substantial fraction of the sample (107 sources; Section~\ref{observations}) -- particularly the smallest and plausibly youngest and least evolved UCHII regions -- remained unresolved or insufficiently resolved, precluding morphological classification of these objects and biasing morphology-based analyses towards the larger, more evolved members of the sample.

In the years since, several new Galactic plane radio continuum surveys have become available. The VLA Sky Survey \citep[VLASS;][]{lacy:2020} covered the entire sky north of $\delta = -40\degr$ at 3~GHz with a $\sim$2.5\arcsec~synthesised beam -- coarser than CORNISH itself -- although its three observing epochs made it a valuable resource for identifying radio variability. The SARAO MeerKAT Galactic Plane Survey \citep[SMGPS;][]{goedhart:2024} reached excellent sensitivity at L band (886--1678~MHz) but with an 8\arcsec~resolution, insufficient to resolve the internal structure of typical UCHII regions. The GLOSTAR survey \citep{brunthaler:2021}, combining the VLA and the Effelsberg 100-m telescope at C band (4--8~GHz), reached a resolution of 1.5\arcsec, though the VLA B-configuration source catalogue \citep{dzib:2023} achieved 1.0\arcsec\ by suppressing short baselines.
None of these more recent surveys reached the sub-arcsecond resolution needed to resolve the internal structure of UCHII regions, which typically have angular sizes of only a few arcseconds at kiloparsec-scale distances (Paper~III).

To address this limitation, we obtained JVLA follow-up observations of the unresolved or insufficiently resolved CORNISH UCHII regions at 10~GHz and sub-arcsecond resolution, described in Section~\ref{observations}.

A further motivation for revisiting these sources is the growing evidence that UCHII regions are not static on human observational timescales. Comparing the 5~GHz CORNISH fluxes to the earlier (1989--1991) MAGPIS 6\,cm survey, tentative evidence was found in Paper~III that $\sim$5 per cent of the CORNISH UCHII sample had brightened by 50 per cent or more over $\sim$15~yr, a rate broadly consistent with, if somewhat higher than, predictions from three-dimensional radiation-hydrodynamic collapse simulations of accreting massive protostars \citep{peters:2010a, galvanmadrid:2011}. More recently, \citet{yang:2025} compared CORNISH and GLOSTAR peak fluxes over a baseline of $\sim$8~yr and found that 44$\pm$5 per cent of the 86 UCHII regions in common between the two surveys showed variability -- substantially higher than earlier estimates, and difficult to reconcile with steady, quiescent accretion alone. Repeated high-resolution imaging of the same targets, of the kind presented here, therefore also lays the groundwork for future studies of morphological and flux variability in individual UCHII regions, complementing the multi-epoch but lower-resolution VLASS data.

In this paper, we present these observations. The primary objective is to determine the morphologies of these smaller, possibly younger UCHII regions, as their morphologies trace the distribution and structure of ionised gas in the immediate vicinity of newly formed massive stars. This will enable morphological classification of nearly the entire CORNISH UCHII sample, providing a more complete view of its morphological diversity. A comprehensive morphological classification will be presented in a follow-up paper (in prep.).

We describe the data reduction, imaging, and quality assessment of the high-angular-resolution observations, with particular emphasis on observations of busy fields and the imaging techniques applied to the W49A star-forming region. We also discuss selected re-observed CORNISH sources in which the higher-resolution data reveal additional structural complexity. Flux density estimates, source angular extents, and derived properties such as physical sizes are presented. Taking advantage of the 10~GHz flux measurements and the available CORNISH UCHII flux densities at lower frequencies (see Paper~III), we derive spectral indices using least-squares fitting and discuss them in the context of the findings for the CORNISH UCHII sample.

\section{The JVLA A-configuration targeted survey of CORNISH UCHIIs}\label{observations}

\subsection{Target selection}
Visual inspection of the original sample of UCHII regions found in CORNISH showed that the morphology of all sources smaller than 2.7\arcsec could not be reliably determined \citep{hoare:2012}.
This subset comprises 107 sources.
A total of 22 of these unresolved CORNISH UCHIIs lie at $\delta < -15\degree$ ($l \lesssim 16\degree$), where CORNISH was observed in the hybrid BnA configuration to limit beam elongation at low elevation, which also provided a somewhat higher resolution along one axis \citep{hoare:2012}. These sources were not included in the follow-up, as at such low declinations A-configuration observations would suffer from a strongly elongated beam and a limited observable time range.
 Excluding all cases where the morphology was clearly established through previous high-resolution radio continuum observations, as well as seven particularly compact sources that required even higher resolution (Q-band) observations, the target sample size was reduced to 69 sources (one of which was later re-classified as a radio star; Section~\ref{detn_stats}).

\subsection{JVLA observational set-up}\label{setup_evla}
\begin{table}
\centering
\caption[Summary of observational parameters]{Summary of observational parameters. The synthesised beam depends on the declination and hour angle. The rms noise was measured in \texttt{CASA} from emission-free regions near field centres (the average for all target fields).}
\begin{tabular}{ll}
\hline
\multicolumn{1}{l}{Parameter} & \multicolumn{1}{l}{Value}  \\
\hline
Band & X (8 -- 12 GHz) \\
Primary beam & $\sim$4.2$'$ \\
Synthesised beamwidth & 0.20\arcsec--0.34\arcsec (median 0.24\arcsec~$\times$~0.17\arcsec) \\
Theoretical rms & 0.011 mJy~beam$^{-1}$  \\
Measured rms & 0.008--0.012 mJy~beam$^{-1}$ \\
Largest angular scale& 5.3\arcsec \\
Image pixel size & 0.05\arcsec -- 0.08\arcsec  \\
\hline
\end{tabular}
\label{table:obs_parameters}
\end{table}

The follow-up targeted survey was observed with the Jansky Very Large Array (JVLA) in continuum mapping mode in A configuration, achieving a resolution 7.5~times higher than that of CORNISH.
Observing at X band (8--12~GHz) provided a further gain in resolution and a sensitivity significantly higher than that of CORNISH, which was observed at C band.

Follow-up observations of unresolved CORNISH sources were undertaken
on 6 and 14 April 2014, in six-hour observing blocks (project code 14A-085, PI: M.~G.~Hoare)\footnote{The CORNISH survey observations were conducted in the 2006 and 2007/2008 VLA observing seasons; see Table 1 in \protect \cite{purcell:2013}.}.
The array had 27 functional antennas at the time. The minimum
antenna baseline for this configuration is 0.68~km, and the maximum is 36.4~km\footnote{See \url{https://science.nrao.edu/facilities/vla/docs/manuals/oss2016A/performance/resolution}}.
The X-band continuum observations covered the targeted CORNISH UCHIIs with $16\degree<l<65\degree$. A total bandwidth of 2~$\times$~2~GHz was used, with each 2~GHz portion divided into 16 spectral windows of 128~MHz each to reduce bandwidth smearing. The restoring beam sizes range from 0.20\arcsec~$\times$~0.14\arcsec\ to 0.34\arcsec~$\times$~0.17\arcsec, with a median of $\sim$0.24\arcsec~$\times$~0.17\arcsec. The largest reliable angular scale (i.e.\ the largest structure that can be imaged reasonably well in full-synthesis observations) for this set-up is $\sim$5.3\arcsec. Three sources found in the imaged fields exceed this limit -- G026.0907$-$00.0575 (deconvolved major axis $\sim$7.9\arcsec), G34.26+0.15 ($\sim$5.75\arcsec\ in CORNISH) and the W49A neighbour G043.1701+00.0078 ($\sim$5.4\arcsec) -- so their flux densities are likely underestimated. All science targets are smaller than this limit: the largest, G038.6465$-$00.2260, has a deconvolved major axis of $3.84 \pm 0.12$\arcsec\ at 10~GHz.

\begin{figure*}
\centering
\begin{tabular}{@{}ccc@{}}
\includegraphics[width=0.30\textwidth]{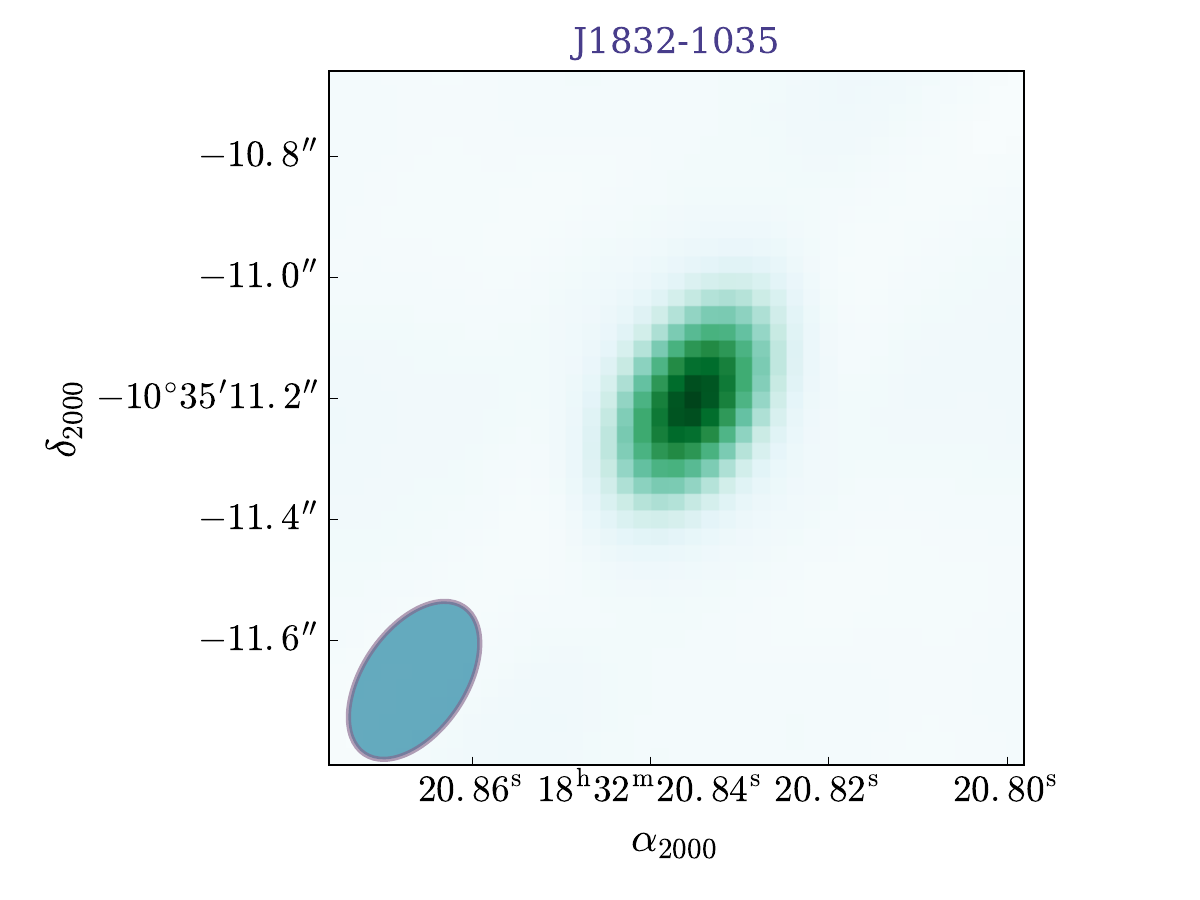} &
\includegraphics[width=0.30\textwidth]{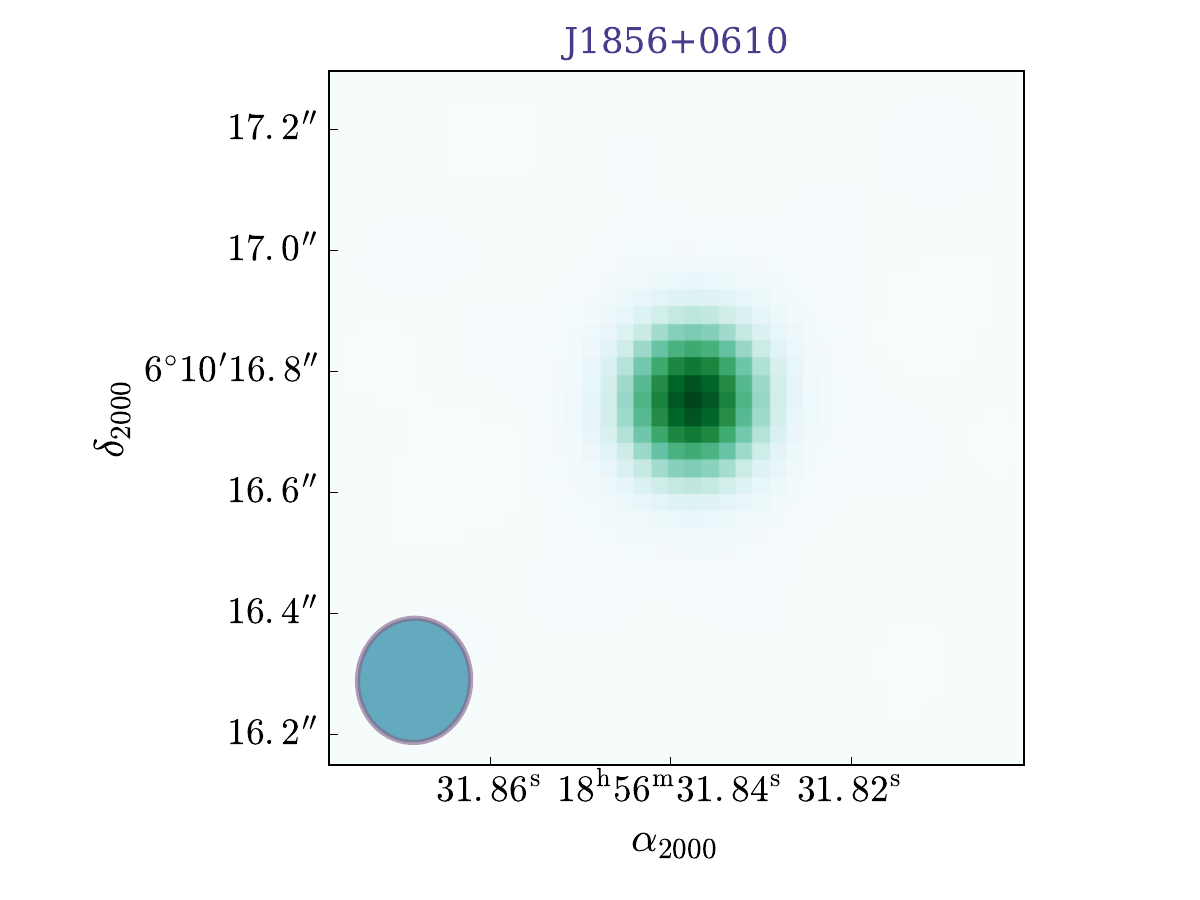} &
\includegraphics[width=0.30\textwidth]{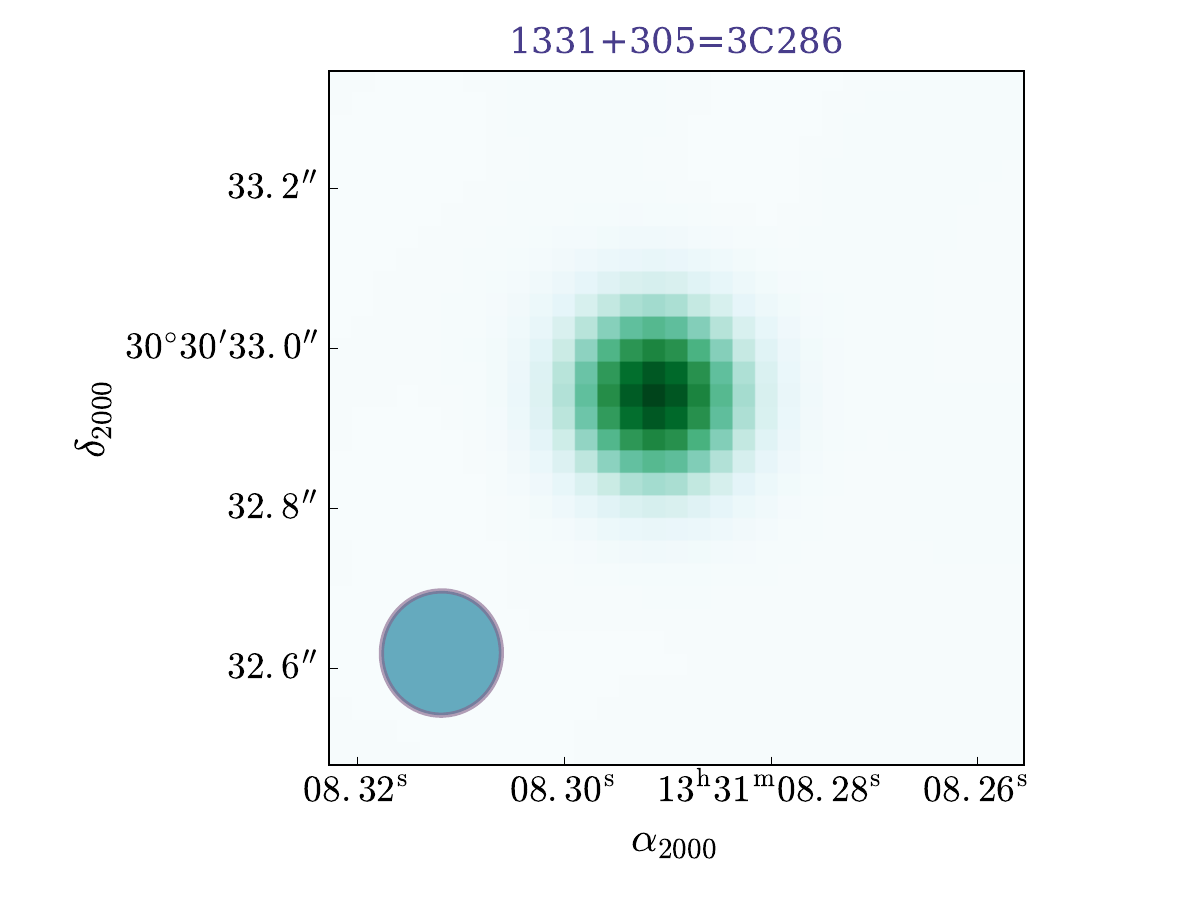} \\[2mm]
\includegraphics[width=0.30\textwidth]{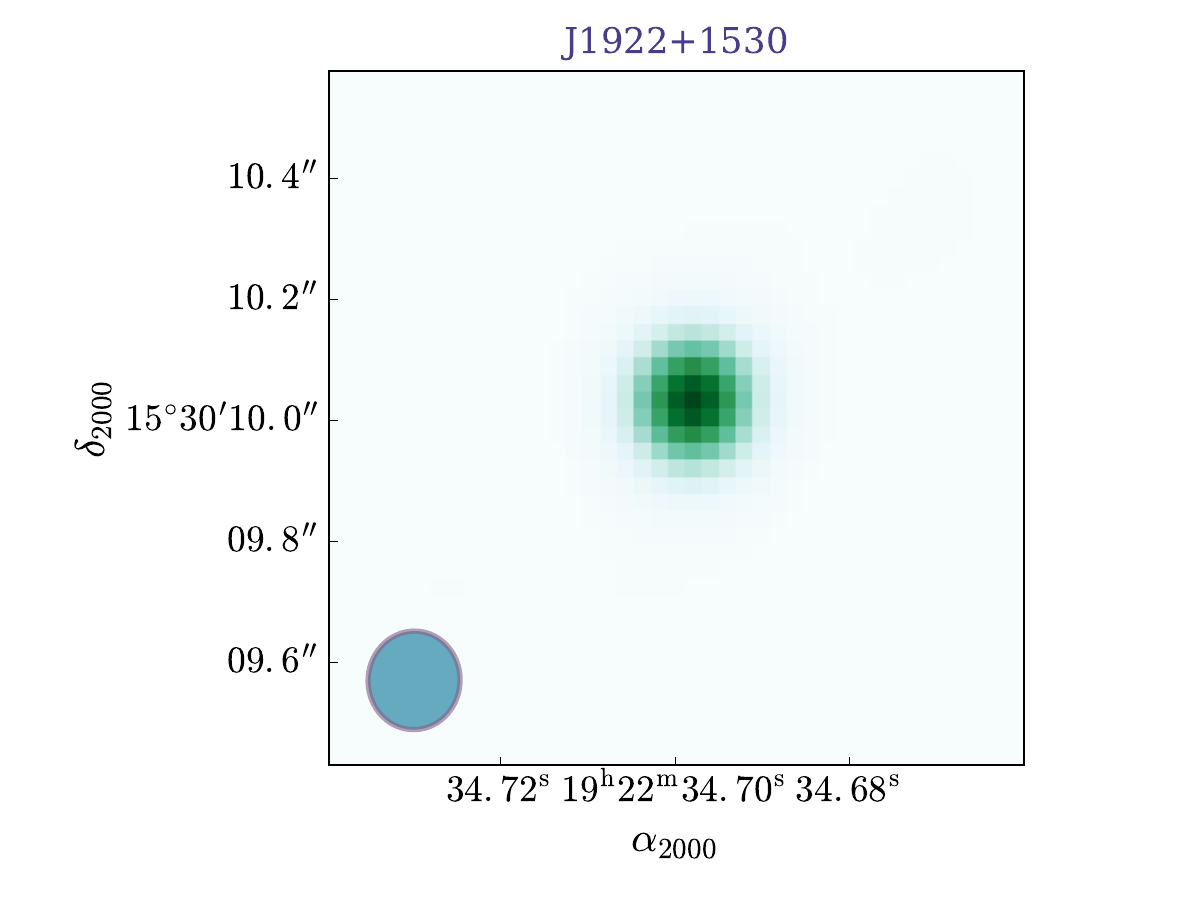} &
\includegraphics[width=0.30\textwidth]{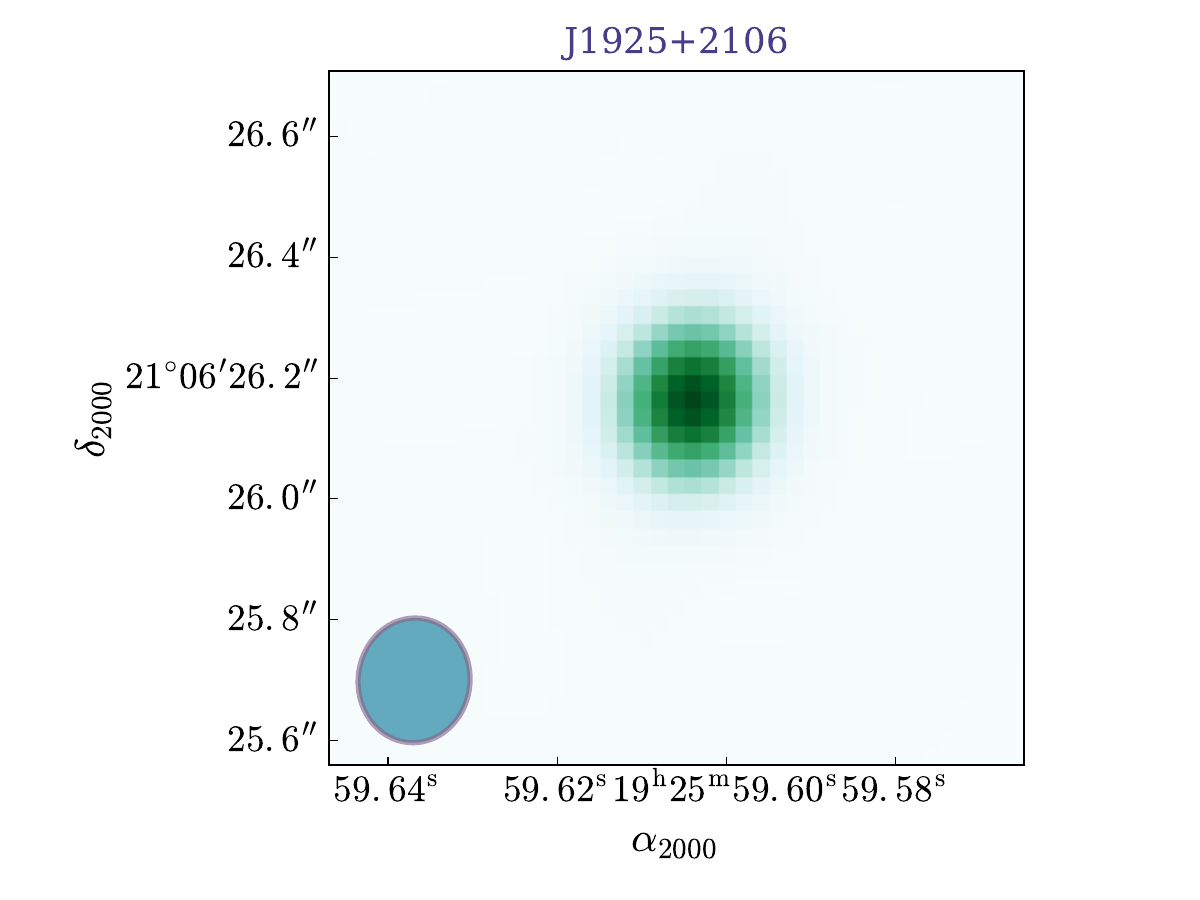} &
\end{tabular}
\caption[Phase calibrators and flux calibrator used in the JVLA follow-up]{Phase calibrators (J1832$-$1035, J1856+0610, J1922+1530, J1925+2106) and flux calibrator (3C 286 = J1331+3030) used in the JVLA follow-up observations. The clean beam is illustrated in the bottom left corner of each panel.}
\label{fig:calibrators}
\end{figure*}

The target UCHIIs were grouped by position into small blocks; a short observation of a nearby  phase calibrator (usually 1--2 minutes, depending on its flux density) was conducted before and after observing each block.
The phase calibrators J1832$-$1035, J1856+0610, J1922+1530, and J1925+2106 were used in the JVLA follow-up to correct for fluctuations in the amplitude and phase due to atmospheric and instrumental effects. As can be seen in Fig.~3 of \cite{hoare:2012}, these sources provide coverage of the entire CORNISH area, as all survey blocks are located within $\sim$ 10$\degree$ of a calibrator.
The quasar 1331+305 (also known as 3C 286\footnote{In the Third Cambridge Catalogue of Radio Sources.}), one of the four primary VLA calibrators, was used for absolute flux density calibration. The primary flux calibrator was observed at the beginning of each of the two observing sessions, with a scan length of 10 minutes each time.
Images of all calibrators are shown in Fig.~\ref{fig:calibrators}.

For the target UCHIIs, the total scan time amounted to about 10 minutes (8 minutes on-source, with additional time for overheads).
This exposure time was needed to obtain a noise level of 11~$\upmu$Jy~beam$^{-1}$. A lower limit to the surface brightness of the targets was estimated from the CORNISH 7$\sigma$ detection limit of 2.8~mJy per 1.5\arcsec\ beam (Paper~III), which corresponds to $\sim$50~$\upmu$Jy per 0.2\arcsec\ beam at 5~GHz. When planning the observations, it was anticipated that, as the most compact and possibly youngest members of the CORNISH UCHII sample, many of the targets could still be partly optically thick at 5~GHz, with turnover frequencies of $\sim$5--15~GHz \citep{kurtz2005hypercompact}, and thus at least as bright at 10~GHz as at 5~GHz. This assumption is tested in Section~\ref{cha3phys}.

\section{Data reduction}\label{data_reduction}

\subsection{Pipeline processing and flagging}

The data (comprising two blocks corresponding to the two observing sessions) were reduced with the NRAO Common Astronomy Software Applications package (\texttt{CASA}, see \citealt{mcmullin:2007}), by using the standard JVLA calibration pipeline and methodology for continuum emission\footnote{See \protect\url{https://casa.nrao.edu/} and \protect\url{https://science.nrao.edu/facilities/vla/data-processing/pipeline}.}.
Known radio frequency interference (RFI), e.g. from commercial broadcasting, and other issues (e.g. shadowed data, edge channels of sub-bands, system problems) were automatically flagged during the pipeline processing. Additional editing was performed to mark and exclude any data still plagued by problems, such as remaining significant phase and amplitude excursions (for example from sporadic natural or human-made RFI, which is typically highly polarised). The upper end of the band was the most strongly affected: in the final five spectral windows (640~MHz in total), up to 84 per cent of the data in the first two and virtually all data ($>$94 per cent) in the final three were flagged. Each of the two six-hour datasets was split into individual target fields prior to the second round of flagging, to expedite the visual inspection, and the pipeline was re-run to apply the new flags. This was done to ensure good imaging sensitivity, as any remaining issues can severely reduce it or even cause calibration to fail.

All viable spectral windows (out of the 32 unique ones)
were combined. As the upper $\sim$640~MHz of the band were largely flagged, the effective frequency of the frequency-averaged images, computed as the mean of the spectral window frequencies weighted by their unflagged fractions, is $\sim$9.75~GHz. Using a nominal frequency of 10~GHz instead changes the derived spectral indices by $\lesssim$1 per cent, and 10~GHz is used hereafter for simplicity.
The computed theoretical rms noise is 11~$\upmu$Jy~beam$^{-1}$ at medium elevation (25$\degree$--50$\degree$)\footnote{The theoretical rms was computed with the \protect\href{https://obs.vla.nrao.edu/ect/}{NRAO JVLA Exposure Calculator} v18A for semester 18A. The tool accounts for the RFI levels characteristic of the 8--12~GHz frequency range, which can be up to 15 per cent.}.

The typical measured 1$\sigma$ rms noise (close to the field centres) was in excellent agreement with predictions -- ranging from $\sim$8--12~$\upmu$Jy~beam$^{-1}$ between fields.
For point sources, the new data are thus $\sim$40 times more sensitive than CORNISH (7$\sigma$ limits of $\sim$0.07~mJy vs 2.8~mJy). The gain in surface-brightness sensitivity is, however, much smaller: scaled to the median JVLA beam, the CORNISH 7$\sigma$ limit corresponds to $\sim$50~$\upmu$Jy~beam$^{-1}$, slightly below the JVLA 7$\sigma$ limit. The detection of the more extended, fainter targets therefore depends on their surface brightness at 10~GHz rather than on the point-source sensitivity, whereas compact sources such as massive young stellar object (MYSO) jets and, in some cases, possibly the stellar winds of the ionising OB stars, are well within reach.
 Towards the edges of fields, the rms noise levels reach 26 $\upmu$Jy~beam$^{-1}$, or a 7$\sigma$ sensitivity limit of $\sim$0.18~mJy -- this is considered the sensitivity limit for non-target sources found
 within the observed fields.
 Such sources were also analysed and are subsequently discussed (the `neighbour group' from here on).

\subsection{Quality check -- calibrator fluxes}

\begin{table*}
\centering
\caption[Comparison of tabulated and newly measured JVLA calibrator fluxes]{Comparison of tabulated (fifth and sixth columns) and newly measured (final column) JVLA calibrator fluxes.}
\begin{tabular*}{\textwidth}{@{\extracolsep{\fill}} ll ll S[table-format=1.2] S[table-format=1.2] S[table-format=1.3(3)]}
\hline
\multicolumn{1}{l}{Calibrator} & \multicolumn{1}{l}{Type} & \multicolumn{1}{c}{$\alpha$} & \multicolumn{1}{c}{$\delta$} & \multicolumn{1}{c}{$F_{\rm 3.7 cm}$}&  \multicolumn{1}{c}{$F_{\rm 2 cm}$}& \multicolumn{1}{c}{Measured $F_{\rm 3 cm}$} \\

\multicolumn{1}{l}{} & \multicolumn{1}{l}{} &\multicolumn{1}{c}{(J2000)} & \multicolumn{1}{c}{(J2000)} & \multicolumn{1}{c}{(Jy)}&  \multicolumn{1}{c}{(Jy)}& \multicolumn{1}{c}{(Jy)} \\

\hline

1832$-$1035 & Phase &  18$^{\rm h}$32$^{\rm m}$20.836$^{\rm s}$ & $-$10$\degree$35$'$11.200$''$ & 1.28  & 1.25 & 1.04 \pm 0.01 \\
1856+0610   & Phase  &   18$^{\rm h}$56$^{\rm m}$31.839$^{\rm s}$ & +06$\degree$10$'$16.754$''$ &  0.3 & {-} & 0.182 \pm 0.001\\
1922+1530    & Phase  &   19$^{\rm h}$22$^{\rm m}$34.699$^{\rm s}$& +15$\degree$30$'$10.032$''$ & 1.0  & 0.7 & 1.04 \pm 0.01\\
1925+2106   & Phase  &    19$^{\rm h}$25$^{\rm m}$59.605$^{\rm s}$& +21$\degree$06$'$26.162$''$ &  1.0 & 1.8 & 0.767 \pm 0.001 \\
\hline
\end{tabular*}
\label{table:calibrators}
\end{table*}

Table \ref{table:calibrators} presents a comparison between the newly measured and the catalogued fluxes of the calibrators\footnote{See \url{https://science.nrao.edu/facilities/vla/observing/callist}}.
The results are comparable, noting that the flux density values included in the VLA calibrator manual are more than 10 years out of date and that calibrators are typically intrinsically variable (and thus differences from current values by more than a factor of 2 may be expected).

\subsection{Imaging}

\subsubsection{Overall method}
The frequency-averaged, split, and calibrated target fields were imaged with \texttt{CASA} and converted to FITS format.
Deconvolution was performed using the standard task \texttt{clean} to remove the artefacts from the dirty beam. Each image was cleaned down to a conservative threshold of 0.15~mJy or, in several cases, 0.1~mJy --
this cutoff level was found to be optimal through visual inspection with the \texttt{CASA~viewer} (to avoid potential \texttt{clean} `bias'). The exact value was determined depending on the individual noise level of each field -- by measuring the rms noise levels in the central portion of the field, as well as near the field edge. Interactive cleaning was performed whenever necessary, to remove artefacts resulting from bright sources in the primary beam -- this entailed `boxing' bright neighbours prior to focusing on the field target. The issue occurred in fields where the target UCHII was relatively weak. The weaker flux was `cleaned' at a later stage, thus incurring larger flux errors unless the
bright neighbour is handled appropriately.
This effect was not prominent in most of the fields containing a multitude of sources, as the data averaged over all (available after the flagging stage) spectral windows were used, ensuring a good signal-to-noise ratio.

Imaging was sensitivity-limited, with the exception of busier fields where bright, evolved \hii regions were also present. In the latter case, which is the case for  $\sim$10 per cent of the maps, results were limited by the dynamic range.
 Imaging was first approached in a more automated manner; the resulting clean maps were examined visually, and the best successfully deconvolved maps were selected. An individual approach was used in all cases with one or more especially bright neighbours in the field, as well as in otherwise challenging cases.

\subsubsection{Choice of \texttt{clean} parameters}

\begin{figure}
\centering
\textbf{W49A -- tapering, centre}\par\medskip
   \includegraphics[width=\columnwidth]{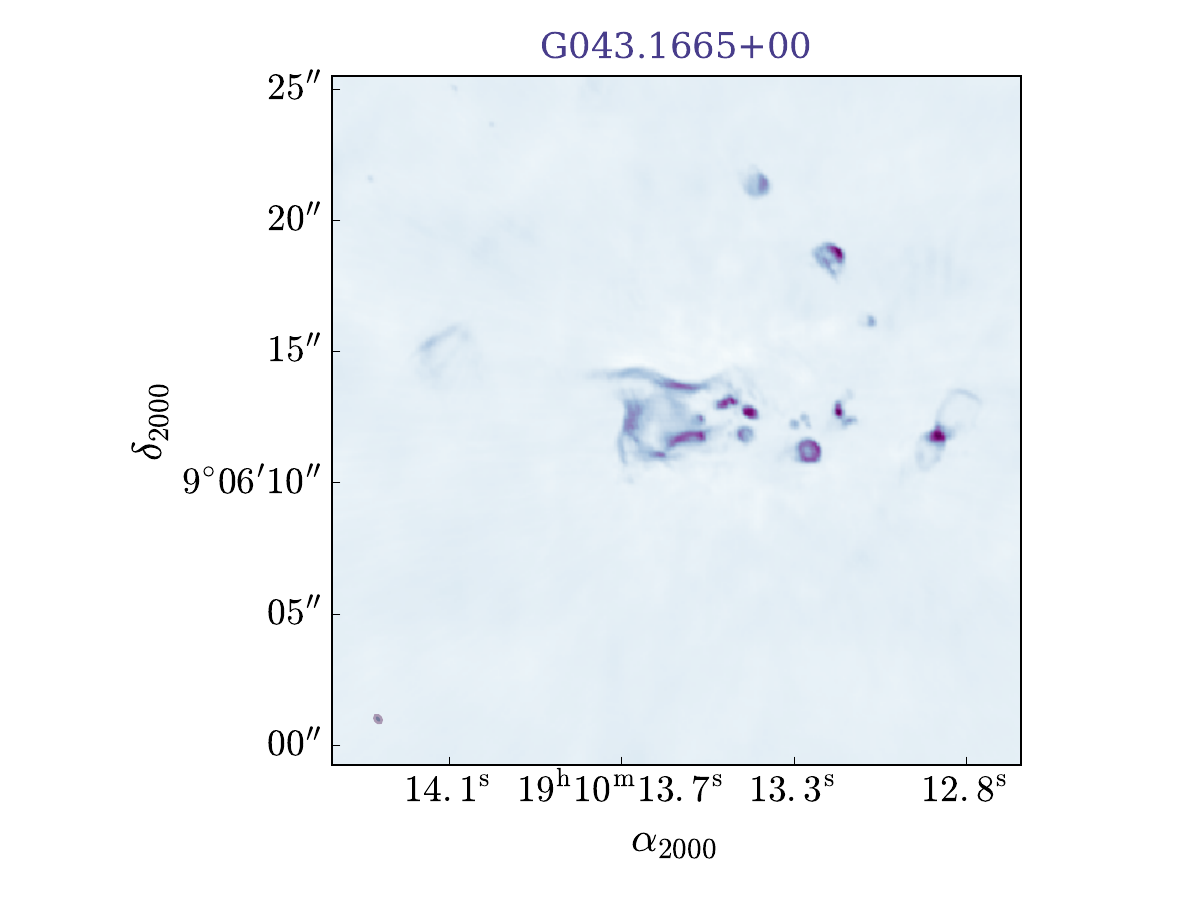}
   \includegraphics[width=\columnwidth]{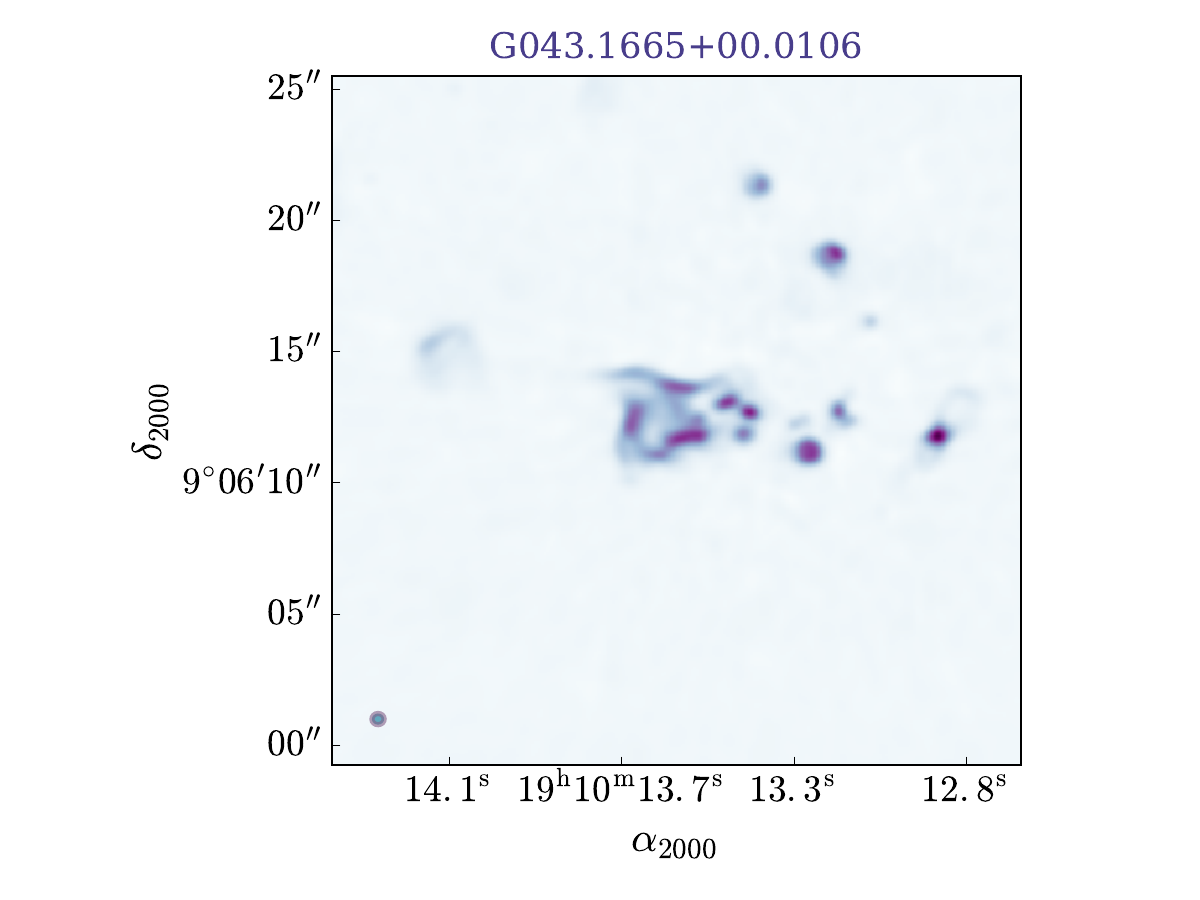}
   \caption[Standard cleaning vs. multi-scale clean -- central portion of W49A]{Comparison between standard cleaning (top panel) and multiscale CLEAN (at 0, 5, 10, 20, and 50 pixels, with a 2.5-pixel beam)  with 0.5\arcsec~tapering of the outer baselines in the $uv$-plane, and small scale bias parameter set to 0.9 (bottom panel), for the central portion of the W49A complex. The standard method performs well close to the field centre and preserves the high resolution.}
\label{fig:centre_taper}
\end{figure}

 \begin{figure}
\centering
\textbf{W49A -- tapering, off-centre}\par\medskip
   \includegraphics[width=\columnwidth]{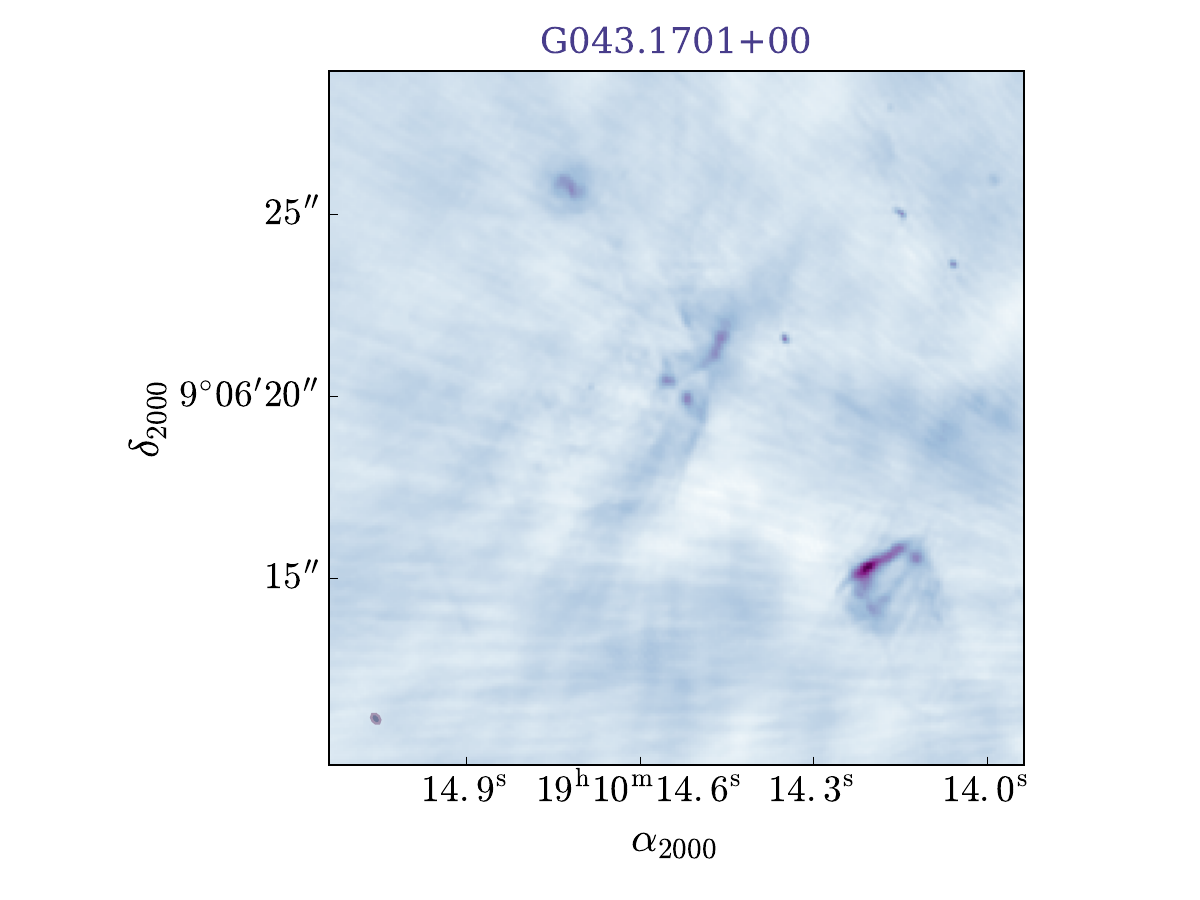}
   \includegraphics[width=\columnwidth]{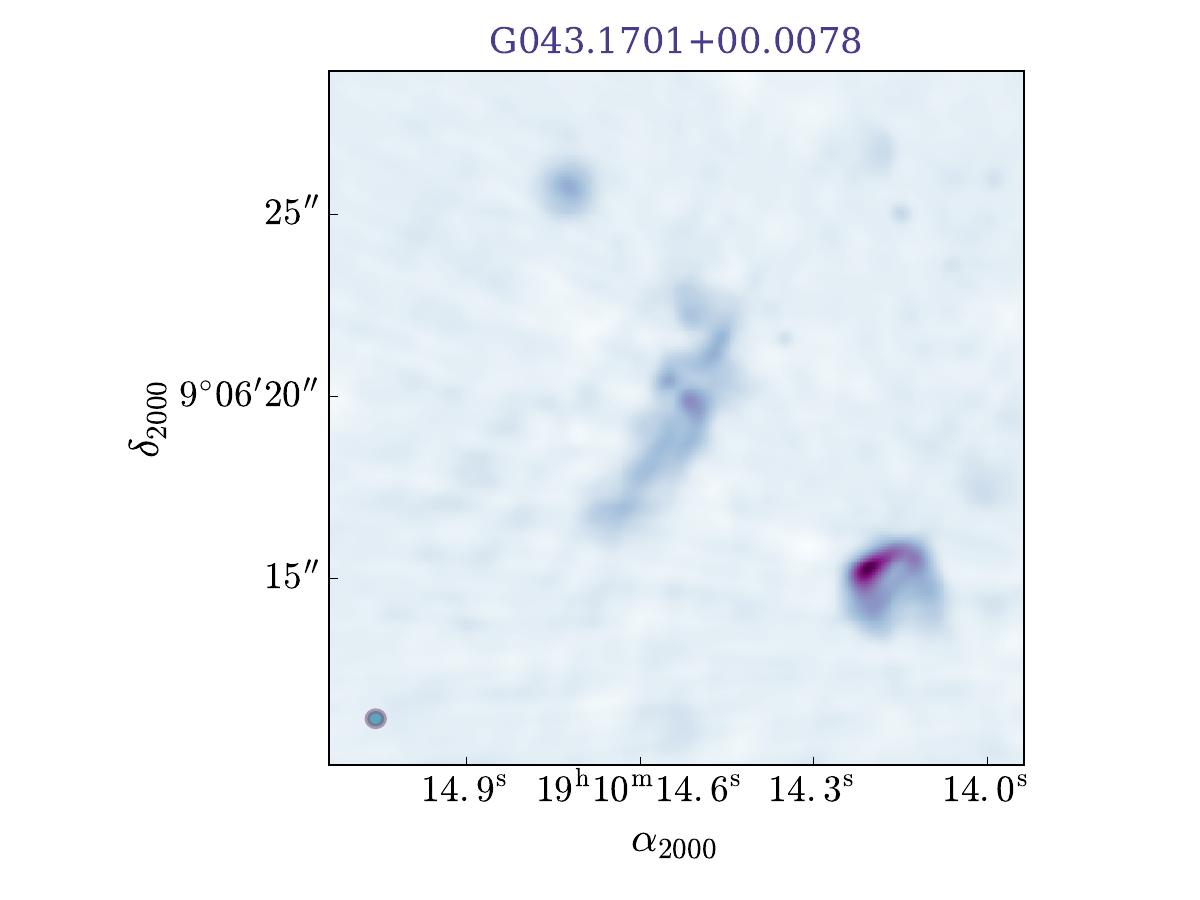}
\caption[Standard cleaning vs. multi-scale clean -- off-centre portion of W49A]{Comparison between standard cleaning (top panel) and multiscale CLEAN (at 0, 5, 10, 20, and 50 pixels, with a 2.5-pixel beam)  with 0.5\arcsec~tapering of the outer baselines in the $uv$-plane, and small scale bias parameter set to 0.9 (bottom panel), for an off-centre portion of the W49A complex. The latter method allows weaker and more extended sources to be imaged better in the vicinity of bright, compact sources.}
\label{fig:multi_taper}
\end{figure}

Unless stated otherwise, all fields were imaged with the \texttt{CASA} task \texttt{clean} in multi-frequency synthesis mode \citep{conway:1990}, combining all spectral windows remaining after flagging into a single continuum image. A single Taylor term was used (\texttt{nterms}~=~1), i.e.\ no spectral dependence was modelled across the band. Briggs weighting with a robustness parameter of 0.5 was applied, as a compromise between natural (high-sensitivity) and uniform (high-resolution) weighting that slightly favours sensitivity \citep{briggs:1995}.

For the minor cycles, the Clark algorithm \citep{clark:1980} was used instead of the H\"ogbom algorithm \citep{hogbom:1974}. It cleans only the brightest pixels in each minor cycle using a truncated beam patch, and corrects the resulting approximation by subtracting the accumulated model from the visibilities in each major cycle; it is therefore considerably faster, with no significant loss of accuracy for compact sources.

Cell sizes of 0.05\arcsec\ or 0.08\arcsec\ were used, corresponding to $\sim$3.5 and $\sim$2 pixels across the minor axis of the median (0.17\arcsec) restoring beam, respectively. The smaller cell size was adopted for the compact targets and the larger one for the more extended targets, for which it reduced the image sizes required and hence the imaging time. A test with a cell size of 0.03\arcsec\ gave no further improvement and was not pursued. The variation in cell sizes and clean thresholds was kept small to preserve the homogeneity of the data set. Most images were 1024~$\times$~1024 or 1280~$\times$~1280 pixels ($\sim$50--100\arcsec\ across, well within the 4.2\arcmin\ primary beam FWHM); their sizes were chosen from the CORNISH 5~GHz images to include the targets and any neighbouring sources.

In the W49A field, which contains extended emission and was imaged with 2560~$\times$~2560 pixels of 0.08\arcsec\ ($\sim$205\arcsec\ across), multiscale CLEAN \citep{cornwell:2008} was used instead, with scales of 0, 5, 10, 20 and 50 pixels and a small-scale bias of 0.9, combined with a 0.5\arcsec\ Gaussian taper of the visibilities. Tapering down-weights the longest baselines, which increases the sensitivity to extended, low-surface-brightness emission at the cost of resolution (the restoring beam increasing to 0.47\arcsec~$\times$~0.43\arcsec) and of point-source sensitivity. As both techniques also slow down imaging considerably, they were applied only to this field. Their effect is illustrated in Figs~\ref{fig:centre_taper} and \ref{fig:multi_taper} for the centre of W49A and for the region to the east of it: standard cleaning performs well for the bright, compact sources and preserves the full resolution, whereas multiscale cleaning with tapering recovers the weaker, more extended emission in their vicinity.

\subsubsection{Note on busy fields with different-brightness neighbours}
A common issue was posed by confusing sources outside the target area, for example in the case of G018.4433$-$00.0056. This is one example of a source with a bright neighbour at the edge of the imaged field (in this case, the four times brighter G018.4614$-$00.0038). In this and similar cases, the maps improved considerably when the confusing source was deconvolved together with the target, by defining `outlier' fields in \texttt{clean} with the outliers' phase centres and mask box sizes. In the case of G020.0797\allowbreak$-$00.1337, with the $>$40 times brighter G020.0809\allowbreak$-$00.1362 only 9\arcsec~away, the map was improved but still left with bright streaks. This was also the case for G045.0694\allowbreak+00.1323, 6.3\arcsec~away from the 14 times brighter G045.0712\allowbreak+00.1321.

\subsubsection{Other imaging hurdles}
  Some targets have no bright sources in their vicinity, but are very bright themselves and cause significant sidelobes in the image. The most extreme example of this effect is G061.7207\allowbreak+00.8630, which required a greater number of cleaning cycles for successful deconvolution (and cleaning down to a slightly lower threshold of 0.1~mJy).

\section{Target detection statistics}\label{detn_stats}

Of the 69 observed targets, G065.2462+00.3505 has since been re-classified as a radio star (Paper~III). It is unresolved in the IR, and its field was not imaged, as it contains no other sources\footnote{The nearest CORNISH source, a radio galaxy, lies $\sim$8.4\arcmin\ from the field centre.}. The remaining 68 target fields were imaged, and all but two of the targets were detected. At the higher resolution, six of the 66 detected targets are revealed to be close pairs of sources (Section~\ref{maps}).

Two of the follow-up targets are detected in neither the dirty nor the clean maps. Both fields are dynamic-range limited rather than sensitivity limited, owing to bright neighbouring sources.

The first is G034.2571+00.1466, with a CORNISH flux density of 47.8~mJy and an angular size of 2.43\arcsec, corresponding to a mean surface brightness of $\sim$0.33~mJy per 0.2\arcsec\ beam at 5~GHz. It lies in the same field as the cometary UCHII G034.2572+00.1535 (G34.26+0.15), which has a flux density of 1.8~Jy at 5~GHz and $\gtrsim$2.4~Jy at 10~GHz (likely underestimated, as its extent exceeds the largest angular scale; Section~\ref{neighbours}). The rms noise near the pointing centre is $\sim$90~$\upmu$Jy~beam$^{-1}$, about nine times higher than in the sensitivity-limited maps ($\sim$10~$\upmu$Jy~beam$^{-1}$). Assuming the mean 10-to-5~GHz flux ratio of the single-component targets (1.12), the target would be expected at only $\sim$4$\sigma$, well below the 7$\sigma$ detection threshold, and at $\sim$1$\sigma$ for the lowest ratio in the sample ($\sim$0.3).

The second is G043.7960$-$00.1286, with a CORNISH flux density of 9.78~mJy and an angular size of 2.38\arcsec, i.e.\ a mean surface brightness of only $\sim$0.07~mJy per 0.2\arcsec\ beam. It lies next to the bright two-component UCHII G043.7954$-$00.1274 ($\sim$25~mJy at 5~GHz for the merged source and $\sim$55~mJy at 10~GHz for its brighter component). The rms noise near the pointing centre is $\sim$40~$\upmu$Jy~beam$^{-1}$, four times higher than in the sensitivity-limited maps, so the target would be expected at only $\sim$2$\sigma$ for a flux ratio of $\sim$1.

Both non-detections are therefore consistent with the low surface brightness of these relatively extended targets combined with the reduced dynamic range of their fields, and do not require the sources to have faded. Self-calibration on the bright neighbours, wide-band spectral modelling ($\texttt{nterms} = 2$), and $uv$ tapering to match the resolution to the source sizes could recover both targets.

MYSO candidates that were not detected by CORNISH due to the lower sensitivity and observation frequency are also seen within the target fields.  A string of four aligned, compact, MYSO-jet-like knots is seen ahead of the cometary UCHII G027.1859\allowbreak$-$00.0816 (see Fig.~\ref{fig:targets1}); none of them is detected in CORNISH.%
The main observables of the MYSO candidates are summarised in Table \ref{table:mysos}.

\begin{table*}
\centering
\caption[Properties of candidate MYSO-jet components]{Observed properties (position, integrated and peak flux, convolved major and minor source axes) of the candidate MYSO-jet components found next to the UCHII G027.1859$-$00.0816 (field clean beam size of 0.33\arcsec $\times$ 0.17\arcsec).}
\begin{tabular*}{\textwidth}{@{\extracolsep{\fill}}lll S[table-format=1.2(2)] S[table-format=1.2(2)] S[table-format=1.2(2)] S[table-format=1.2(2)]}
\hline
\multicolumn{1}{l}{Comp.}& \multicolumn{1}{c}{$\alpha$} & \multicolumn{1}{c}{$\delta$} & \multicolumn{1}{c}{$S_{\rm int}$}&  \multicolumn{1}{c}{$S_{\rm peak}$}& \multicolumn{1}{c}{$\theta_{\rm M}$} & \multicolumn{1}{c}{$\theta_{\rm m}$}  \\
\multicolumn{1}{l}{} & \multicolumn{1}{c}{(J2000)} & \multicolumn{1}{c}{(J2000)} & \multicolumn{1}{c}{(mJy)}&  \multicolumn{1}{c}{(mJy/beam)}& \multicolumn{1}{c}{(\arcsec)}&\multicolumn{1}{c}{(\arcsec)}  \\
\hline
(A) & 18$^{\rm h}$41$^{\rm m}$13.32$^{\rm s}$& $-$05$\degree$08$'$59.80\arcsec  & 0.45 \pm 0.04 & 0.18 \pm 0.01  & 0.46 \pm 0.04  & 0.30 \pm 0.02  \\
(B) & 18$^{\rm h}$41$^{\rm m}$13.23$^{\rm s}$& $-$05$\degree$09$'$00.54\arcsec  & 0.35 \pm 0.04 & 0.18 \pm 0.01  & 0.32 \pm 0.02  & 0.22 \pm 0.01  \\
(C) & 18$^{\rm h}$41$^{\rm m}$13.20$^{\rm s}$& $-$05$\degree$09$'$01.06\arcsec  & 0.79 \pm 0.13 & 0.26 \pm 0.03  & 0.45 \pm 0.08  & 0.17 \pm 0.01  \\
(D) & 18$^{\rm h}$41$^{\rm m}$13.15$^{\rm s}$& $-$05$\degree$09$'$01.39\arcsec  & 0.61 \pm 0.05 & 0.51 \pm 0.02  & 0.31 \pm 0.02  & 0.22 \pm 0.01  \\
\hline
\end{tabular*}
\label{table:mysos}
\end{table*}

\section{Target maps} \label{maps}

All target
maps are presented in Appendix~\ref{app:appendixA}, zoomed in on the relevant CORNISH source (Fig.~\ref{fig:targets1}). The panels
are centred on the JVLA pointing centres, except for G043.1489+00.0130, G043.1520+00.0115 and G043.1720+00.0080 in the W49A field, which are centred on their CORNISH positions.
Contours are overplotted to aid the morphological classification (presented in a follow-up paper). The levels run from 7$\sigma$ to the peak flux of each source (or of the brightest source, if several occupy the same image), with the number of levels chosen according to the dynamic range of each field.

The flux densities and angular sizes of the targets are summarised in Table \ref{table:fluxes_evla} in Appendix \ref{app:appendixD}. The vast majority of target UCHIIs were sufficiently resolved to be morphologically classified (to be presented in a follow-up paper).
One target UCHII is still unresolved at the higher resolution -- this is G027.9352\allowbreak+00.2056, at a distance of 12.1~kpc \citep{urquhart:2012} (giving an upper limit to its physical diameter of 0.016~pc; as \texttt{imfit} could not deconvolve it from the beam, its convolved size is given as an upper limit in Table~\ref{table:fluxes_evla}). Target UCHIIs that are only just resolved, and whose morphological classification is therefore ambiguous, are G026.0083\allowbreak+00.1369, G026.1094\allowbreak$-$00.0937, G030.0096\allowbreak$-$00.2734, and G038.6529\allowbreak+00.0875.

As mentioned in Section~\ref{detn_stats}, six of the CORNISH targets appear at the higher resolution to be merged pairs of sources.
Three CORNISH UCHIIs consist of one extended and one unresolved source (G018.6654\allowbreak+00.0294, G030.5887\allowbreak$-$00.0428, and G044.3103\allowbreak+00.0410). Two targets were resolved into two extended components (G033.8100\allowbreak$-$00.1864 and G043.7954\allowbreak$-$00.1274). One target, G020.9636\allowbreak$-$00.0744, was revealed at the higher resolution to consist of a bright and a dim extended component.
The unresolved components could be younger or less luminous sources, particularly in the cases where the flux of the unresolved component is quite low (relative to the resolved companion), and are discussed below.

G018.6654\allowbreak+00.0294 (see Fig.~\ref{fig:targets1}) belongs to the massive protocluster G18.67+0.03, explored by \cite{cyganowski2012protocluster}\footnote{See also \citealt{cyganowski2009class, cyganowski2011deep} for a VLA radio continuum survey toward a sample of GLIMPSE Extended Green Objects (EGOs) with maser associations (in particular, 6.7~GHz Class II CH$_{3}$OH masers, which are thought to be associated exclusively with MYSOs).}.
Figure~1d in \cite{cyganowski2012protocluster} shows the closest association, the UCHII G18.67+0.03--CM1, which lies $\sim$0.5\arcsec\ from both 10~GHz components (separated by 0.47\arcsec). At the resolution of their 24.64~GHz continuum data, the two components are not resolved, and CM1 corresponds to their combined emission; the compact component G018.6654\allowbreak+00.0294b is revealed only in the 10~GHz map (see panels b and d of Fig.~1 in \citealt{cyganowski2012protocluster}).
The source G18.67+0.03 -- CM1 is associated only with Class I CH$_{3}$OH masers (no hot core line emission),
on the basis of which the authors concluded that Class I masers may be excited not exclusively by young (hot core) MYSOs, but also by older `UCHII MYSO' stages within the same protocluster, in contrast with other studies (see the references in \citealt{cyganowski2012protocluster}). If the masers are instead excited by the compact component G018.6654\allowbreak+00.0294b, and this represents a younger evolutionary phase, this conclusion may need to be revisited.

G030.5887$-$00.0428, also with a compact component ahead of a cometary UCHII, is associated with an interstellar hydroxyl (OH) maser -- another signpost of recent massive star formation \citep{argon2000interstellar}. Given the positional accuracy of the maser data ($\sim$0.3\arcsec\ E--W and $\sim$0.5\arcsec\ N--S), the maser lies significantly west of both 10~GHz components, $\sim$1.0\arcsec\ from the cometary component G030.5887$-$00.0428b and $\sim$1.3\arcsec\ from the compact component G030.5887$-$00.0428a. It is therefore not coincident with the compact component, and is more likely associated with the dense gas surrounding the cometary UCHII.

The two-component G044.3103\allowbreak+00.0410 appears very red in the three-colour GLIMPSE image and is associated with a 22~GHz water maser \citep{urquhart2011red}. Water masers occur  in  both  high-  and  low-mass star-forming regions and are thought to be associated with molecular outflows (\citealt{urquhart2011red} and the references therein).

A water maser associated with IRAS~18385$-$0512 \citep{beuther2002high} lies along the chain of MYSO-jet-like knots ahead of the cometary UCHII G027.1859$-$00.0816 (Table~\ref{table:mysos}, Fig.~\ref{fig:targets1}). Its nominal position is within 0.1\arcsec\ of component C, although, given the absolute positional accuracy of the maser data ($\lesssim$1\arcsec), it could be associated with any of components B--D. The maser lies $\sim$3\arcsec\ from the UCHII itself, which makes an association with the \hii\ region less likely and supports the interpretation of the knots as a jet from a young massive protostar.%

Compact components are thus found ahead of the cometary UCHIIs G018.6654\allowbreak+00.0294 and G030.5887\allowbreak$-$00.0428, and a chain of MYSO-jet-like knots ahead of G027.1859\allowbreak$-$00.0816 (see also the neighbour G020.0809\allowbreak$-$00.1362, Section~\ref{neighbours}). The presence of compact, possibly younger, sources ahead of cometary UCHIIs is consistent with the sequential structure formation hypothesis of \citet{kim2001radio, kim2002interaction, kim2003molecular}, although higher-resolution, multi-wavelength observations are needed to establish whether these components are physically associated with the UCHIIs.

\section{Neighbour maps}\label{neighbours}
A total of 15 CORNISH UCHIIs that were not observation targets lie within the imaged fields (the neighbour group). Two of them are resolved into multiple components at 10~GHz -- G032.7966+00.1909 into four and G036.4062+00.0221 into two -- so that the neighbour group comprises 19 individual components, whose properties are listed in Table~\ref{table:fluxes_evla_n} in Appendix~\ref{app:appendixD}. The neighbour maps are presented in Fig.~\ref{fig:maps_neighbours}, centred on the CORNISH coordinates of each source, with contours plotted as for the targets.

The neighbour CORNISH UCHIIs found in the target fields have a greater variation in (CORNISH) angular size than the targets, and at the higher resolution some reveal complex morphology and a multitude of components.  This includes newly resolved pairs and multiples that are the counterparts to CORNISH UCHII regions, as well as clumpy extended CORNISH UCHIIs. A search was performed for high-resolution maps of the neighbour sources, and, when available,
those were inspected to help separate (by assigning a component label and measuring the component fluxes and sizes, as shown in Table \ref{table:fluxes_evla_n}) the individual components comprising each merged CORNISH source.
A notable example is G032.7966\allowbreak+00.1909, which appears in CORNISH as a $\sim$10\arcsec\ UCHII of irregular morphology but is resolved at 10~GHz into four bright components with distinct morphologies (three of them cometary); it is thus likely a cluster of young sources rather than a single, over-resolved source. Another source, G036.4062\allowbreak+00.0221, neighbour to G036.4057\allowbreak+00.0226, appears as two individual compact sources.

Notably, the bright region G043.1665\allowbreak+00.0106 is a well-explored source in the Welch ring of  UCHII regions in W49A. This is source G in Fig.~2 of \cite{depree:1997}, where it can be seen in the multi-frequency, multi-configuration maps as a collection of sub-clumps, despite the slightly larger beam size (0.8\arcsec).
Another merged CORNISH source in the W49A map is G043.1657\allowbreak+00.0116, corresponding to the cometary-shaped components B and D in the same figure in \cite{depree:1997}. Component E is also seen as two unresolved knots.

G020.0809\allowbreak$-$00.1362 presents an intriguing nested, `Russian-doll'-like structure: a bright knot ahead of two successively larger shells. The same curious pattern continues in the 5~GHz maps and the mid-IR images, as the source is ahead of a larger bubble (corresponding to the $\sim$20\arcsec\ CORNISH \hii region G020.0789\allowbreak$-$00.1383).

  \begin{figure*}
\centering
   \includegraphics[width=0.8\columnwidth]{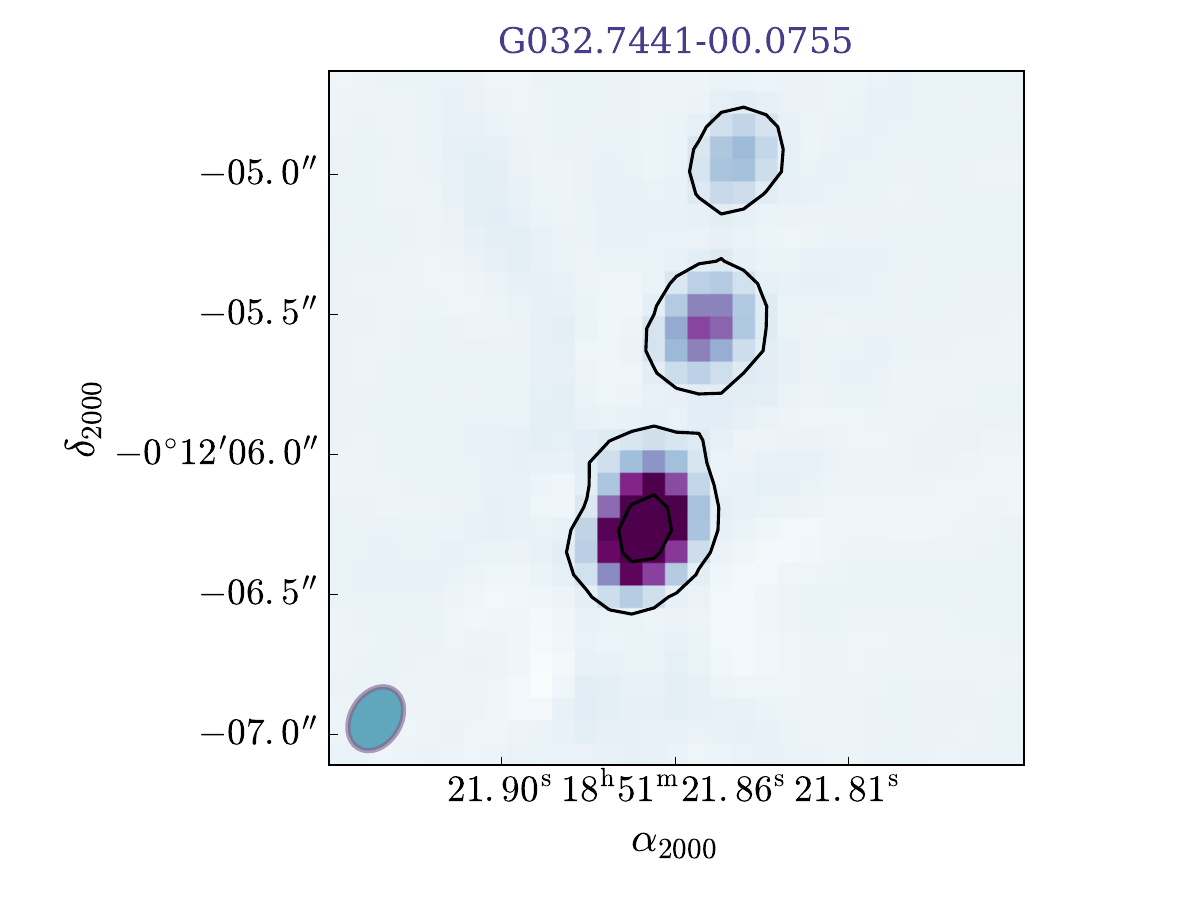}
\caption[G032.7441$-$00.0755 at higher resolution]{G032.7441$-$00.0755 was classified as a candidate IR-dark \hii region in the CORNISH survey.
At 10~GHz it is resolved into three aligned compact knots -- possibly a very young (hyper-)compact \hii\ region.}
\label{fig:jet_knots}
\end{figure*}

\begin{table*}
\begin{footnotesize}
\centering
\caption[Properties of a dark \hii region]{Observed properties (position, integrated and peak flux, convolved major and minor source axes) of the components corresponding to the candidate IR-dark \hii region G032.7441$-$00.0755 (field clean beam size of 0.24$''$ $\times$ 0.18$''$).}
\begin{tabular*}{\textwidth}{@{\extracolsep{\fill}} l ll S[table-format=2.2(2)] S[table-format=2.2(2)] S[table-format=1.2(2)] S[table-format=1.2(2)]}
\hline
\multicolumn{1}{l}{Comp.}& \multicolumn{1}{c}{$\alpha$} & \multicolumn{1}{c}{$\delta$} & \multicolumn{1}{c}{$S_{\rm int}$}&  \multicolumn{1}{c}{$S_{\rm peak}$}& \multicolumn{1}{c}{$\theta_{\rm M}$} & \multicolumn{1}{c}{$\theta_{\rm m}$}  \\
\multicolumn{1}{l}{} & \multicolumn{1}{c}{(J2000)} & \multicolumn{1}{c}{(J2000)} & \multicolumn{1}{c}{(mJy)}&  \multicolumn{1}{c}{(mJy~beam$^{-1}$)}& \multicolumn{1}{c}{(\arcsec)}&\multicolumn{1}{c}{(\arcsec)}  \\
\hline
(A) & 18$^{\rm h}$51$^{\rm m}$21.86$^{\rm s}$& $-$00$\degree$12$'$06.26$''$  & 10.67 \pm 0.22 & 8.15 \pm 0.10  & 0.28 \pm 0.01  & 0.20 \pm 0.01  \\
(B) & 18$^{\rm h}$51$^{\rm m}$21.85$^{\rm s}$& $-$00$\degree$12$'$05.54$''$  & 2.58 \pm 0.10 & 2.20 \pm 0.05  & 0.25 \pm 0.01  & 0.20 \pm 0.01  \\
(C) & 18$^{\rm h}$51$^{\rm m}$21.84$^{\rm s}$& $-$00$\degree$12$'$04.95$''$  & 1.18 \pm 0.04 & 1.08 \pm 0.02 & 0.25 \pm 0.01  & 0.19 \pm 0.01  \\
\hline
\end{tabular*}
\label{table:darkhii}
\end{footnotesize}
\end{table*}

Three further CORNISH sources in the imaged fields are not included in the neighbour group. G026.0907\allowbreak$-$00.0575 is classified in CORNISH as a diffuse \hii\ region, and its extent ($\sim$7.9\arcsec) exceeds the largest angular scale of the JVLA data (Section~\ref{setup_evla}), so its 10~GHz flux density is unreliable. G032.7441\allowbreak$-$00.0755 is classified as a candidate IR-dark \hii\ region and is discussed below. The bright, well-studied cometary UCHII G034.2572\allowbreak+00.1535 (G34.26+0.15), together with its neighbouring components G034.2581\allowbreak+00.1533 and G034.2573\allowbreak+00.1523 (unresolved here), is part of the CORNISH UCHII sample, but its extent in CORNISH ($\sim$5.75\arcsec) also exceeds the largest angular scale. Neither G026.0907\allowbreak$-$00.0575 nor G34.26+0.15 was therefore analysed further.
At 10~GHz, G032.7441$-$00.0755 is resolved into three aligned compact knots (Fig.~\ref{fig:jet_knots}, Table~\ref{table:darkhii}). Consistent with its original classification, it has no counterpart in the near-IR \citep{lucas:2008} or in the GLIMPSE mid-IR images \citep{benjamin:2003, churchwell:2009}, but it has strong counterparts at 24~$\upmu$m in MIPSGAL \citep{carey:2009} and in the Hi-GAL far-IR images \citep{molinari:2010}, indicating a deeply embedded, luminous source. It is not detected in the MAGPIS 20~cm image \citep{helfand:2006} ($S_{1.4\,\rm GHz} \lesssim$1--2~mJy), implying a rising spectrum between 1.4 and 10~GHz, i.e.\ emission that is still optically thick at 1.4~GHz. Such spectra are characteristic of both hyper-compact \hii\ regions and MYSO jets, and unlike those of most CORNISH UCHIIs, the vast majority of which have 20~cm counterparts (216 of 239) and spectral indices close to the optically thin value (Paper~III). The brightest knot is most likely a very young, still optically thick (hyper-)compact \hii\ region, although the linear arrangement of the knots also allows a MYSO-jet interpretation, which cannot be excluded without a distance estimate.

\section{Flux densities and source sizes}\label{sn:fluxes}

The new photometric results for the targets are presented in Table \ref{table:fluxes_evla} in Appendix \ref{app:appendixD}. The sources were identified first through their CORNISH coordinates and angular extents. With the higher-resolution data, more precise positions of the source centres, particularly for CORNISH sources comprising multiple components, were obtained.
Flux densities, positions and sizes were measured with the \texttt{CASA} (v5.1.1) task \texttt{imfit}, using the interactive fitting tool of the \texttt{CASA} viewer. For each source, a region enclosing its emission was defined manually, and a two-dimensional Gaussian was fitted together with a constant zero-level offset; sources resolved into several components were fitted one component per region. Each fit was inspected together with its residual image and repeated with adjusted regions where the initial fit was poor (e.g.\ in the presence of negative sidelobes from bright neighbouring sources). The angular diameter is the geometric mean of the deconvolved major and minor axes; the clean beam axes are also listed in Table~\ref{table:fluxes_evla}.

The quoted uncertainties are the statistical errors returned by \texttt{imfit}; an additional absolute flux-scale uncertainty of 5 per cent should be added in quadrature. For sources with non-Gaussian morphologies a single Gaussian does not fully describe the emission: for eight of the tabulated fits, the residual rms exceeds 15 per cent of the peak flux density (marked with $^{\dagger}$ in Tables~\ref{table:fluxes_evla} and \ref{table:fluxes_evla_n}), and their flux densities should be treated with caution. More generally, Gaussian fits may underestimate the total flux density of extended sources with non-Gaussian morphologies, whereas the CORNISH 5~GHz flux densities of extended sources were measured within hand-drawn polygons (Paper~III); this should be borne in mind when comparing the two.
Similarly, for sources in dynamic-range-limited fields close to bright neighbours, the fitted flux densities depend on the choice of fitting region and on the fitted background level; for G018.4433$-$00.0056, for example, fits with different regions differ by up to a factor of $\sim$2, whereas repeated fits of G018.7106+00.0002 agree to within $\sim$5 per cent. For overlapping components the uncertainties may additionally be underestimated, as correlations between the parameters of separate components are not accounted for\footnote{See the \texttt{CASA} \texttt{imfit} task documentation for a detailed discussion of the fitting limitations.}.%

Individual components of CORNISH UCHIIs are treated as separate sources in the table in the cases where the morphology of at least one of the components is distinct, and, more importantly, the components are not found within a common outermost (7$\sigma$) contour, which is assumed to be the source boundary.

The images were not corrected for primary-beam attenuation during imaging. For the targets outside W49A, located at the pointing centres, and for the neighbours in these fields, which lie within $\sim$13\arcsec\ of the pointing centres, the attenuation is $<$1 per cent and was neglected. The W49A complex was observed with a single pointing, centred on G043.1706$-$00.0003. The flux densities of all other sources in this field -- including the targets G043.1489+00.0130, G043.1520+00.0115 and G043.1720+00.0080, located up to $\sim$100\arcsec\ from the pointing centre -- were corrected using the primary-beam response at their positions, approximated by a Gaussian of FWHM 4.2\arcmin\ (correction factors of 1.04--1.59; Tables~\ref{table:fluxes_evla} and \ref{table:fluxes_evla_n}).

\section{Observed and derived properties}

\subsection{10 GHz sample properties}\label{cha3phys}
 \begin{figure}
\centering
   \includegraphics[width=0.9\columnwidth]{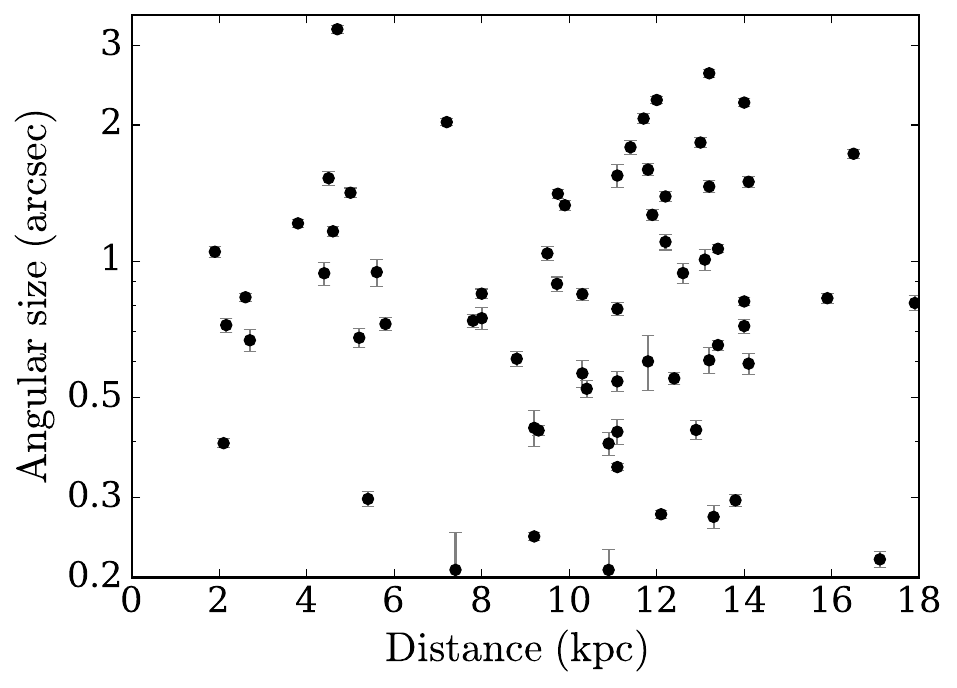}
\caption[JVLA angular sizes against distance]{Measured JVLA 10 GHz angular diameters for the target UCHIIs against distances.}
\label{fig:jvla_angsize}
\end{figure}

\begin{figure}
\centering
   \includegraphics[width=0.9\columnwidth]{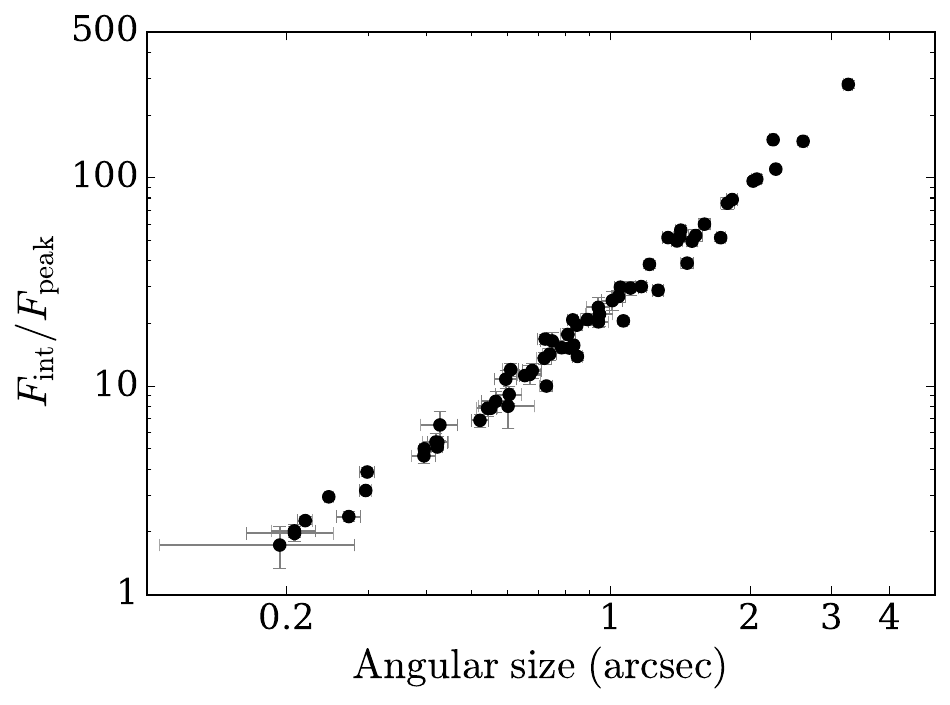}
\caption[Ratio of integrated to peak flux against angular size of the JVLA targets]{Ratio of measured integrated to peak flux of the target UCHII regions, against deconvolved angular size (computed as the geometric average of the deconvolved major and minor source axes).}
\label{fig:int_peak10}
\end{figure}

As expected for Gaussian fits, the ratio of integrated to peak 10~GHz flux density of the target UCHII regions increases with angular size (Fig.~\ref{fig:int_peak10}).

\begin{figure*}
\centering
\begin{subfigure}[b]{0.4\textwidth}
\includegraphics[width=\linewidth]{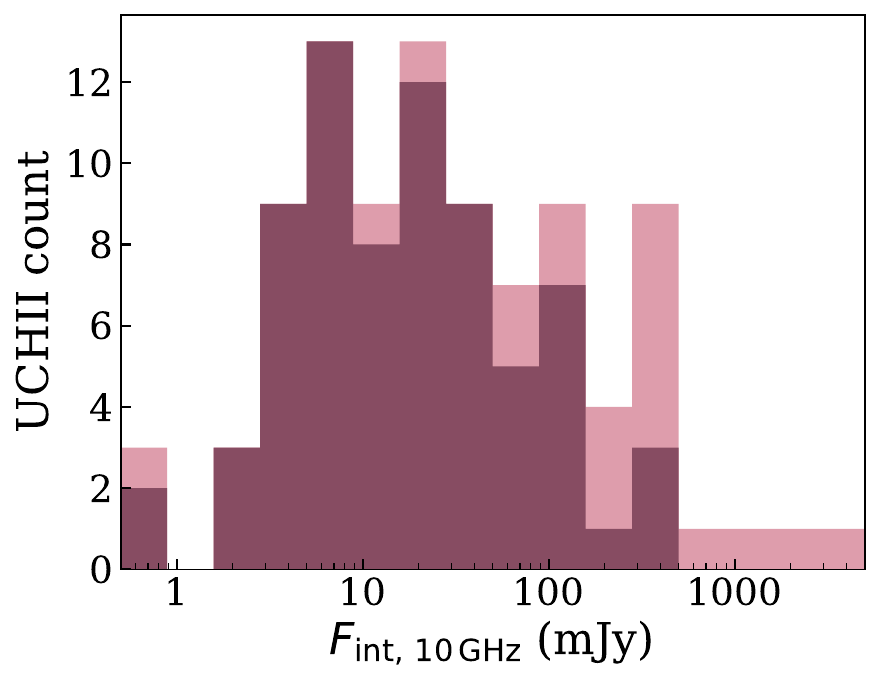}
\caption{}
\label{fig:10GHz_int}
\end{subfigure}\hspace{1em}
\begin{subfigure}[b]{0.4\textwidth}
\includegraphics[width=\linewidth]{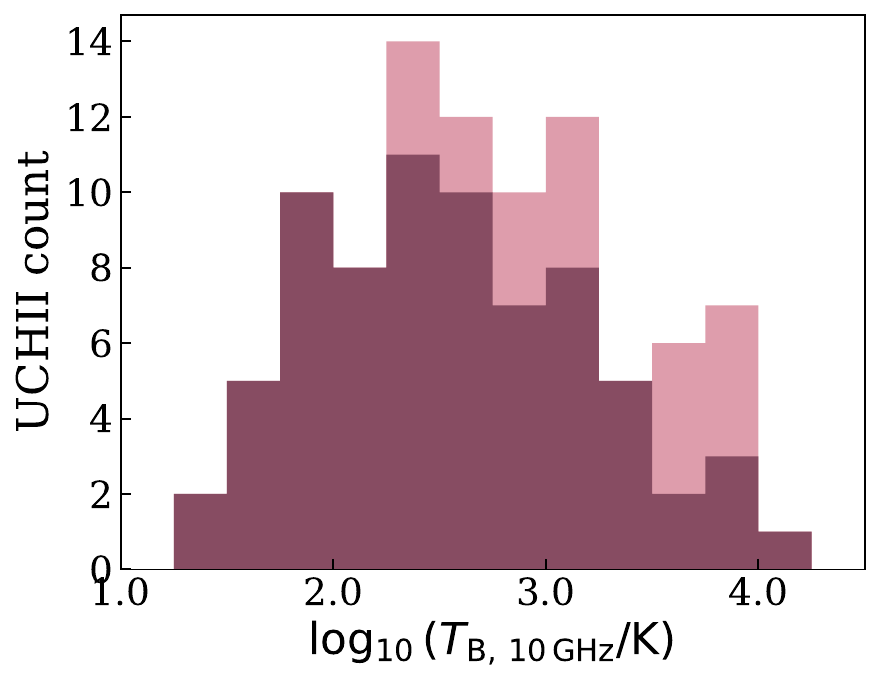}
\caption{}
\label{fig:newTb}
\end{subfigure}

\vspace{-1mm}

\begin{subfigure}[b]{0.4\textwidth}
\includegraphics[width=\linewidth]{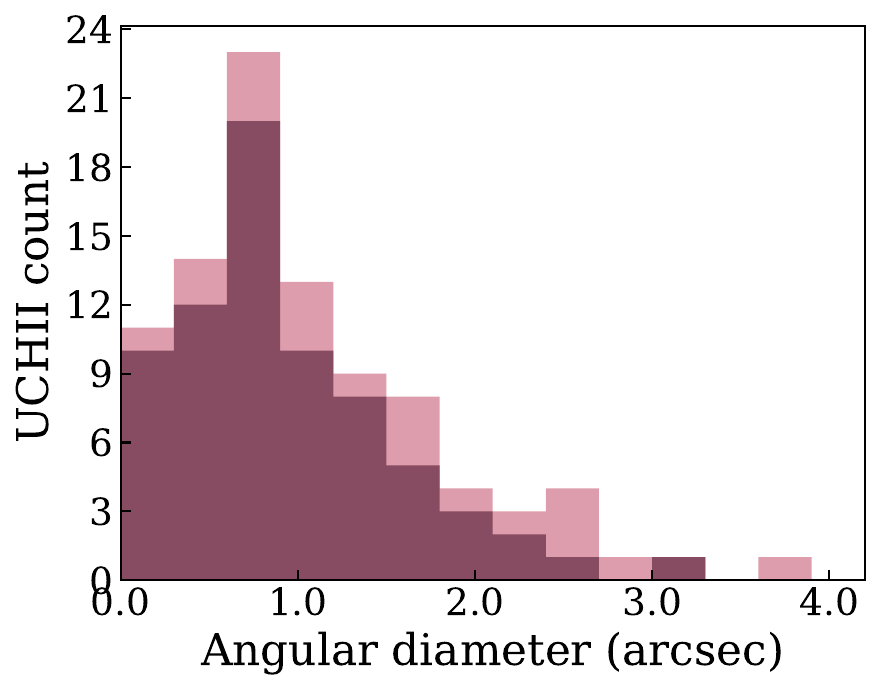}
\caption{}
\label{fig:new_angsize}
\end{subfigure}\hspace{1em}
\begin{subfigure}[b]{0.4\textwidth}
\includegraphics[width=\linewidth]{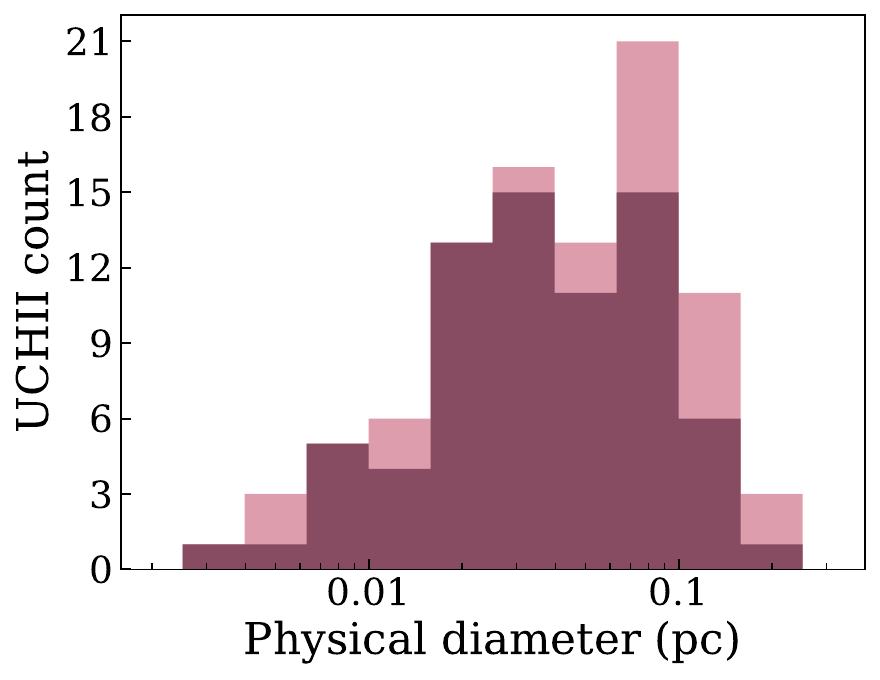}
\caption{}
\label{fig:new_phsize}
\end{subfigure}

\vspace{-1mm}

\begin{subfigure}[b]{0.4\textwidth}
\includegraphics[width=\linewidth]{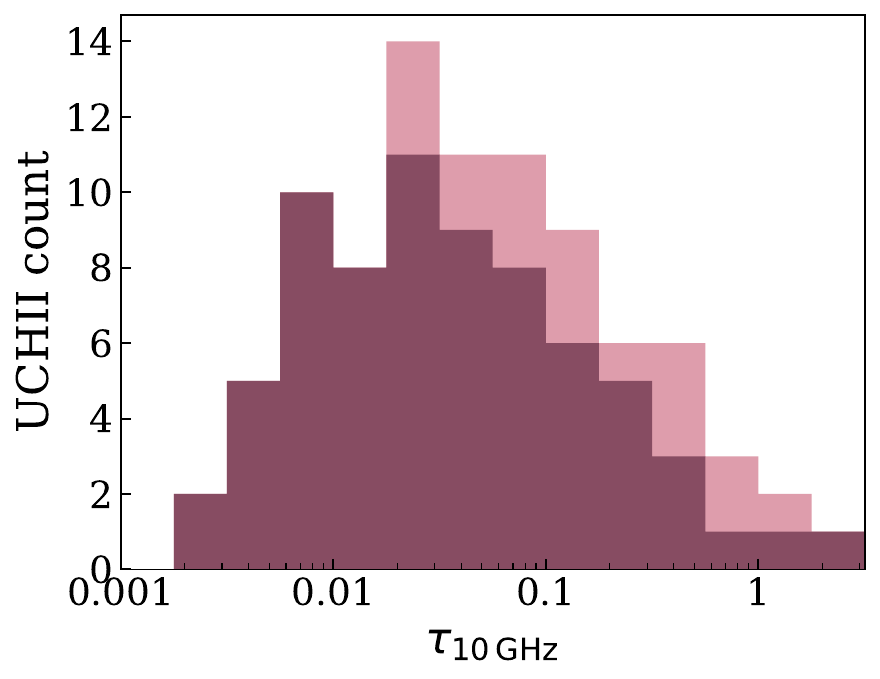}
\caption{}
\label{fig:new_tau10GHz}
\end{subfigure}

\caption[Histograms of the JVLA sample properties]{Histograms of the observational properties, brightness temperatures, optical depths, and physical sizes of the target (in wine red) and neighbour (in pink) UCHIIs at 10 GHz.}
\label{fig:10GHz_param}
\end{figure*}

Figure \ref{fig:10GHz_param} provides a summary of the observational properties, brightness temperatures, optical depths and physical sizes of the
 UCHIIs at 10~GHz (computed as in Section~5 of Paper~III).
The median 10~GHz flux density of the single-component targets is slightly lower than at 5~GHz for the same sub-sample ($\sim$16.3~mJy vs. $\sim$18.7~mJy; Fig.~\ref{fig:10GHz_int}), and the median brightness temperature is the same as at 5~GHz ($\log_{10}(T_{\rm B}/{\rm K}) \approx 2.5$, Fig.~\ref{fig:newTb}). Although many of the targets were expected to be young and optically thick at 5~GHz (Section~\ref{setup_evla}), $\sim$45 per cent of them are fainter at 10~GHz than at 5~GHz, and their brightness temperatures ($T_{\rm B} \sim 300$~K, far below typical electron temperatures of $\sim$10$^{4}$~K) indicate largely optically thin emission at 10~GHz, as found for the bulk of the CORNISH UCHII sample at 5~GHz (Paper~III). Part of the lower 10~GHz flux densities is likely due to flux on scales of a few arcseconds being filtered out at the higher resolution, or not recovered by the Gaussian fits (Section~\ref{sn:fluxes}).%
The optical depths at 10~GHz
are presented in Fig.~\ref{fig:new_tau10GHz}. For $\sim$76 per cent of the targets, the difference between the measured and the theoretical unattenuated flux is below 10 per cent.
The deconvolved angular size histogram (Fig.~\ref{fig:new_angsize}) peaks at $\sim$0.9\arcsec. From visual inspection, sources cannot be morphologically classified if their convolved major axis is $\lesssim$1.5 times the clean beam size
(i.e.\ $\lesssim$0.35--0.5\arcsec, or $\sim$0.7\arcsec\ in the tapered W49A field).
The largest target, G038.6465$-$00.2260, is $\sim$3.3\arcsec in diameter at 10~GHz (slightly larger than its corresponding size of 2.5$''$ at 5~GHz).

The distributions of physical sizes of the UCHIIs
are shown in Fig.~\ref{fig:new_phsize}. They were obtained from the distances in Table D.1 in Paper~III and the measured angular diameters (as presented in Table \ref{table:fluxes_evla} in Appendix \ref{app:appendixD}). The median physical diameter of the targets, which are the smallest CORNISH UCHIIs, is $\sim$0.03~pc, and their diameters range from $\sim$0.004~pc to $\sim$0.17~pc (with an upper limit of 0.003~pc for the unresolved component of G044.3103\allowbreak+00.0410 being the smallest value).
All target sources have an available distance estimate (assuming the same distance for each source in multi-component CORNISH UCHIIs).

Angular diameter is plotted against distance in Fig.~\ref{fig:jvla_angsize}; the two components with deconvolved sizes consistent with zero, G030.5887$-$00.0428a and G044.3103+00.0410b, are not shown (Section~\ref{maps}). No correlation is found (Pearson $r = 0.01$, $p = 0.93$, for the 68 plotted targets and components; $r = 0.02$, $p = 0.85$, using only the largest component of each CORNISH source), whereas sources of similar physical size would show a clear anti-correlation. The compactness of the targets is therefore not primarily a distance effect, consistent with the targets being intrinsically compact, and possibly young, UCHII regions.

The results computed for the neighbour subset are also plotted in the histograms. These have a larger span, reaching up to larger values than the target subset, as expected for a subset encompassing a random, broader range of sizes and ages. The largest observed neighbour, G043.1665\allowbreak+00.0106 -- a clumpy source in the Welch ring in the W49A complex \citep{depree:1997}, discussed in Section~\ref{neighbours}, is $\sim$3.8$''$ in diameter. The physical diameters of the neighbours range from $\sim$0.004~pc to $\sim$0.20~pc, with a median of $\sim$0.07~pc.

\subsection{10--1.4~GHz spectral indices}\label{sn:spind}

 \begin{figure}
\centering
   \includegraphics[width=0.9\columnwidth]{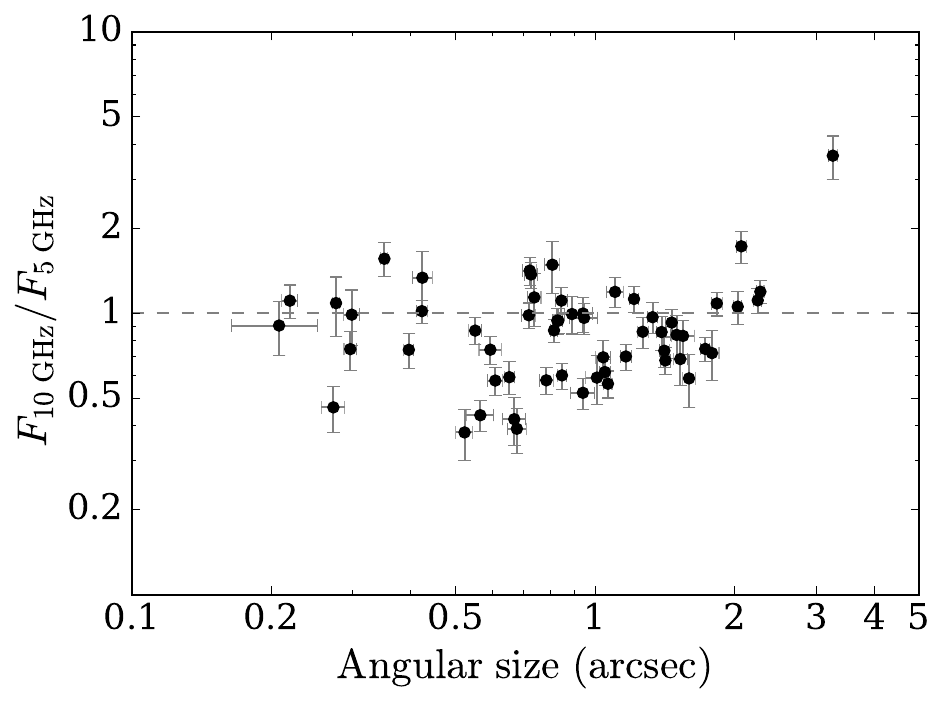}
\caption[10 GHz over 5 GHz integrated flux against angular size at 10 GHz]{Measured JVLA 10 GHz integrated flux over CORNISH 5 GHz integrated flux of the target UCHII regions, against angular size at 10 GHz (excluding resolved pairs). }
\label{fig:fevla_fc}
\end{figure}

  \begin{figure}
\centering
   \includegraphics[width=0.9\columnwidth]{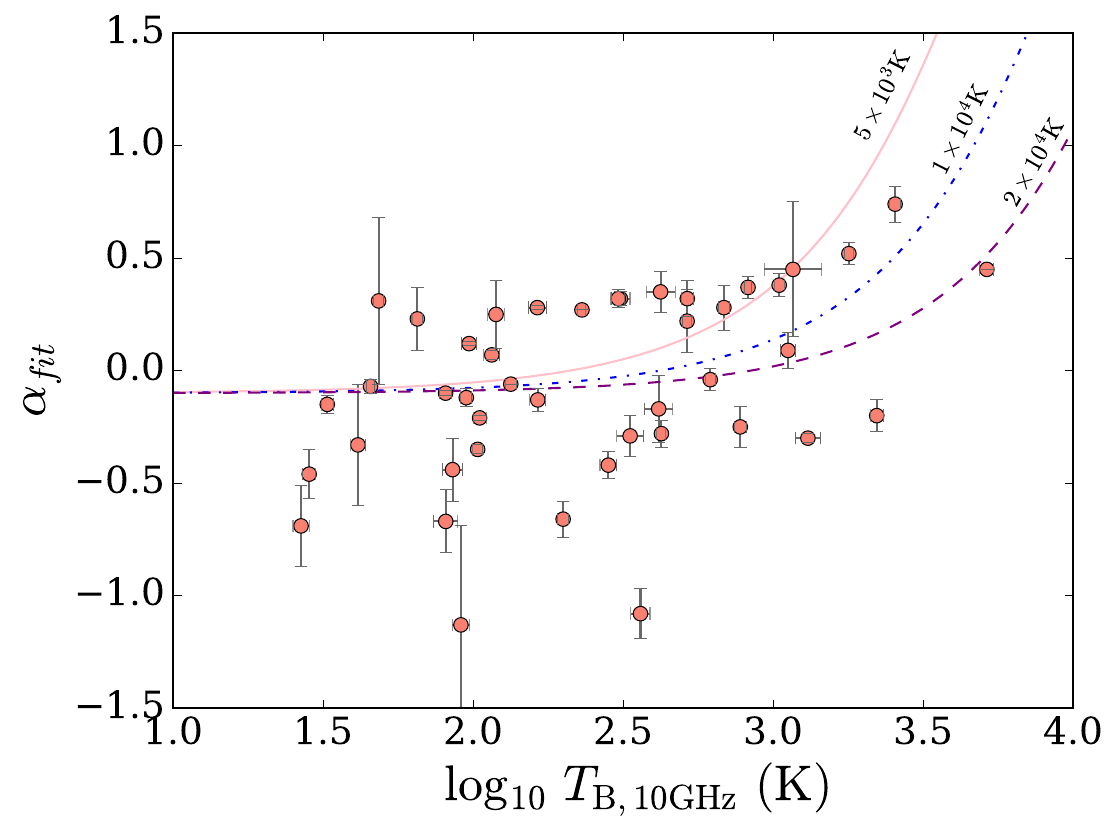}
\caption[Spectral indices from fits to the 10~GHz, 5~GHz and 1.4~GHz integrated flux of the target UCHIIs]{Spectral indices estimated from least-squares fits to the available 10~GHz, 5~GHz and 1.4~GHz integrated flux of the target UCHII regions (excluding MAGPIS~6~cm), against brightness temperature at 10 GHz (excluding sources merged at 1.4~GHz and/or 5~GHz). }
\label{fig:spind_fit}
\end{figure}

Figure \ref{fig:spind_fit} shows spectral indices obtained from least-squares fits to the 10, 5 and 1.4~GHz flux densities available for the JVLA targets
(Table \ref{table:fluxes_evla} in the Appendix and Table B.1 in Paper~III), against brightness temperature at 10~GHz. The plotted theoretical models were adapted from \cite{siodmiak:2001} and \cite{bojicic:2011}, similarly to Section~3.2 of Paper~III, in this case showing the model predictions for the 1.4--10~GHz range at electron temperatures of $5\times10^{3}$, $10^{4}$ and $2\times10^{4}$~K.
The individual linear fits are shown in Appendix \ref{app:appendixC}.
Flux upper limits for all non-detections of
20~cm counterparts in the MAGPIS images are included (in four cases), with the upper limit used as the fitted value and an uncertainty equal to its value.
Note that the MAGPIS~6~cm fluxes are shown for comparison only, and were not taken into account in the fits.
CORNISH UCHIIs that comprise distinct pairs or multiples of sources at the higher resolution were not fitted.

As discussed in Section~3.2 of Paper~III, individual spectral indices are reliable only in cases where the source is unresolved at all fitted frequencies. This is clearly not the case here, as the fitted UCHIIs are compact at 1.4 and 5~GHz but extended at 10~GHz. The different VLA configurations used at each frequency further limit the reliability of the results. Therefore, the estimated 10~GHz--1.4~GHz spectral indices should be considered lower limits.
A method to improve the \textit{uv} coverage and include zero-spacing flux is to combine interferometric data with single-dish data (see e.g. \citealt{stanimirovic:2002, taylor:2003, tian2005radio}). However, this is beyond the scope of this paper and not feasible with the available archival data.
The 10--1.4~GHz spectral indices from the linear fits were not used for detailed spectral analysis of individual sources, but only to compare the properties of the sample of most compact CORNISH UCHIIs to those of the general sample (Section~3.2 of Paper~III).

A well-sampled radio SED could indicate whether the turnover frequency of the source is within the available spectral range. Flat or falling spectral indices and optically thin emission are characteristic of UCHIIs (e.g. \citealt{yang2018search}, Section~5 of Paper~III).
The youngest UCHIIs and hyper-compact \hii\ regions (HCHIIs), however, have higher turnover frequencies and rising spectra between 1.4 and 5~GHz, which suggests a mix of optically thin and thick emission due to a much less uniform environment \citep{kurtz2005hypercompact, yang2018search}.
If the targets are particularly young UCHIIs (Section~\ref{cha3phys}), some of them could have turnover frequencies between 5 and 15~GHz.
Given the very limited number of data points, the plots only hint at the shape of the radio SED in the available range.
Only about 41 per cent of the radio SEDs appear to be well described by a single straight line. In this subset,
about 23 per cent appear to have rising spectra.
Thus, at least $\sim$9 per cent ($0.23 \times 0.41$) of the targets may have a turnover frequency above 10~GHz and could be particularly young.

As expected, the majority of sources scatter around the optically thin limit
in Fig.~\ref{fig:spind_fit} at low $T_{\rm b}$. A larger fraction of indices lies below $-$0.1 than for the 5~GHz--1.4~GHz spectral indices (Fig. 6 in Paper~III) -- 32 per cent in the former and 18 per cent in the latter, taking errors into account.
This reflects the fact that in nearly 45 per cent of the cases (including errors), the 10~GHz flux is lower than the 5~GHz flux.
 The effect does not increase systematically with angular size (Fig.~\ref{fig:fevla_fc}), so filtering of extended emission (due to the difference in \textit{uv} coverage and also since the 5~GHz values in the 10~GHz--5~GHz plot belong predominantly to unresolved sources) may not be the primary cause in all cases. In fact, sources smaller than (a deconvolved angular diameter of) $\sim$0.7\arcsec have the smallest 10-to-5~GHz flux ratios (around 0.5). These are G017.0299\allowbreak$-$00.0696, G023.4835\allowbreak+00.0964,
and G031.1590\allowbreak+00.0465. Two of these, G023.4835\allowbreak+00.0964 and G031.1590\allowbreak+00.0465, are known variable sources that faded between the CORNISH and GLOSTAR epochs \citep{yang:2025}, which may account for part of their low flux ratios (see below). At the other extreme, the target with the largest flux ratio of $\sim$4.3, G038.6465\allowbreak$-$00.2260, also has one of the largest angular extents -- about 3.3\arcsec. In this case, the angular extent is also larger than the one measured in CORNISH (2.51\arcsec). However, the 5~GHz flux measured in the MAGPIS image is comparable to the 10~GHz result (Tables B.1 in Paper~III and \ref{table:fluxes_evla} in this paper, respectively). Its extended, non-Gaussian morphology is, however, poorly described by a single Gaussian (residual rms $\sim$18 per cent of the peak), so its 10~GHz flux density is uncertain.

The three steepest positive spectral indices ($\gtrsim$~0.5) are found at the largest brightness temperatures and belong to G018.7106\allowbreak+00.0002, G019.7549\allowbreak$-$00.1282, and G028.2003\allowbreak$-$00.0494. These therefore lie in the optically thick region of the model shown in Fig.~\ref{fig:spind_fit} (see \citealt{siodmiak:2001}, \citealt{bojicic:2011}, Paper~III for details), between 1.4~GHz and 10~GHz.
The 10~GHz images of G018.7106\allowbreak+00.0002 and G028.2003\allowbreak$-$00.0494 show hints of two-component structure, which could become more apparent at even higher resolution. G019.7549\allowbreak$-$00.1282 is a compact cometary-shaped source at 10~GHz.

The UCHIIs with steep negative fitted spectral indices ($\lesssim-0.4$, from good fits) are G017.0299\allowbreak$-$00.0696, G020.7619\allowbreak$-$00.0646, G023.4835\allowbreak+00.0964, G026.1094\allowbreak$-$00.0937, G028.4518\allowbreak+00.0027, G031.0595\allowbreak+00.0922, G031.1590\allowbreak+00.0465, G034.0901\allowbreak+00.4365, G042.1090\allowbreak$-$00.4469 and G043.1520\allowbreak+00.0115. G017.0299\allowbreak$-$00.0696 and G026.1094\allowbreak$-$00.0937 are undetected at 20~cm, so a 1.4~GHz upper limit was used in their fits and their spectral indices are not well constrained. Three of the sources, G023.4835\allowbreak+00.0964, G028.4518\allowbreak+00.0027 and G031.1590\allowbreak+00.0465, are known to be variable (see below). In the remaining cases, the 1.4~GHz--5~GHz slopes are also steep and negative. As these sources are compact at both 1.4 and 5~GHz due to the lower resolution at these frequencies, their steep spectra are unlikely to be entirely the effect of filtering of extended flux with the change in \textit{uv} coverage, and may instead indicate a non-thermal contribution or time variability.
Most of these sources have cometary morphologies. G043.1520\allowbreak+00.0115 and the relatively extended G034.0901\allowbreak+00.4365 could be parts of, or superimposed on, a much larger bubble,
 as hinted by their 10~GHz morphologies.

The UCHIIs in the 10~GHz sub-sample are likely younger than the rest of the CORNISH UCHII sample (Section~\ref{cha3phys}).
It is clear from a comparison between the radio maps that improving the resolution reveals at each step further morphological complexity, and, in some cases ($\sim$9 per cent of the targets), multiplicity of source components. The computed negative spectral indices could be evidence of an underlying non-thermal component, e.g.\ from bow shocks from runaway massive stars (see \citealt{pereira2016modeling, delvalle2018}),
or stellar wind interactions in massive binaries \citep{de2007non}.
More complete radio SEDs could be studied in the future to look for an underlying physical mechanism causing the apparently non-thermal spectral indices.

Time variability of the continuum flux density is another possible explanation, as the radio data are from different observational epochs (see the discussion in Section~3.3 in Paper~III). About 5 per cent of the CORNISH UCHIIs could be time-variable, based on the 5~GHz flux densities from two epochs $\sim$15 years apart (Section~3.3 in Paper~III), and none of the follow-up targets are among these candidates. However, from a comparison of CORNISH with the GLOSTAR survey, \citet{yang:2025} identified 38 variable UCHIIs, 12 of which are among our targets: G017.5549\allowbreak+00.1654, G021.3571\allowbreak$-$00.1766, G023.4835\allowbreak+00.0964, G025.7157\allowbreak+00.0487, G026.8304\allowbreak$-$00.2067, G027.1859\allowbreak$-$00.0816, G027.9352\allowbreak+00.2056, G028.4518\allowbreak+00.0027, G031.0495\allowbreak+00.4697, G031.1590\allowbreak+00.0465, G032.7398\allowbreak+00.1940 and G037.8209\allowbreak+00.4125. Ten of these faded by 30--55 per cent between the two surveys, while G027.9352\allowbreak+00.2056 and G032.7398\allowbreak+00.1940 brightened by 36 and 52 per cent, respectively. The JVLA follow-up was conducted $\sim$7~years after CORNISH, and a flux-density change of this size between the 5 and 10~GHz epochs alone would change the 5--10~GHz spectral index by $\sim$0.5--1.2. Three of the sources with steep negative spectral indices (G023.4835\allowbreak+00.0964, G028.4518\allowbreak+00.0027 and G031.1590\allowbreak+00.0465), two of which also have among the smallest 10-to-5~GHz flux ratios, are fading variables. Their low 10~GHz flux densities and steep spectral indices may therefore be due, at least in part, to variability rather than to non-thermal emission. Moreover, as the 1.4, 5 and 10~GHz data were obtained at different epochs, a source that faded steadily over this period and was observed at successively higher frequencies would mimic a steep negative spectrum over the whole frequency range. Multi-epoch, or ideally simultaneous, observations at 1.4, 5 and 10~GHz are needed to separate variability from intrinsic spectral effects.

\section{Conclusions}

High-resolution 10~GHz continuum observations were presented, following up on CORNISH UCHII regions smaller than the
limit at which sufficient morphological information is available from the 5~GHz maps (2.7\arcsec).

The observed fields were successfully reduced and deconvolved, resulting in good-quality images. All but two of the target UCHII regions were detected; the two non-detections are most likely due to the bright neighbours in their fields, which limit the dynamic range, combined with the low expected surface brightness of the targets.

Fifteen further CORNISH UCHIIs, and candidate MYSOs/MYSO jet components, were also detected as neighbours to the targets within the observed fields.
All target UCHIIs and their components are smaller than the largest angular scale recovered in A configuration, and only three non-target sources exceed it. The flux densities of sources with non-Gaussian morphologies, or in dynamic-range-limited fields, are less certain, and those of the W49A sources were corrected for primary-beam attenuation.

At the $\sim$0.2\arcsec resolution, about 9 per cent of the target UCHIIs were shown to consist of at least two components which are likely distinct sources -- either UCHII pairs or an UCHII with an unresolved companion. Some of the CORNISH UCHIIs with extents exceeding 2.7\arcsec could also be hiding multiplicity at the 1.5\arcsec resolution, as was the case for the four UCHIIs merged within G032.7966\allowbreak+00.1909 at 5~GHz.

Photometric measurements of the 10~GHz flux densities of all target and neighbour sources of interest were presented. The spectral indices corresponding to the target subset
were estimated from least-squares fits to all available radio flux densities (this paper and Paper~III) at three frequencies.
About 32 per cent of the fitted spectral indices of the target UCHIIs were found below the lower limit for thermal free-free emission ($-$0.1). Some of these, including three of the steepest, belong to known variable sources \citep{yang:2025}, so both a possible non-thermal contribution to the radio continuum emission and time variability should be investigated in the future.

The
angular sizes were used to compute the physical extents of all target and neighbour sources of interest (using the distances listed in Table D.1 in Paper~III). As expected, the follow-up targets are very compact, with physical diameters ranging from $\sim$0.004~pc to $\sim$0.17~pc and a median of $\sim$0.03~pc.

The observed and derived properties of the high-resolution sub-sample suggest that the UCHIIs comprising it are
likely the youngest CORNISH UCHII regions.
The presented maps therefore provide morphological information on the immediate environments of the newborn massive stars powering these radio continuum sources. To take advantage of this, the morphologies of the CORNISH UCHII sample, including the JVLA follow-up presented here, will be classified and explored in more detail in a future paper (in prep.).

\section*{Acknowledgements}
The National Radio Astronomy Observatory is a facility of the National Science Foundation operated under cooperative agreement by Associated Universities, Inc.

\section*{Data Availability}
The JVLA data underlying this article are available from the NRAO Science Data Archive (\url{https://data.nrao.edu}) under project code 14A-085.

\bibliographystyle{mnras}
\bibliography{references_all}

\begin{thebibliography}{}
\makeatletter
\relax
\def\mn@urlcharsother{\let\do\@makeother \do\$\do\&\do\#\do\^\do\_\do\%\do\~}
\def\mn@doi{\begingroup\mn@urlcharsother \@ifnextchar [ {\mn@doi@}
  {\mn@doi@[]}}
\def\mn@doi@[#1]#2{\def\@tempa{#1}\ifx\@tempa\@empty \href
  {http://dx.doi.org/#2} {doi:#2}\else \href {http://dx.doi.org/#2} {#1}\fi
  \endgroup}
\def\mn@eprint#1#2{\mn@eprint@#1:#2::\@nil}
\def\mn@eprint@arXiv#1{\href {http://arxiv.org/abs/#1} {{\tt arXiv:#1}}}
\def\mn@eprint@dblp#1{\href {http://dblp.uni-trier.de/rec/bibtex/#1.xml}
  {dblp:#1}}
\def\mn@eprint@#1:#2:#3:#4\@nil{\def\@tempa {#1}\def\@tempb {#2}\def\@tempc
  {#3}\ifx \@tempc \@empty \let \@tempc \@tempb \let \@tempb \@tempa \fi \ifx
  \@tempb \@empty \def\@tempb {arXiv}\fi \@ifundefined
  {mn@eprint@\@tempb}{\@tempb:\@tempc}{\expandafter \expandafter \csname
  mn@eprint@\@tempb\endcsname \expandafter{\@tempc}}}

\bibitem[\protect\citeauthoryear{Argon, Reid  \& Menten}{Argon
  et~al.}{2000}]{argon2000interstellar}
Argon A.,  Reid M.,   Menten K.~M.,  2000, \apjs, 129, 159

\bibitem[\protect\citeauthoryear{Benjamin et~al.,}{Benjamin
  et~al.}{2003}]{benjamin:2003}
Benjamin R.~A.,  et~al., 2003, \pasp, 115, 953

\bibitem[\protect\citeauthoryear{Beuther, Schilke, Menten, Motte, Sridharan  \&
  Wyrowski}{Beuther et~al.}{2002}]{beuther2002high}
Beuther H.,  Schilke P.,  Menten K.,  Motte F.,  Sridharan T.,   Wyrowski F.,
  2002, \apj, 566, 945

\bibitem[\protect\citeauthoryear{Boji{\v{c}}i{\'c}, Parker, Filipovi{\'c}  \&
  Frew}{Boji{\v{c}}i{\'c} et~al.}{2011}]{bojicic:2011}
Boji{\v{c}}i{\'c} I.~S.,  Parker Q.~A.,  Filipovi{\'c} M.,   Frew D.~J.,  2011,
  \mnras, 412, 223

\bibitem[\protect\citeauthoryear{Briggs}{Briggs}{1995}]{briggs:1995}
Briggs D.,  1995, BAAS, 27, 1444

\bibitem[\protect\citeauthoryear{{Brunthaler} et~al.,}{{Brunthaler}
  et~al.}{2021}]{brunthaler:2021}
{Brunthaler} A.,  et~al., 2021, \mn@doi [\aap] {10.1051/0004-6361/202039856},
  \href {https://ui.adsabs.harvard.edu/abs/2021A&A...651A..85B} {651, A85}

\bibitem[\protect\citeauthoryear{Carey et~al.,}{Carey
  et~al.}{2009}]{carey:2009}
Carey S.,  et~al., 2009, \pasp, 121, 76

\bibitem[\protect\citeauthoryear{Churchwell}{Churchwell}{2002}]{churchwell:2002}
Churchwell E.,  2002, \araa, 40, 27

\bibitem[\protect\citeauthoryear{Churchwell et~al.,}{Churchwell
  et~al.}{2009}]{churchwell:2009}
Churchwell E.,  et~al., 2009, \pasp, 121, 213

\bibitem[\protect\citeauthoryear{{Clark}}{{Clark}}{1980}]{clark:1980}
{Clark} B.~G.,  1980, \aap, 89, 377

\bibitem[\protect\citeauthoryear{{Conway}, {Cornwell}  \& {Wilkinson}}{{Conway}
  et~al.}{1990}]{conway:1990}
{Conway} J.~E.,  {Cornwell} T.~J.,   {Wilkinson} P.~N.,  1990, \mnras, 246, 490

\bibitem[\protect\citeauthoryear{{Cornwell}}{{Cornwell}}{2008}]{cornwell:2008}
{Cornwell} T.~J.,  2008, IEEE Journal of Selected Topics in Signal Processing,
  2, 793

\bibitem[\protect\citeauthoryear{Cyganowski, Brogan, Hunter  \&
  Churchwell}{Cyganowski et~al.}{2009}]{cyganowski2009class}
Cyganowski C.,  Brogan C.,  Hunter T.,   Churchwell E.,  2009, \apj, 702, 1615

\bibitem[\protect\citeauthoryear{Cyganowski, Brogan, Hunter  \&
  Churchwell}{Cyganowski et~al.}{2011}]{cyganowski2011deep}
Cyganowski C.,  Brogan C.,  Hunter T.,   Churchwell E.,  2011, \apj, 743, 56

\bibitem[\protect\citeauthoryear{Cyganowski, Brogan, Hunter, Zhang, Friesen,
  Indebetouw  \& Chandler}{Cyganowski
  et~al.}{2012}]{cyganowski2012protocluster}
Cyganowski C.,  Brogan C.,  Hunter T.,  Zhang Q.,  Friesen R.,  Indebetouw R.,
   Chandler C.,  2012, \apjl, 760, L20

\bibitem[\protect\citeauthoryear{De~Becker}{De~Becker}{2007}]{de2007non}
De~Becker M.,  2007, The Astronomy and Astrophysics Review, 14, 171

\bibitem[\protect\citeauthoryear{De~Pree, Mehringer  \& Goss}{De~Pree
  et~al.}{1997}]{depree:1997}
De~Pree C.,  Mehringer D.~M.,   Goss W.,  1997, \apj, 482, 307

\bibitem[\protect\citeauthoryear{De~Pree, Wilner, Deblasio, Mercer  \&
  Davis}{De~Pree et~al.}{2005}]{depree:2005}
De~Pree C.~G.,  Wilner D.,  Deblasio J.,  Mercer A.,   Davis L.,  2005, \apjl,
  624, L101

\bibitem[\protect\citeauthoryear{Dzib, Yang, Urquhart, Medina, Brunthaler,
  Menten, Wyrowski  et~al.}{Dzib et~al.}{2023}]{dzib:2023}
Dzib S.~A.,  Yang A.~Y.,  Urquhart J.~S.,  Medina S.-N.~X.,  Brunthaler A.,
  Menten K.~M.,  Wyrowski F.,   et~al., 2023, \mn@doi [\aap]
  {10.1051/0004-6361/202143019}, \href
  {https://ui.adsabs.harvard.edu/abs/2023A%26A...670A...9D} {670, A9}

\bibitem[\protect\citeauthoryear{Galv{\'a}n-Madrid, Peters, Keto, Mac~Low,
  Banerjee  \& Klessen}{Galv{\'a}n-Madrid et~al.}{2011}]{galvanmadrid:2011}
Galv{\'a}n-Madrid R.,  Peters T.,  Keto E.~R.,  Mac~Low M.-M.,  Banerjee R.,
  Klessen R.~S.,  2011, \mn@doi [\mnras] {10.1111/j.1365-2966.2011.19101.x},
  416, 1033

\bibitem[\protect\citeauthoryear{{Goedhart} et~al.,}{{Goedhart}
  et~al.}{2024}]{goedhart:2024}
{Goedhart} S.,  et~al., 2024, \mn@doi [\mnras] {10.1093/mnras/stae1166}, \href
  {https://ui.adsabs.harvard.edu/abs/2024MNRAS.531..649G} {531, 649}

\bibitem[\protect\citeauthoryear{Helfand, Becker, White, Fallon  \&
  Tuttle}{Helfand et~al.}{2006}]{helfand:2006}
Helfand D.~J.,  Becker R.~H.,  White R.~L.,  Fallon A.,   Tuttle S.,  2006,
  \aj, 131, 2525

\bibitem[\protect\citeauthoryear{Hoare, Kurtz, Lizano, Keto  \& Hofner}{Hoare
  et~al.}{2007}]{hoare:2007}
Hoare M.,  Kurtz S.,  Lizano S.,  Keto E.,   Hofner P.,  2007, Protostars and
  Planets, 5, 181

\bibitem[\protect\citeauthoryear{Hoare et~al.,}{Hoare
  et~al.}{2012}]{hoare:2012}
Hoare M.,  et~al., 2012, \pasp, 124, 939

\bibitem[\protect\citeauthoryear{H{\"o}gbom}{H{\"o}gbom}{1974}]{hogbom:1974}
H{\"o}gbom J.,  1974, \aaps, 15, 417

\bibitem[\protect\citeauthoryear{Irabor et~al.,}{Irabor
  et~al.}{2023}]{irabor:2023}
Irabor T.,  et~al., 2023, \mn@doi [\mnras] {10.1093/mnras/stad005}, 520, 1073

\bibitem[\protect\citeauthoryear{Kalcheva, Hoare, Urquhart, Kurtz, Lumsden,
  Purcell  \& Zijlstra}{Kalcheva et~al.}{2018}]{kalcheva:2018}
Kalcheva I.~E.,  Hoare M.~G.,  Urquhart J.~S.,  Kurtz S.,  Lumsden S.~L.,
  Purcell C.~R.,   Zijlstra A.~A.,  2018, \mn@doi [\aap]
  {10.1051/0004-6361/201832734}, 615, A103

\bibitem[\protect\citeauthoryear{Kim \& Koo}{Kim \& Koo}{2001}]{kim2001radio}
Kim K.-T.,  Koo B.-C.,  2001, \apj, 549, 979

\bibitem[\protect\citeauthoryear{Kim \& Koo}{Kim \&
  Koo}{2002}]{kim2002interaction}
Kim K.-T.,  Koo B.-C.,  2002, \apj, 575, 327

\bibitem[\protect\citeauthoryear{Kim \& Koo}{Kim \&
  Koo}{2003}]{kim2003molecular}
Kim K.-T.,  Koo B.-C.,  2003, \apj, 596, 362

\bibitem[\protect\citeauthoryear{Kurtz}{Kurtz}{2005}]{kurtz2005hypercompact}
Kurtz S.,  2005, Proceedings of the International Astronomical Union, 1, 111

\bibitem[\protect\citeauthoryear{{Lacy} et~al.,}{{Lacy}
  et~al.}{2020}]{lacy:2020}
{Lacy} M.,  et~al., 2020, \mn@doi [\pasp] {10.1088/1538-3873/ab63eb}, \href
  {https://ui.adsabs.harvard.edu/abs/2020PASP..132c5001L} {132, 035001}

\bibitem[\protect\citeauthoryear{Lucas et~al.,}{Lucas
  et~al.}{2008}]{lucas:2008}
Lucas P.,  et~al., 2008, \mnras, 391, 136

\bibitem[\protect\citeauthoryear{{McMullin}, {Waters}, {Schiebel}, {Young}  \&
  {Golap}}{{McMullin} et~al.}{2007}]{mcmullin:2007}
{McMullin} J.~P.,  {Waters} B.,  {Schiebel} D.,  {Young} W.,   {Golap} K.,
  2007, in {Shaw} R.~A.,  {Hill} F.,   {Bell} D.~J.,  eds,  Astronomical
  Society of the Pacific Conference Series Vol. 376, Astronomical Data Analysis
  Software and Systems XVI. p.~127

\bibitem[\protect\citeauthoryear{{Molinari} et~al.,}{{Molinari}
  et~al.}{2010}]{molinari:2010}
{Molinari} S.,  et~al., 2010, \mn@doi [\pasp] {10.1086/651314}, \href
  {https://ui.adsabs.harvard.edu/abs/2010PASP..122..314M} {122, 314}

\bibitem[\protect\citeauthoryear{Pereira, L{\'o}pez-Santiago, Miceli, Bonito
  \& De~Castro}{Pereira et~al.}{2016}]{pereira2016modeling}
Pereira V.,  L{\'o}pez-Santiago J.,  Miceli M.,  Bonito R.,   De~Castro E.,
  2016, \aap, 588, A36

\bibitem[\protect\citeauthoryear{Peters, Banerjee, Klessen, Mac~Low,
  Galv{\'a}n-Madrid  \& Keto}{Peters et~al.}{2010}]{peters:2010a}
Peters T.,  Banerjee R.,  Klessen R.~S.,  Mac~Low M.-M.,  Galv{\'a}n-Madrid R.,
    Keto E.~R.,  2010, \apj, 711, 1017

\bibitem[\protect\citeauthoryear{Purcell et~al.,}{Purcell
  et~al.}{2013}]{purcell:2013}
Purcell C.,  et~al., 2013, \apjs, 205, 1

\bibitem[\protect\citeauthoryear{Si{\'o}dmiak \& Tylenda}{Si{\'o}dmiak \&
  Tylenda}{2001}]{siodmiak:2001}
Si{\'o}dmiak N.,  Tylenda R.,  2001, \aap, 373, 1032

\bibitem[\protect\citeauthoryear{{Stanimirovi{\'c}}}{{Stanimirovi{\'c}}}{2002}]{stanimirovic:2002}
{Stanimirovi{\'c}} S.,  2002, in {Stanimirovi{\'c}} S.,  {Altschuler} D.,
  {Goldsmith} P.,   {Salter} C.,  eds, ASP Conf. Ser., Vol.~278, Single-Dish
  Radio Astronomy: Techniques and Applications.
Astronomical Society of the Pacific, p.~375

\bibitem[\protect\citeauthoryear{{Taylor}, {Gibson}, {Peracaula}
  et~al.}{{Taylor} et~al.}{2003}]{taylor:2003}
{Taylor} A.~R.,  {Gibson} S.~J.,  {Peracaula} M.,   et~al., 2003, \aj, 125,
  3145

\bibitem[\protect\citeauthoryear{Tian \& Leahy}{Tian \&
  Leahy}{2005}]{tian2005radio}
Tian W.,  Leahy D.,  2005, \aap, 436, 187

\bibitem[\protect\citeauthoryear{Urquhart et~al.,}{Urquhart
  et~al.}{2011}]{urquhart2011red}
Urquhart J.,  et~al., 2011, \mnras, 418, 1689

\bibitem[\protect\citeauthoryear{Urquhart et~al.,}{Urquhart
  et~al.}{2012}]{urquhart:2012}
Urquhart J.,  et~al., 2012, \mnras, 420, 1656

\bibitem[\protect\citeauthoryear{Wood \& Churchwell}{Wood \&
  Churchwell}{1989}]{churchwell:1989a}
Wood D.~O.,  Churchwell E.,  1989, \apjs, 69, 831

\bibitem[\protect\citeauthoryear{Yang, Thompson, Tian, Bihr, Beuther  \&
  Hindson}{Yang et~al.}{2018}]{yang2018search}
Yang A.,  Thompson M.,  Tian W.,  Bihr S.,  Beuther H.,   Hindson L.,  2018,
  \mnras, 482, 2681

\bibitem[\protect\citeauthoryear{Yang et~al.,}{Yang et~al.}{2025}]{yang:2025}
Yang A.~Y.,  et~al., 2025, \mn@doi [\aap] {10.1051/0004-6361/202452078}, 694,
  A26

\bibitem[\protect\citeauthoryear{Zinnecker \& Yorke}{Zinnecker \&
  Yorke}{2007}]{yorke:2007}
Zinnecker H.,  Yorke H.~W.,  2007, \araa, 45, 481

\bibitem[\protect\citeauthoryear{del Valle \& Pohl}{del Valle \&
  Pohl}{2018}]{delvalle2018}
del Valle M.~V.,  Pohl M.,  2018, \apj, 864, 19

\makeatother
\end{thebibliography}

\appendix

\clearpage

\begin{figure*}
\centering
\phantomsection
\refstepcounter{section}\label{app:appendixA}
\addcontentsline{toc}{section}{\thesection\ Target maps}
{\bfseries APPENDIX A: TARGET MAPS}\par\vspace{4mm}
\includegraphics[width=0.30\textwidth]{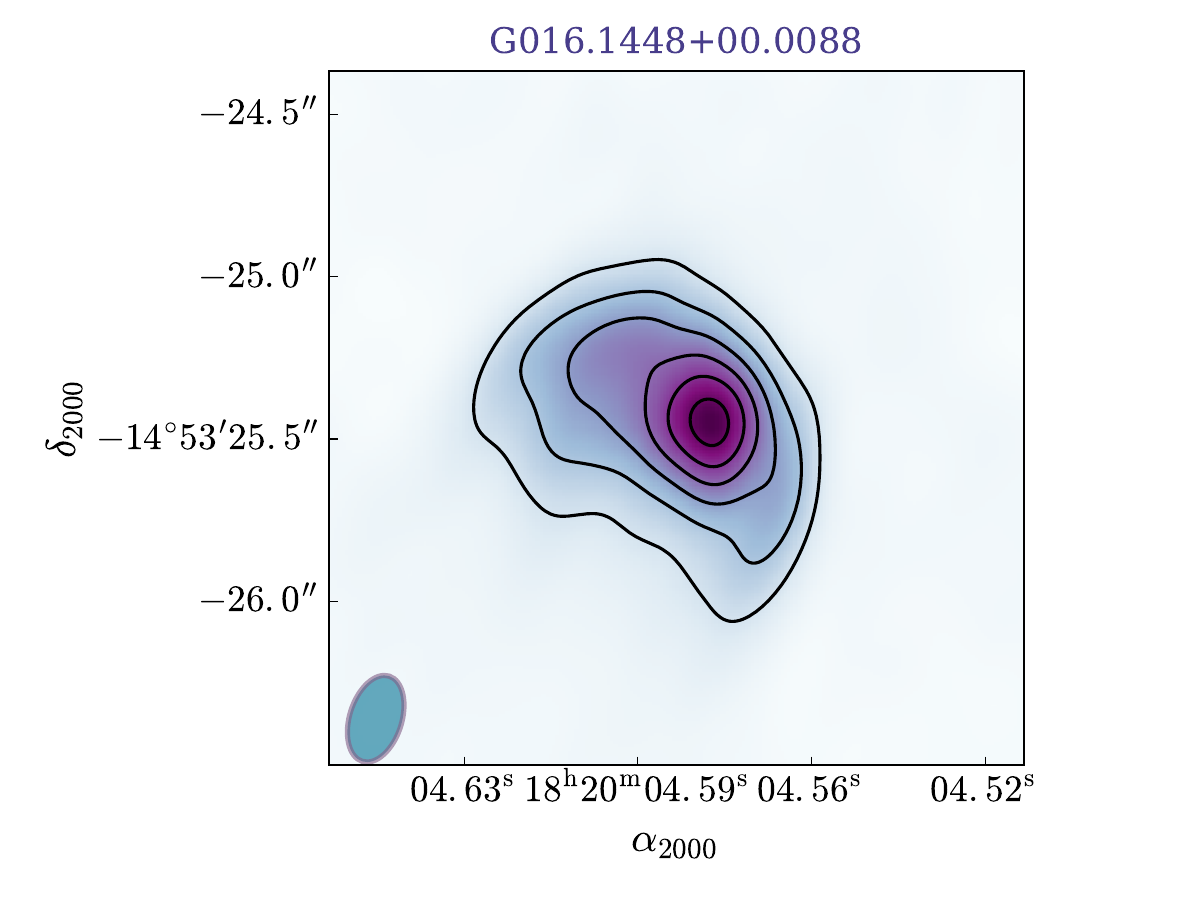} \hfill
\includegraphics[width=0.30\textwidth]{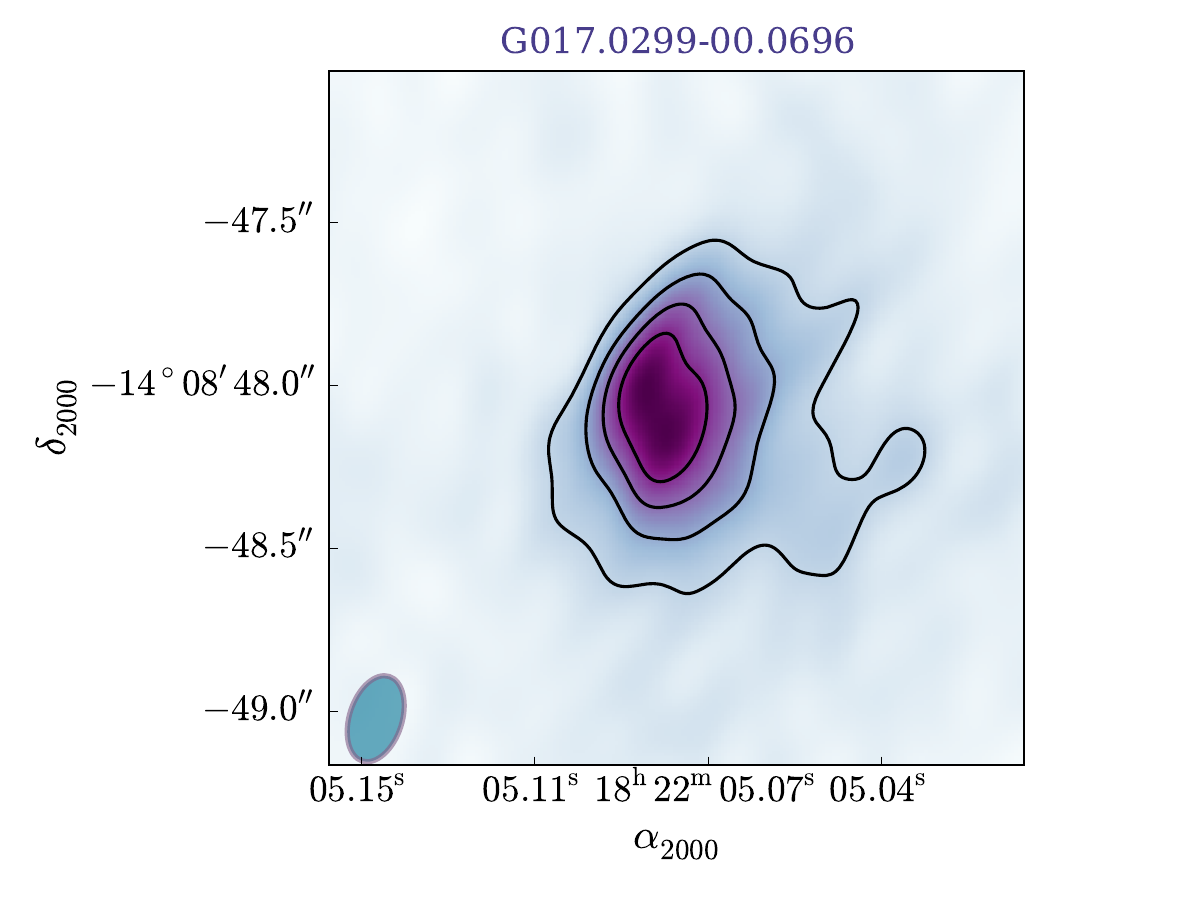} \hfill
\includegraphics[width=0.30\textwidth]{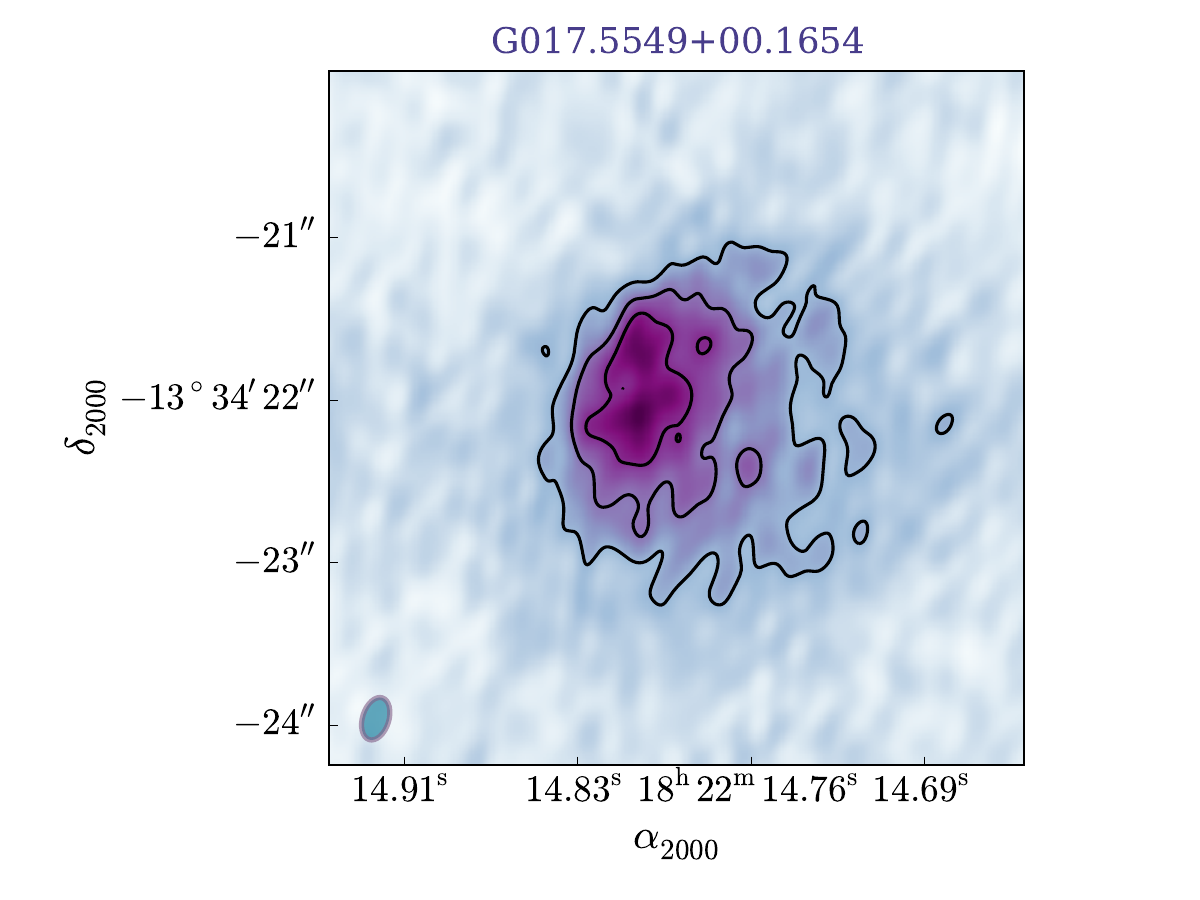}
\\[2mm]
\includegraphics[width=0.30\textwidth]{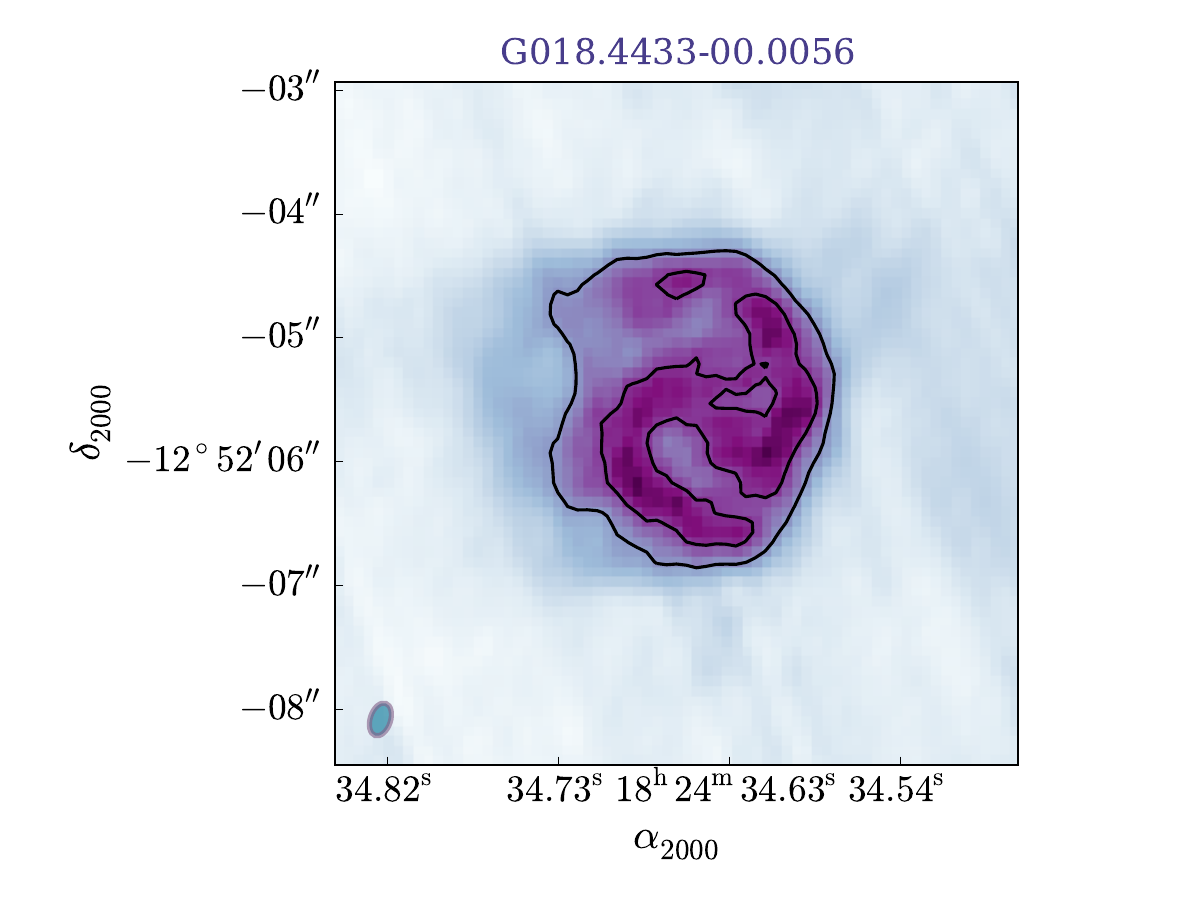} \hfill
\includegraphics[width=0.30\textwidth]{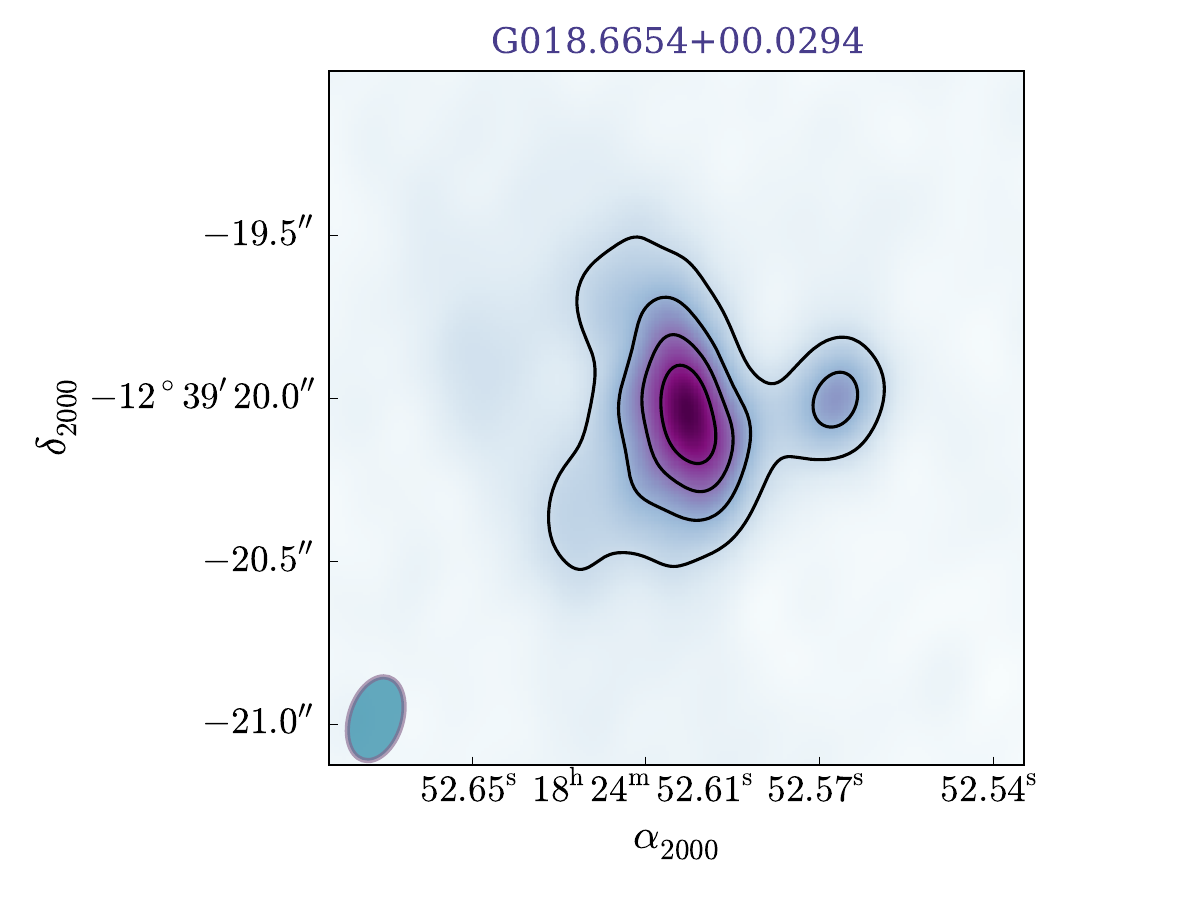} \hfill
\includegraphics[width=0.30\textwidth]{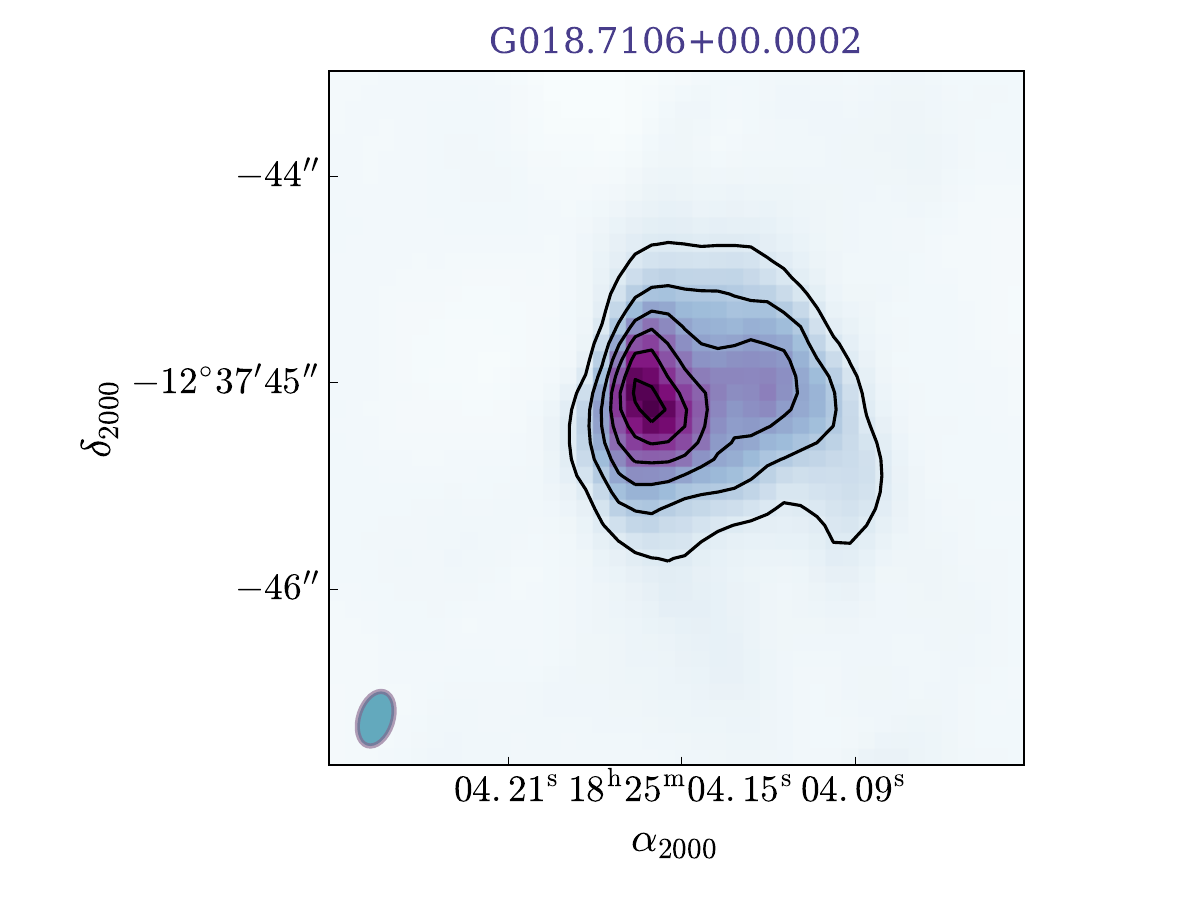}
\\[2mm]
\includegraphics[width=0.30\textwidth]{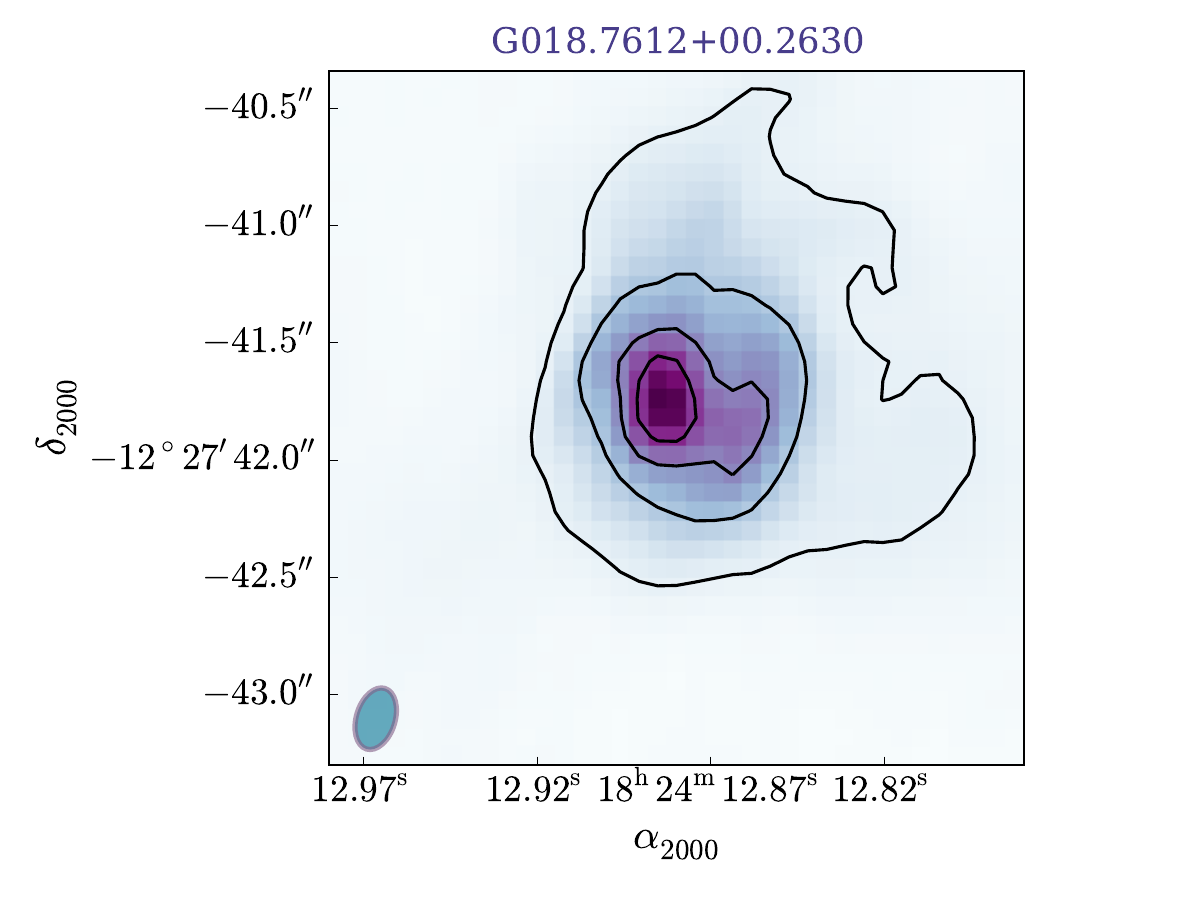} \hfill
\includegraphics[width=0.30\textwidth]{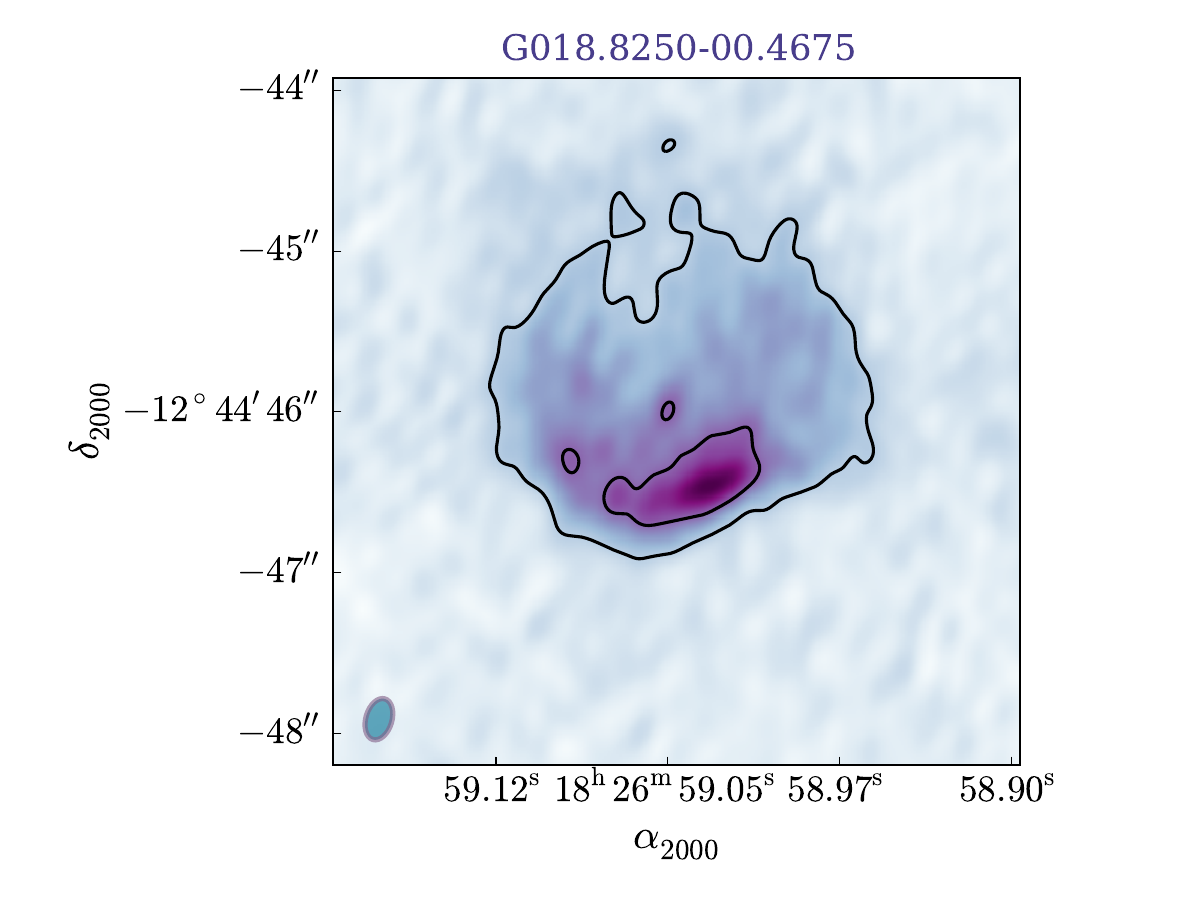} \hfill
\includegraphics[width=0.30\textwidth]{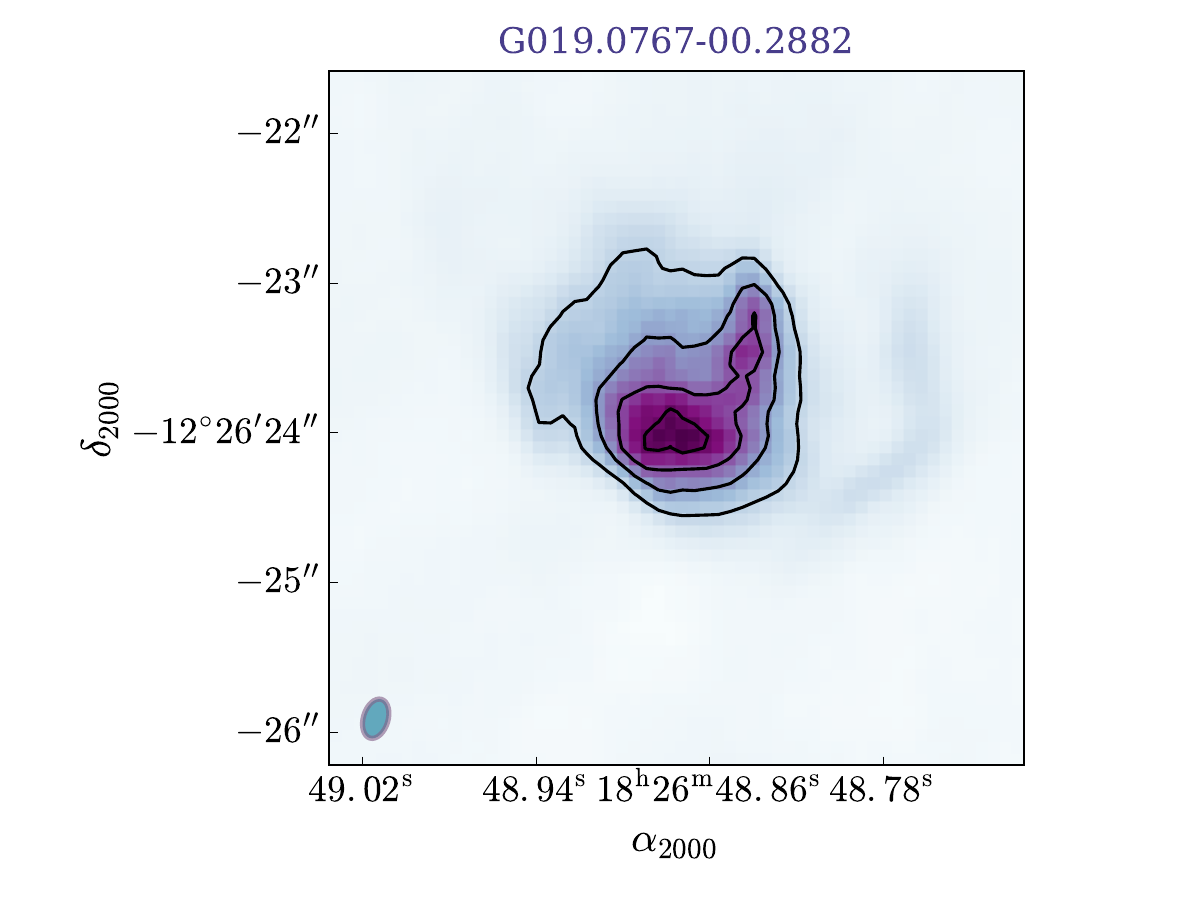}
\\[2mm]
\includegraphics[width=0.30\textwidth]{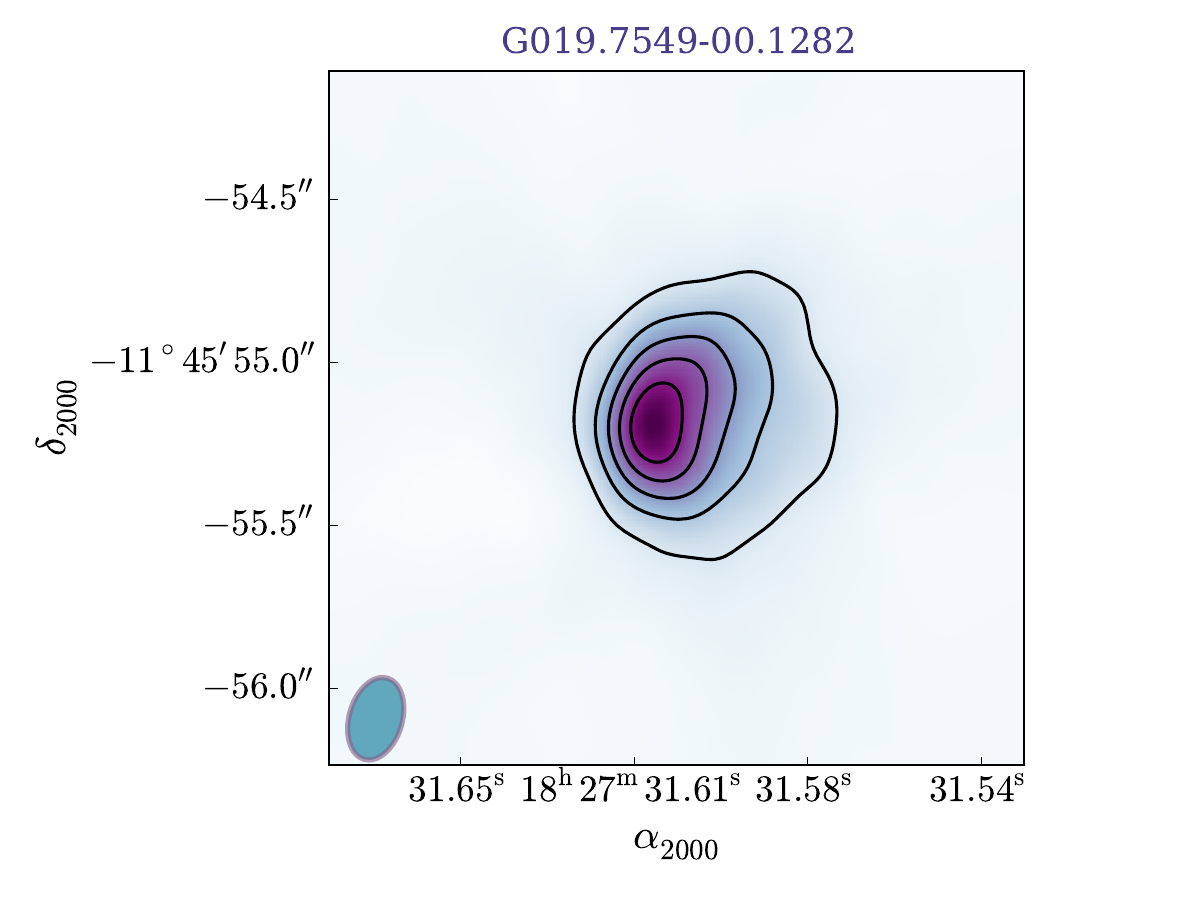} \hfill
\includegraphics[width=0.30\textwidth]{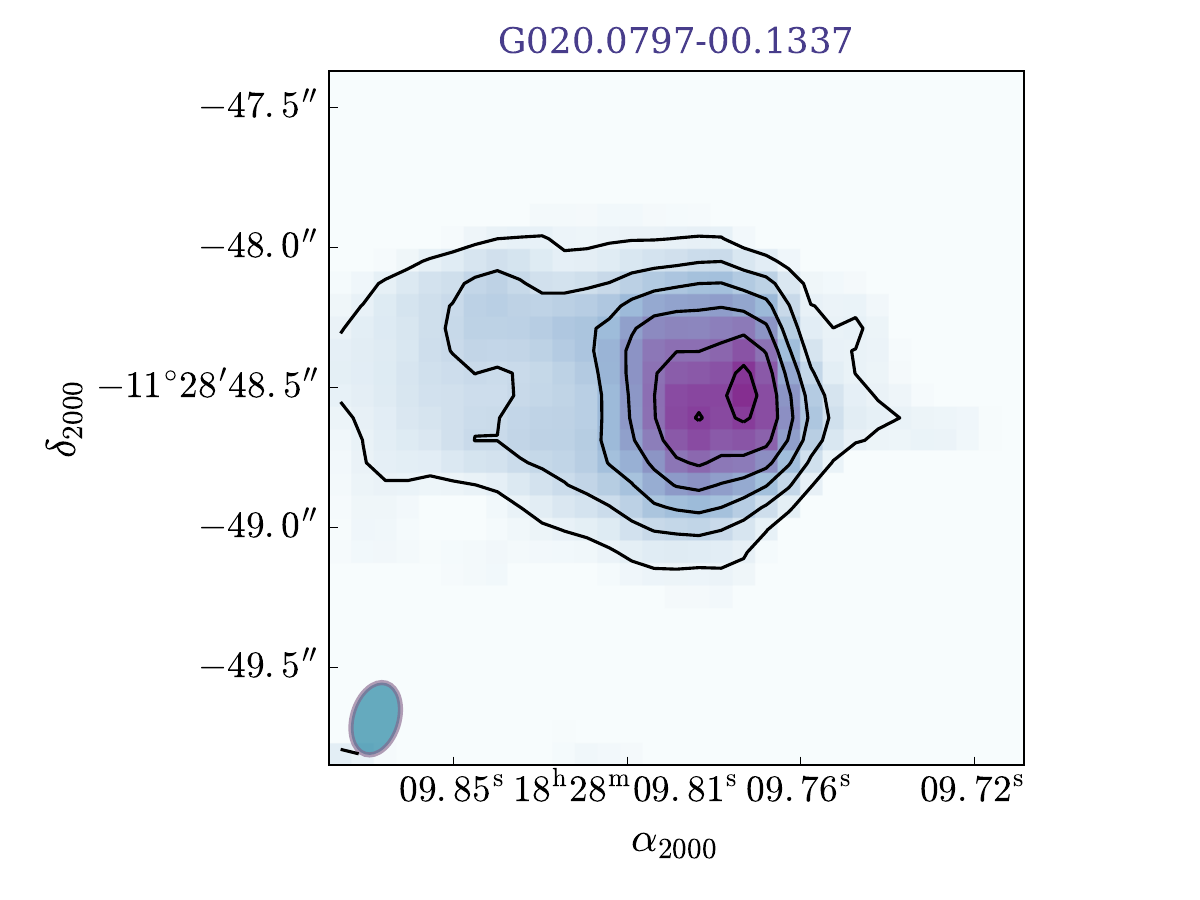} \hfill
\includegraphics[width=0.30\textwidth]{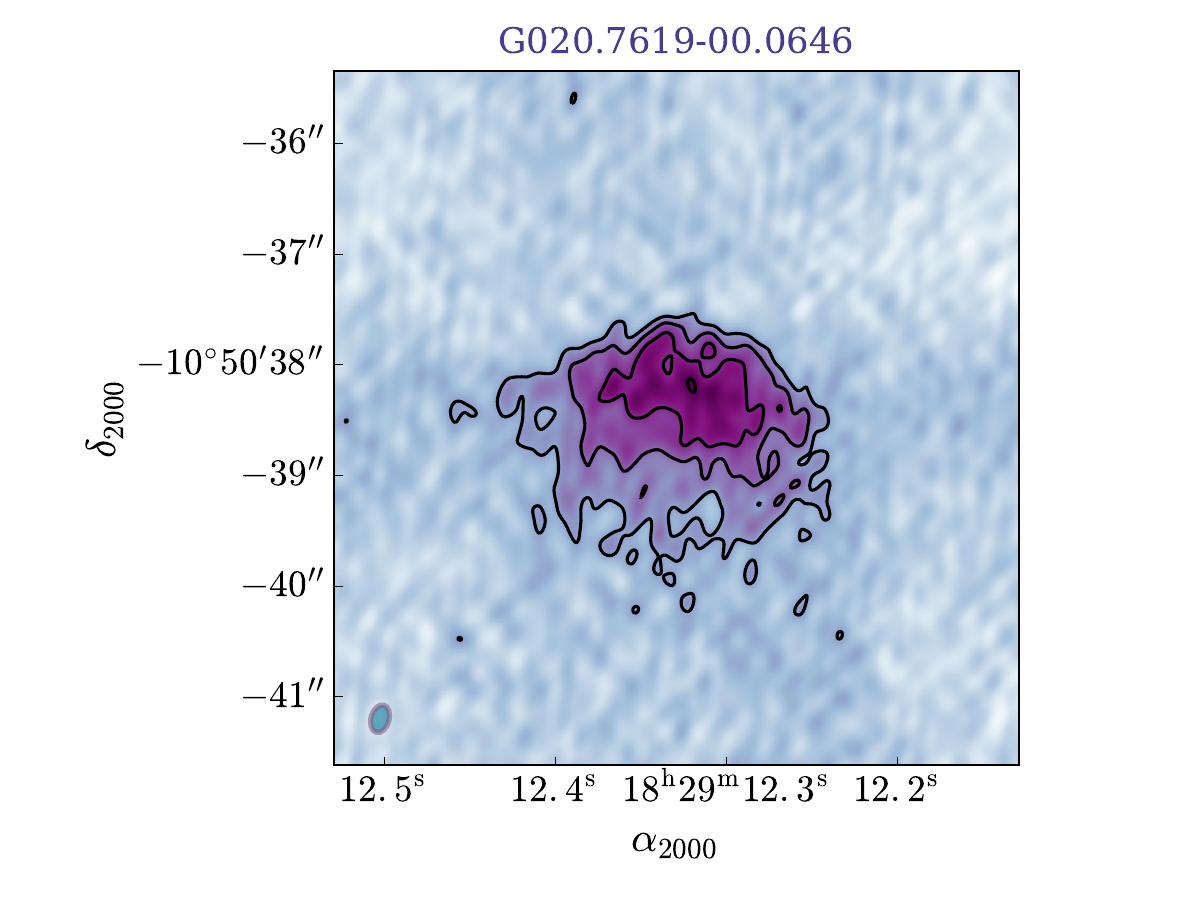}
\\[2mm]
\includegraphics[width=0.30\textwidth]{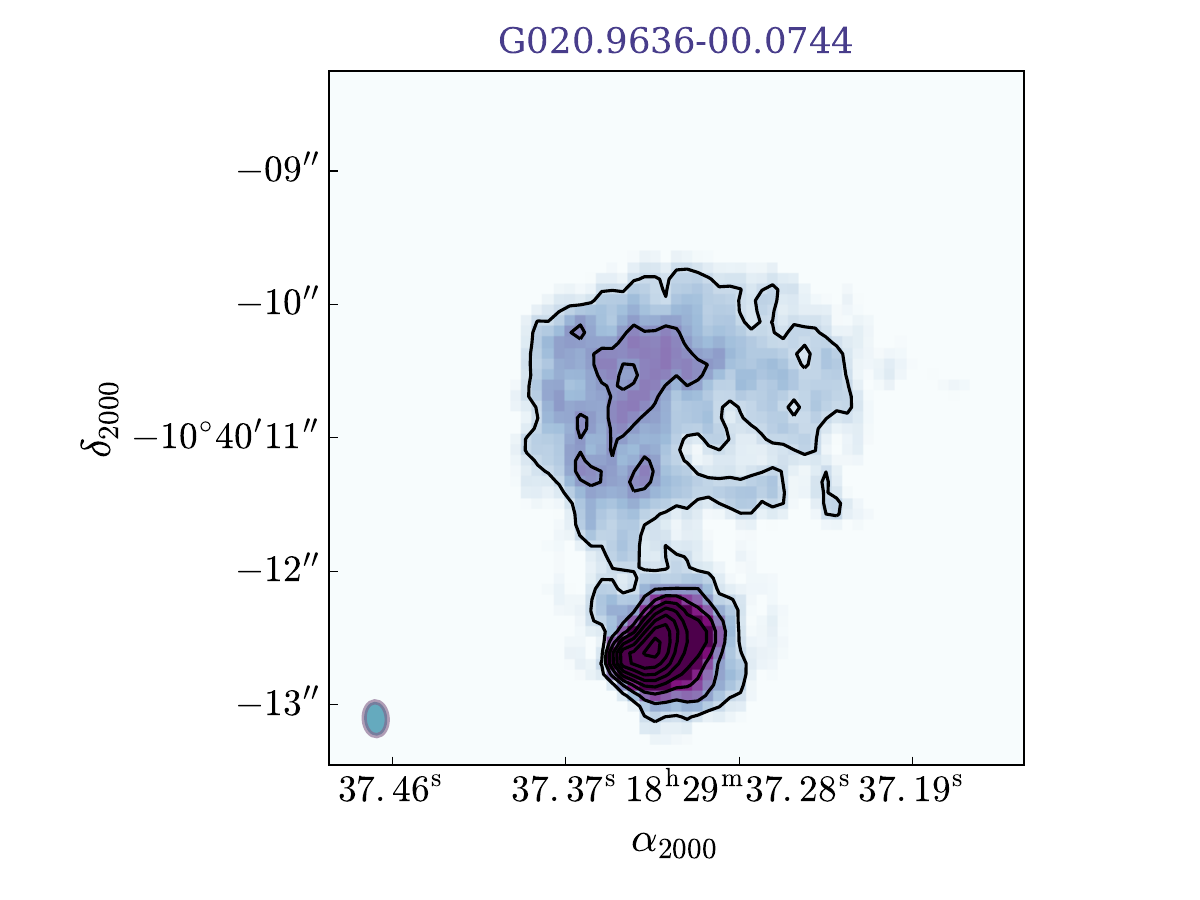} \hfill
\includegraphics[width=0.30\textwidth]{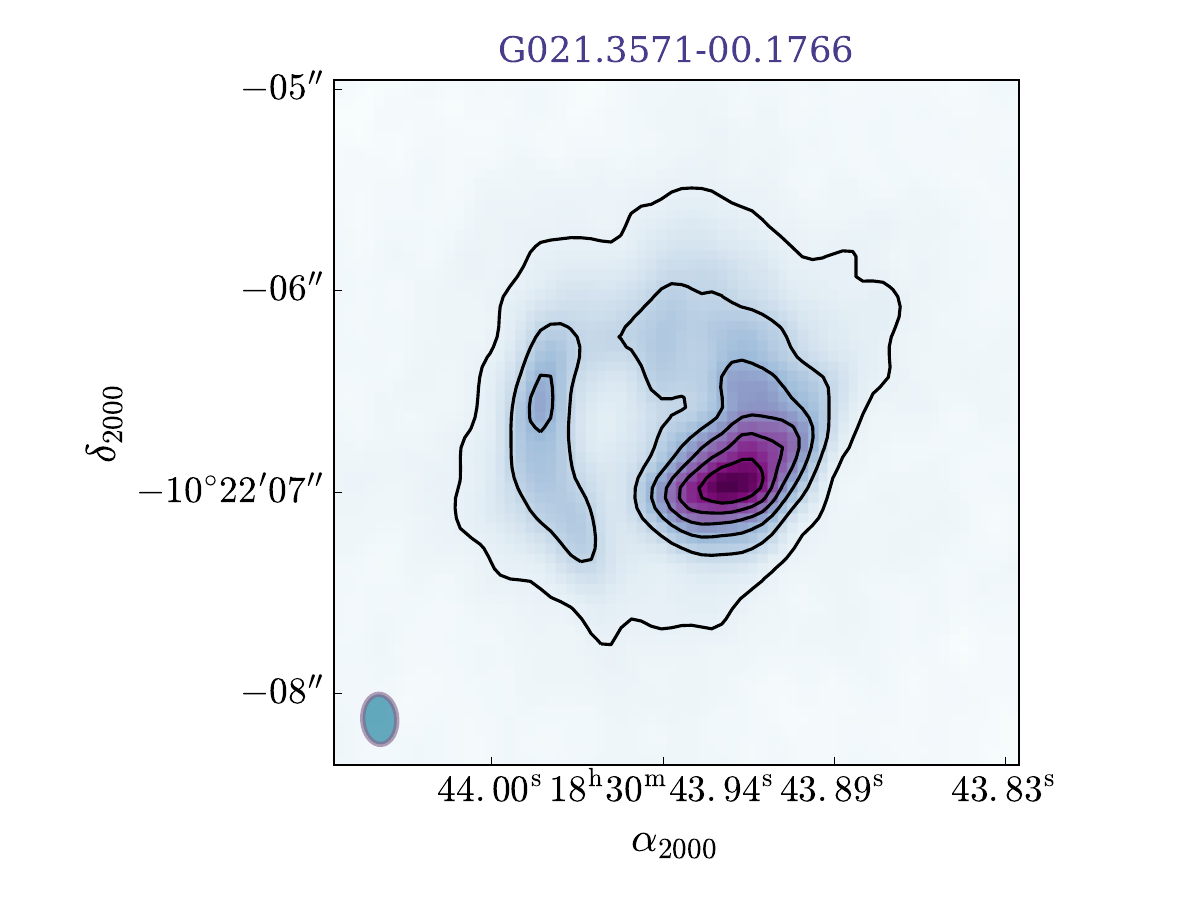} \hfill
\includegraphics[width=0.30\textwidth]{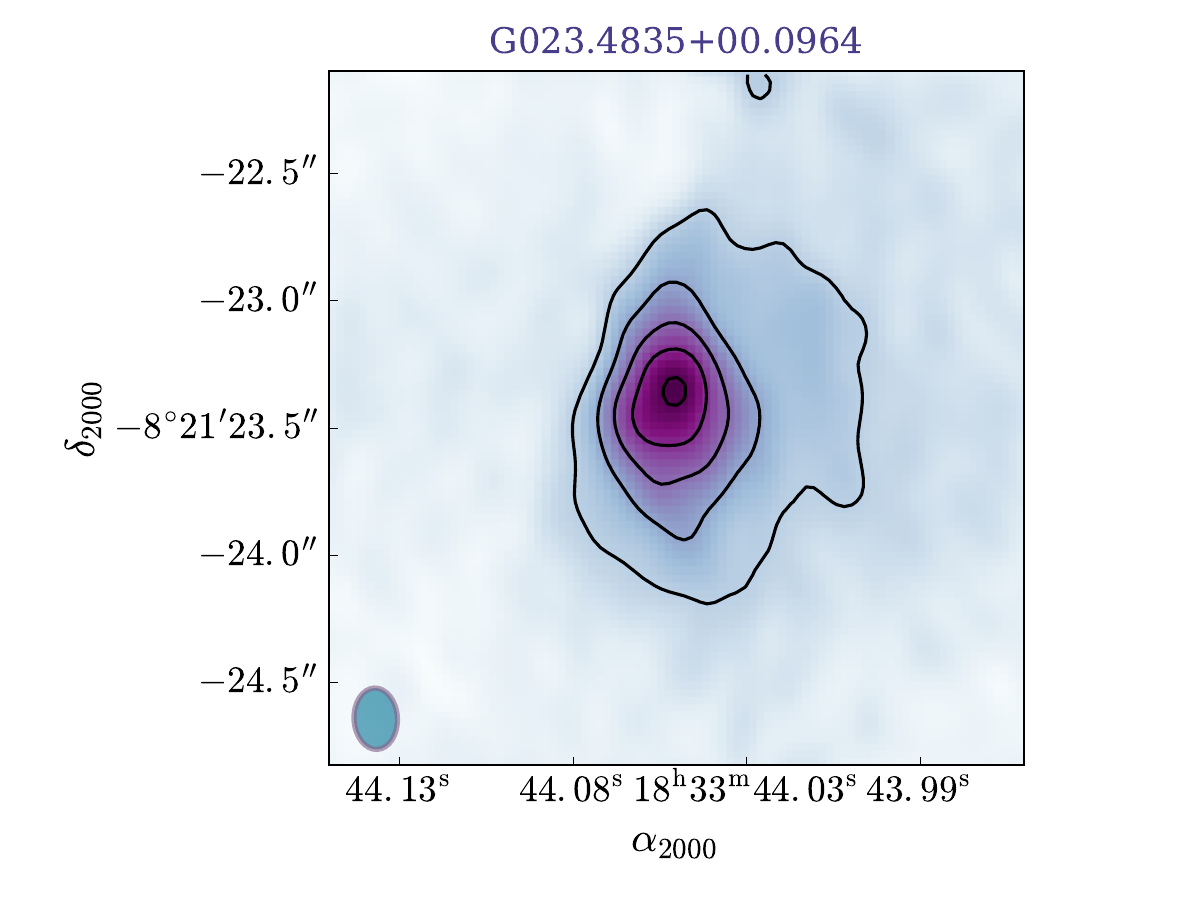}
\\[2mm]
\caption[Target maps]{Target maps. The clean beam is shown in the bottom-left corners. The contour levels start at 7$\sigma$, up to the measured peak flux (darkest purple).}
\label{fig:targets1}
\end{figure*}
\clearpage

\begin{figure*}
\ContinuedFloat
\centering
\includegraphics[width=0.30\textwidth]{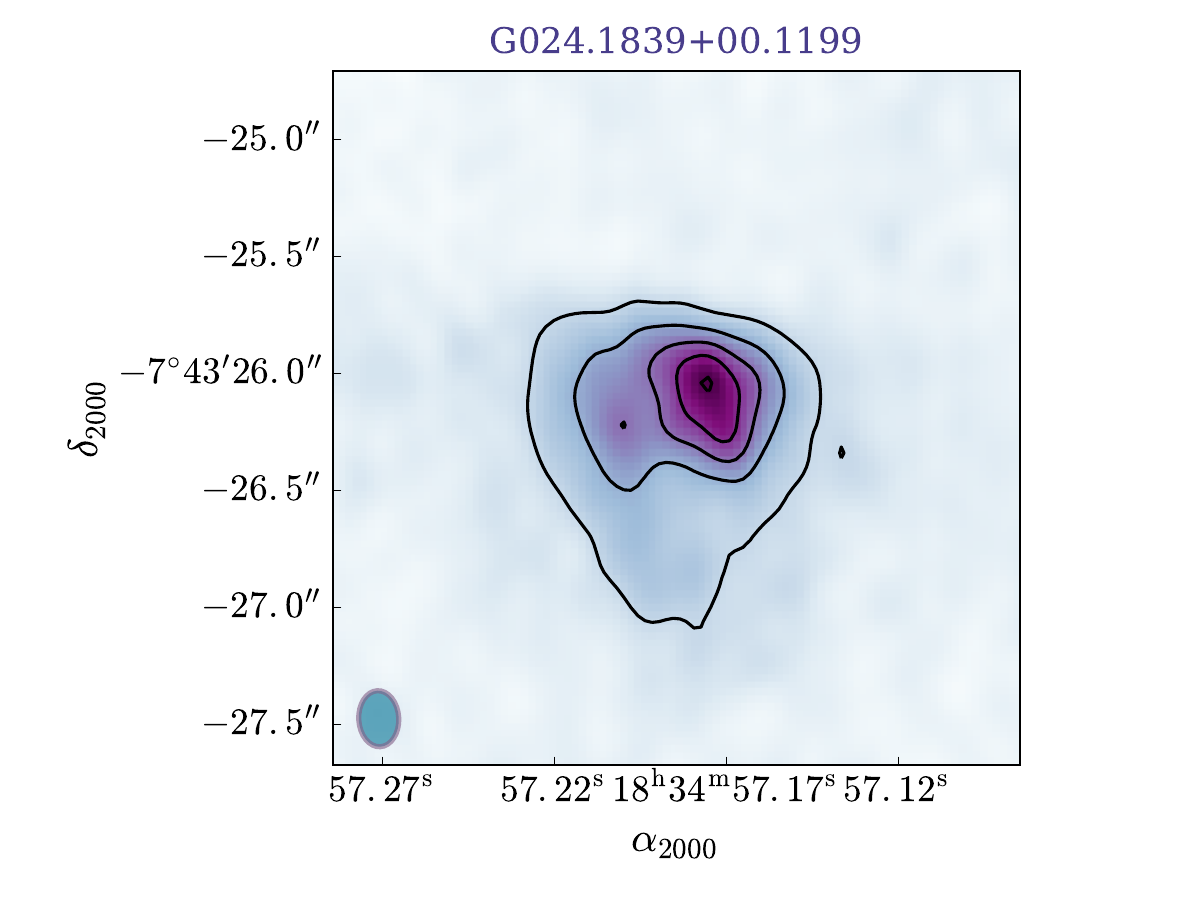} \hfill
\includegraphics[width=0.30\textwidth]{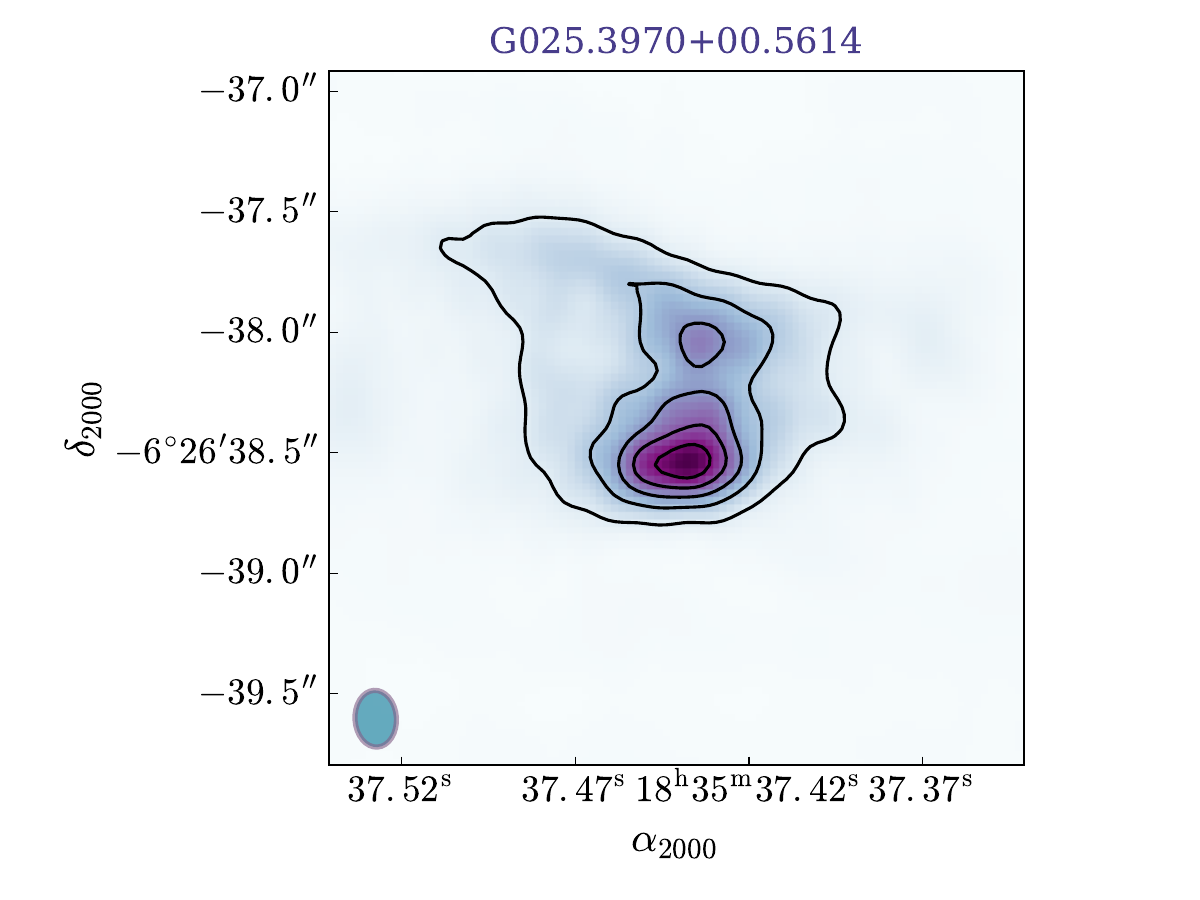} \hfill
\includegraphics[width=0.30\textwidth]{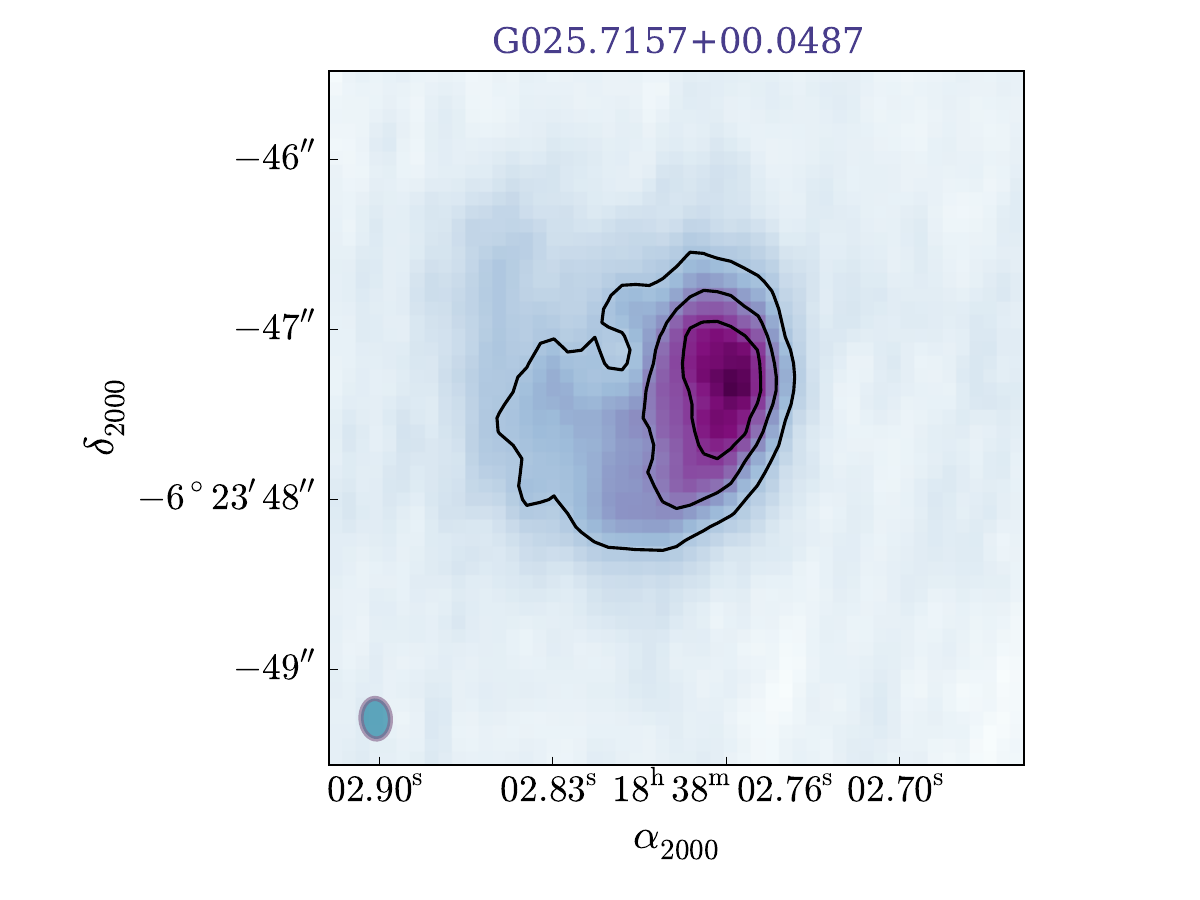}
\\[2mm]
\includegraphics[width=0.30\textwidth]{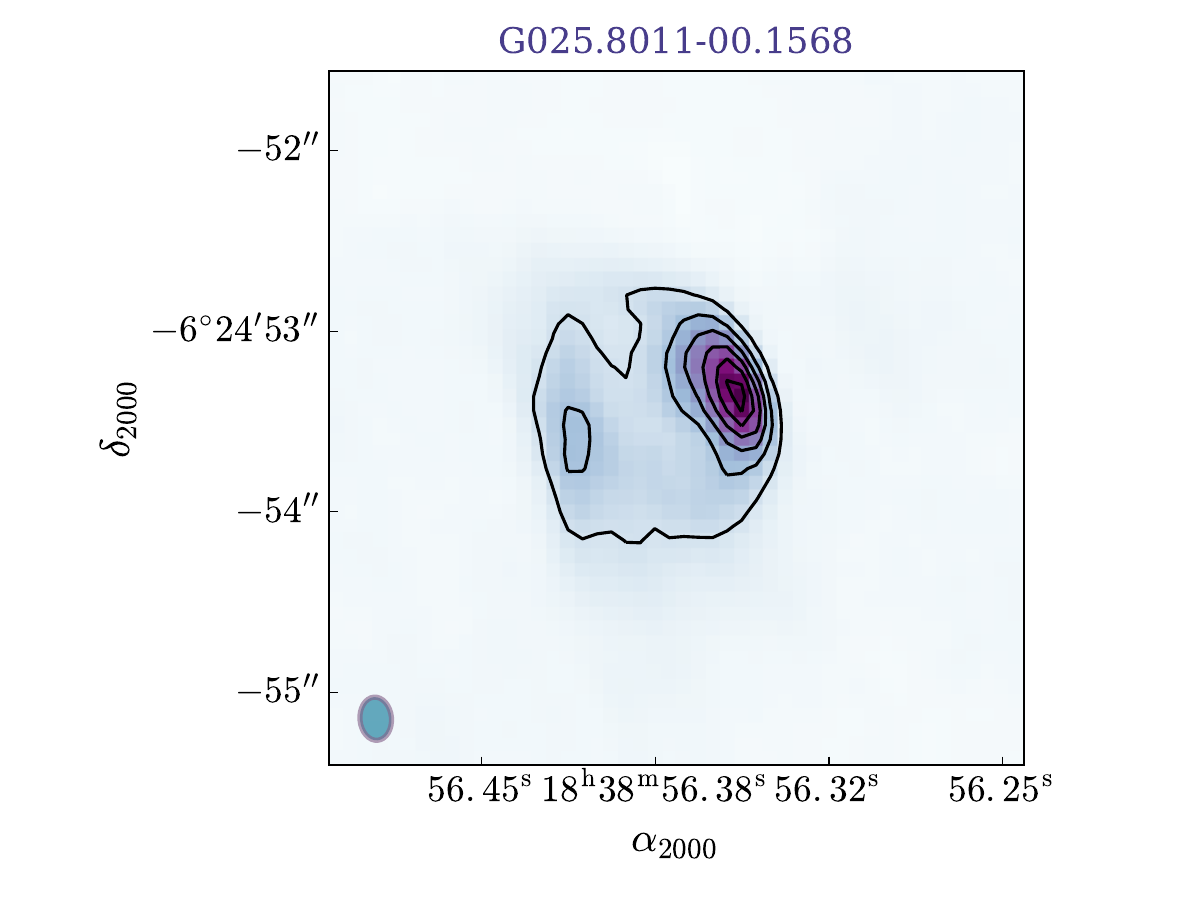} \hfill
\includegraphics[width=0.30\textwidth]{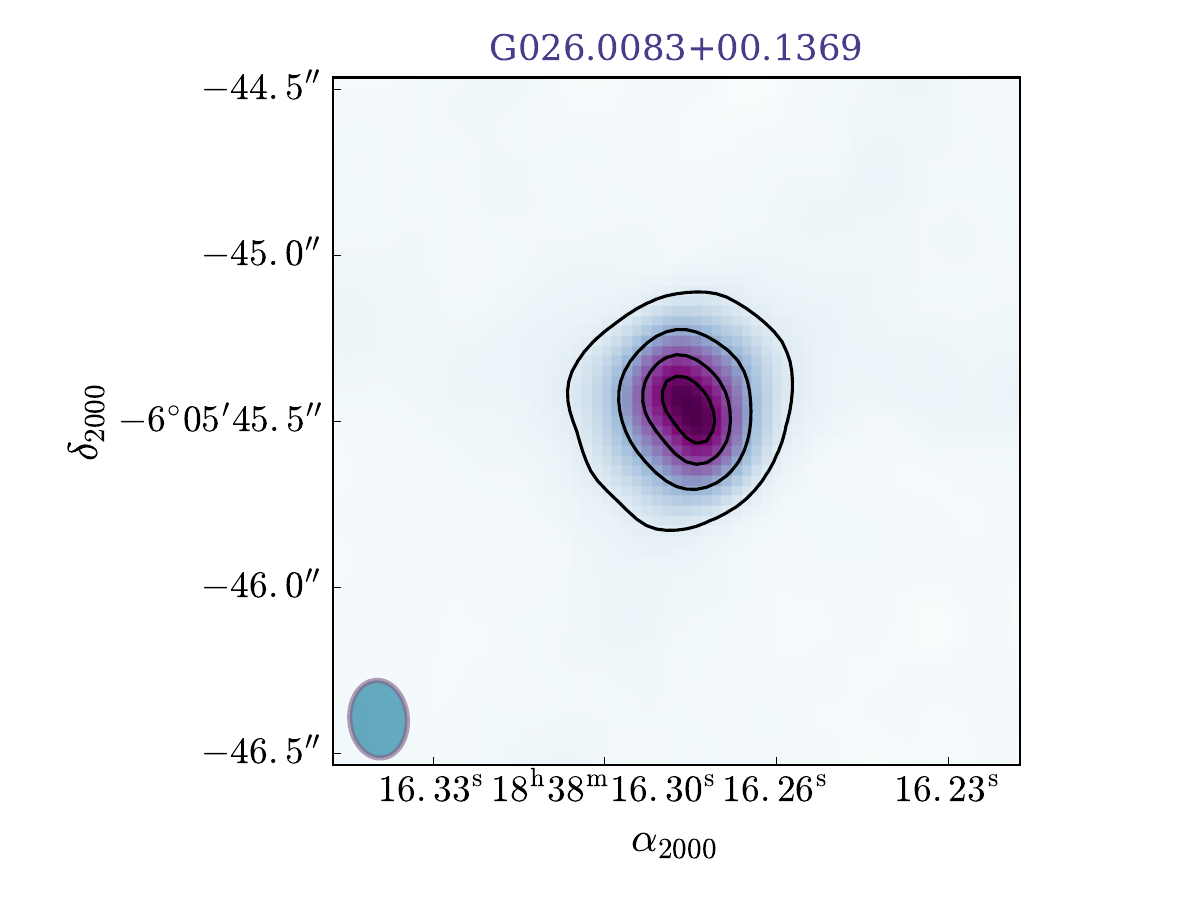} \hfill
\includegraphics[width=0.30\textwidth]{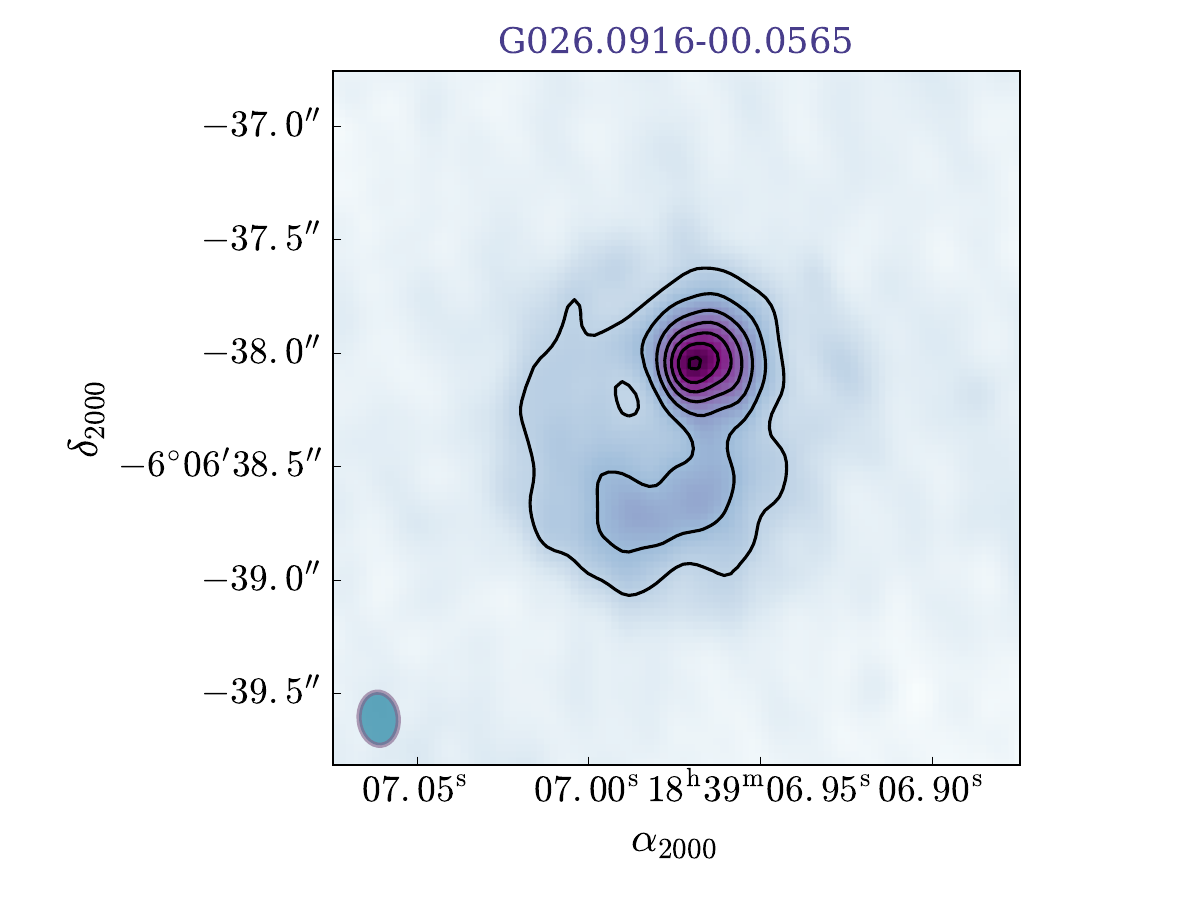}
\\[2mm]
\includegraphics[width=0.30\textwidth]{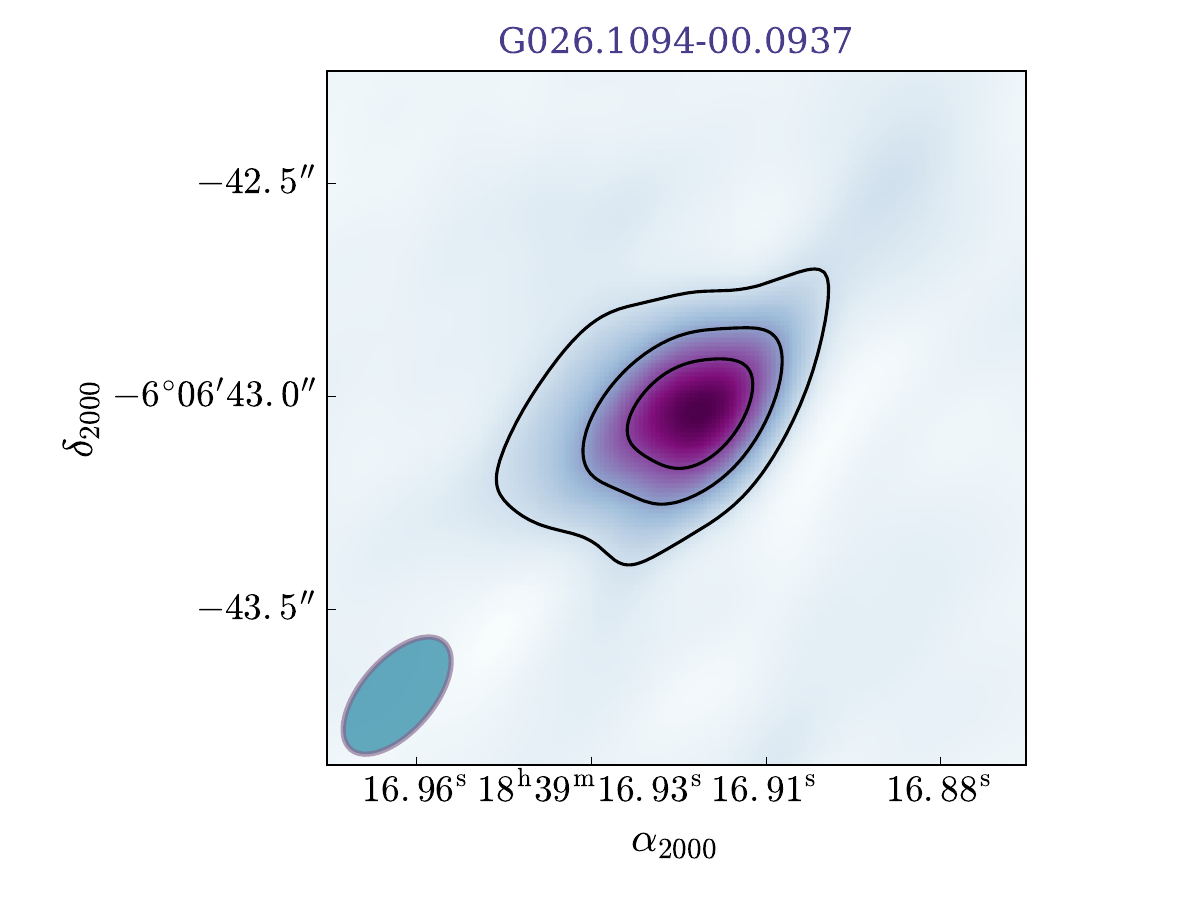} \hfill
\includegraphics[width=0.30\textwidth]{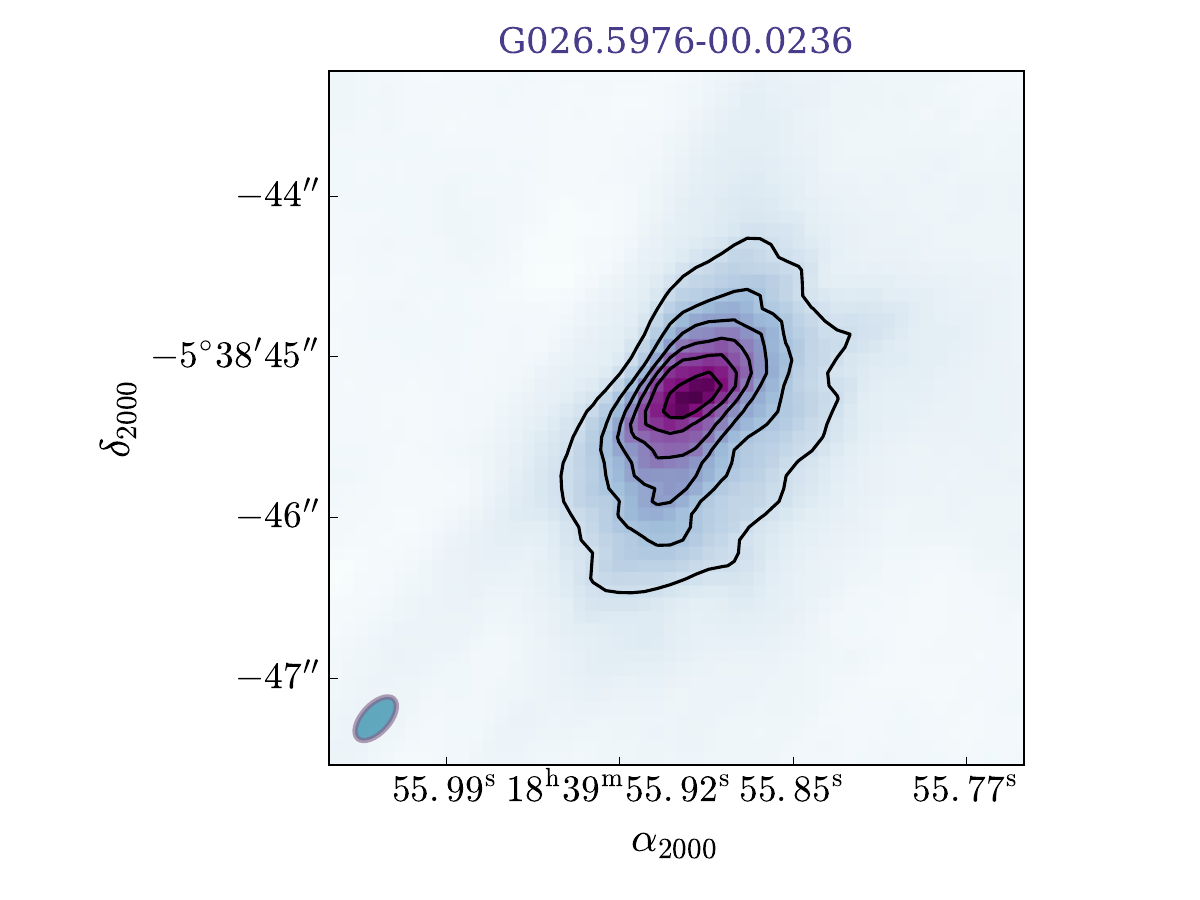} \hfill
\includegraphics[width=0.30\textwidth]{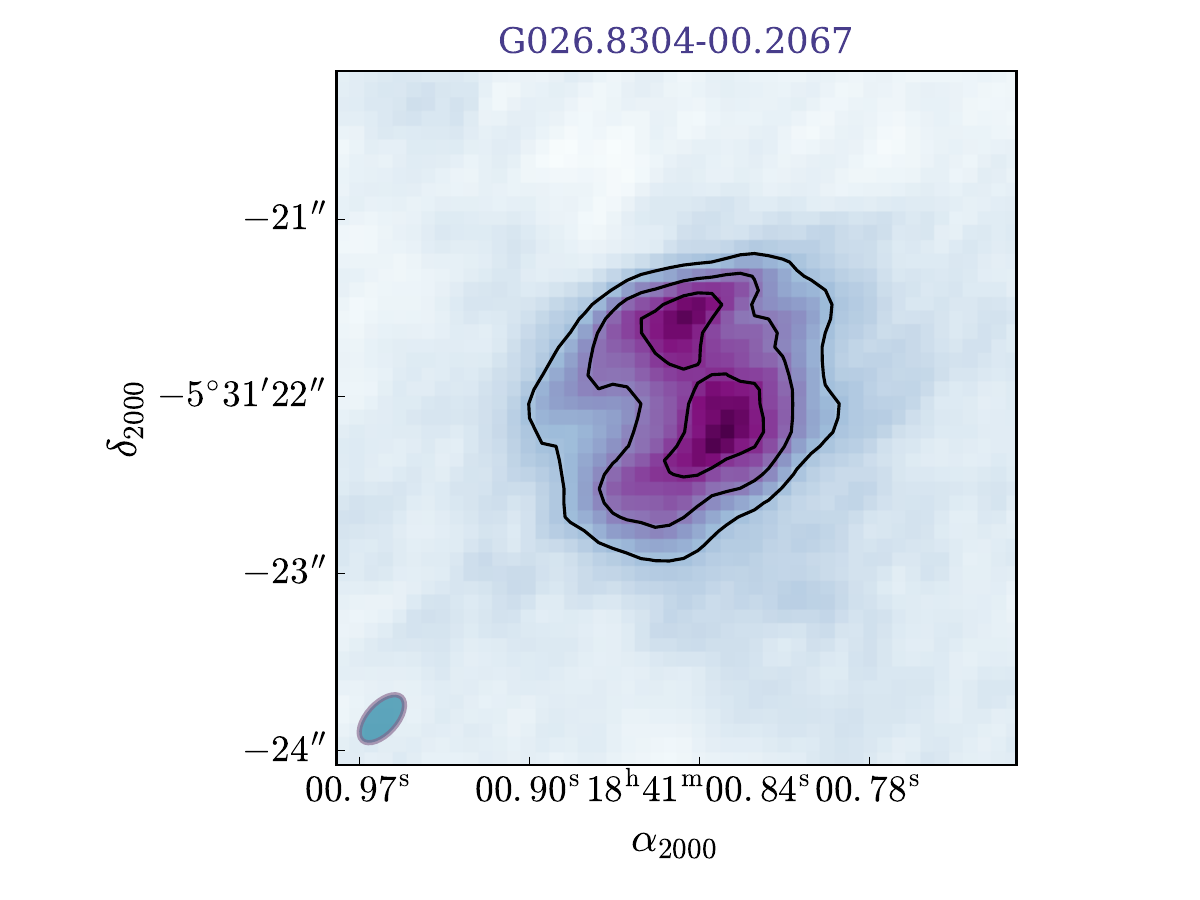}
\\[2mm]
\includegraphics[width=0.30\textwidth]{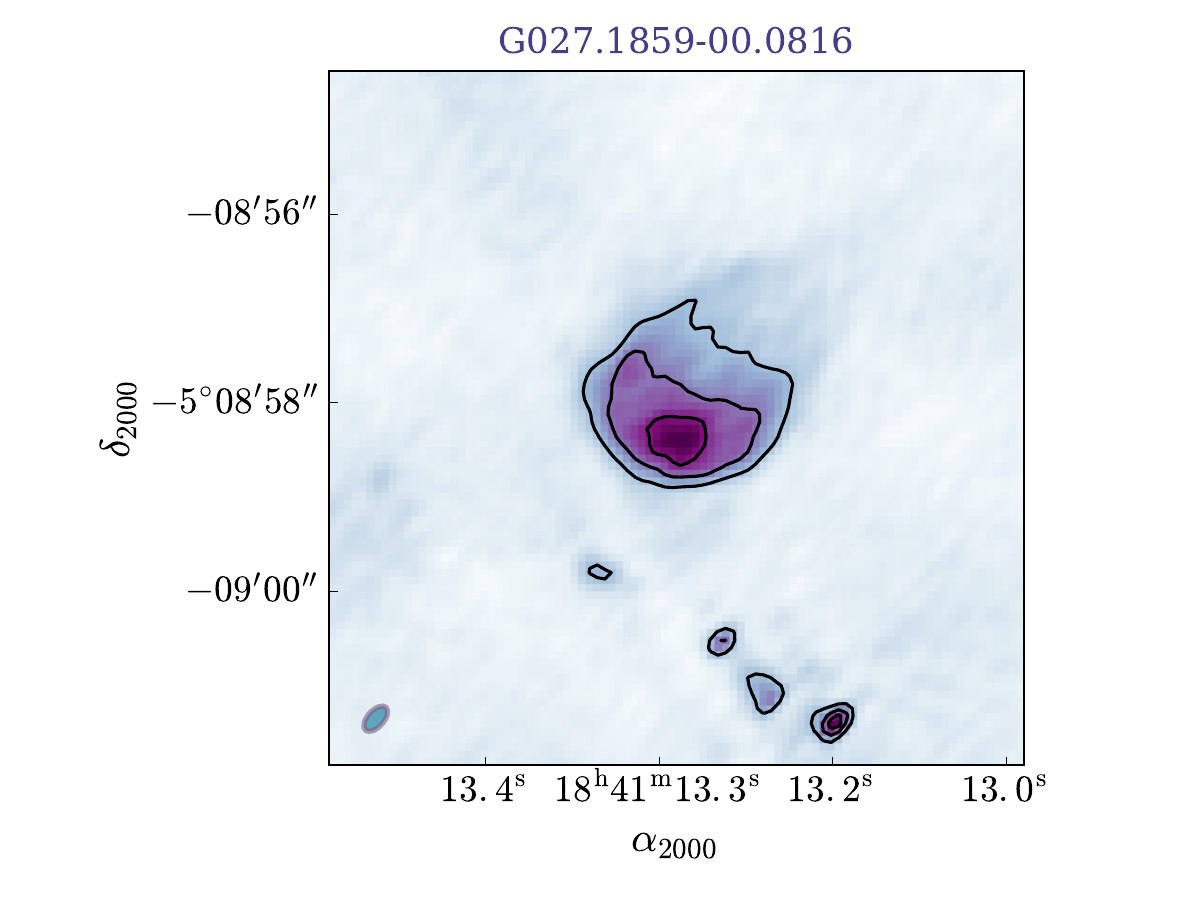} \hfill
\includegraphics[width=0.30\textwidth]{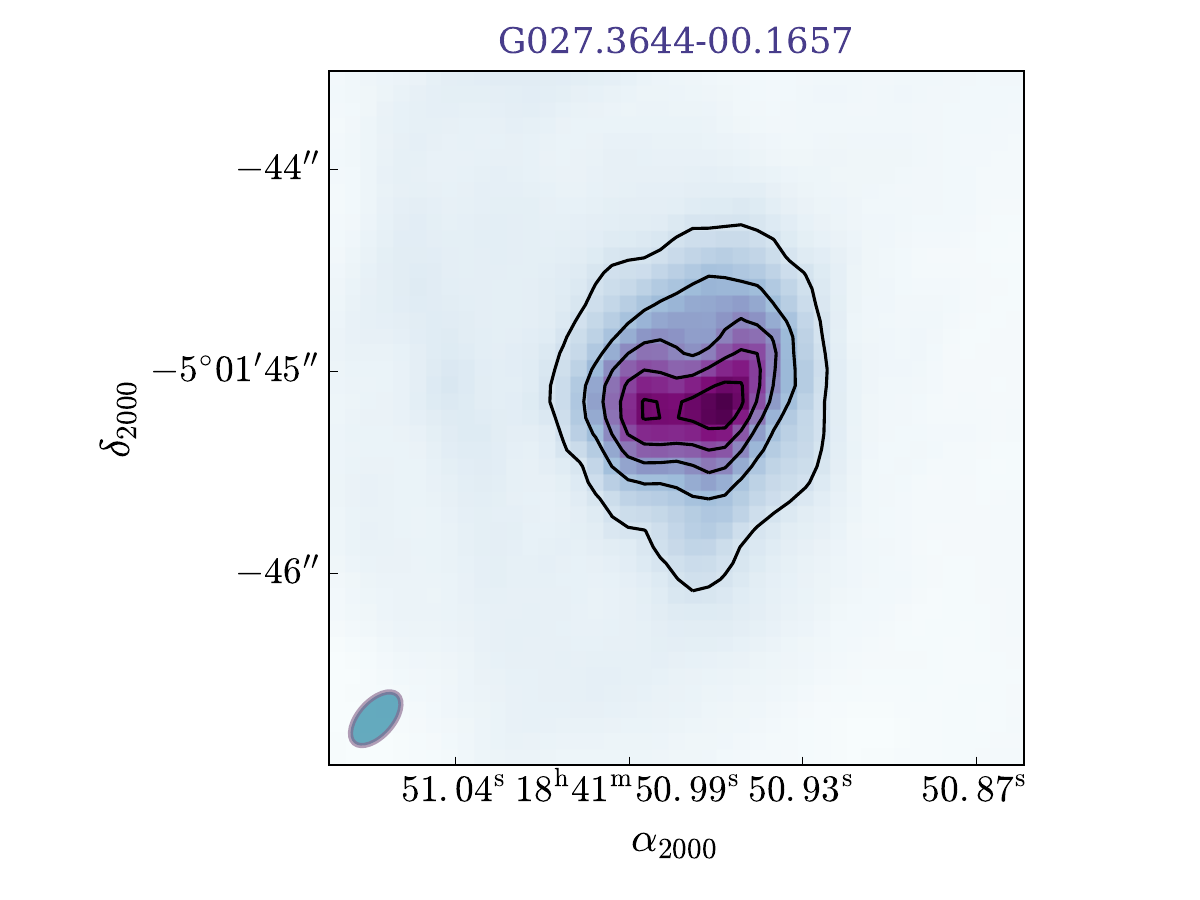} \hfill
\includegraphics[width=0.30\textwidth]{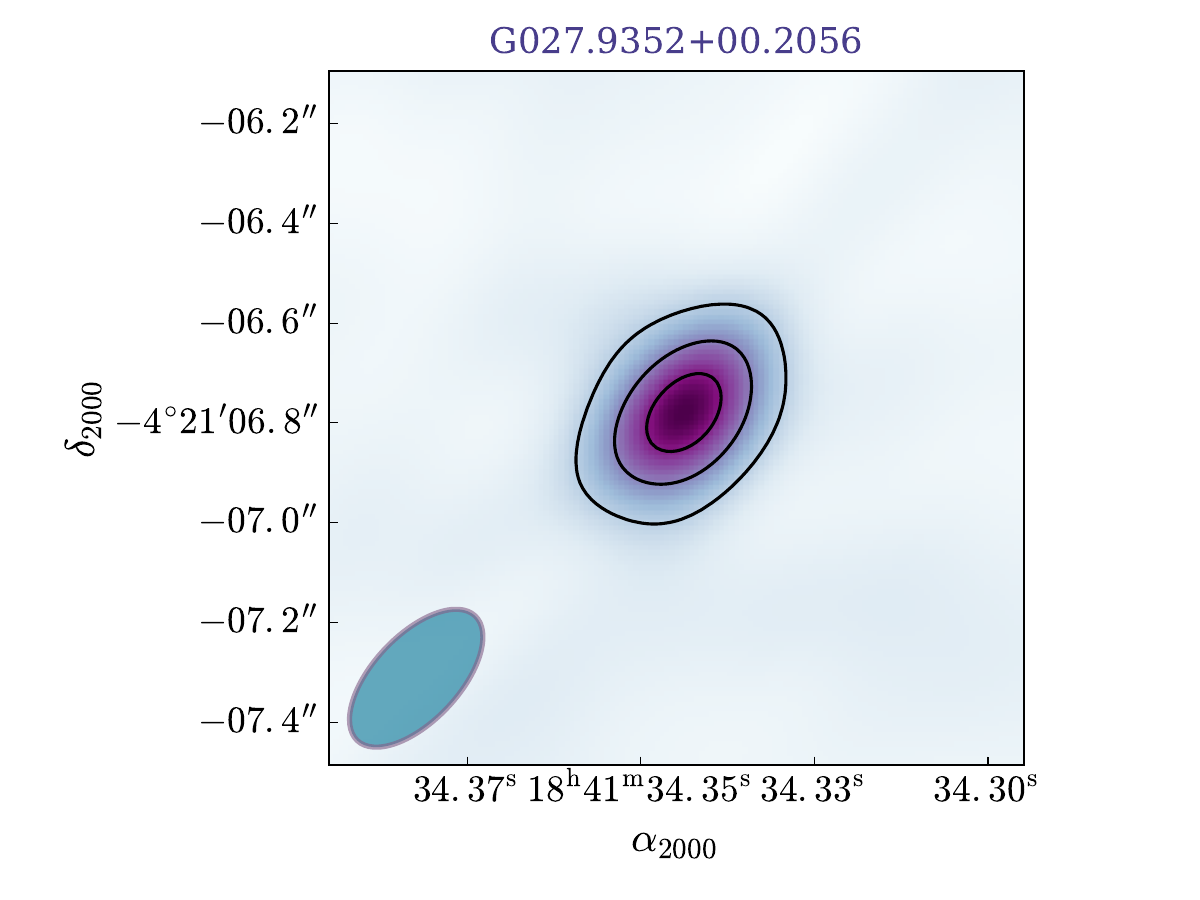}
\\[2mm]
\includegraphics[width=0.30\textwidth]{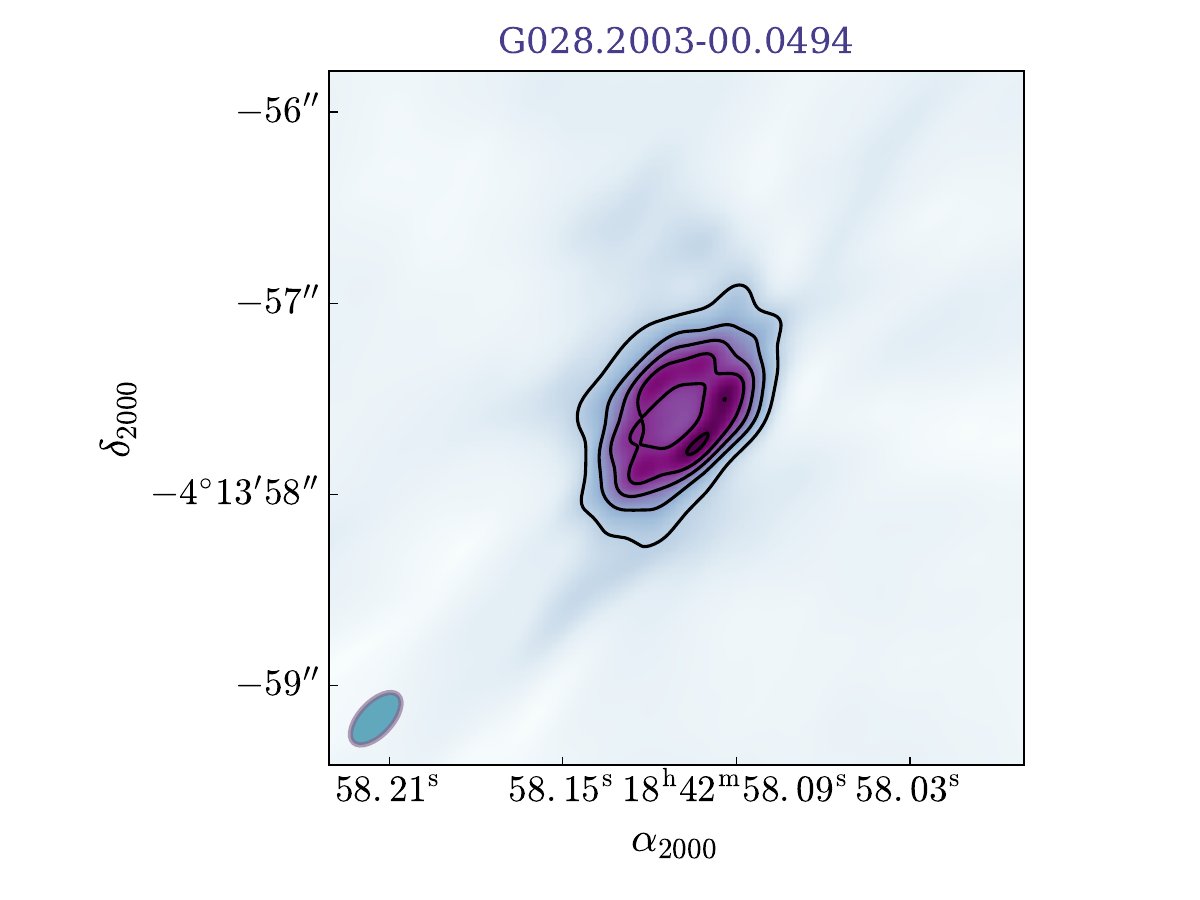} \hfill
\includegraphics[width=0.30\textwidth]{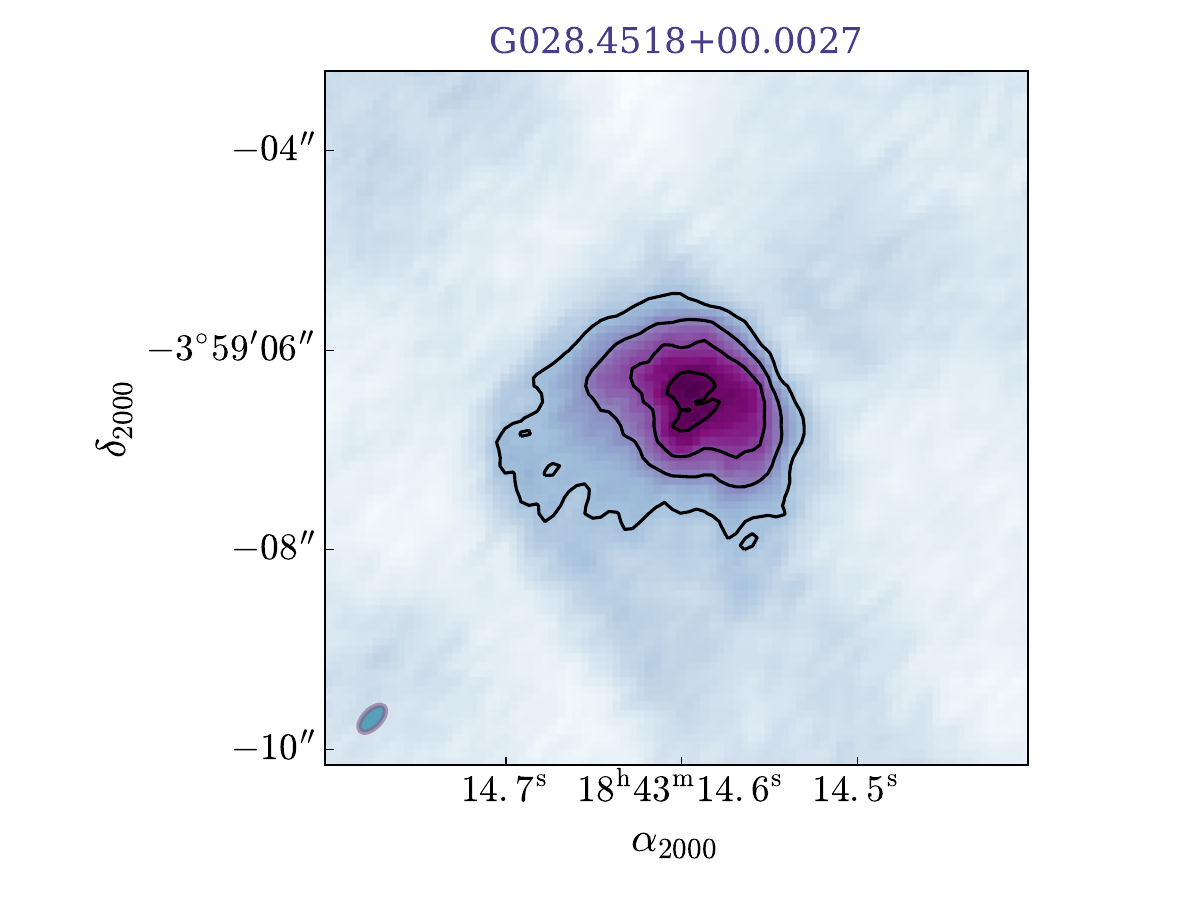} \hfill
\includegraphics[width=0.30\textwidth]{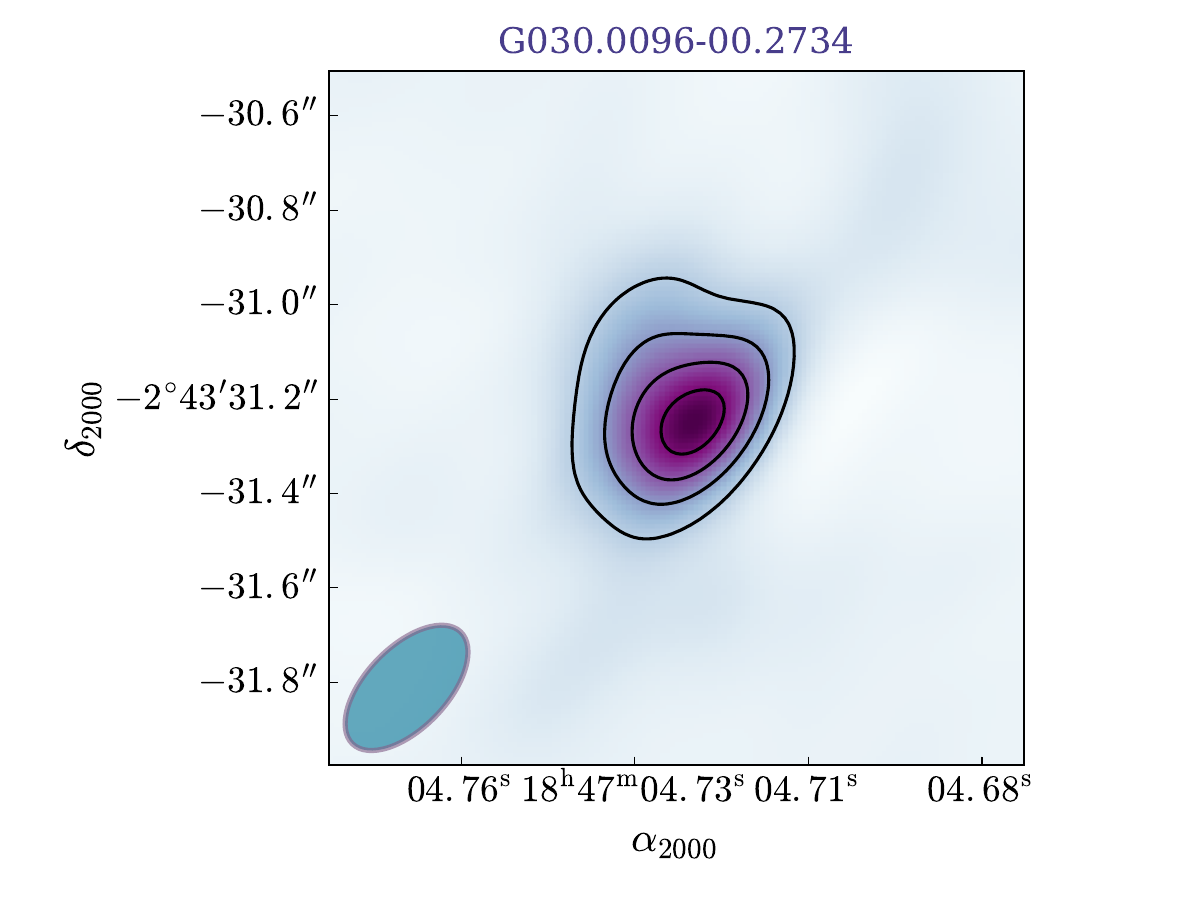}
\\[2mm]
\caption[Target maps, continued]{Target maps, continued.}
\end{figure*}
\clearpage

\begin{figure*}
\ContinuedFloat
\centering
\includegraphics[width=0.30\textwidth]{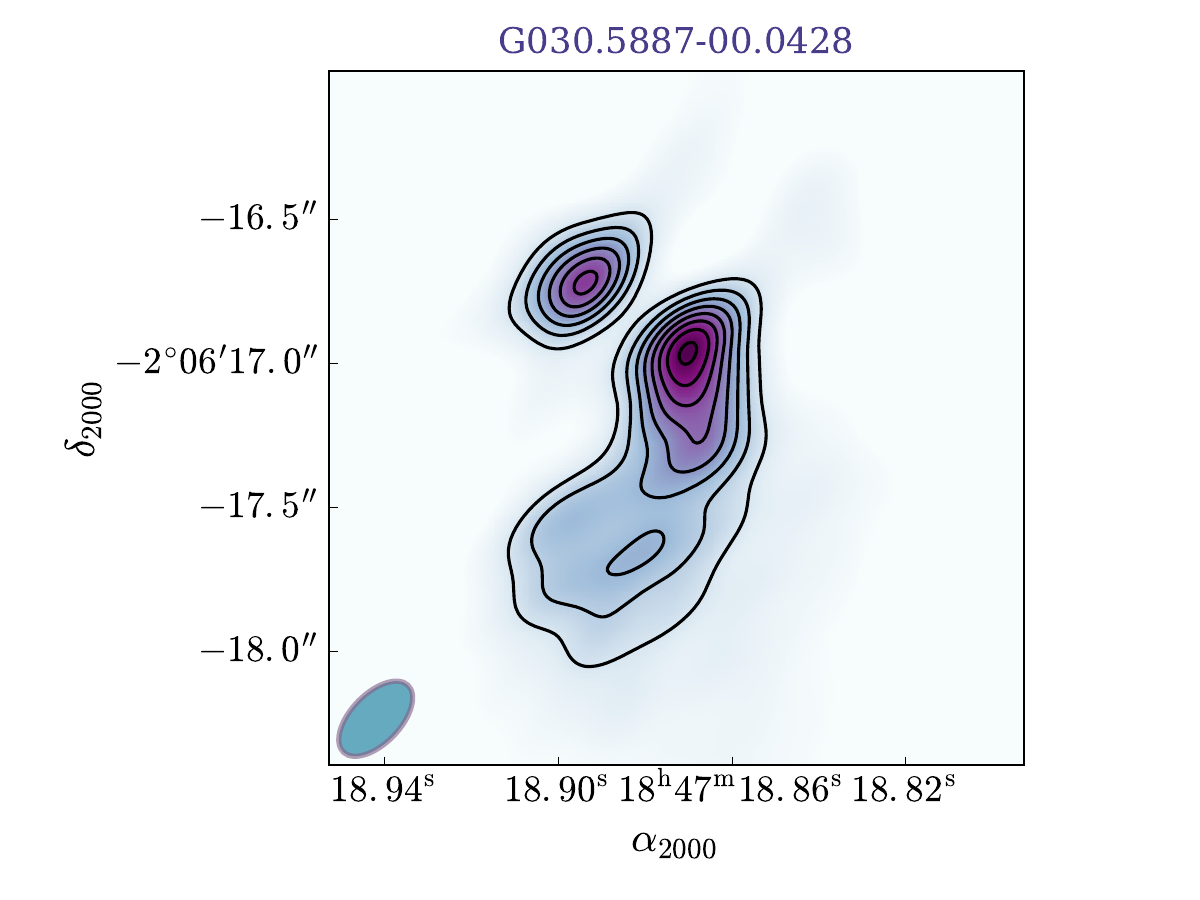} \hfill
\includegraphics[width=0.30\textwidth]{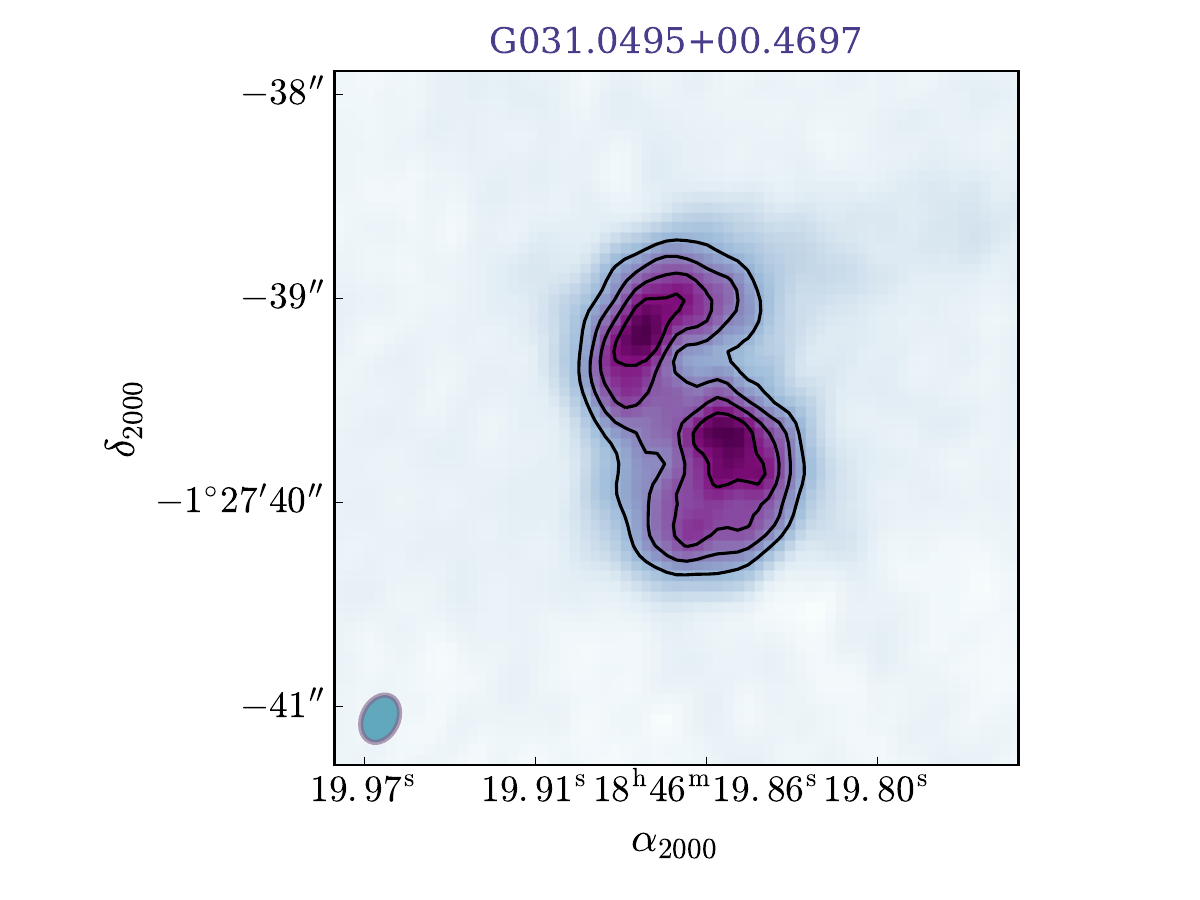} \hfill
\includegraphics[width=0.30\textwidth]{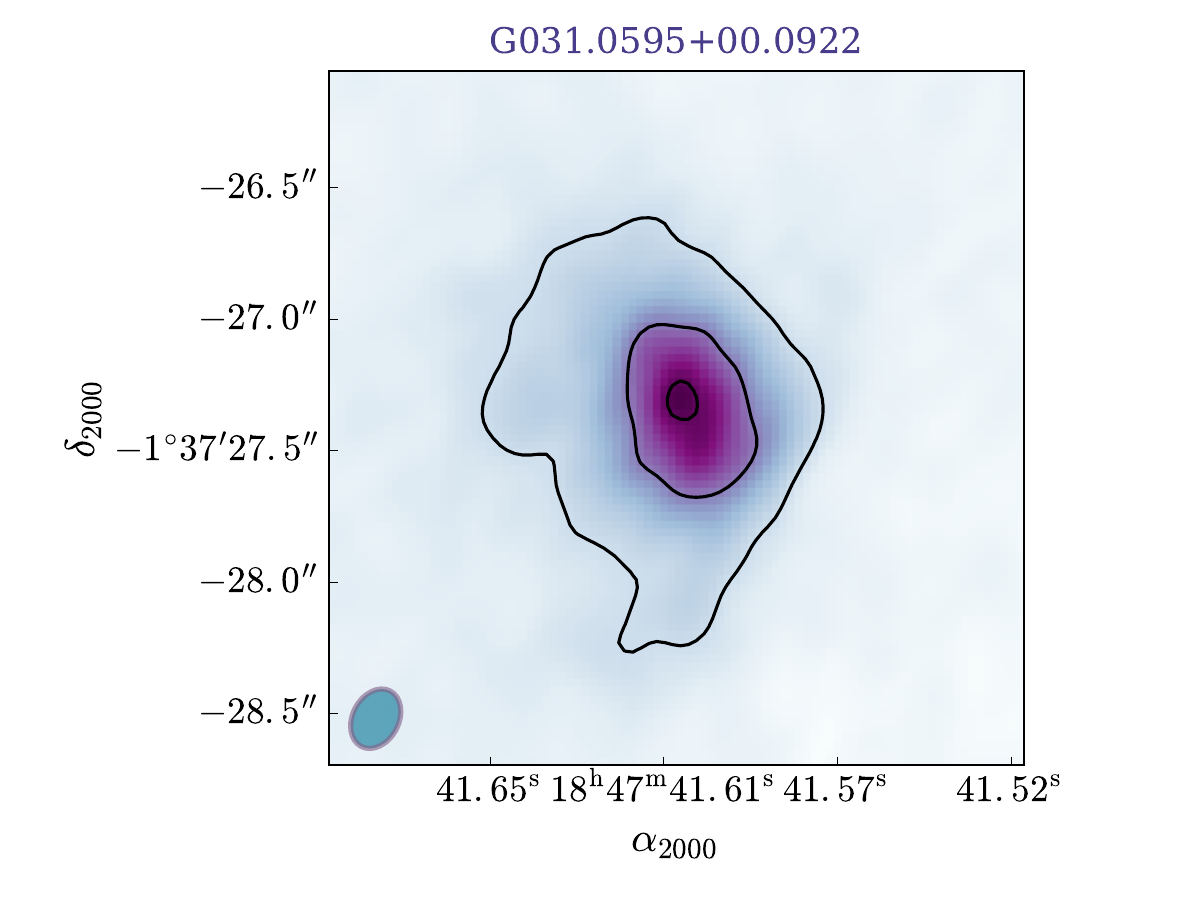}
\\[2mm]
\includegraphics[width=0.30\textwidth]{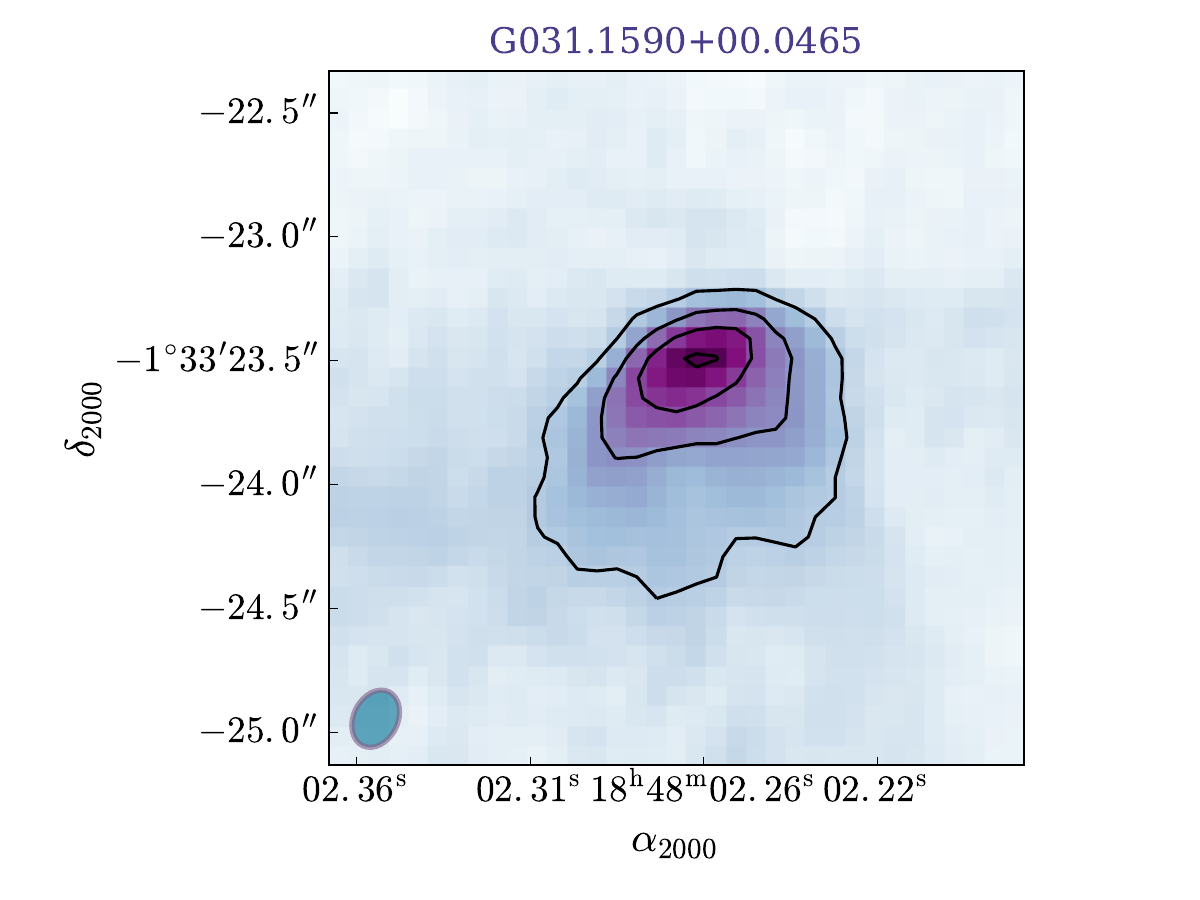} \hfill
\includegraphics[width=0.30\textwidth]{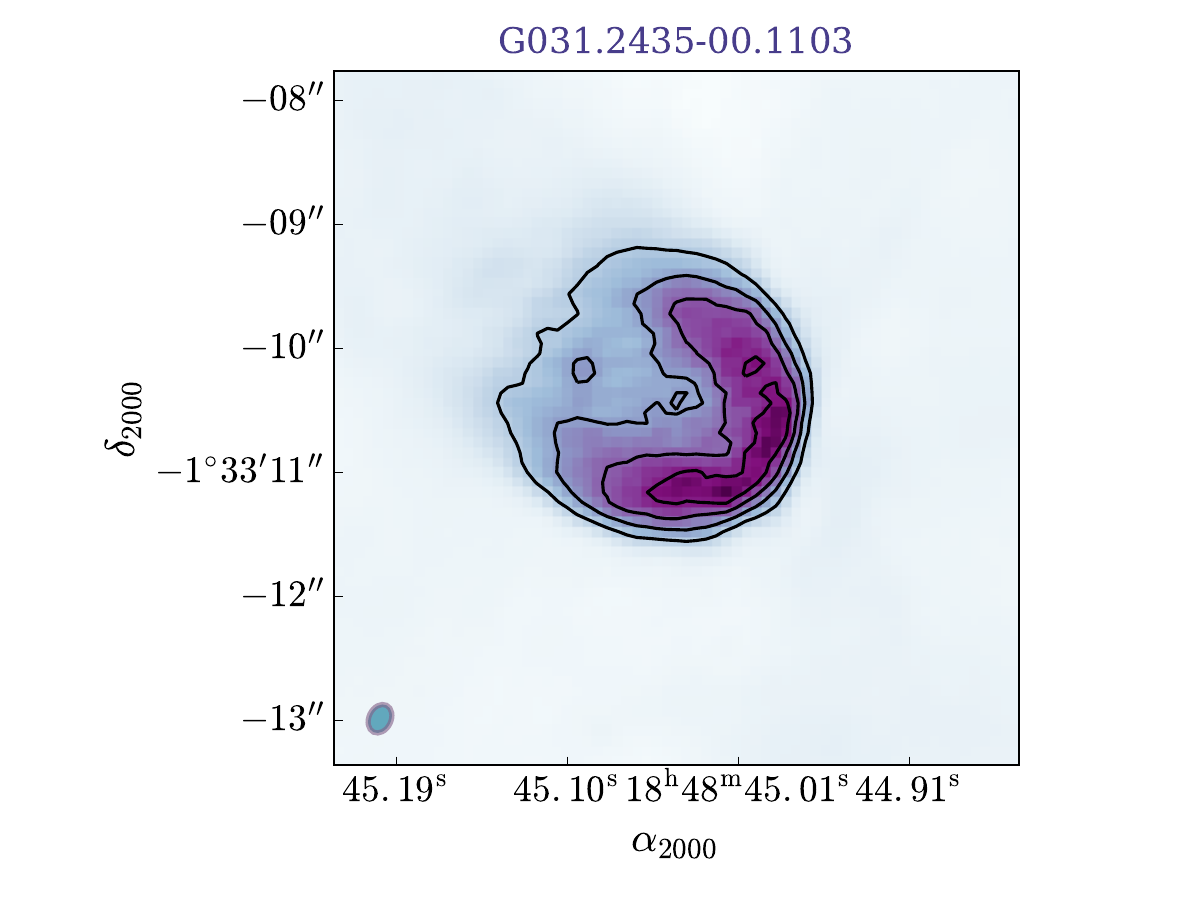} \hfill
\includegraphics[width=0.30\textwidth]{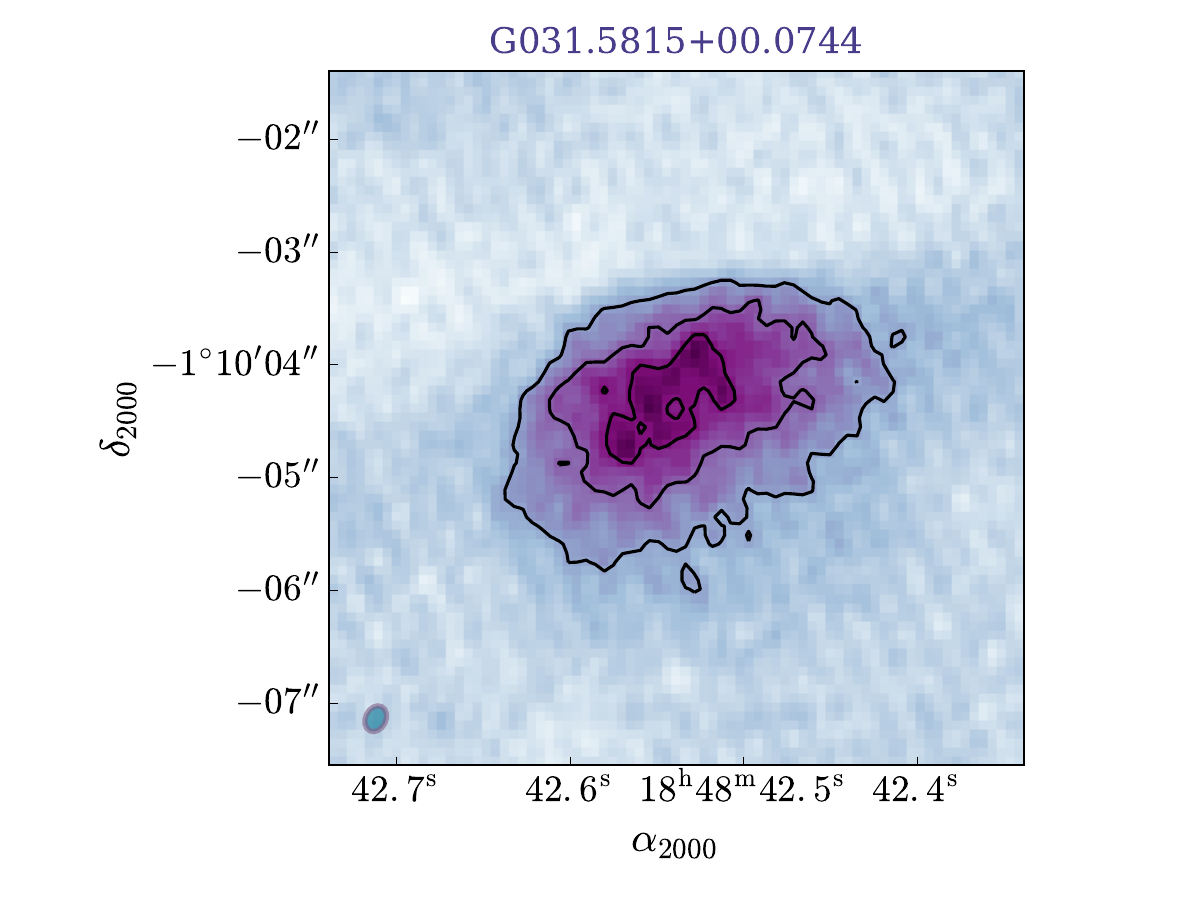}
\\[2mm]
\includegraphics[width=0.30\textwidth]{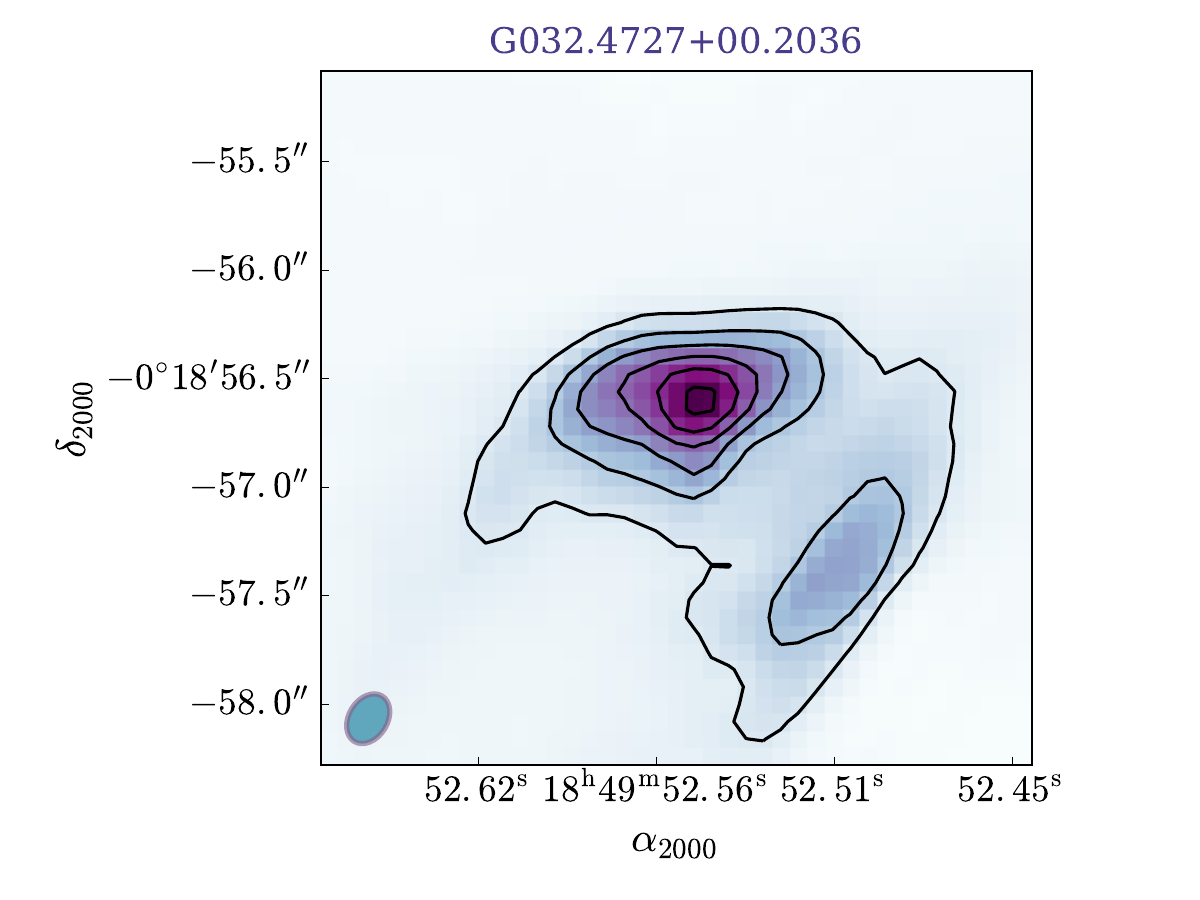} \hfill
\includegraphics[width=0.30\textwidth]{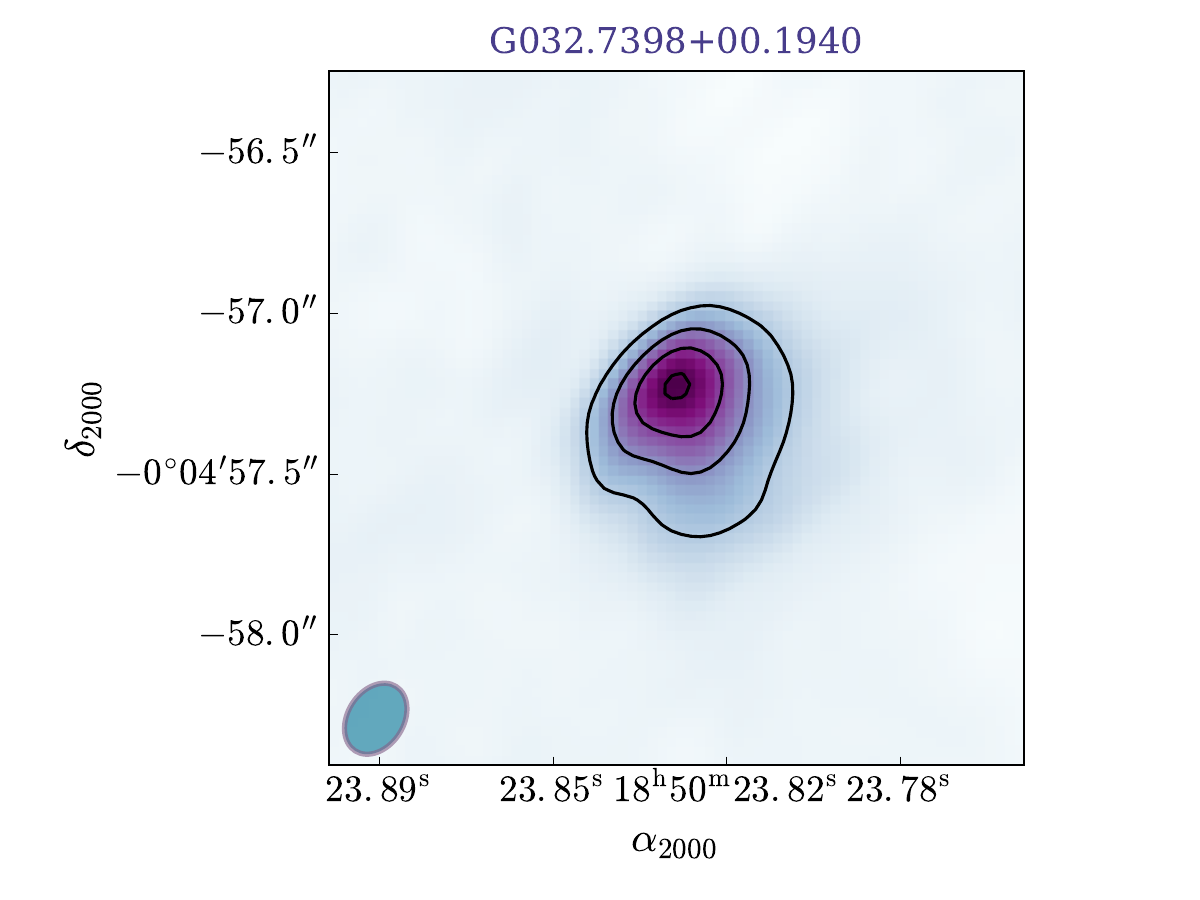} \hfill
\includegraphics[width=0.30\textwidth]{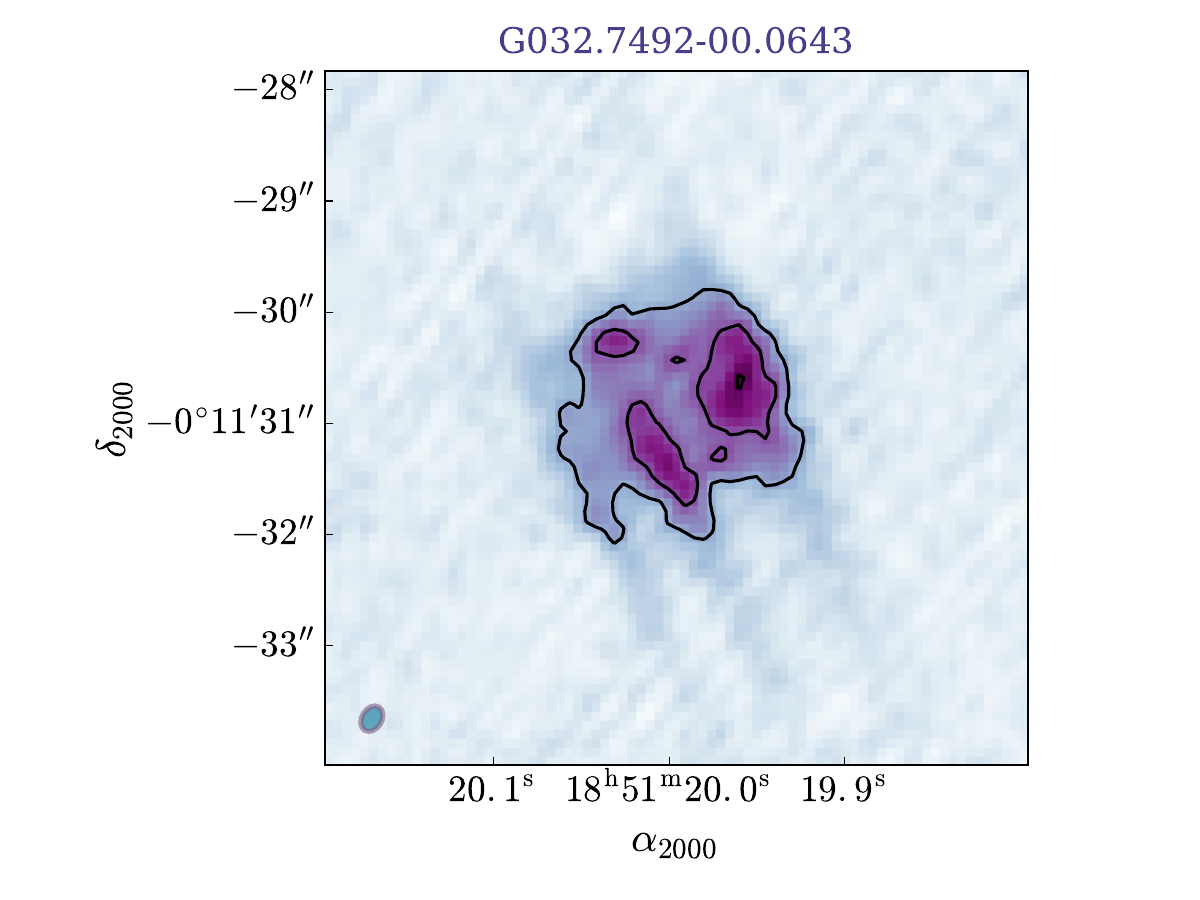}
\\[2mm]
\includegraphics[width=0.30\textwidth]{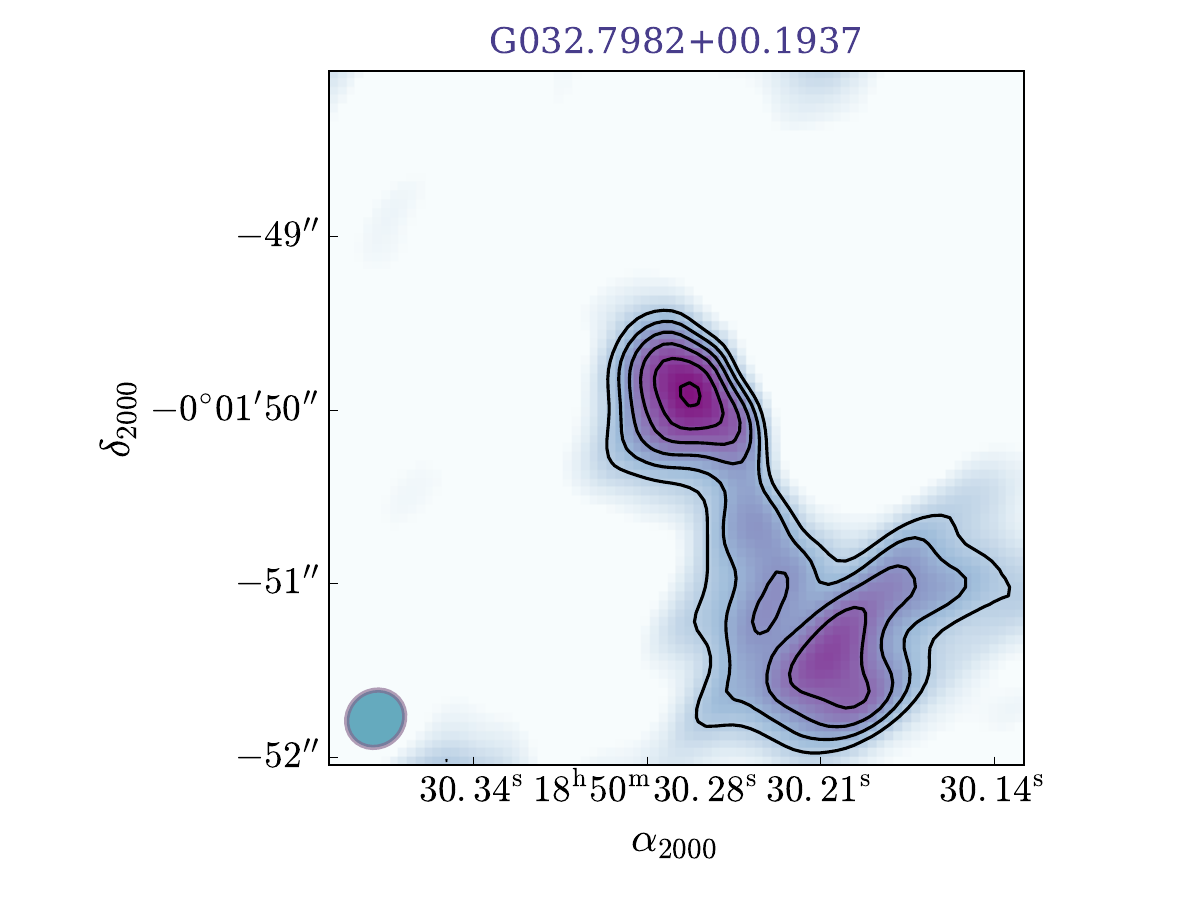} \hfill
\includegraphics[width=0.30\textwidth]{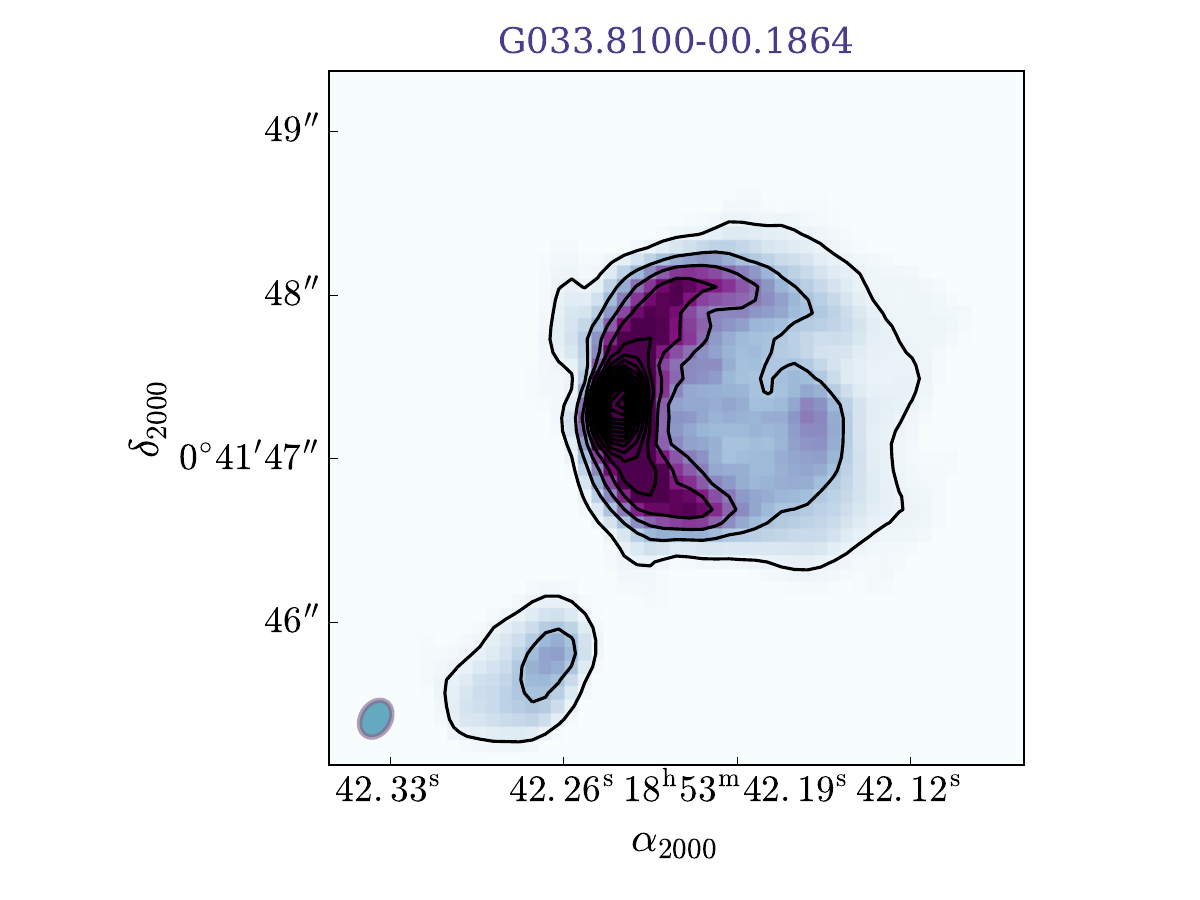} \hfill
\includegraphics[width=0.30\textwidth]{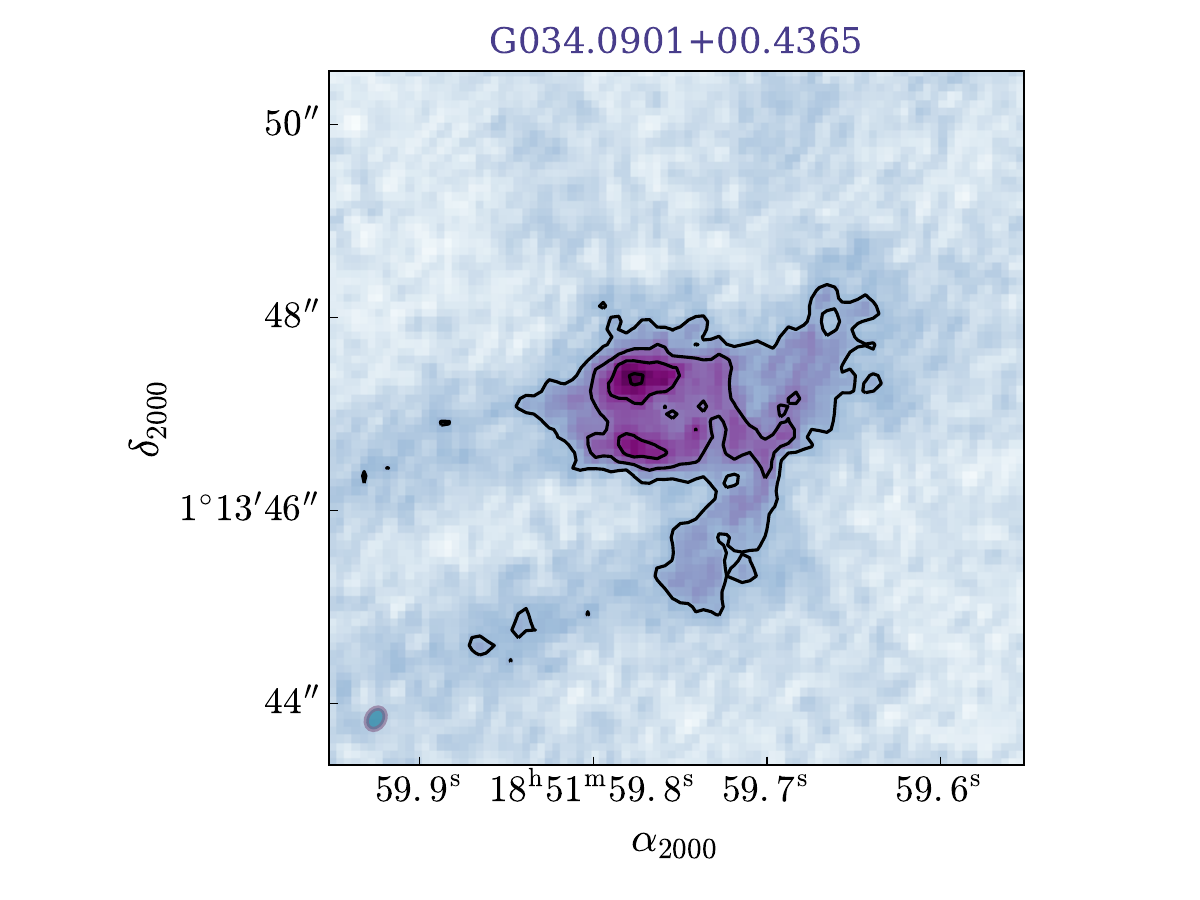}
\\[2mm]
\includegraphics[width=0.30\textwidth]{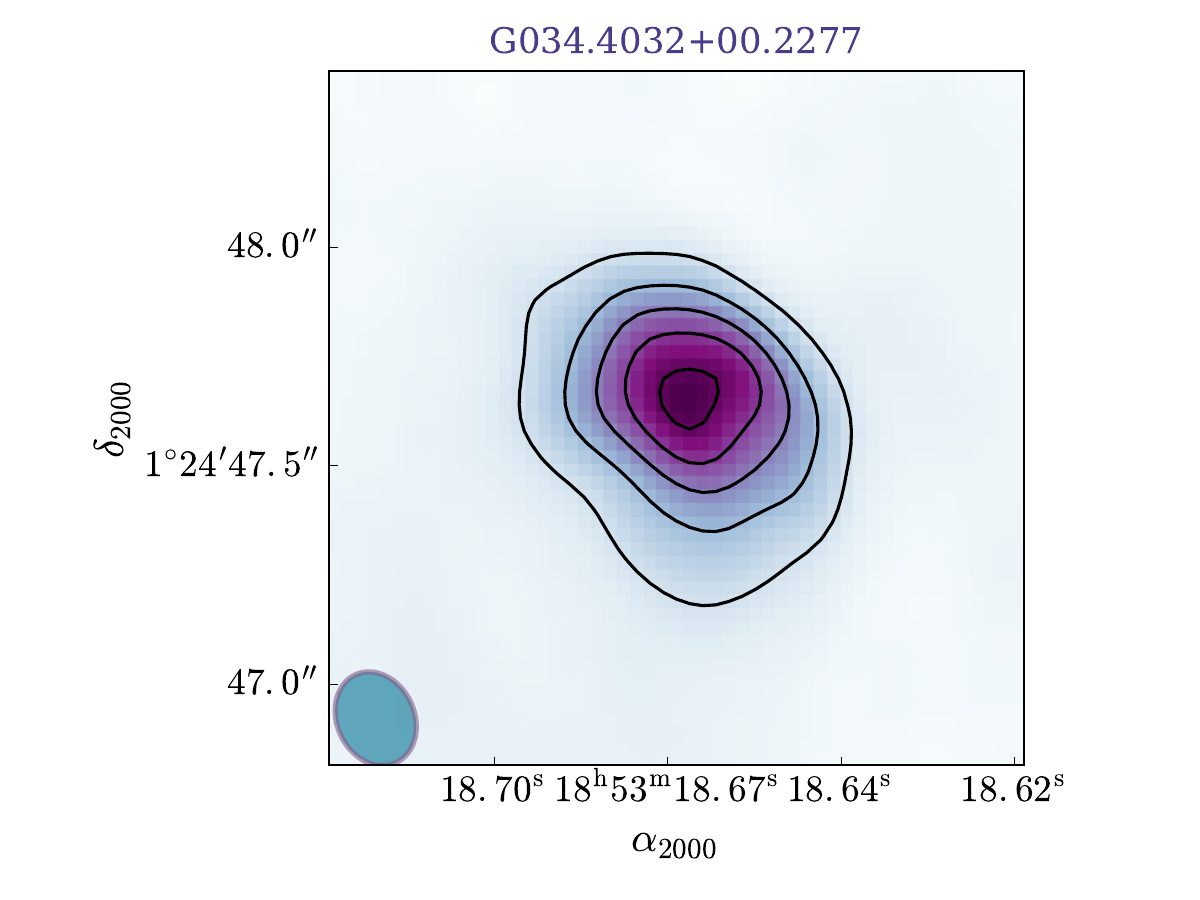} \hfill
\includegraphics[width=0.30\textwidth]{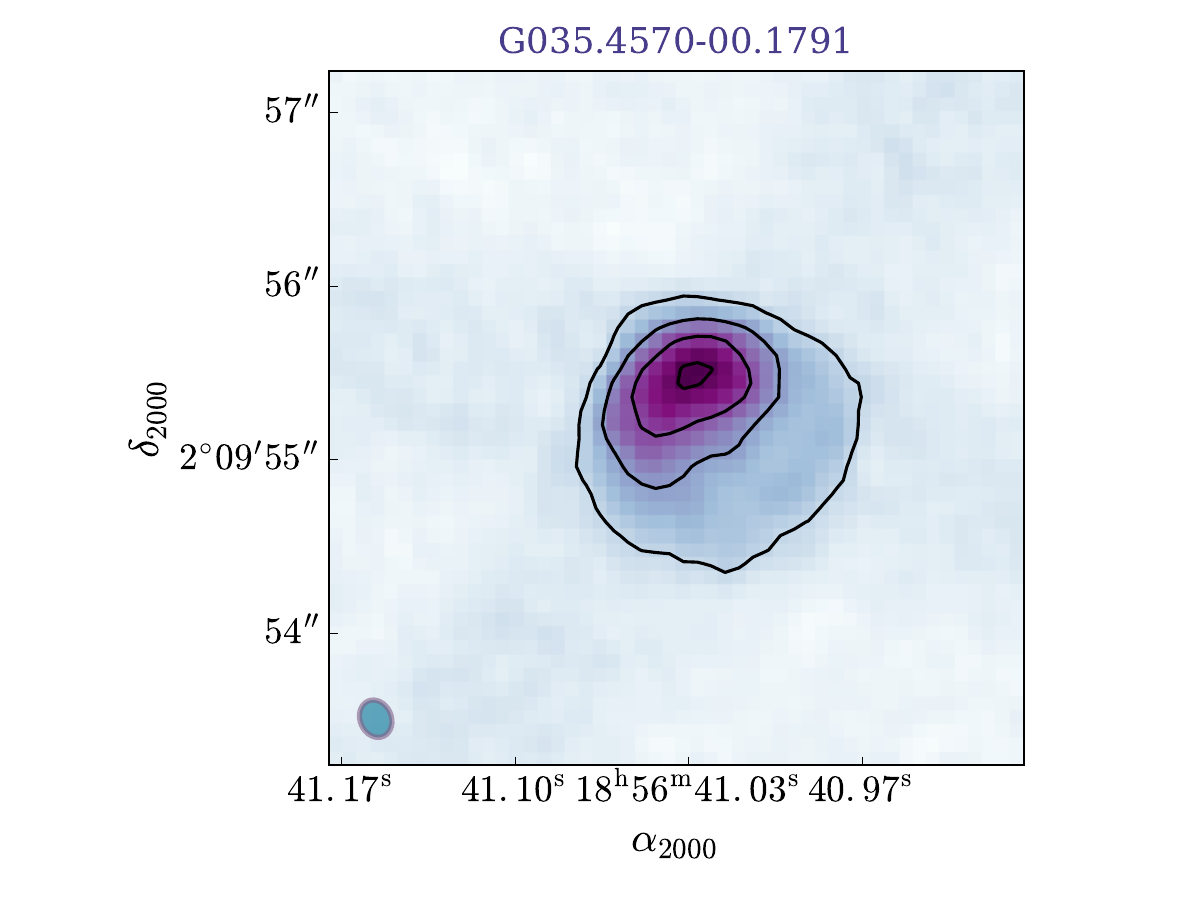} \hfill
\includegraphics[width=0.30\textwidth]{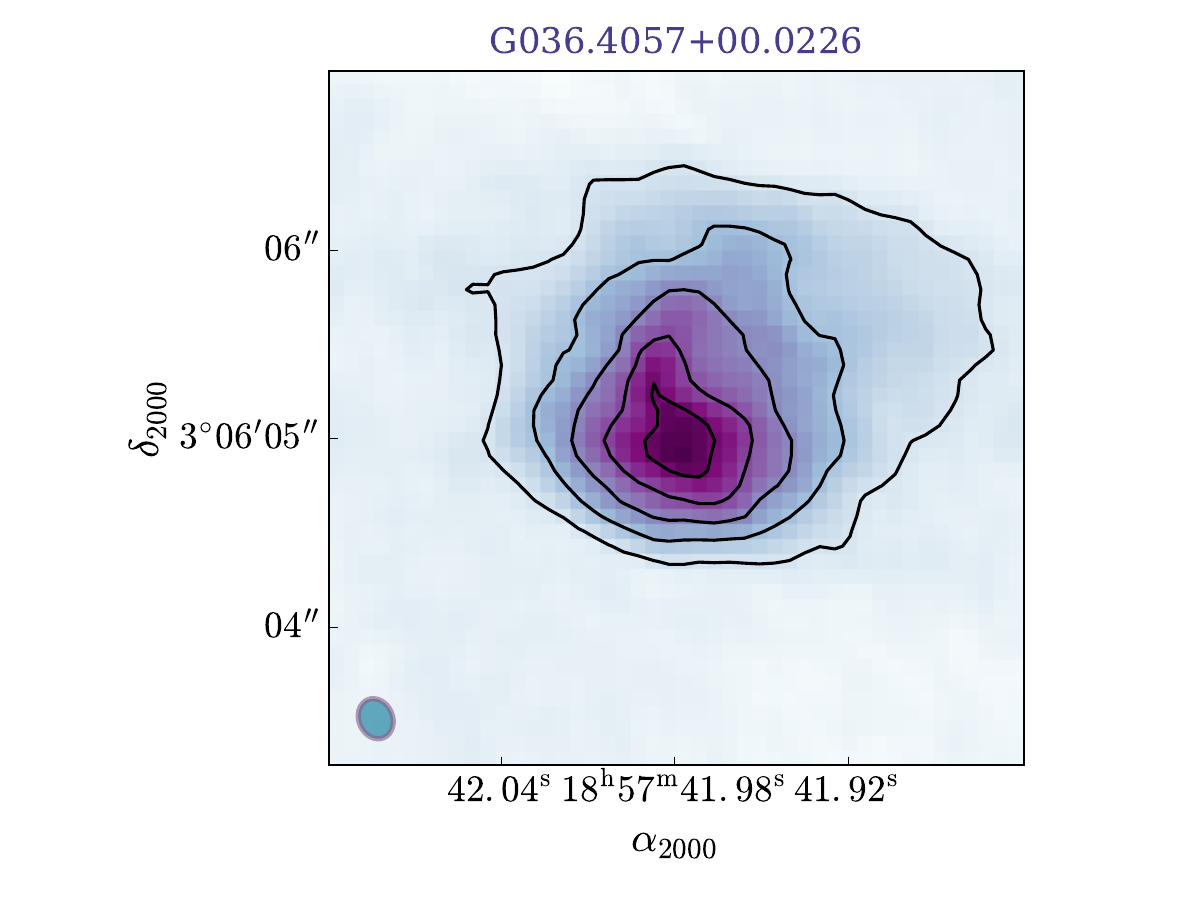}
\\[2mm]
\caption[Target maps, continued]{Target maps, continued.}
\end{figure*}
\clearpage

\begin{figure*}
\ContinuedFloat
\centering
\includegraphics[width=0.30\textwidth]{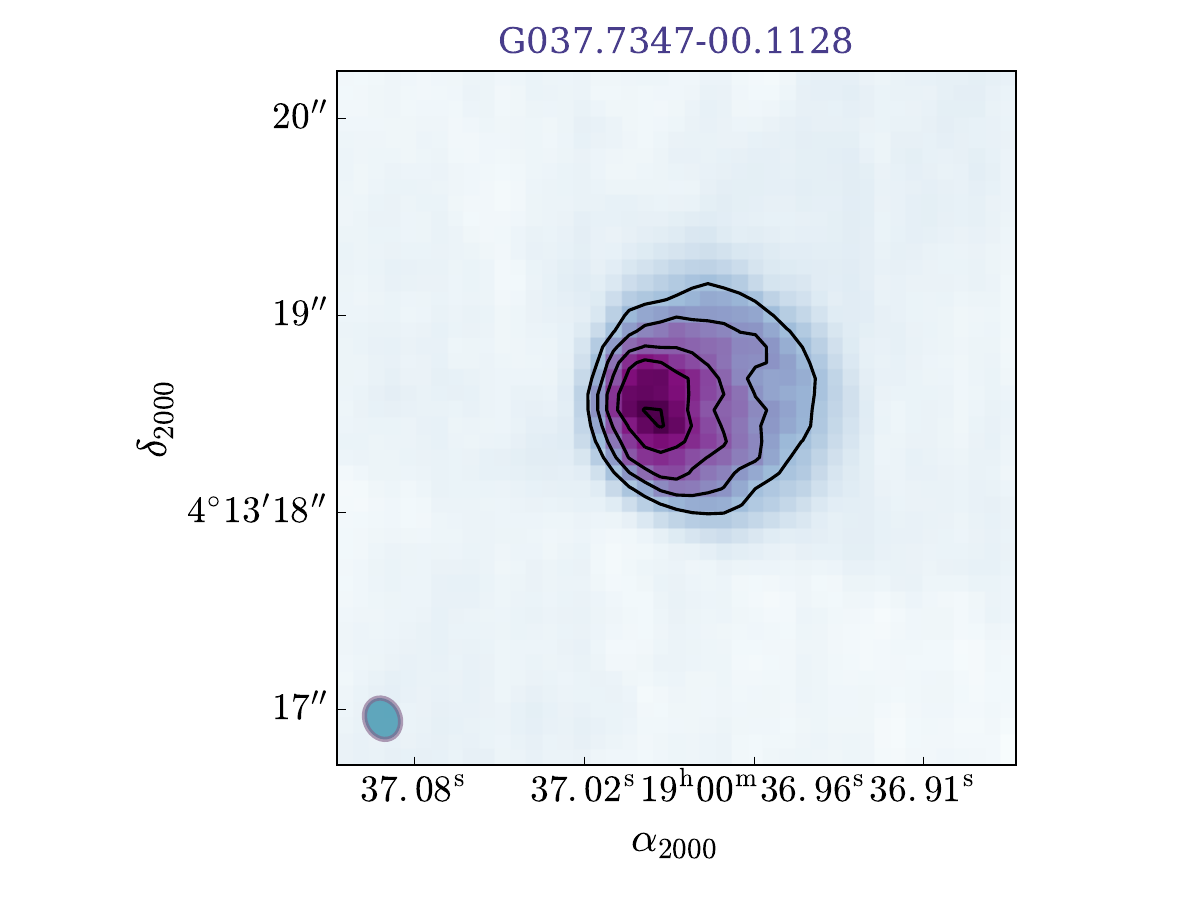} \hfill
\includegraphics[width=0.30\textwidth]{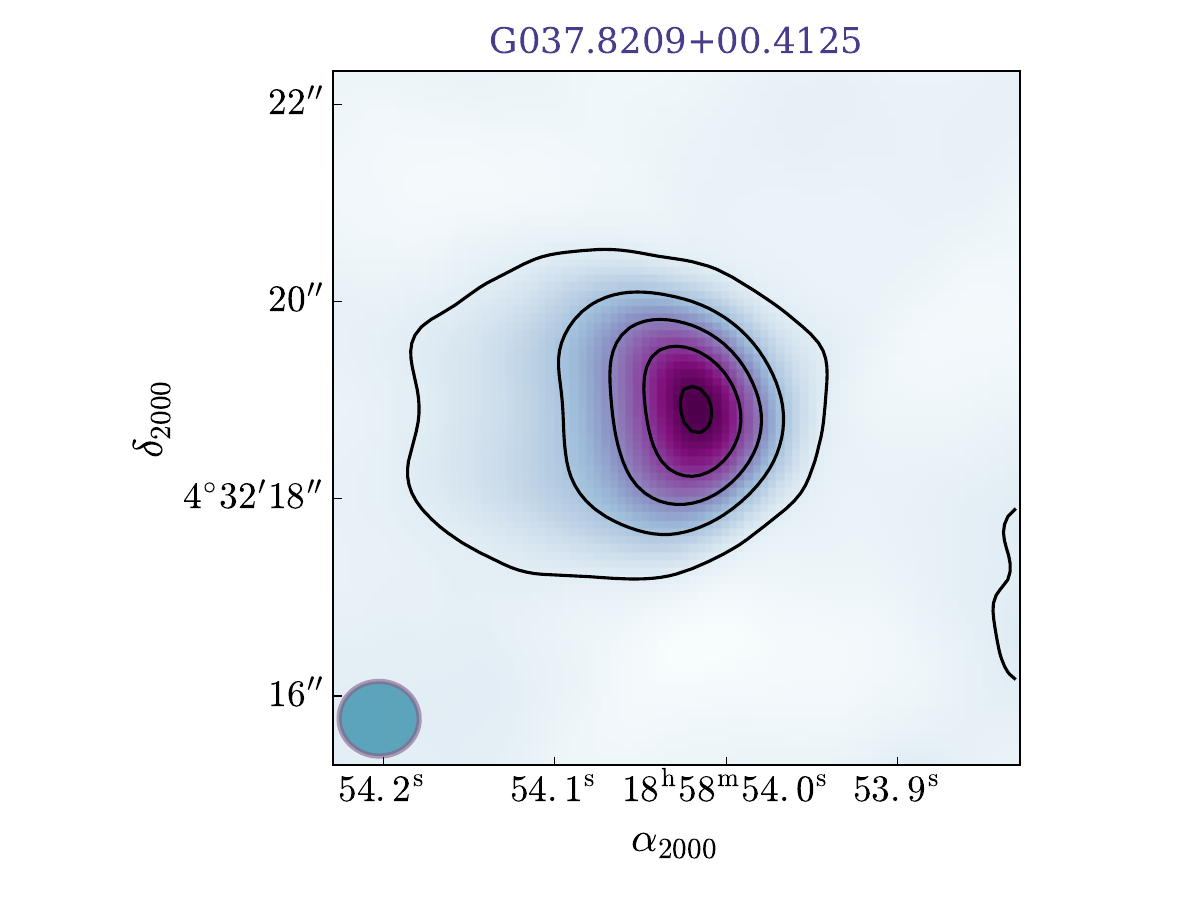} \hfill
\includegraphics[width=0.30\textwidth]{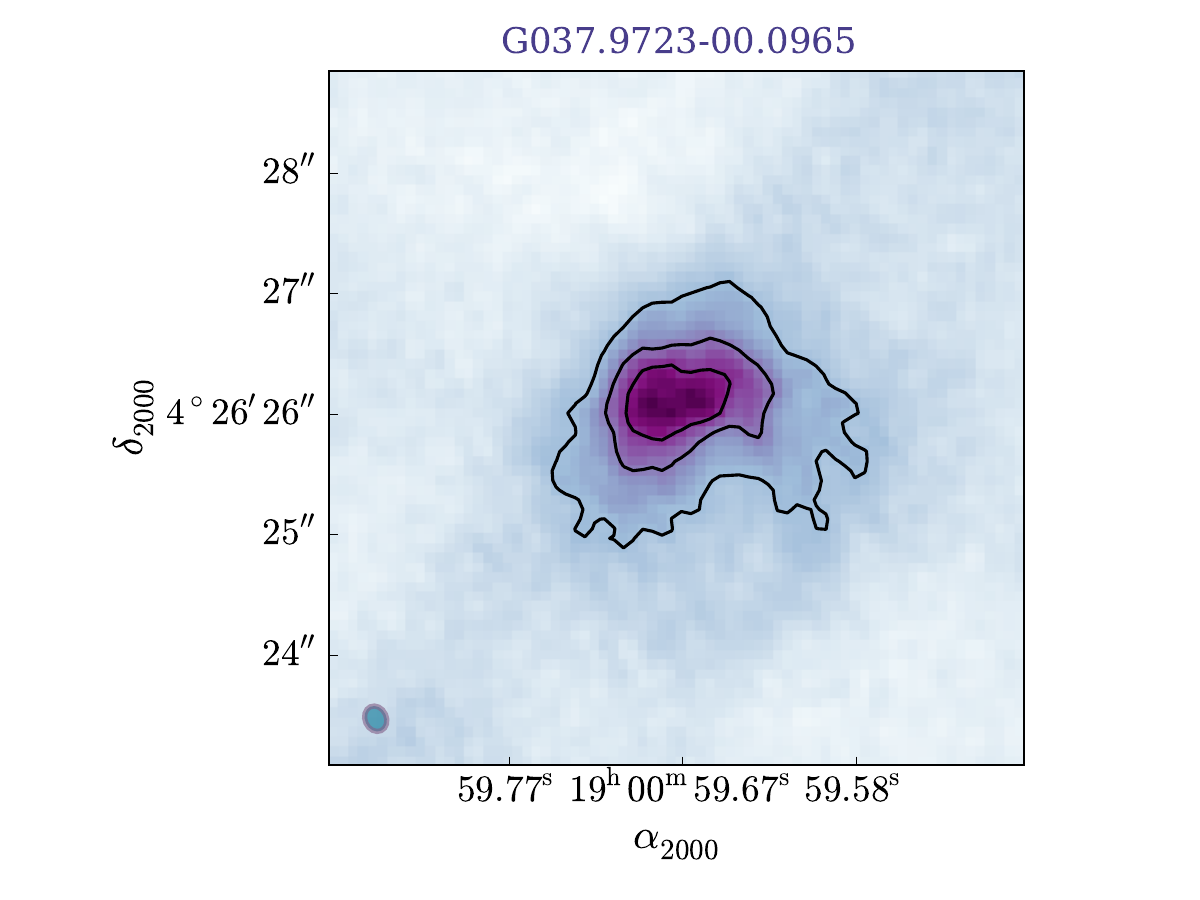}
\\[2mm]
\includegraphics[width=0.30\textwidth]{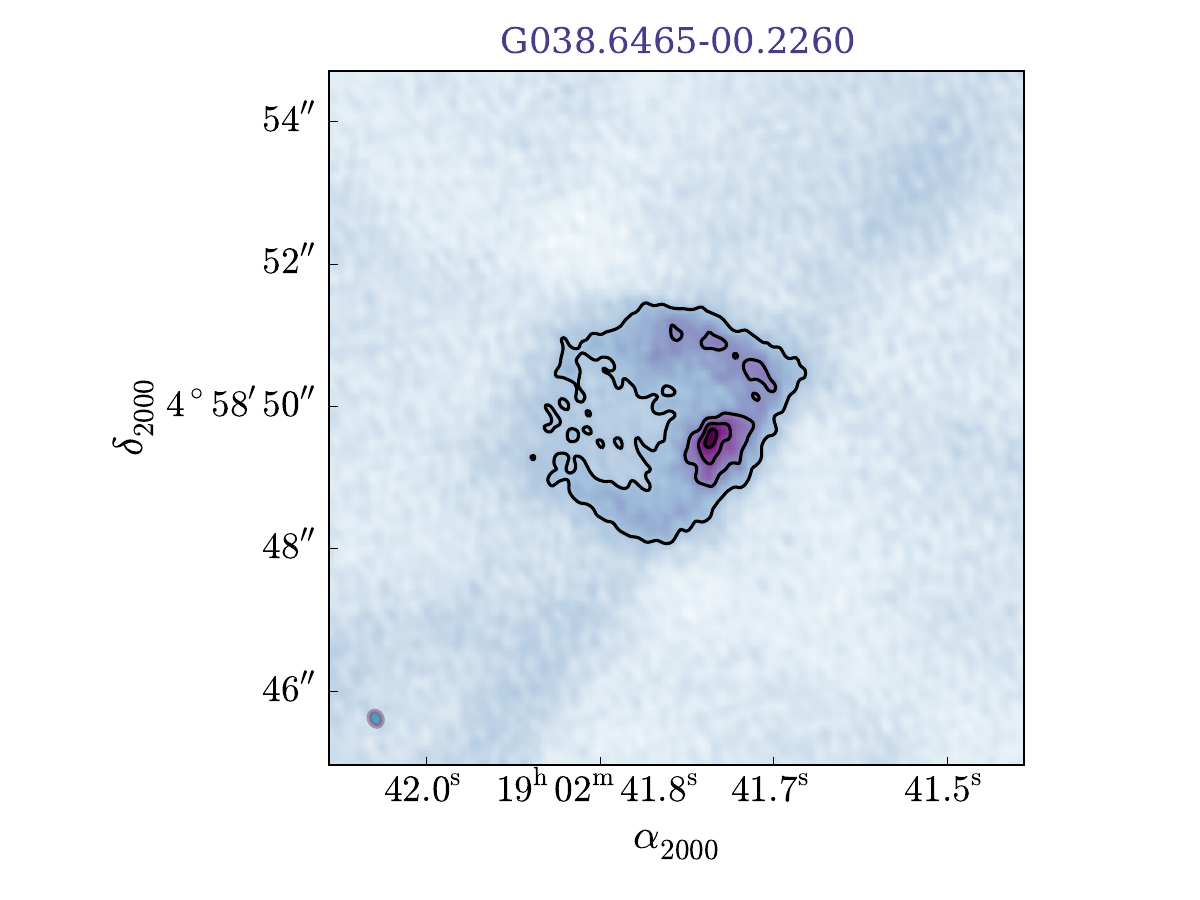} \hfill
\includegraphics[width=0.30\textwidth]{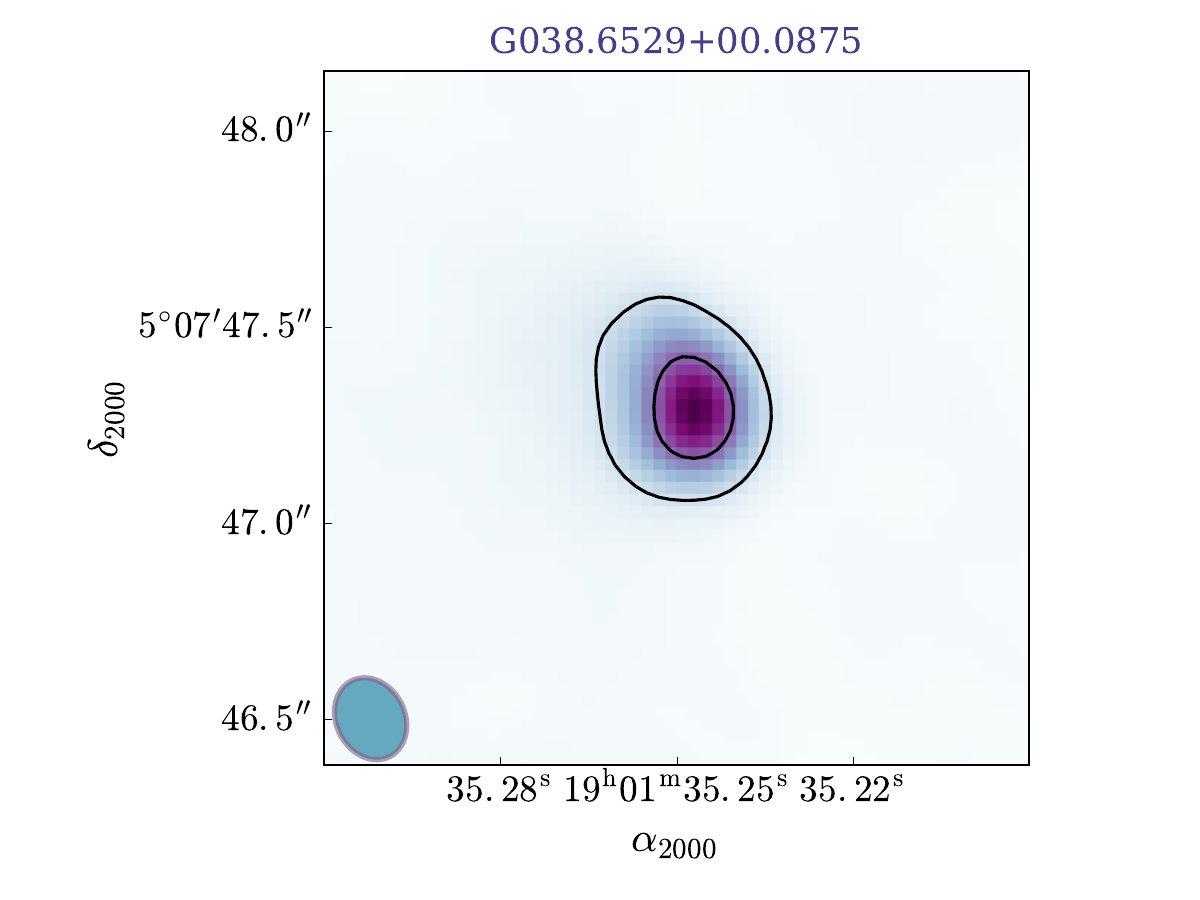} \hfill
\includegraphics[width=0.30\textwidth]{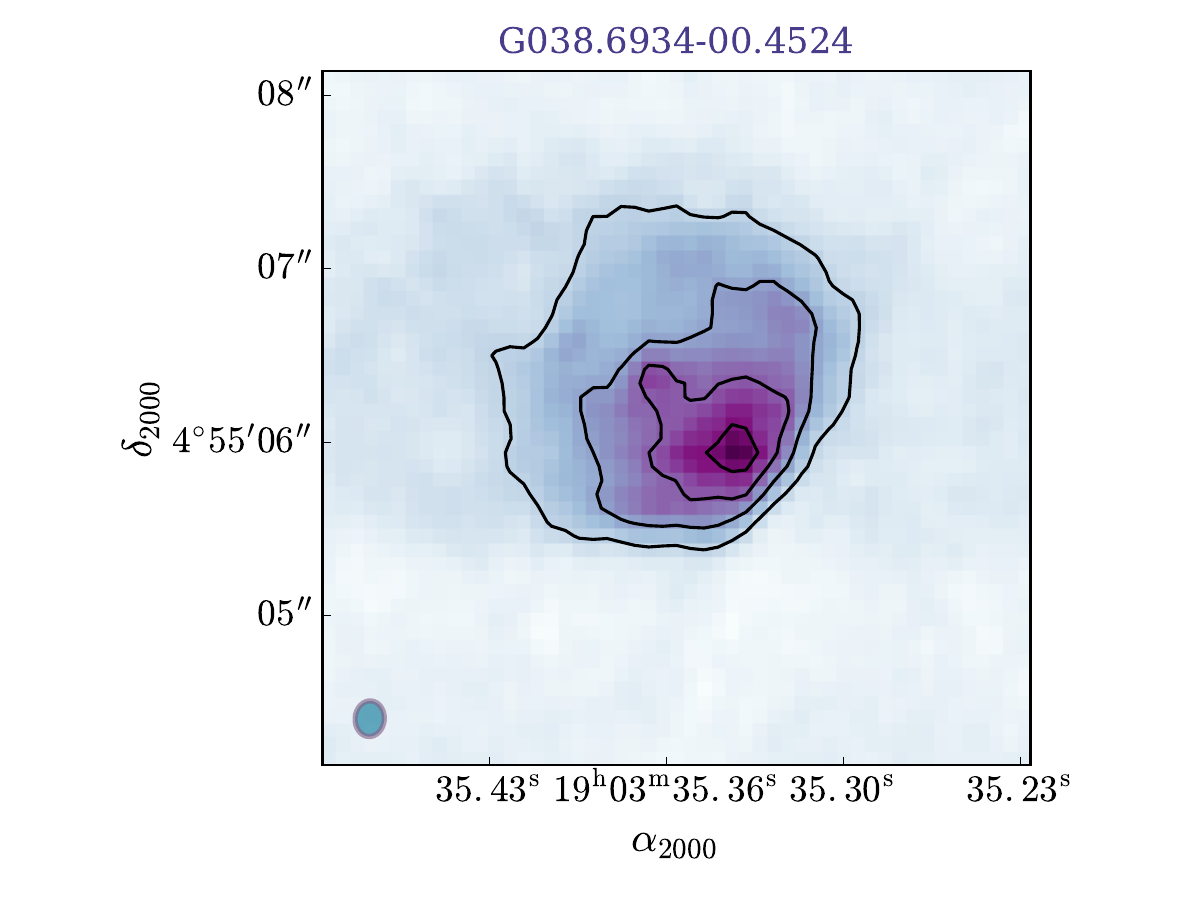}
\\[2mm]
\includegraphics[width=0.30\textwidth]{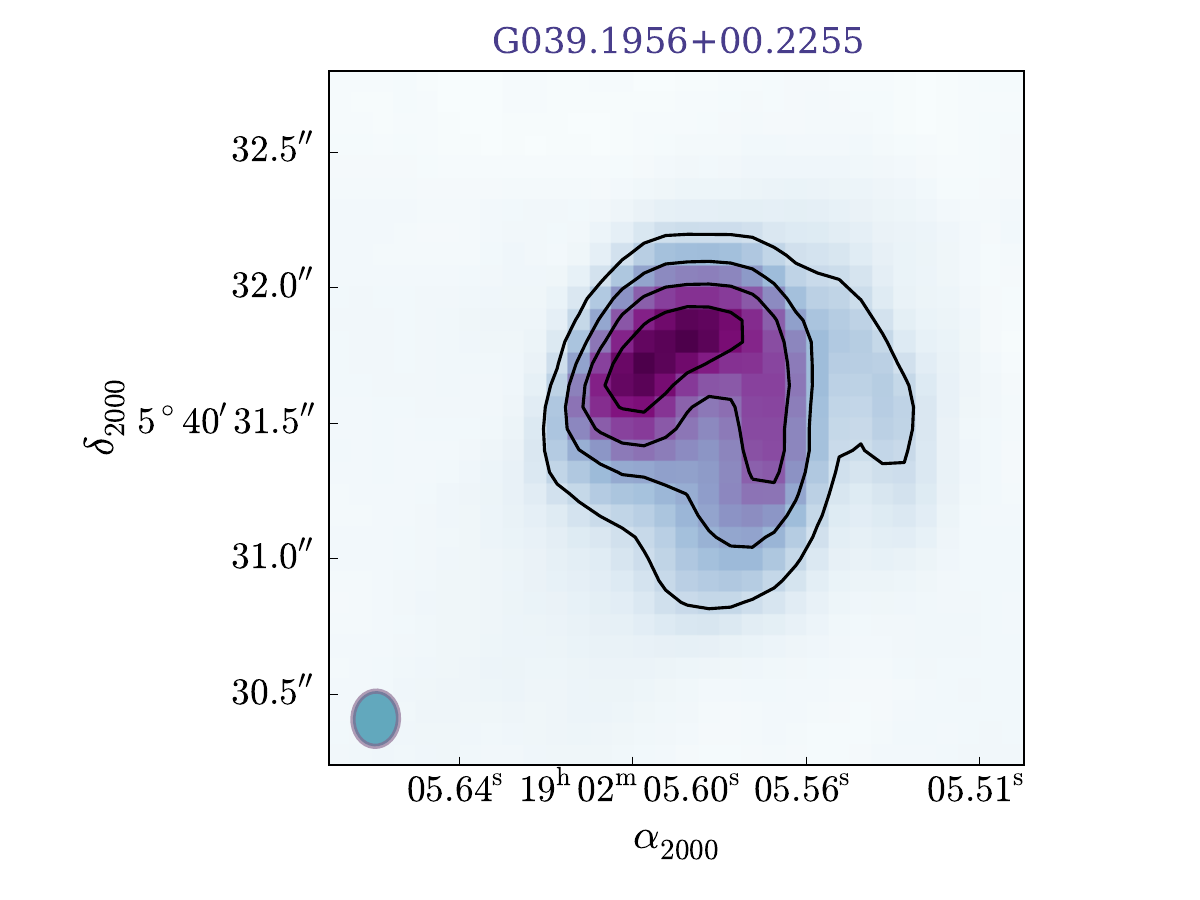} \hfill
\includegraphics[width=0.30\textwidth]{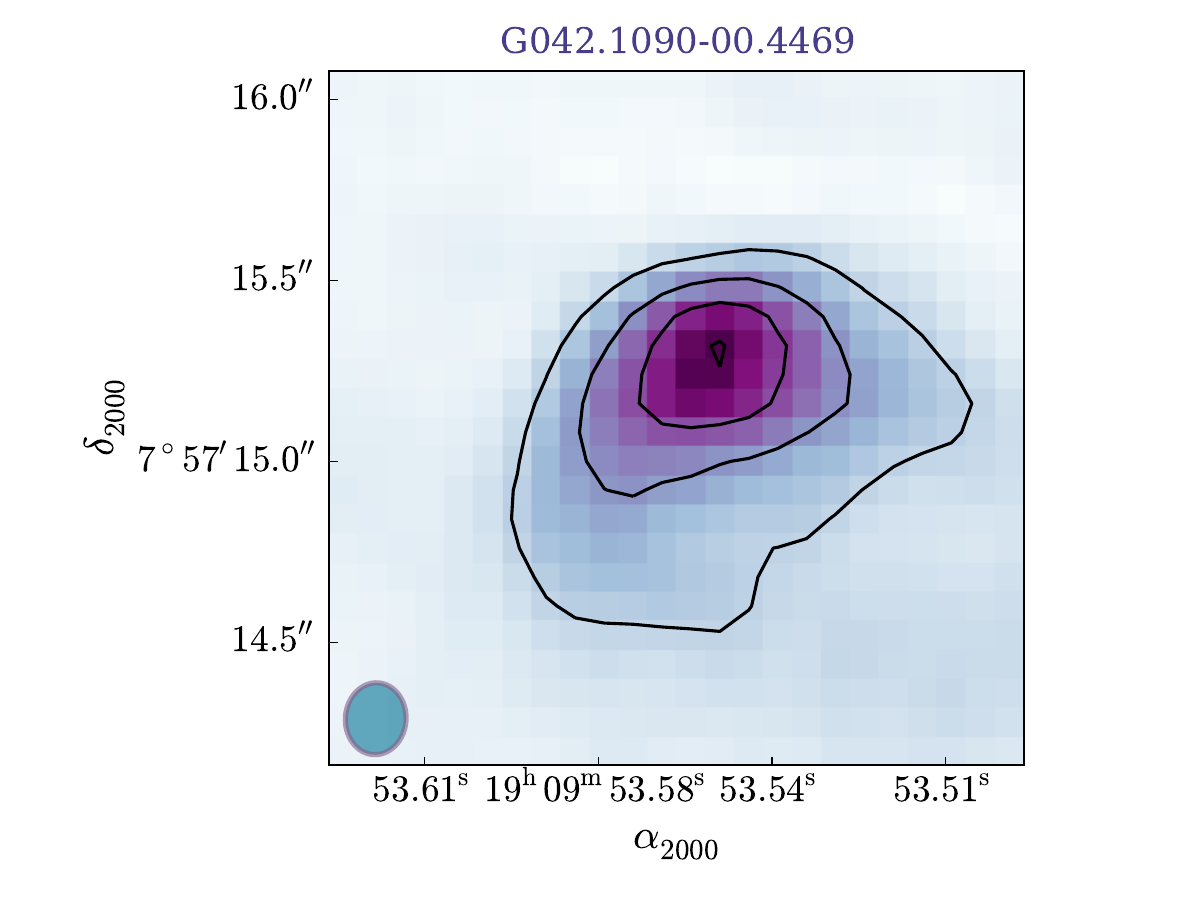} \hfill
\includegraphics[width=0.30\textwidth]{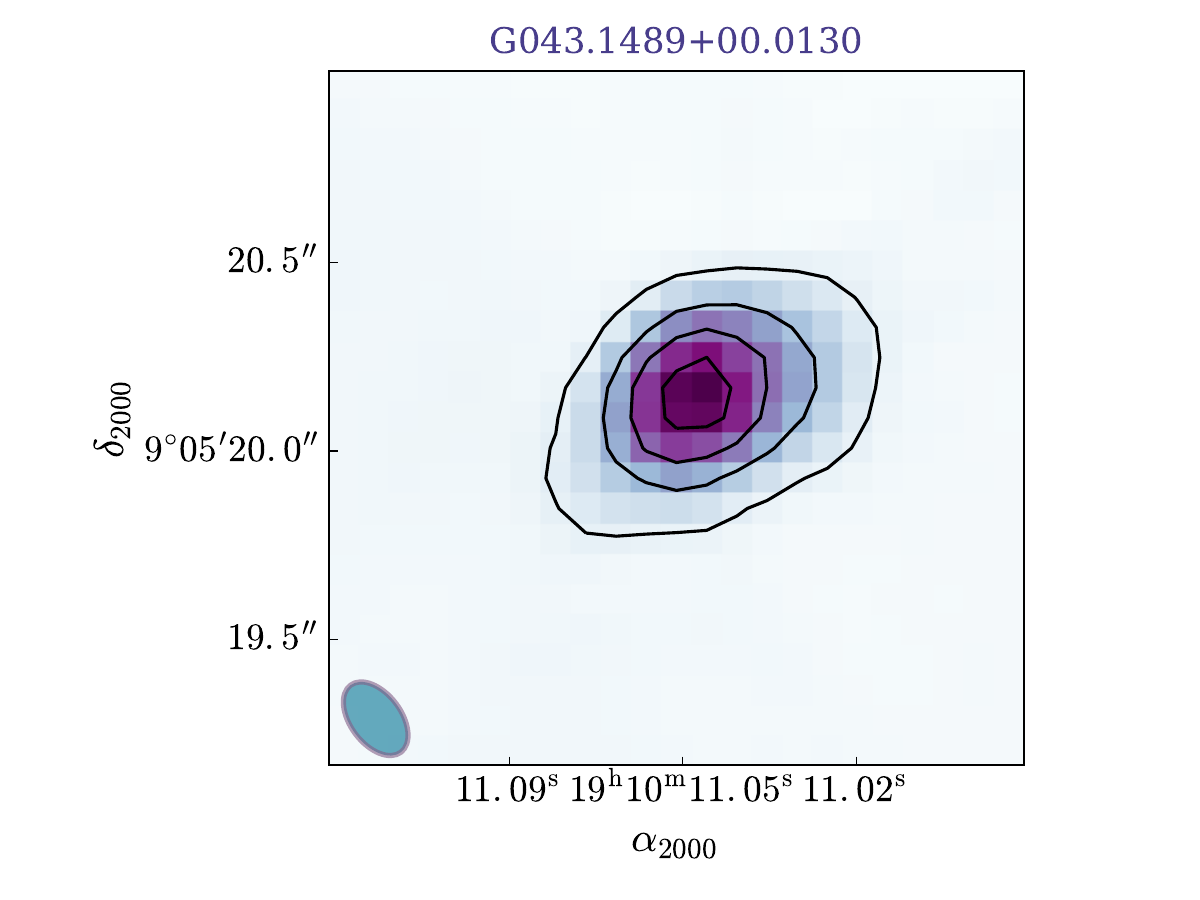}
\\[2mm]
\includegraphics[width=0.30\textwidth]{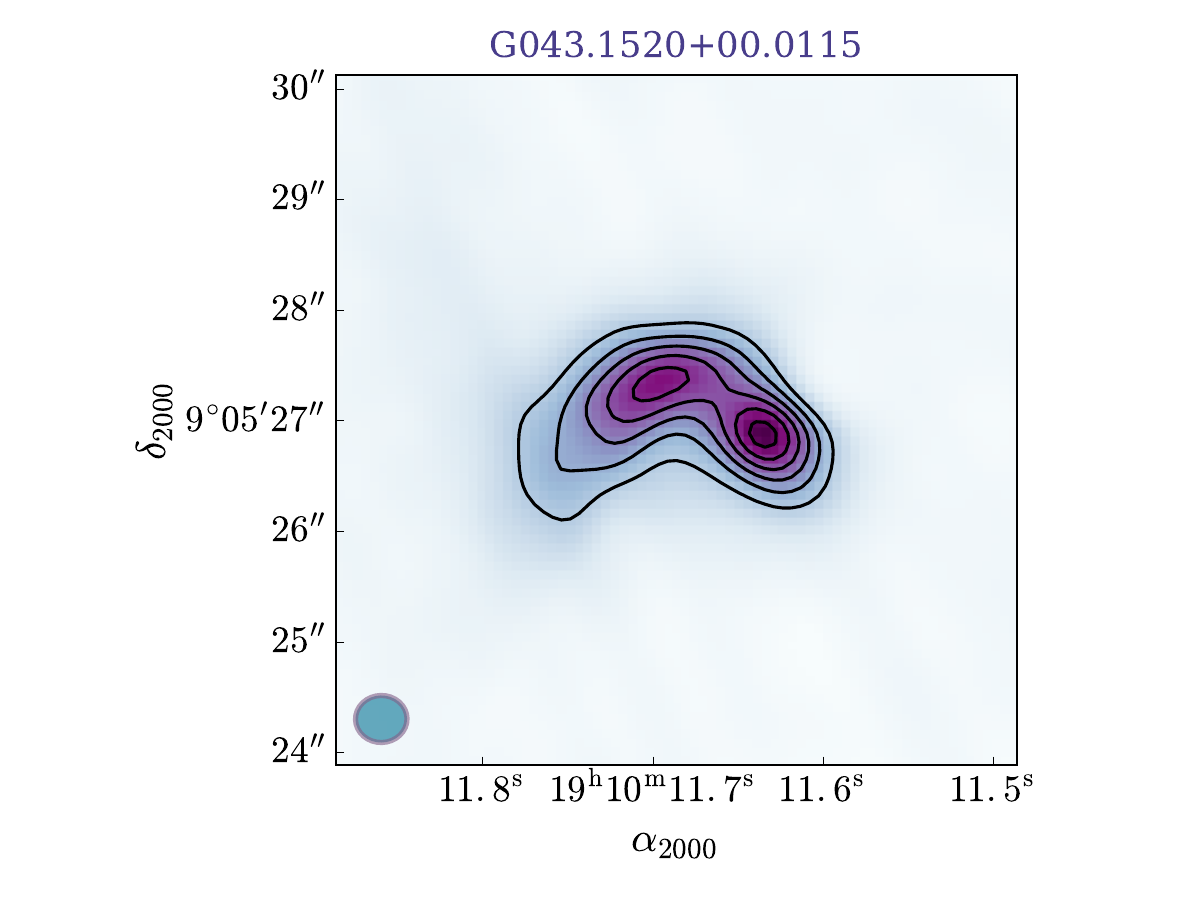} \hfill
\includegraphics[width=0.30\textwidth]{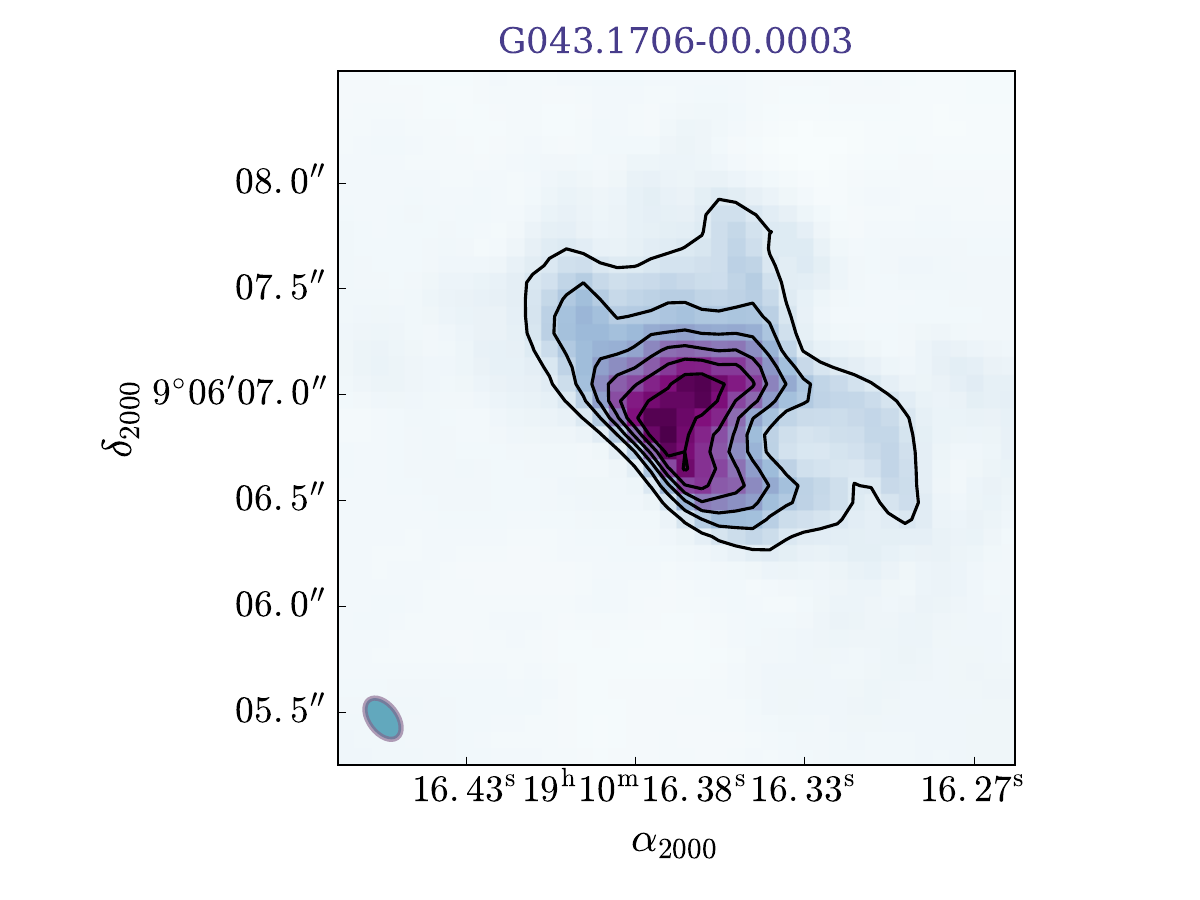} \hfill
\includegraphics[width=0.30\textwidth]{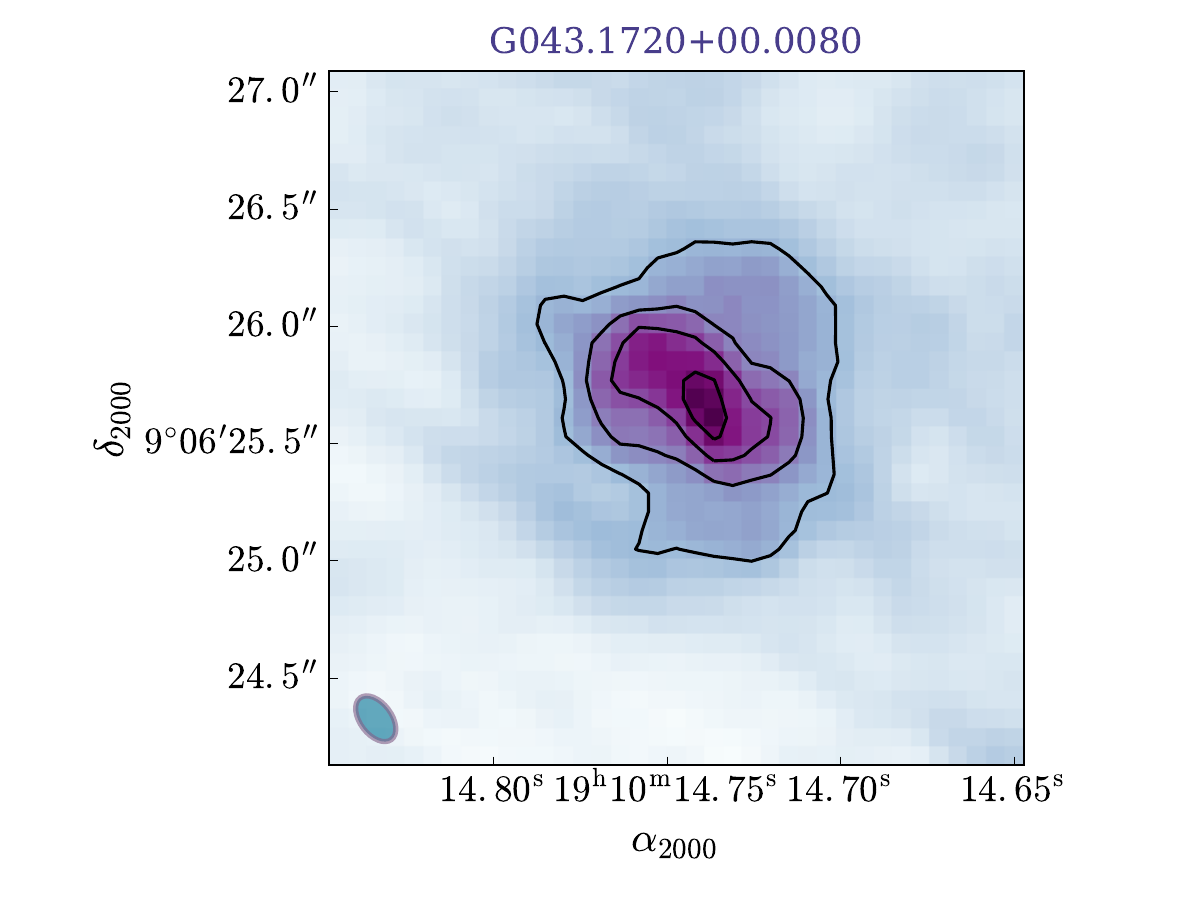}
\\[2mm]
\includegraphics[width=0.30\textwidth]{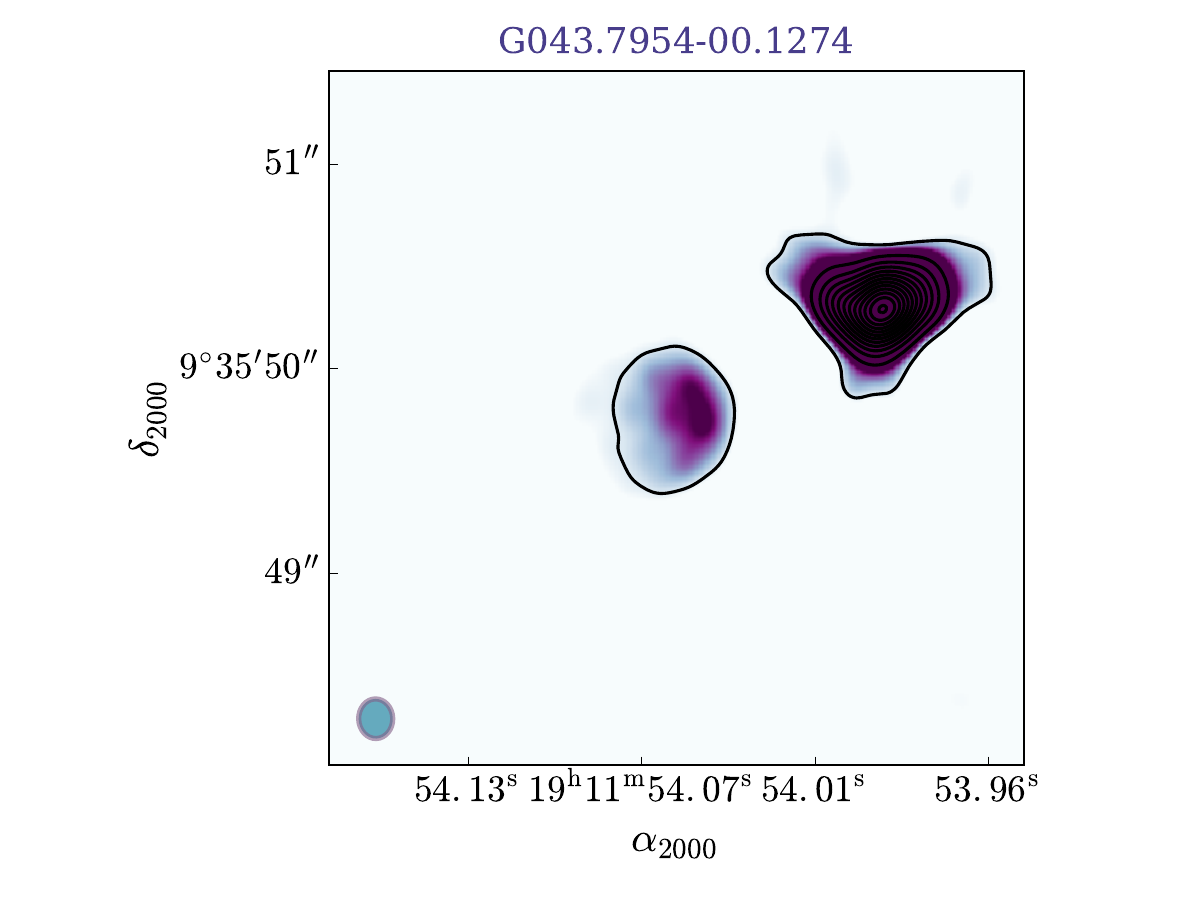} \hfill
\includegraphics[width=0.30\textwidth]{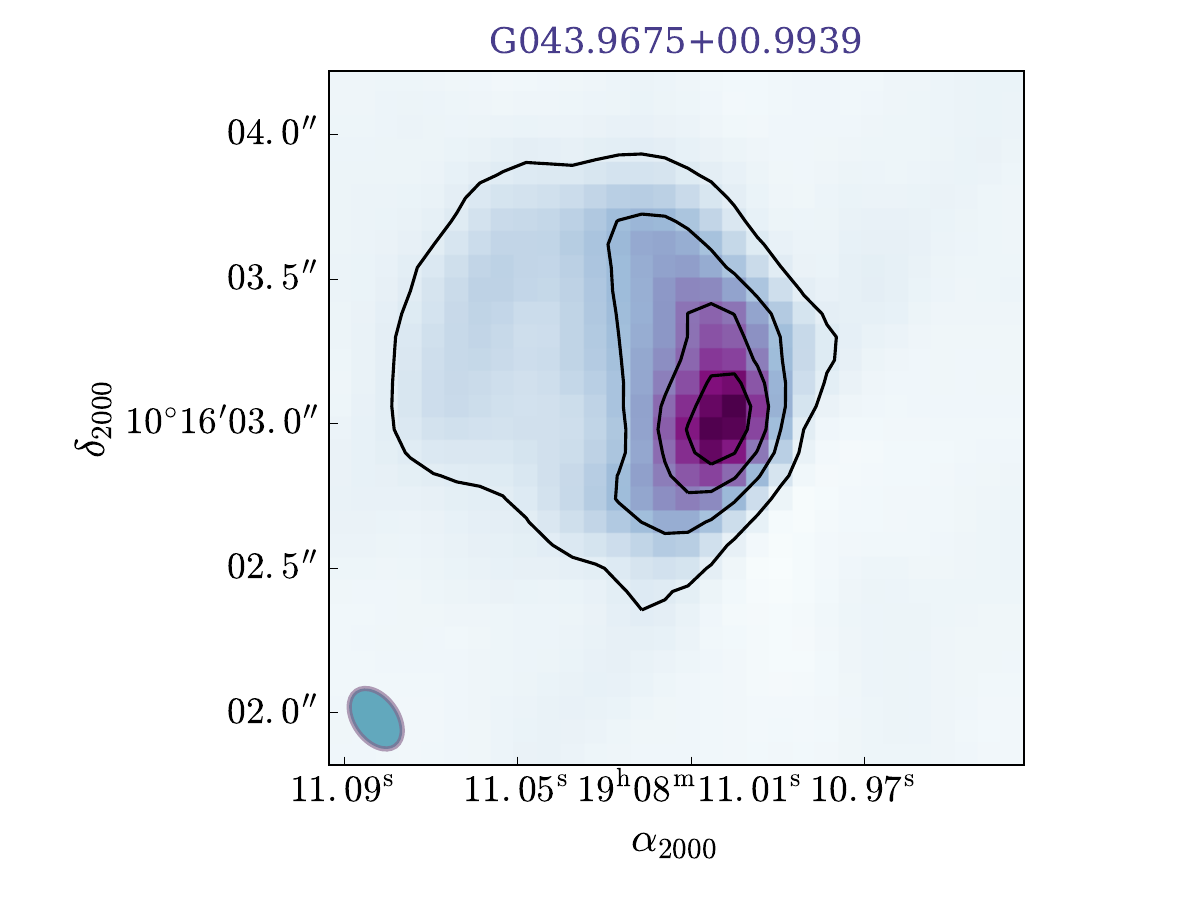} \hfill
\includegraphics[width=0.30\textwidth]{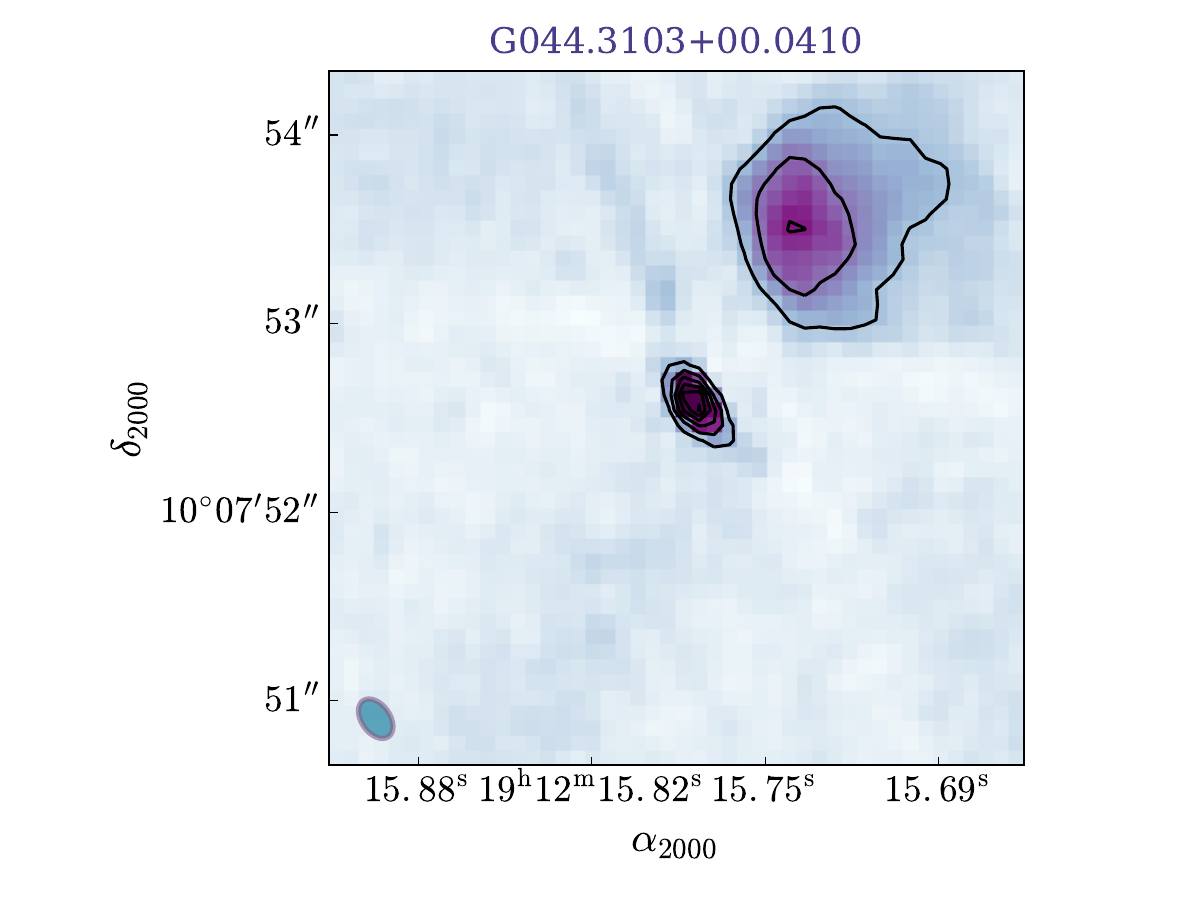}
\\[2mm]
\caption[Target maps, continued]{Target maps, continued.}
\end{figure*}
\clearpage

\begin{figure*}
\ContinuedFloat
\centering
\includegraphics[width=0.30\textwidth]{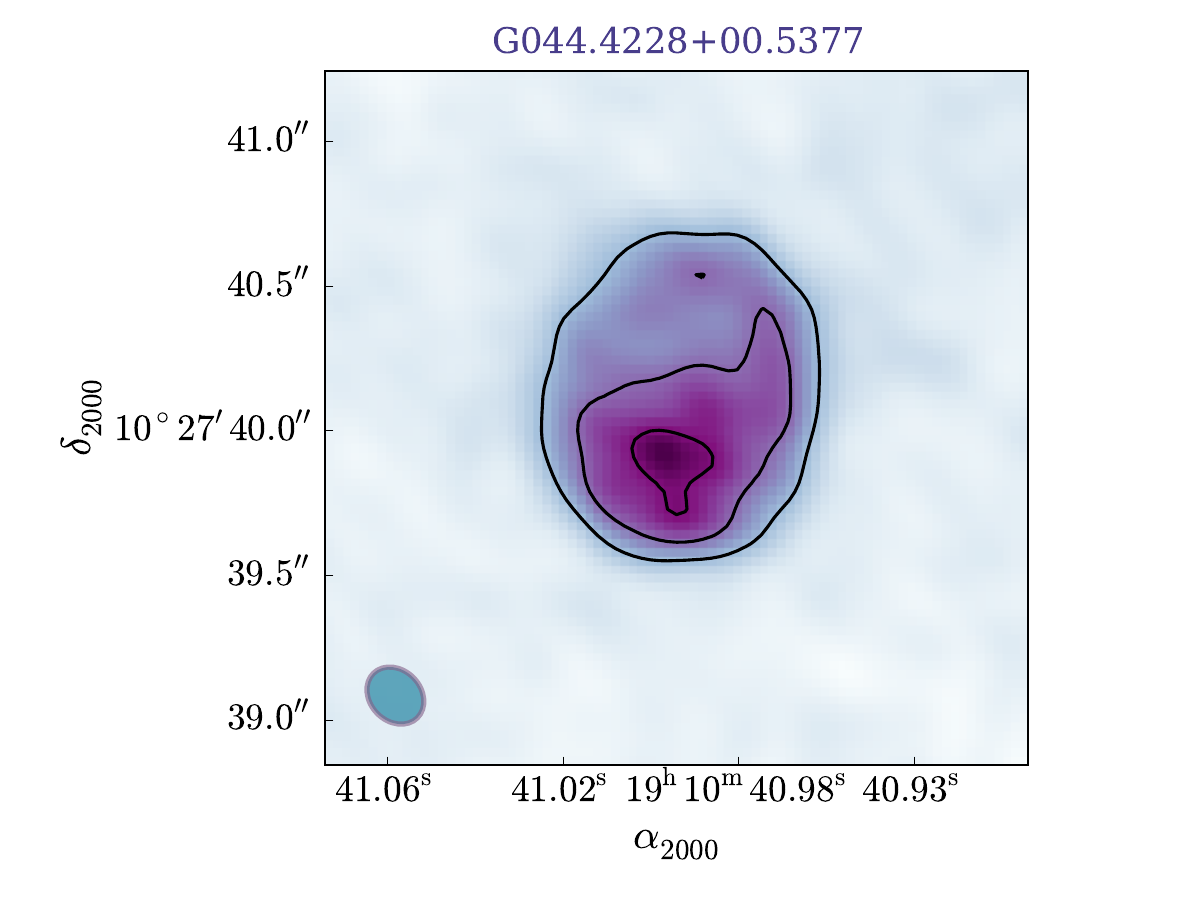} \hfill
\includegraphics[width=0.30\textwidth]{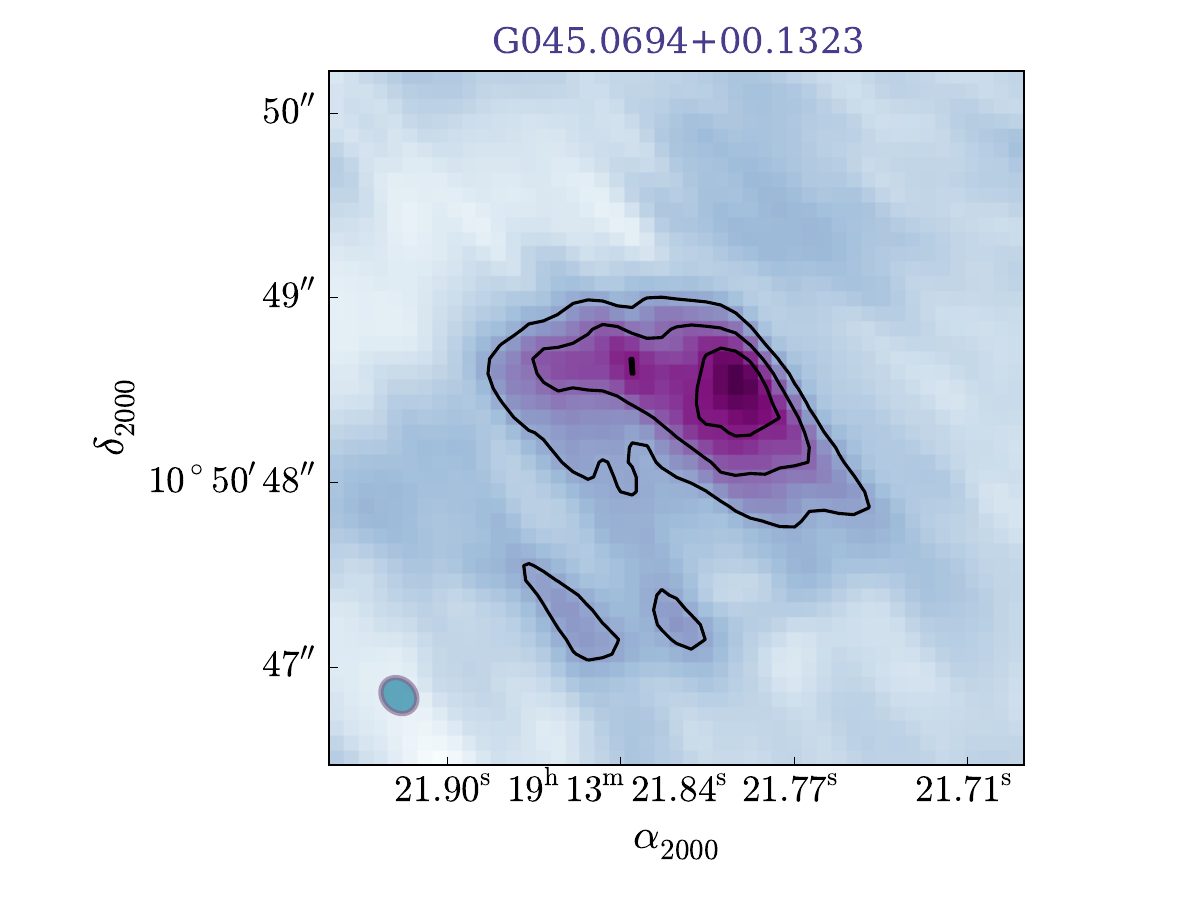} \hfill
\includegraphics[width=0.30\textwidth]{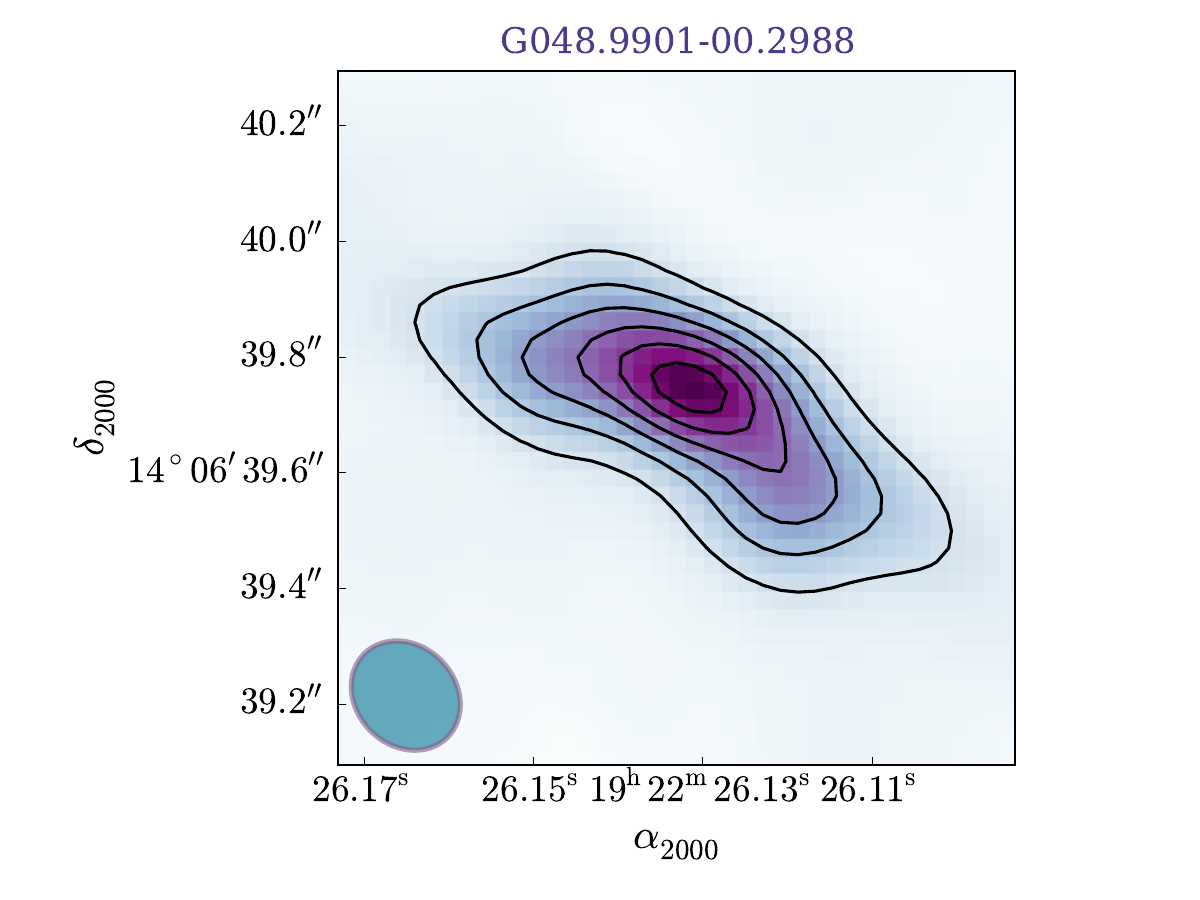}
\\[2mm]
\includegraphics[width=0.30\textwidth]{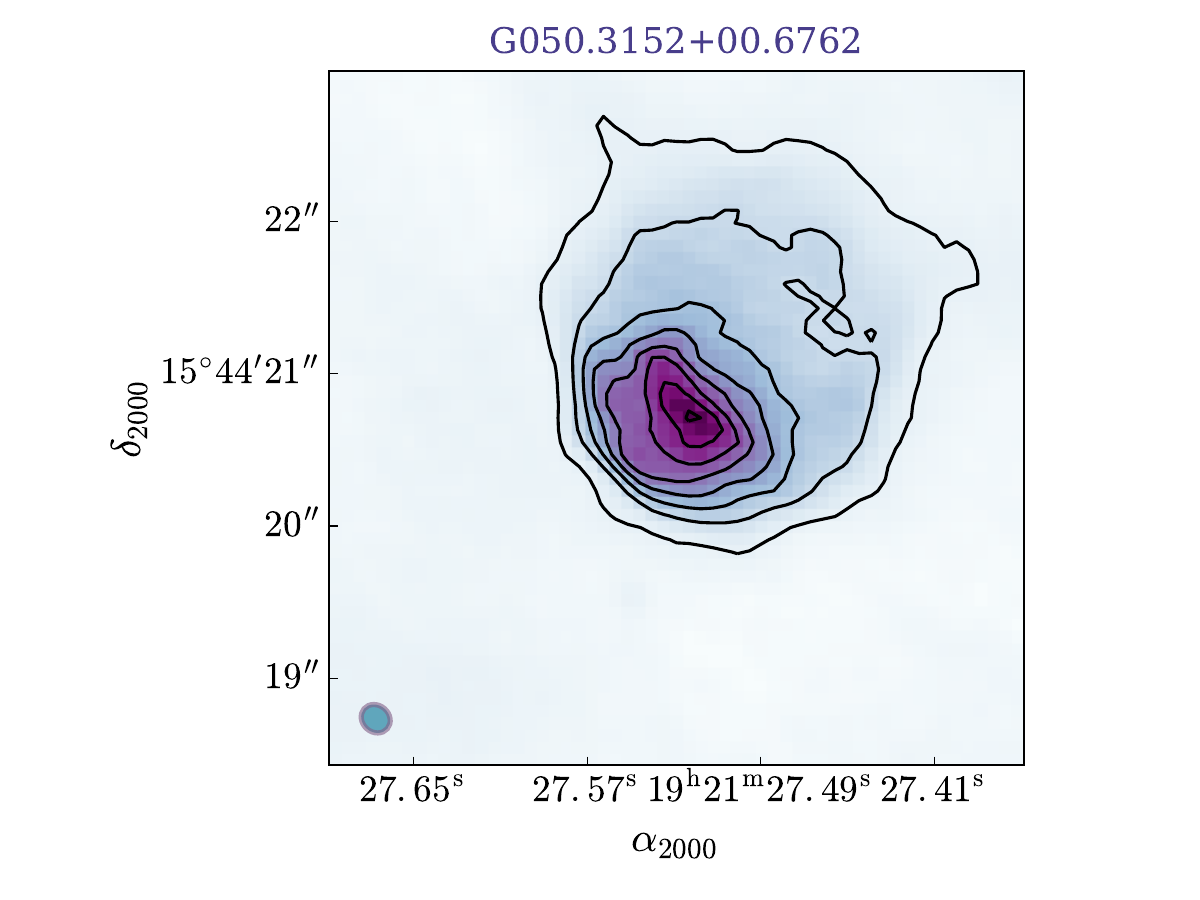} \hfill
\includegraphics[width=0.30\textwidth]{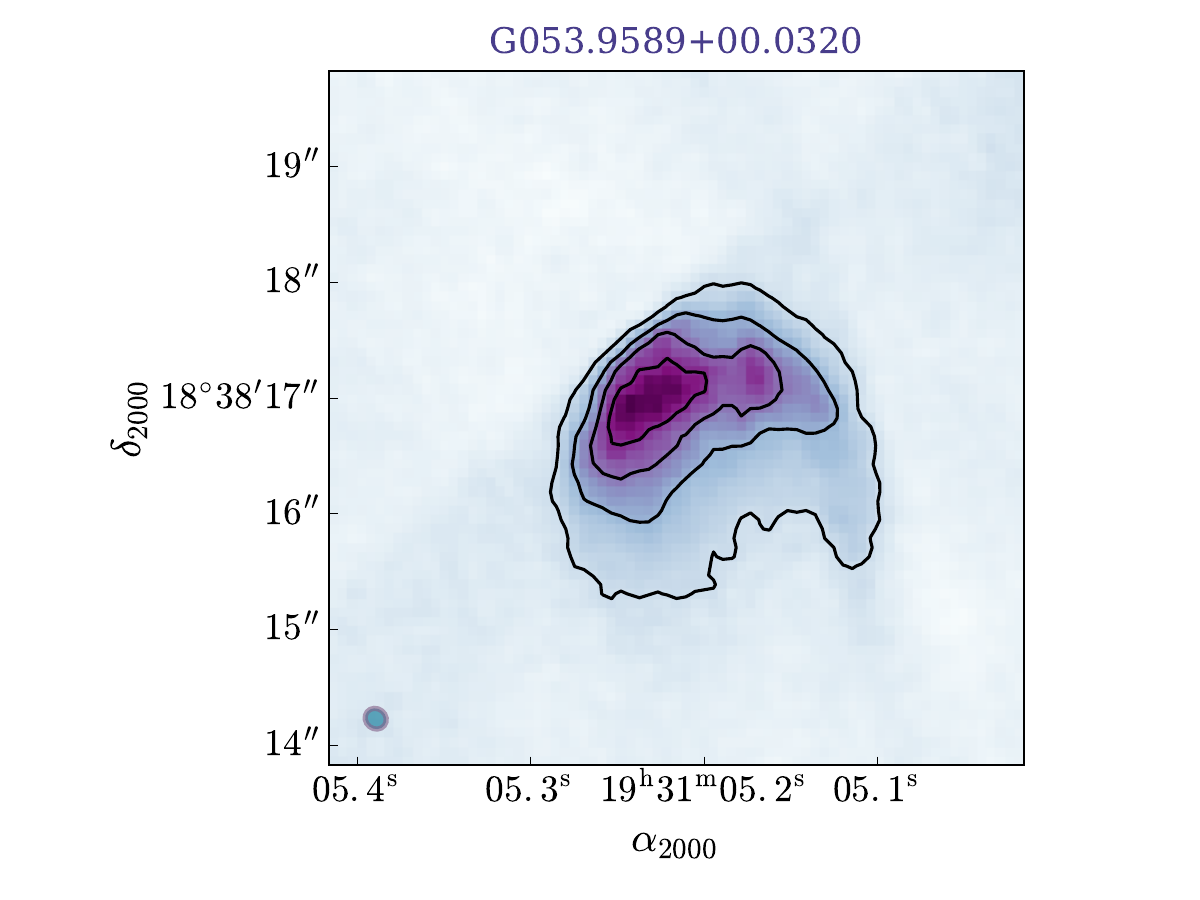} \hfill
\includegraphics[width=0.30\textwidth]{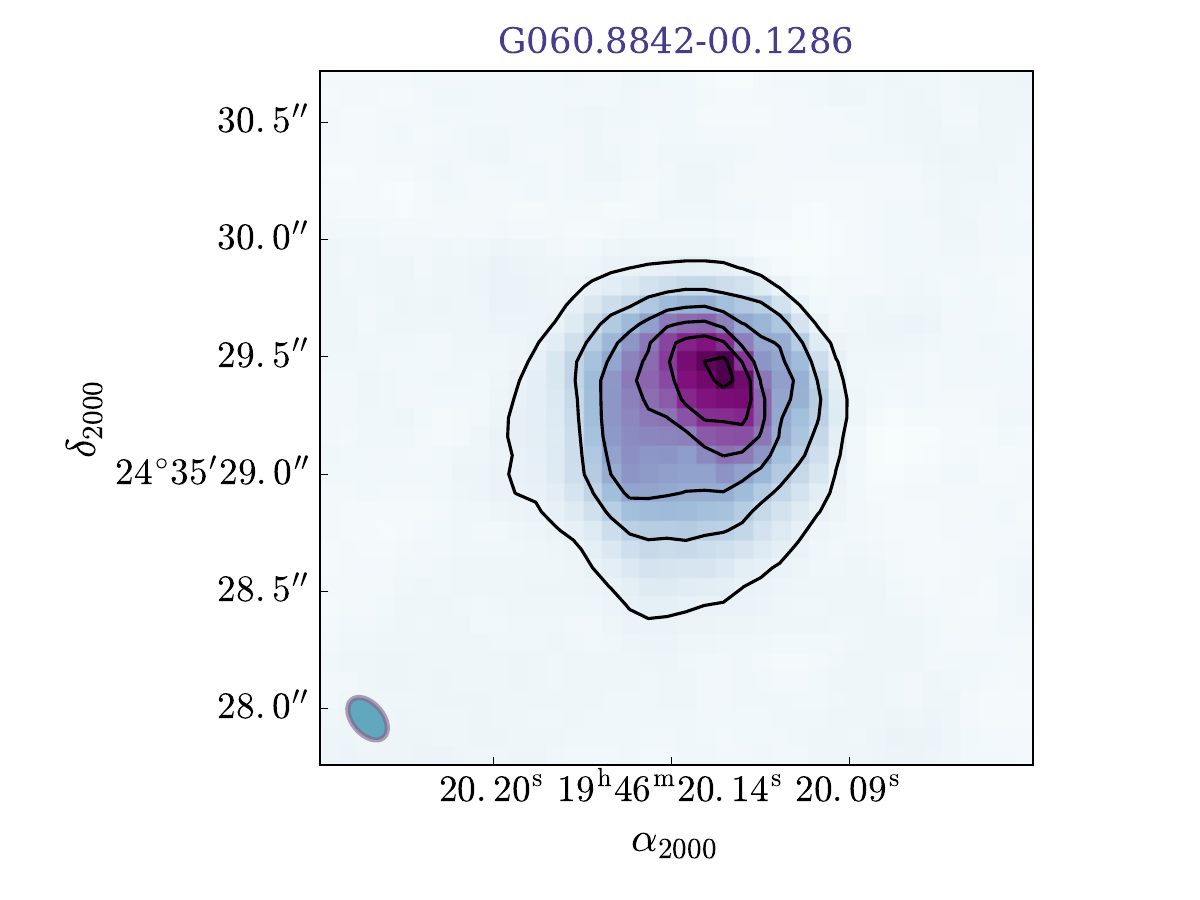}
\\[2mm]
\includegraphics[width=0.30\textwidth]{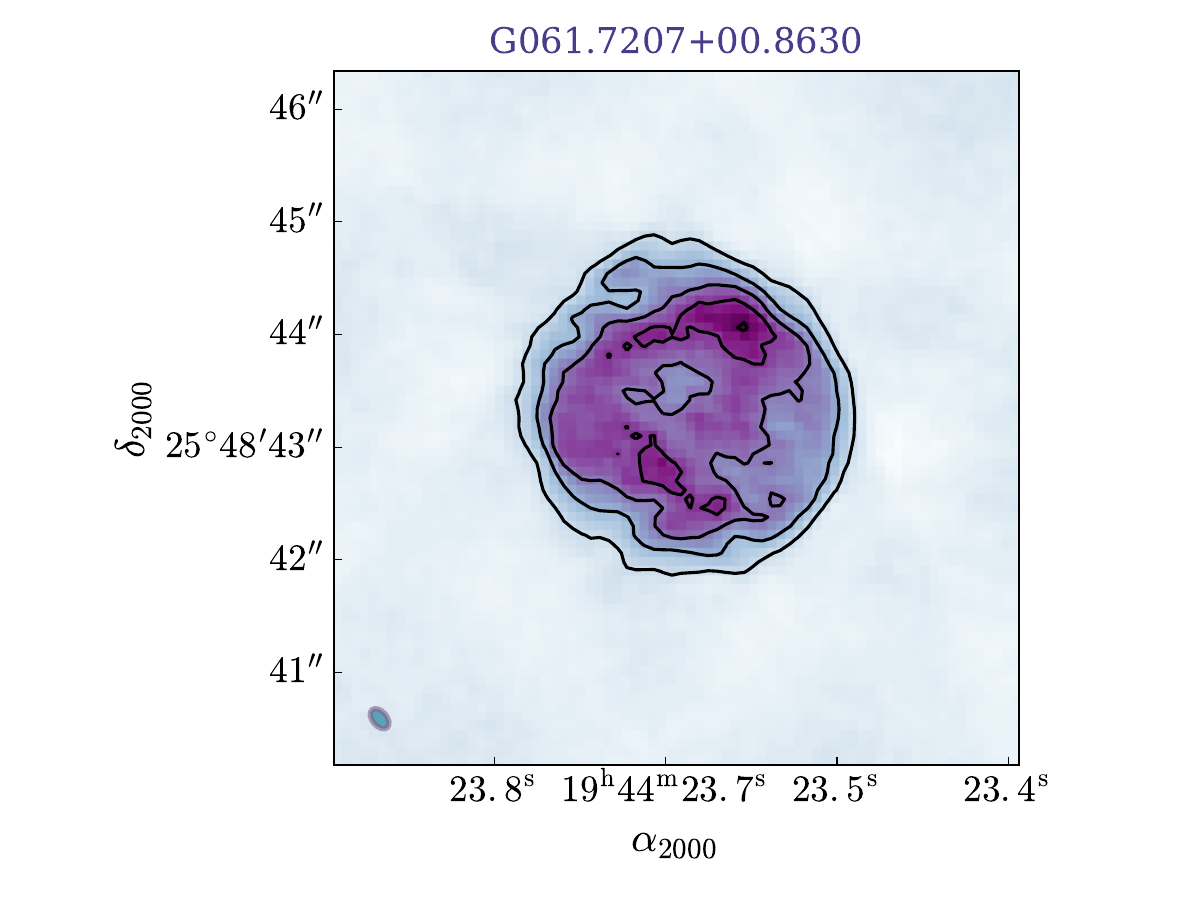}
\\[2mm]
\caption[Target maps, continued]{Target maps, continued.}
\end{figure*}

\clearpage

\begin{figure*}
\centering
\phantomsection
\refstepcounter{section}\label{app:appendixB}
\addcontentsline{toc}{section}{\thesection\ Neighbour maps}
{\bfseries APPENDIX B: NEIGHBOUR MAPS}\par\vspace{4mm}
\includegraphics[width=0.30\textwidth]{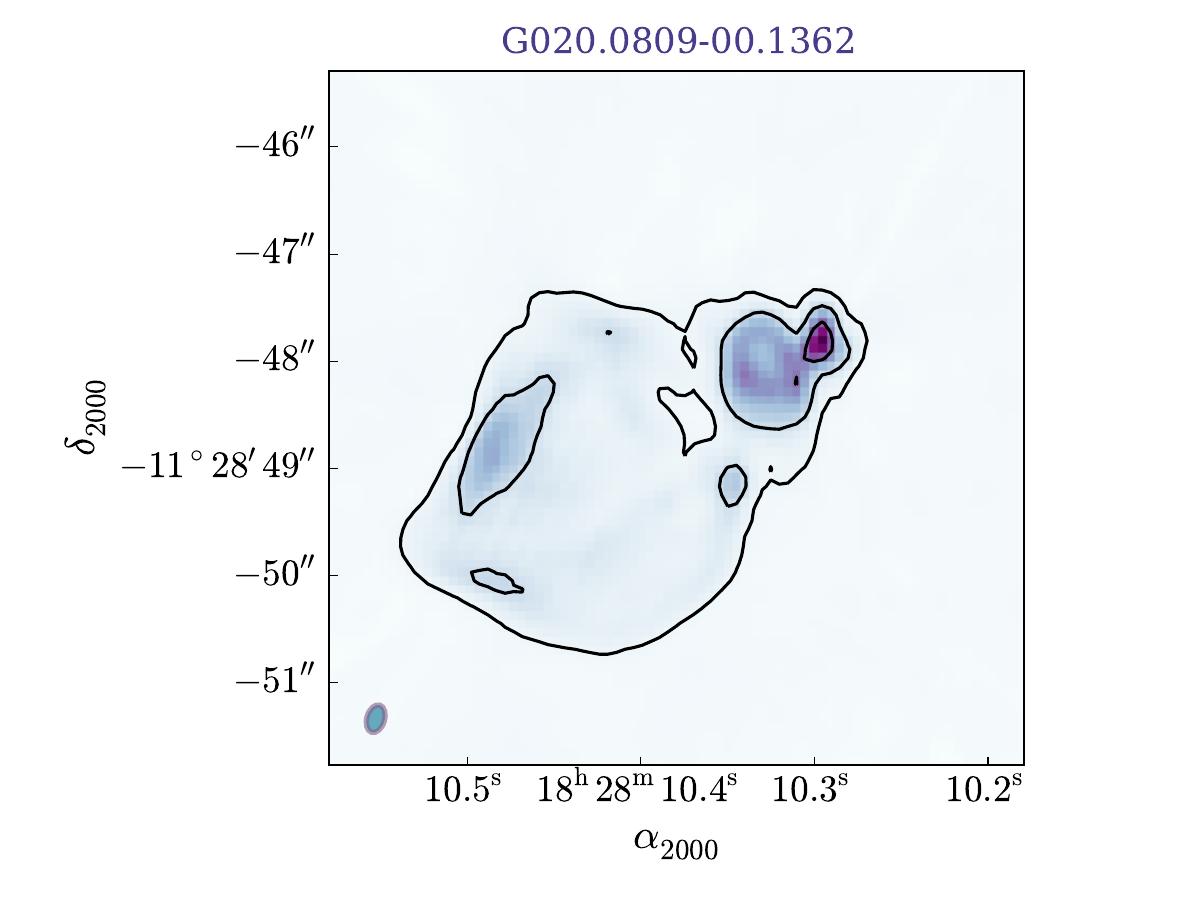} \hfill
\includegraphics[width=0.30\textwidth]{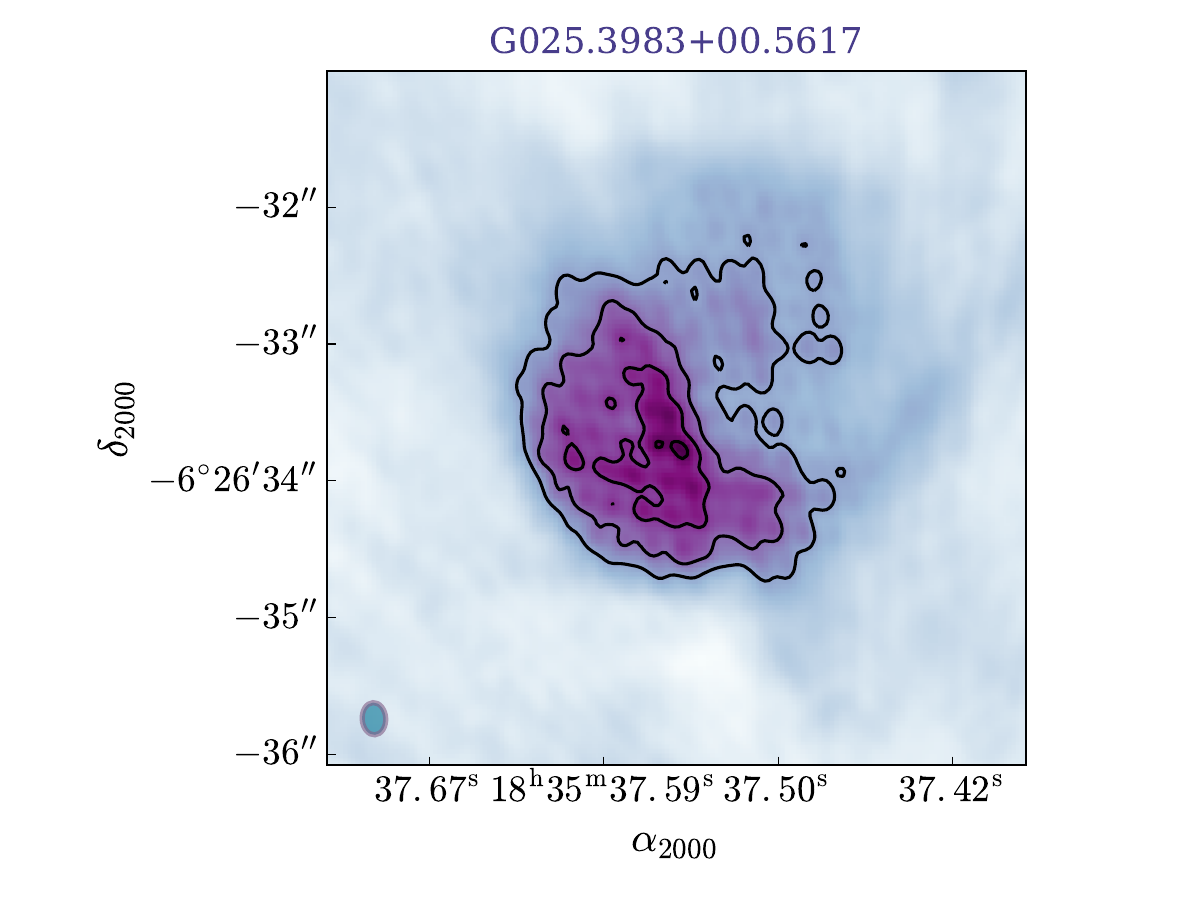} \hfill
\includegraphics[width=0.30\textwidth]{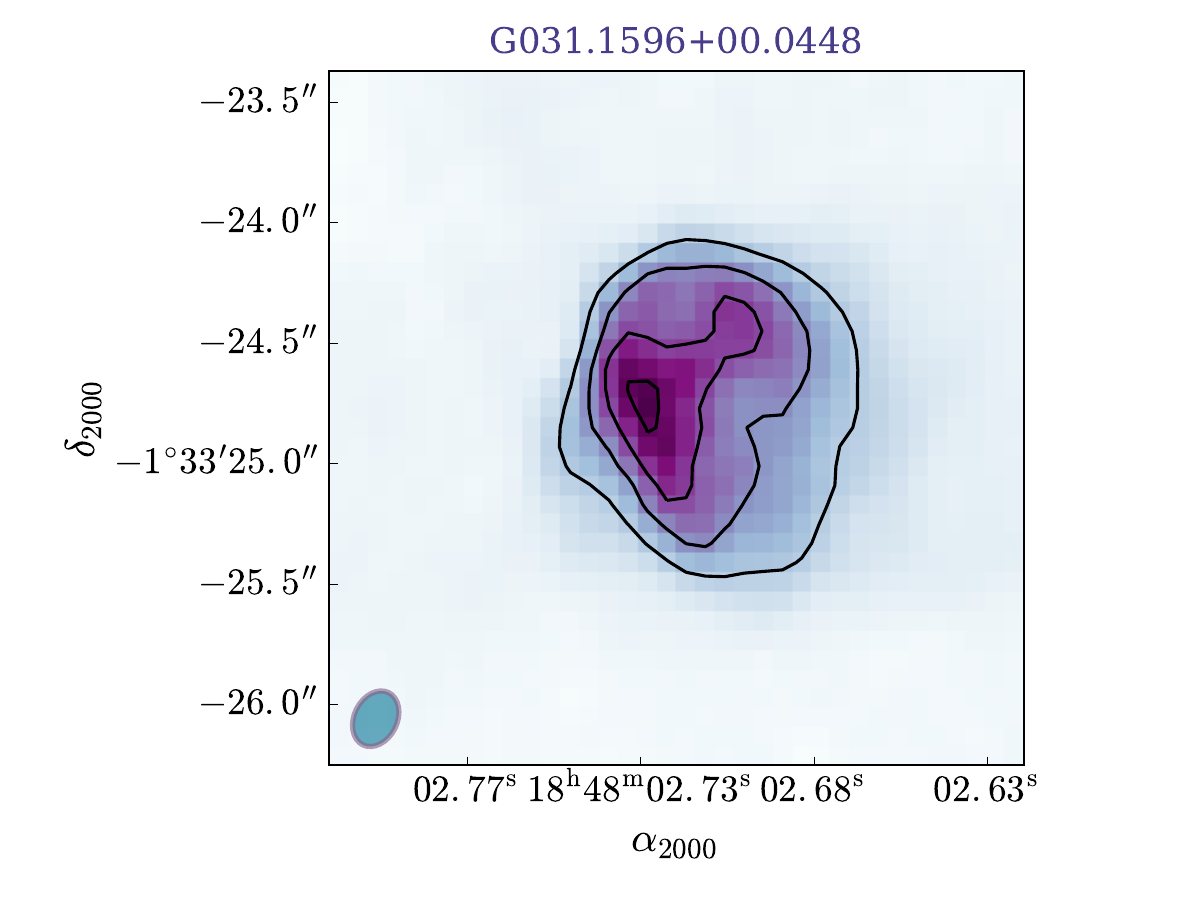}
\\[2mm]
\includegraphics[width=0.30\textwidth]{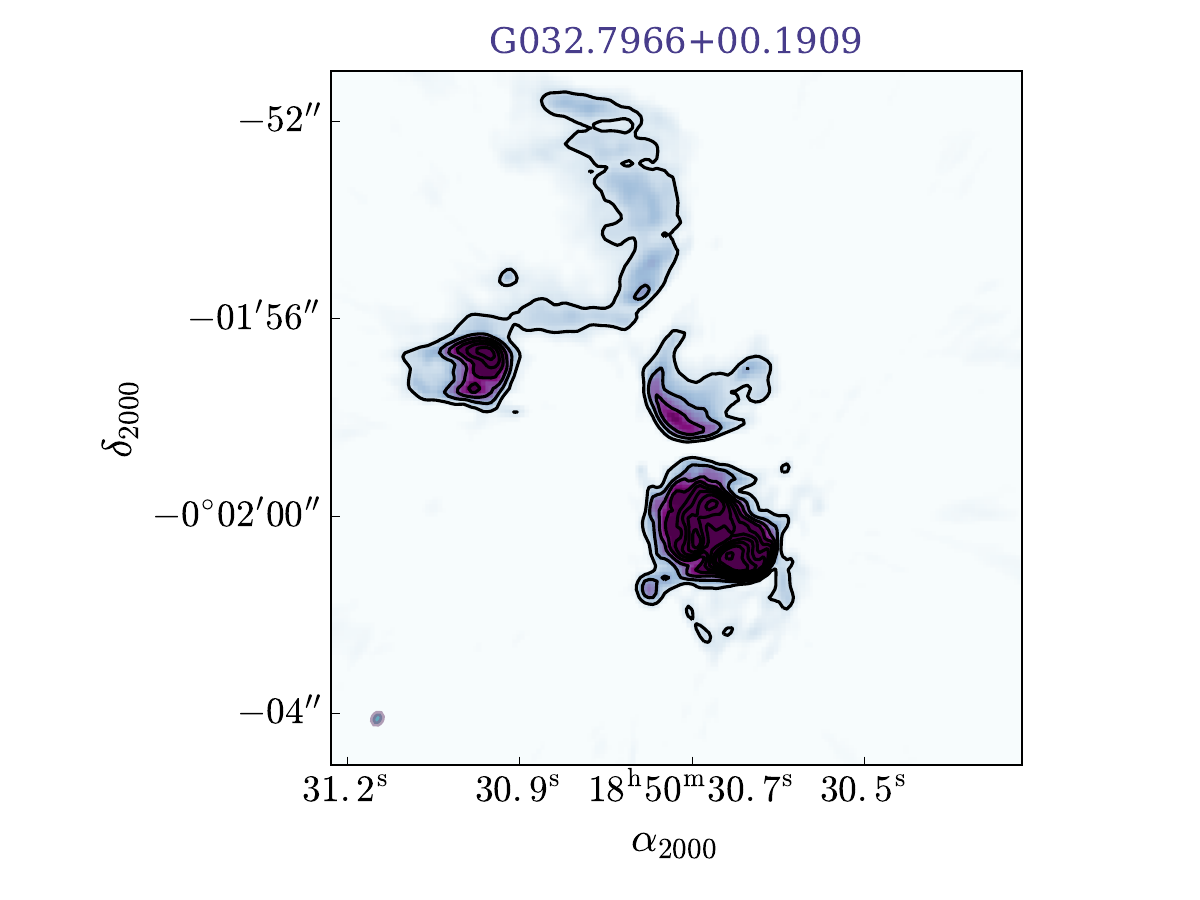} \hfill
\includegraphics[width=0.30\textwidth]{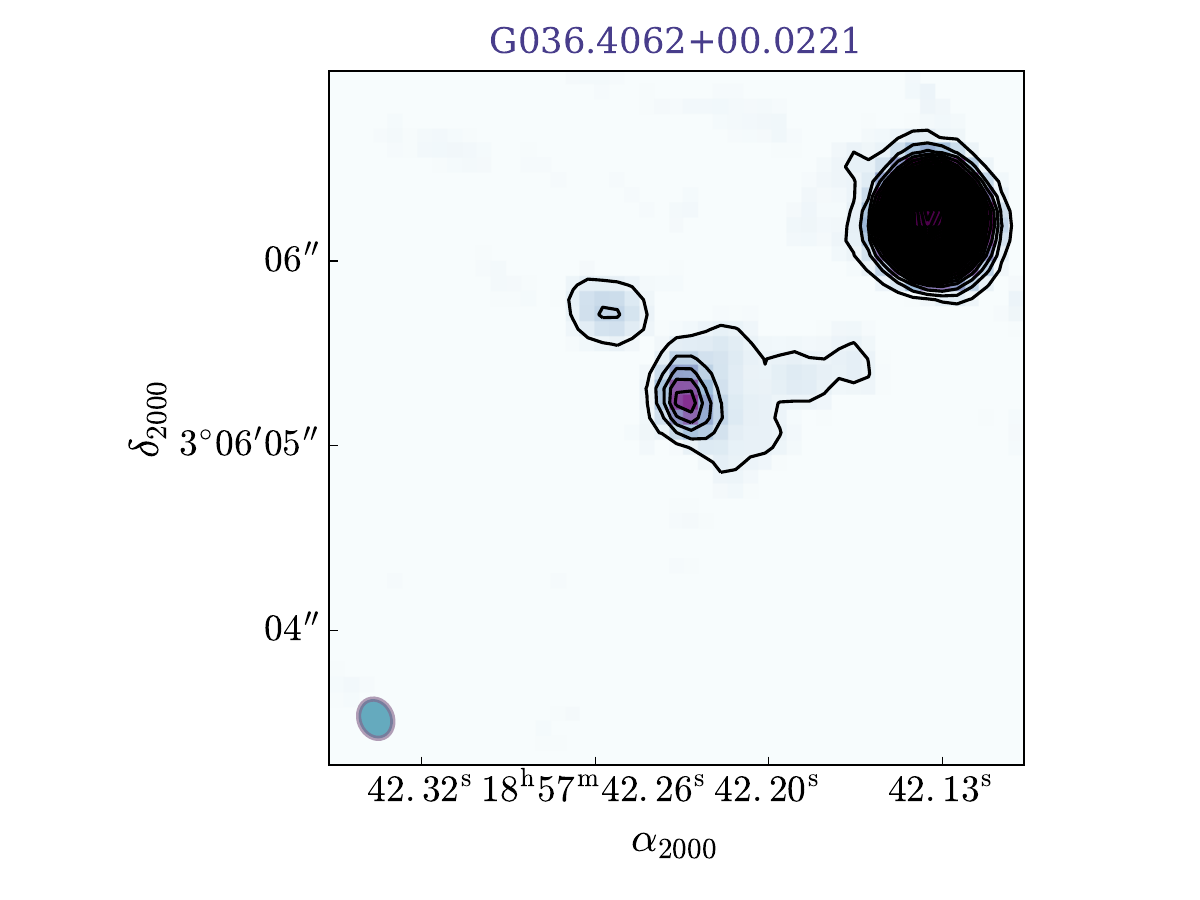} \hfill
\includegraphics[width=0.30\textwidth]{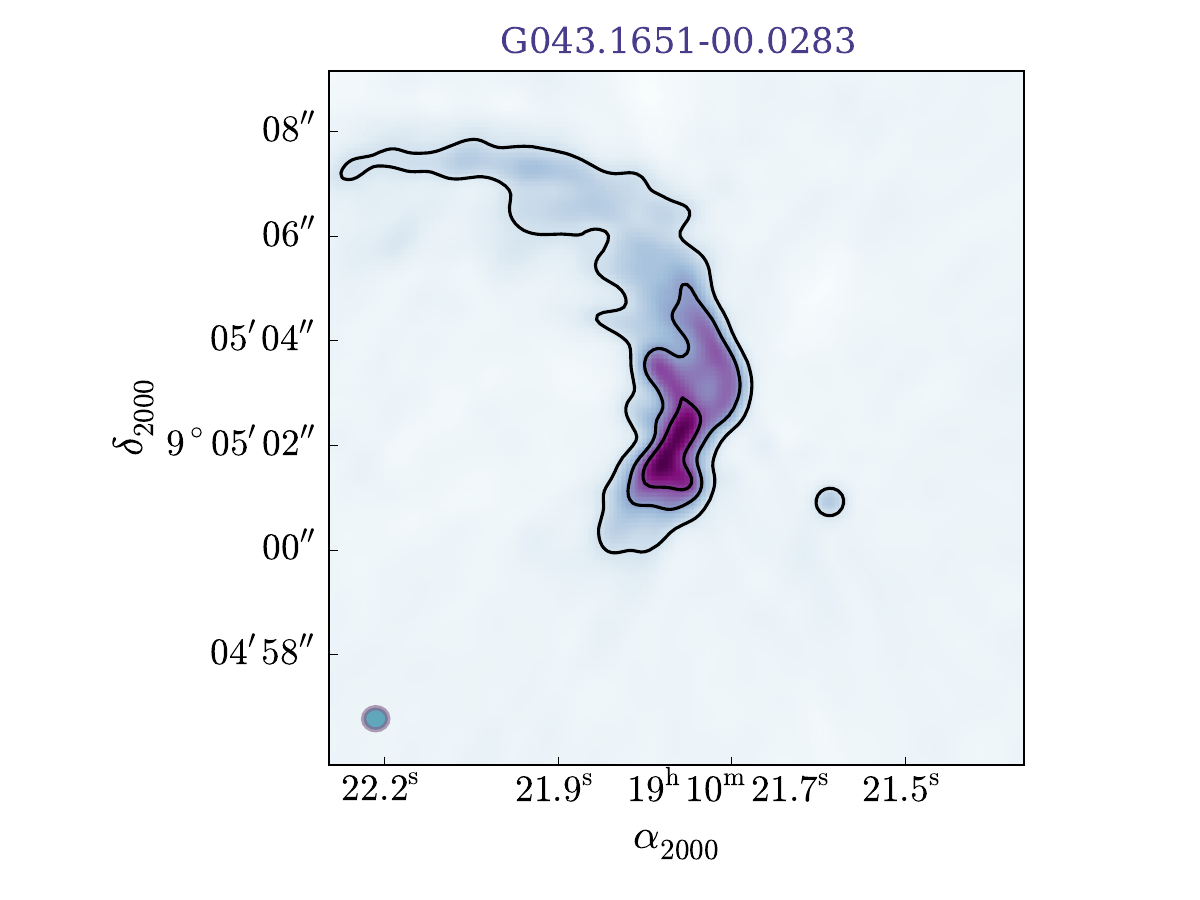}
\\[2mm]
\includegraphics[width=0.30\textwidth]{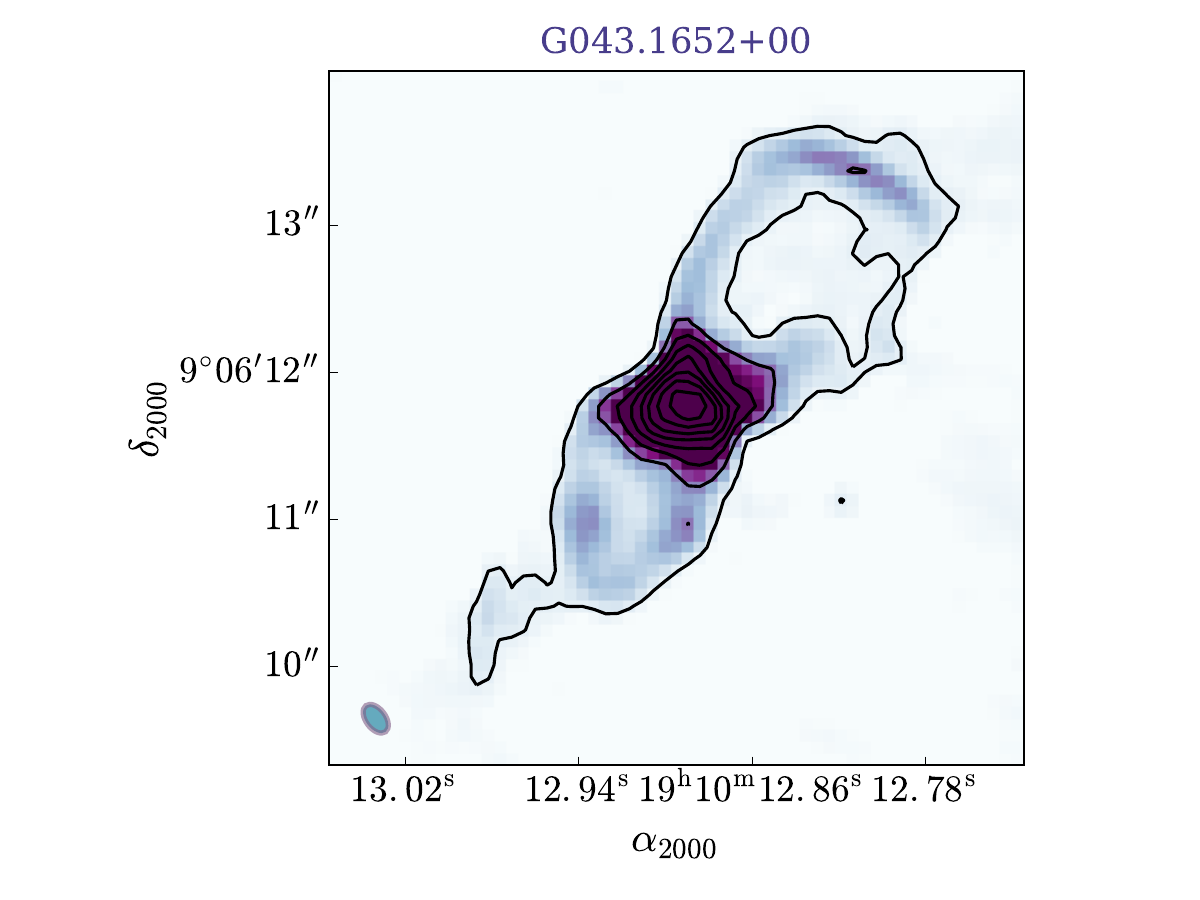} \hfill
\includegraphics[width=0.30\textwidth]{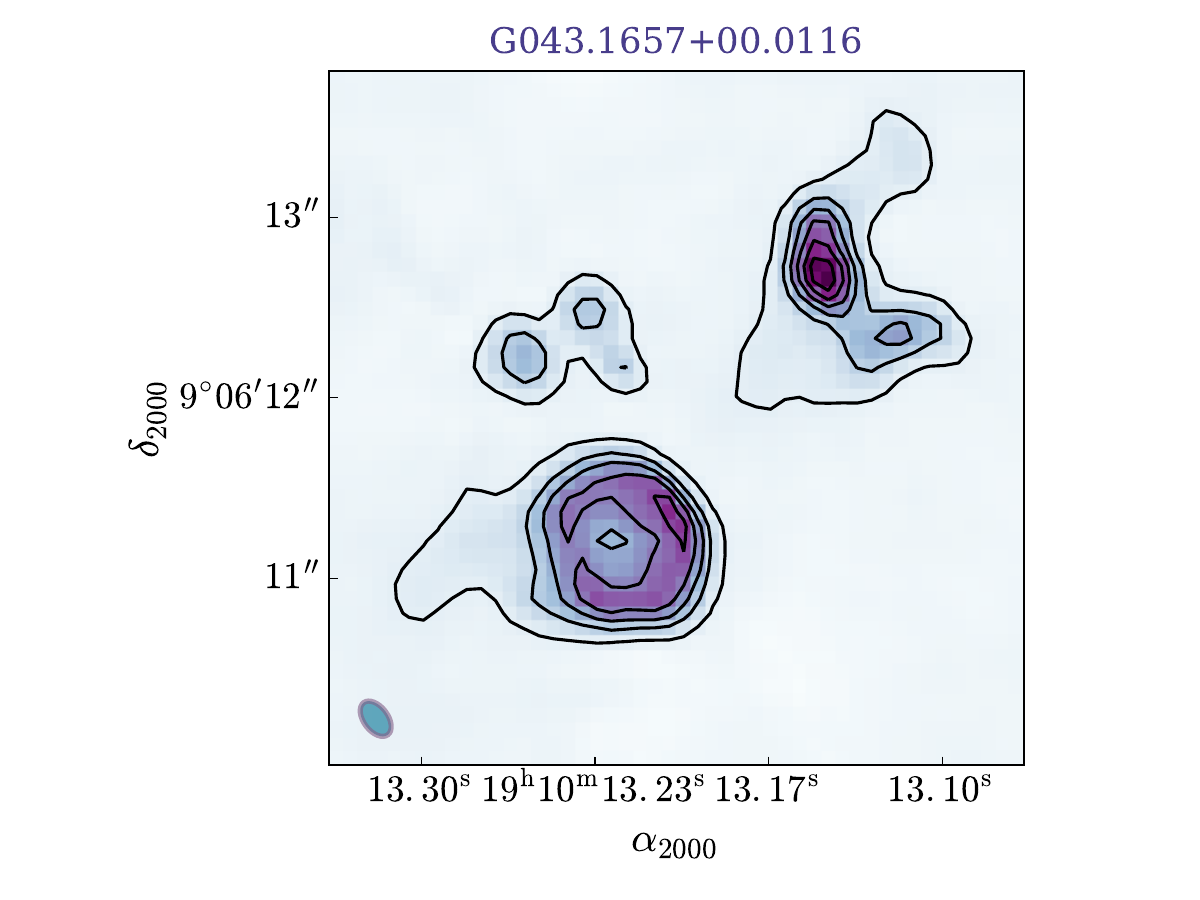} \hfill
\includegraphics[width=0.30\textwidth]{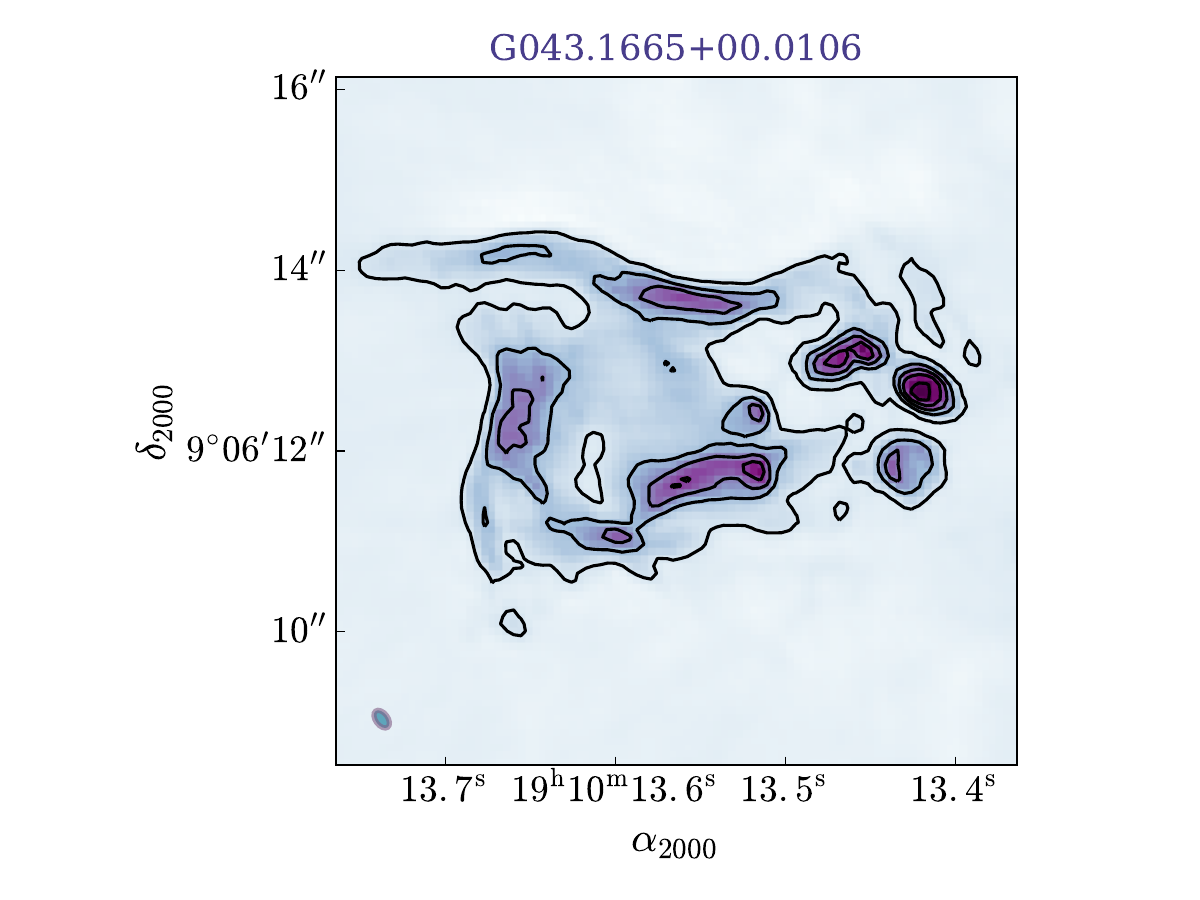}
\\[2mm]
\includegraphics[width=0.30\textwidth]{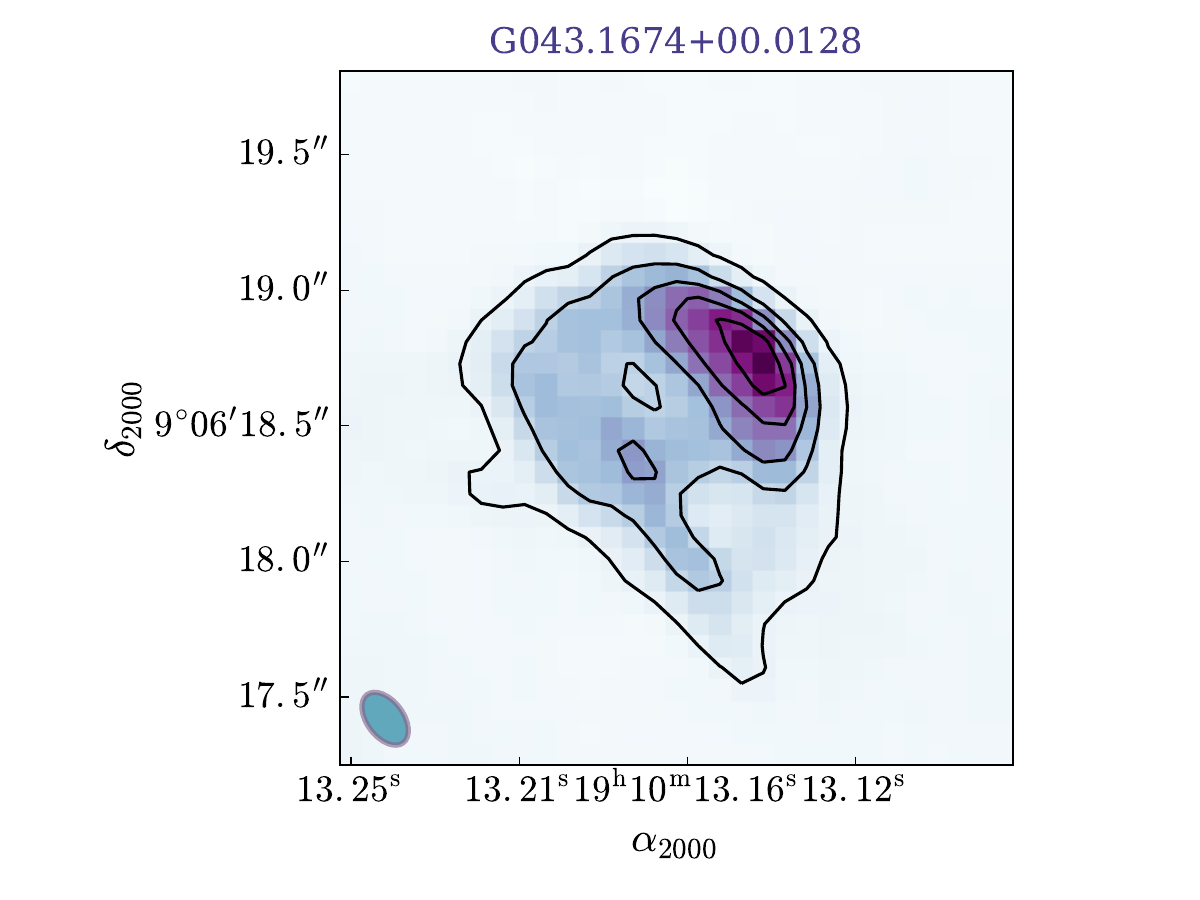} \hfill
\includegraphics[width=0.30\textwidth]{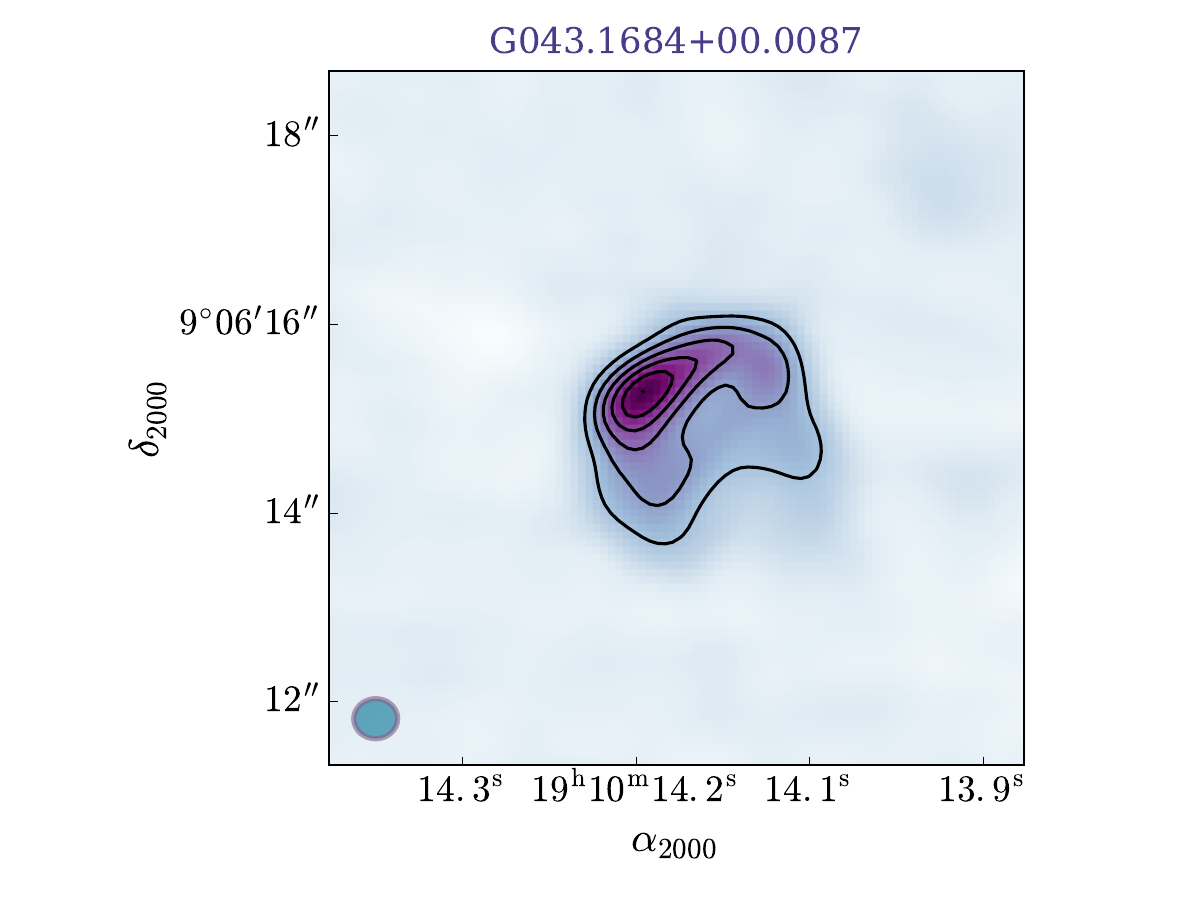} \hfill
\includegraphics[width=0.30\textwidth]{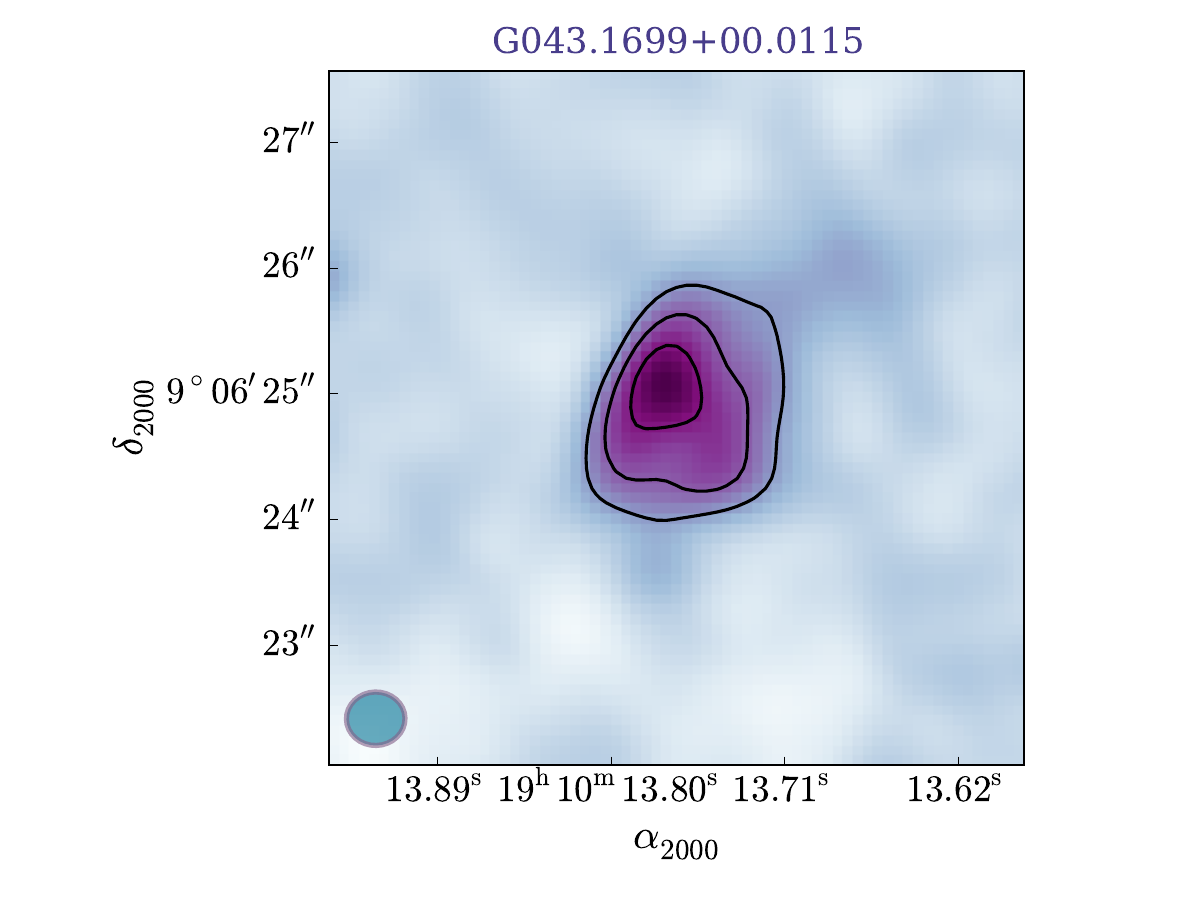}
\\[2mm]
\includegraphics[width=0.30\textwidth]{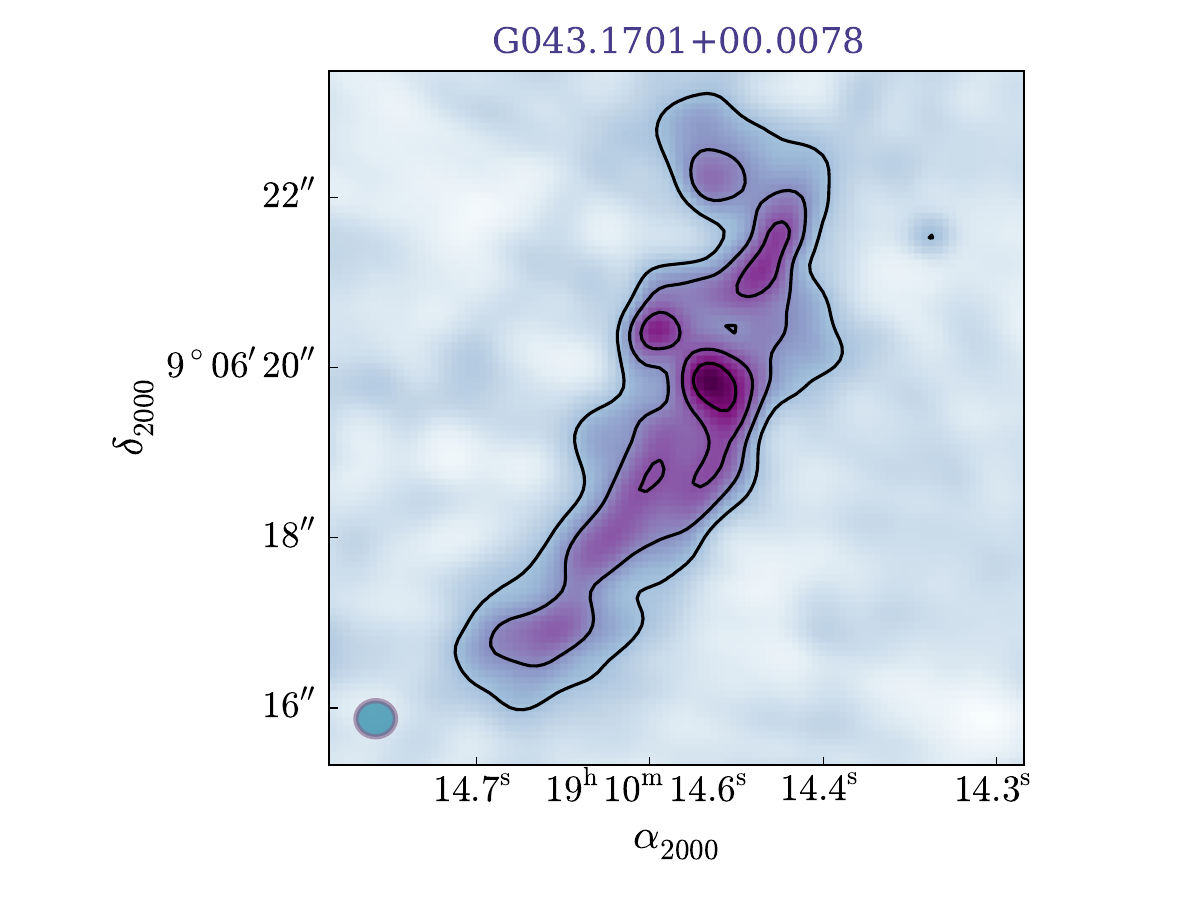} \hfill
\includegraphics[width=0.30\textwidth]{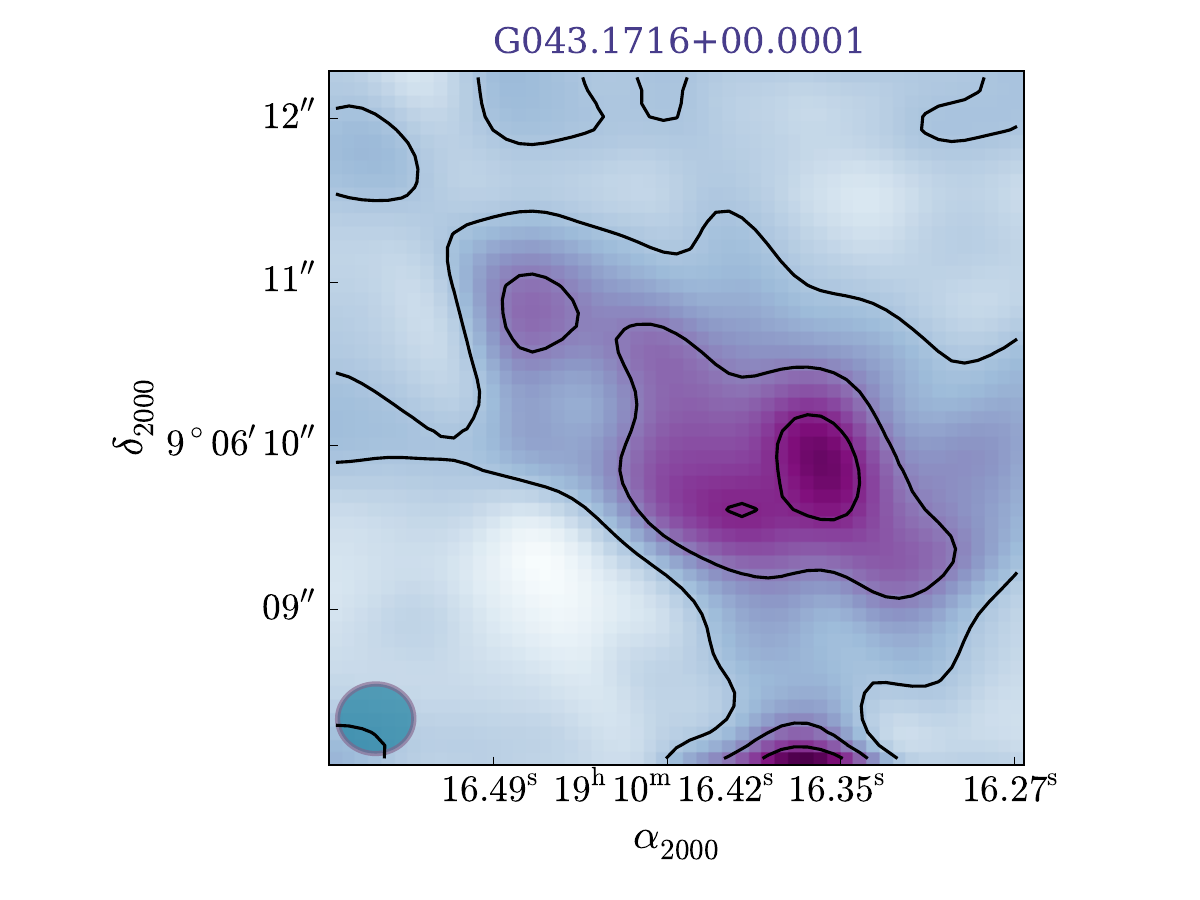} \hfill
\includegraphics[width=0.30\textwidth]{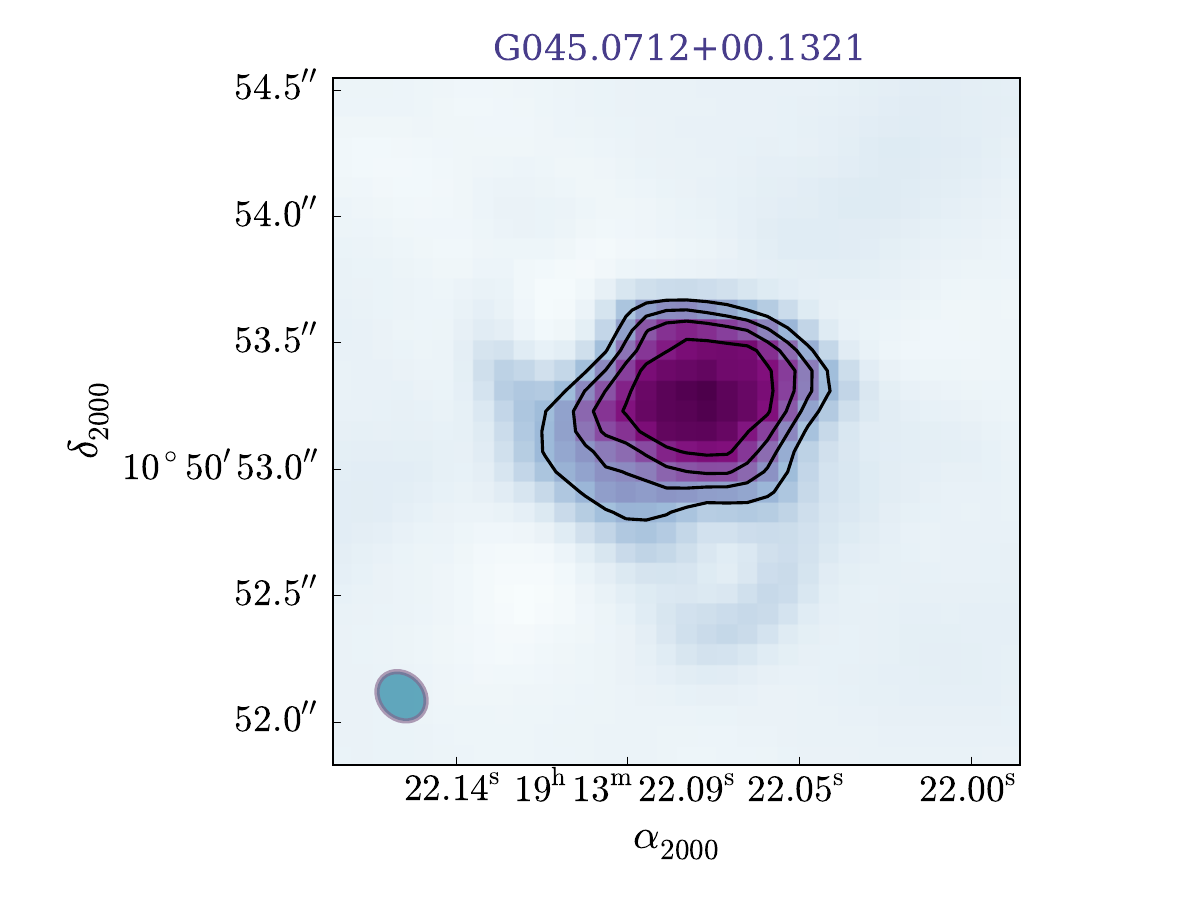}
\\[2mm]
\caption[Neighbour maps]{Neighbour maps. Multiple resolved components of a single CORNISH UCHII are shown together.}
\label{fig:maps_neighbours}
\end{figure*}

\clearpage

\begin{figure*}
\centering
\phantomsection
\refstepcounter{section}\label{app:appendixC}
\addcontentsline{toc}{section}{\thesection\ Radio SEDs of JVLA targets}
{\bfseries APPENDIX C: RADIO SEDs OF JVLA TARGETS}\par\vspace{4mm}
\includegraphics[width=0.26\textwidth]{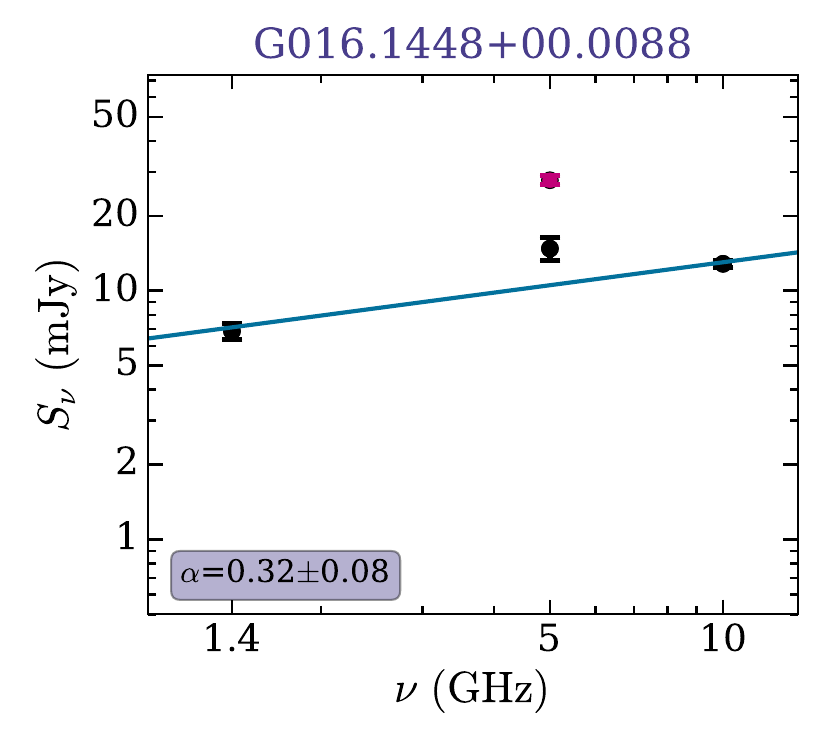} \hfill
\includegraphics[width=0.26\textwidth]{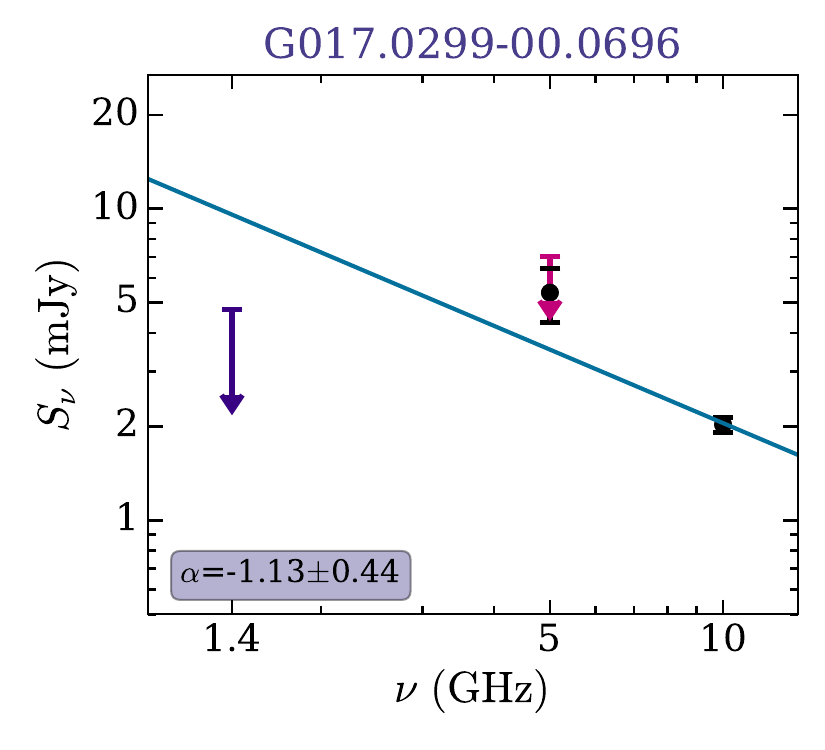} \hfill
\includegraphics[width=0.26\textwidth]{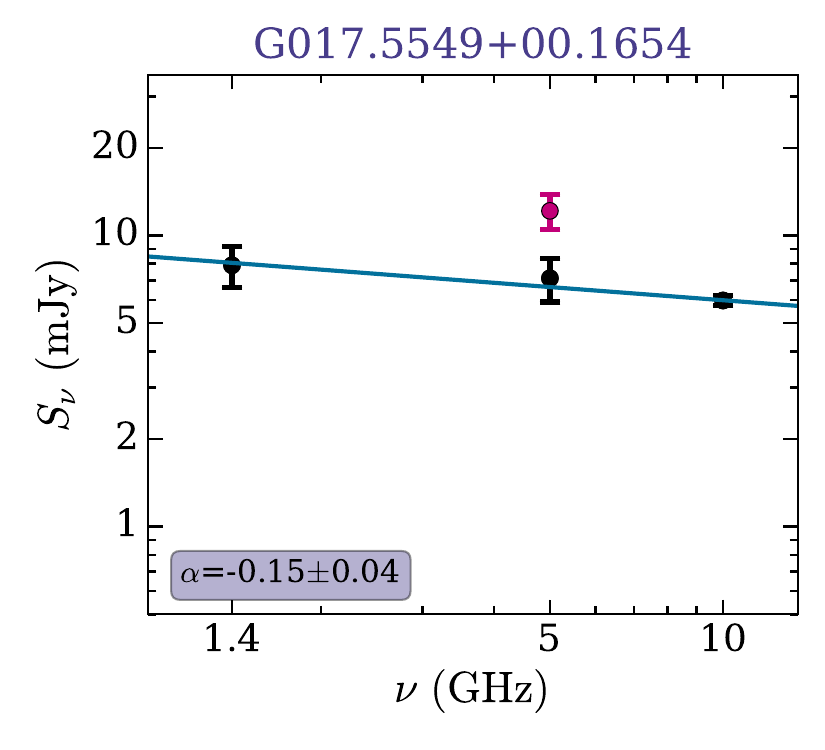}
\\[2mm]
\includegraphics[width=0.26\textwidth]{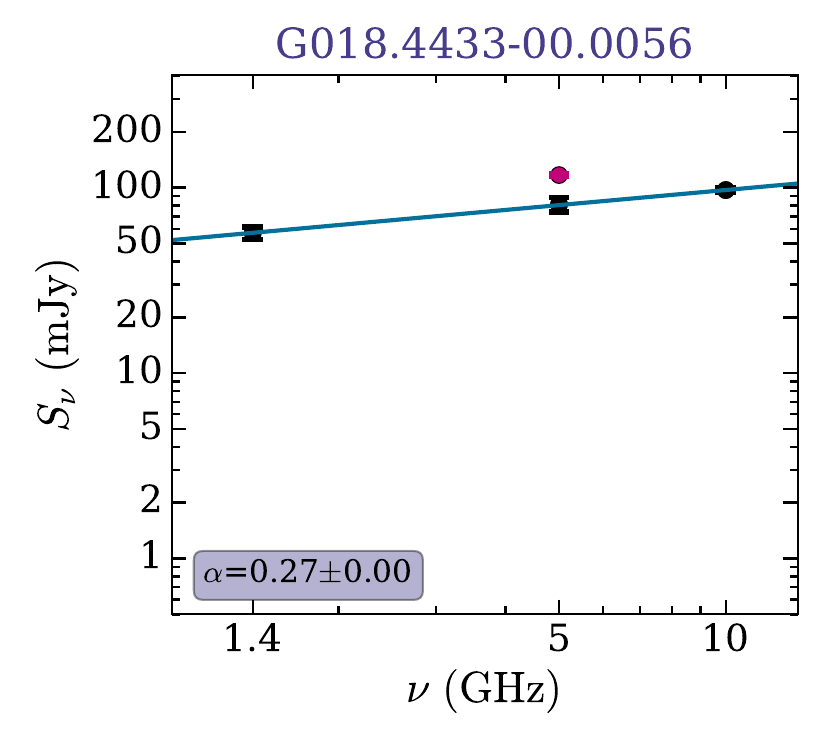} \hfill
\includegraphics[width=0.26\textwidth]{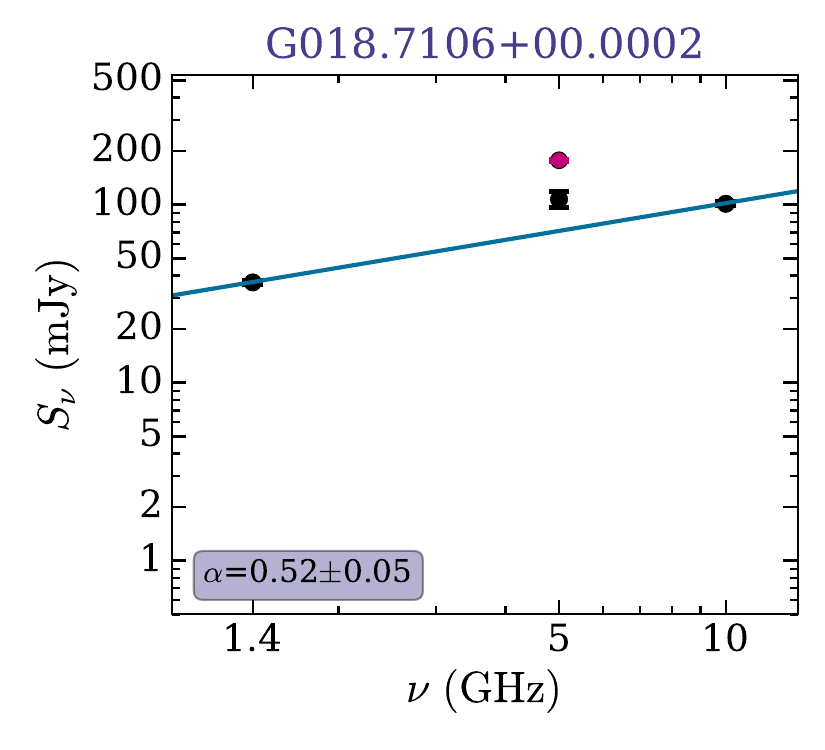} \hfill
\includegraphics[width=0.26\textwidth]{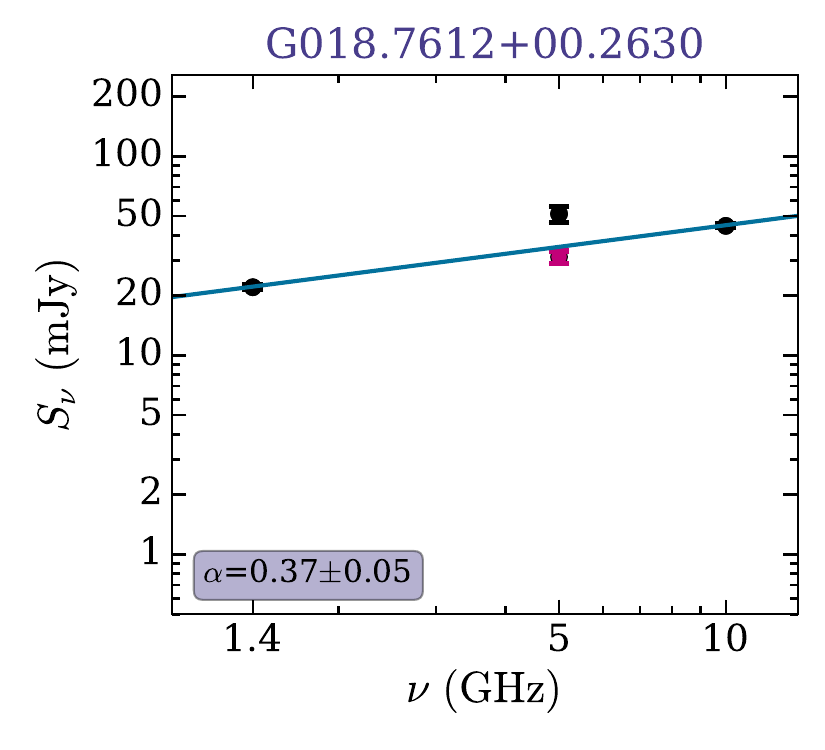}
\\[2mm]
\includegraphics[width=0.26\textwidth]{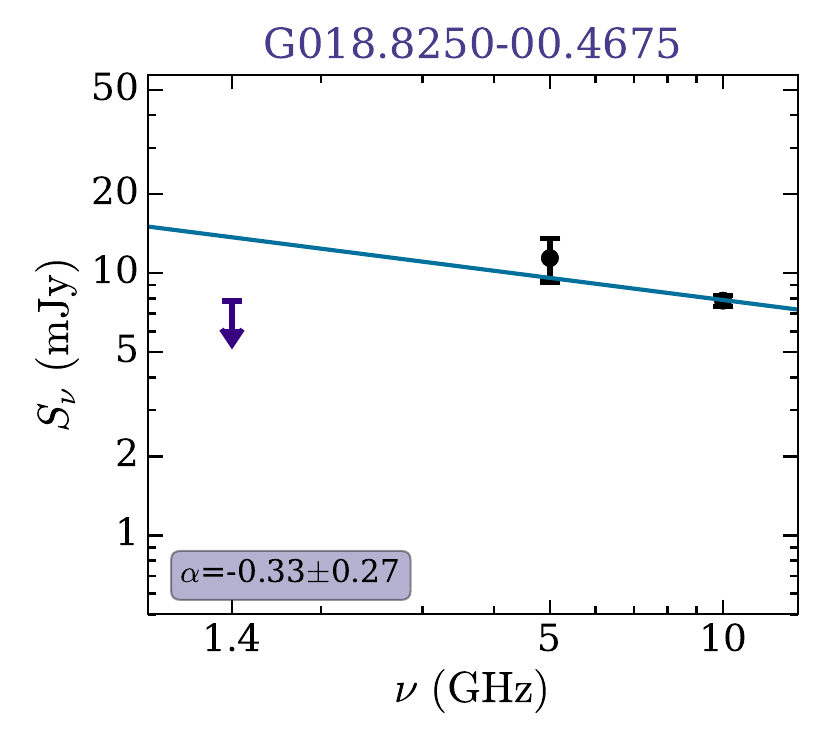} \hfill
\includegraphics[width=0.26\textwidth]{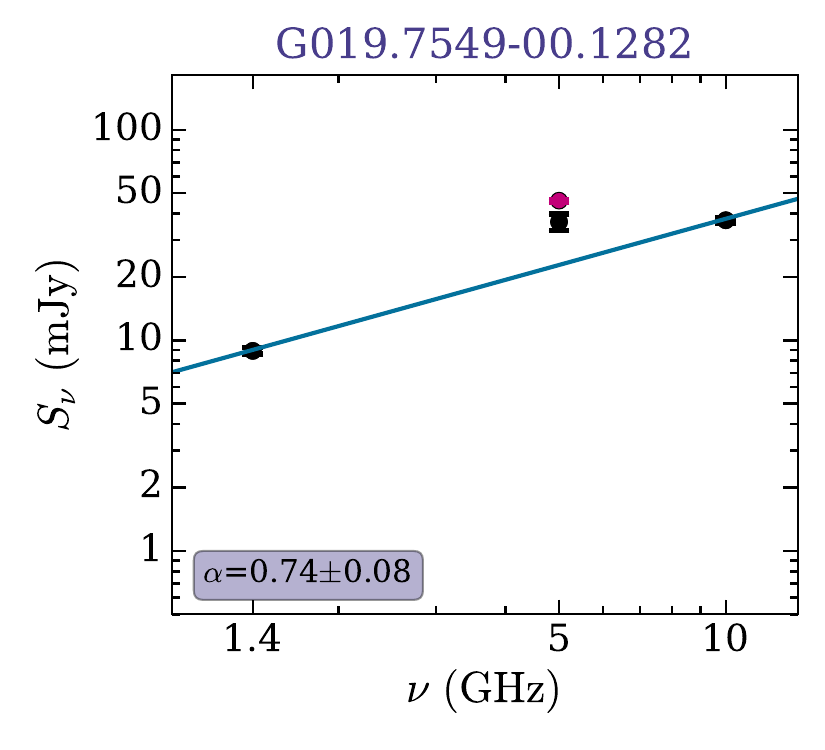} \hfill
\includegraphics[width=0.26\textwidth]{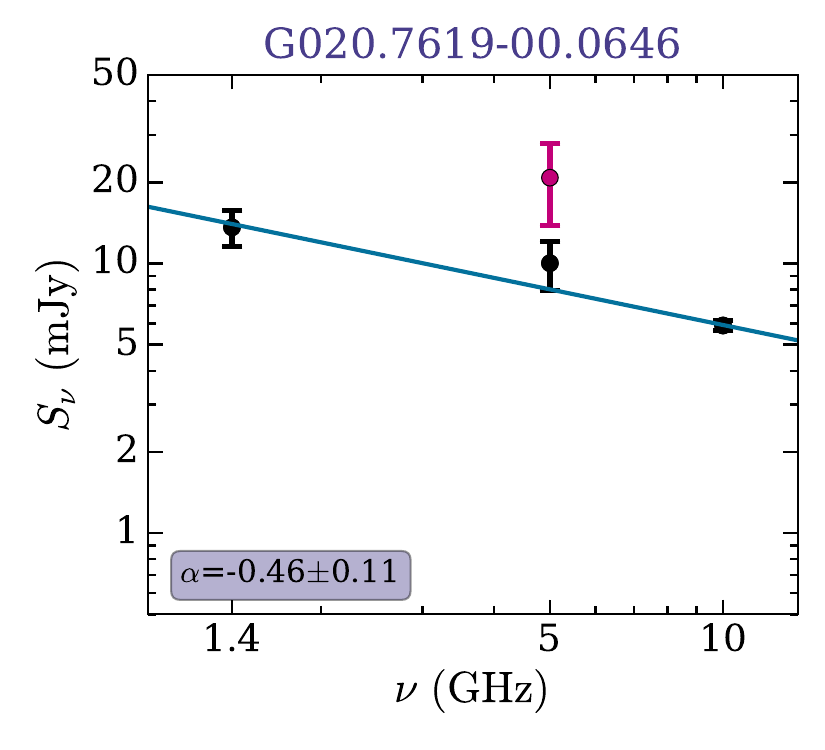}
\\[2mm]
\includegraphics[width=0.26\textwidth]{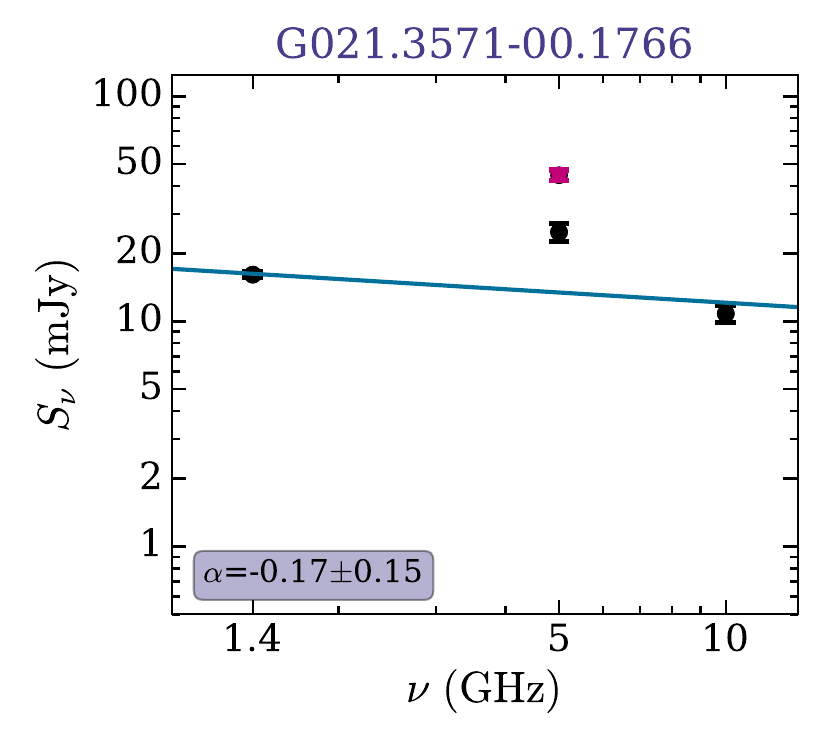} \hfill
\includegraphics[width=0.26\textwidth]{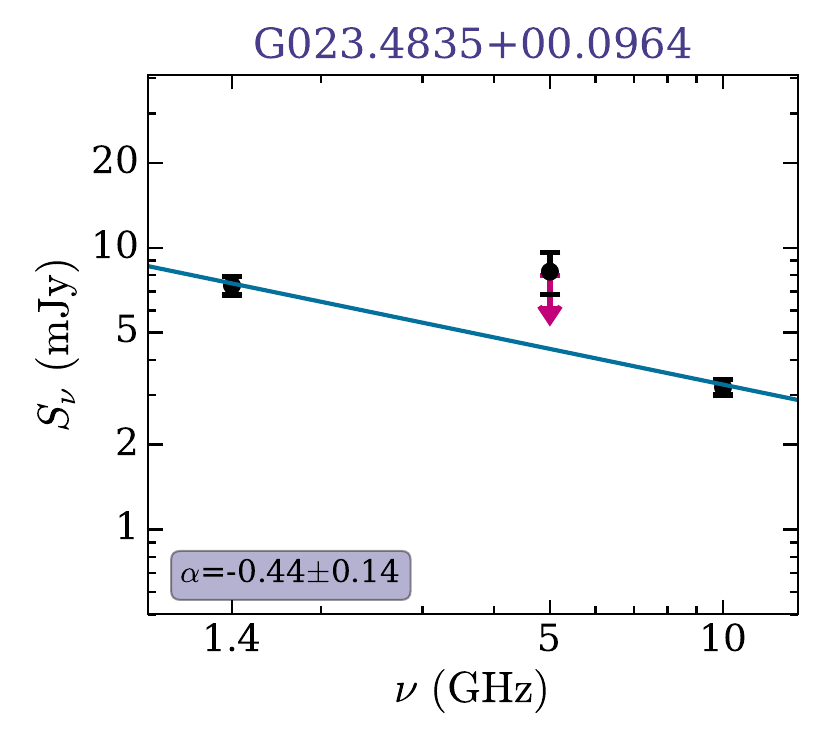} \hfill
\includegraphics[width=0.26\textwidth]{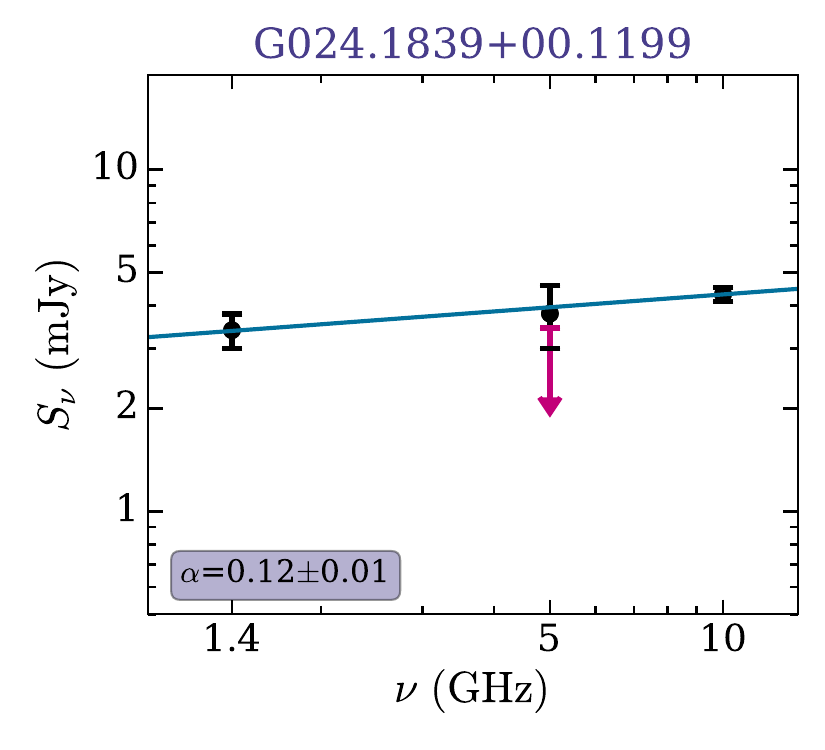}
\\[2mm]
\includegraphics[width=0.26\textwidth]{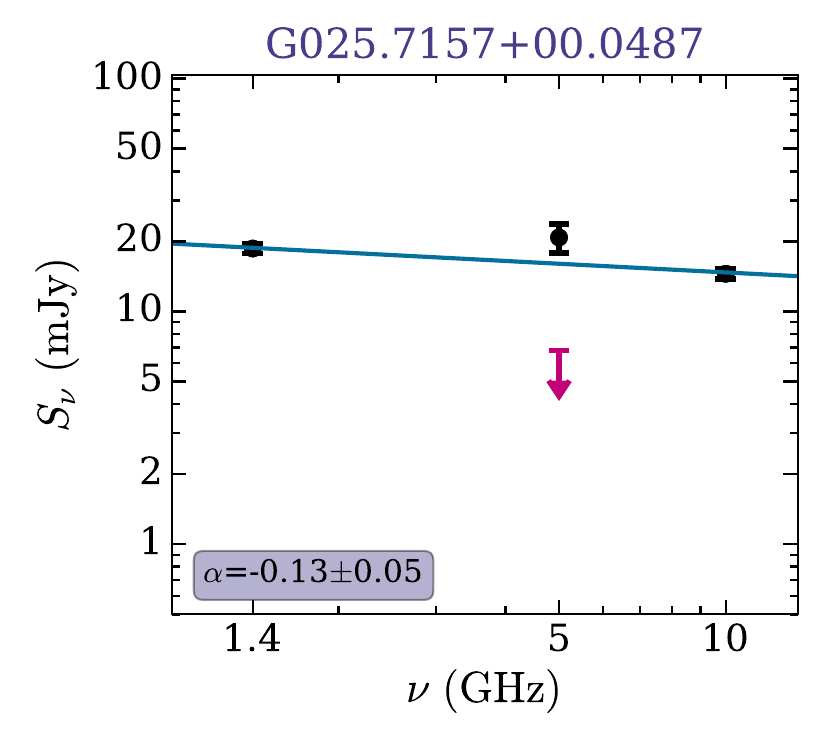} \hfill
\includegraphics[width=0.26\textwidth]{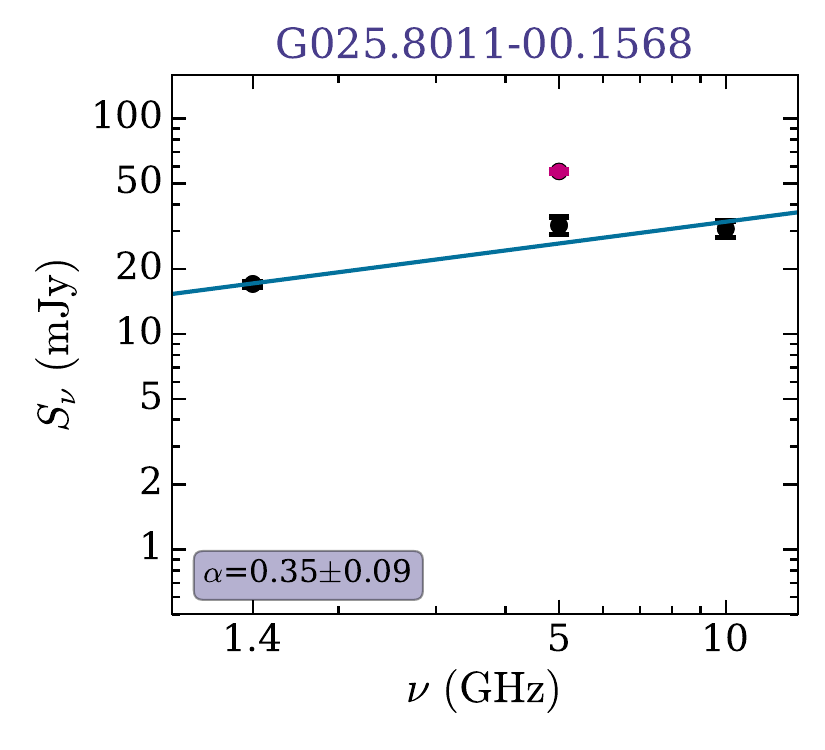} \hfill
\includegraphics[width=0.26\textwidth]{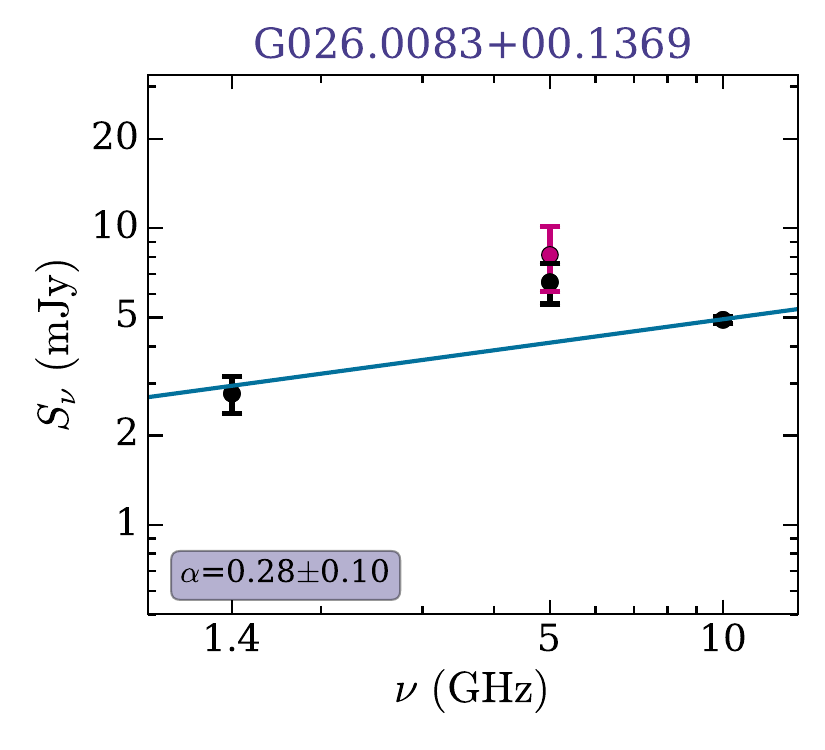}
\\[2mm]
\caption[Radio SEDs of JVLA targets]{Radio spectral energy distributions and fitted spectral indices of UCHIIs with available 1.4~GHz, 5~GHz (where the CORNISH fluxes are marked in black and the MAGPIS 6~cm results -- in red) and 10~GHz flux densities. The least-squares fitted line is plotted. Only 10~GHz targets which appear as single sources at all three frequencies are included.}
\label{fig:lsqfits}
\end{figure*}

\clearpage

\begin{figure*}
\ContinuedFloat
\centering
\includegraphics[width=0.26\textwidth]{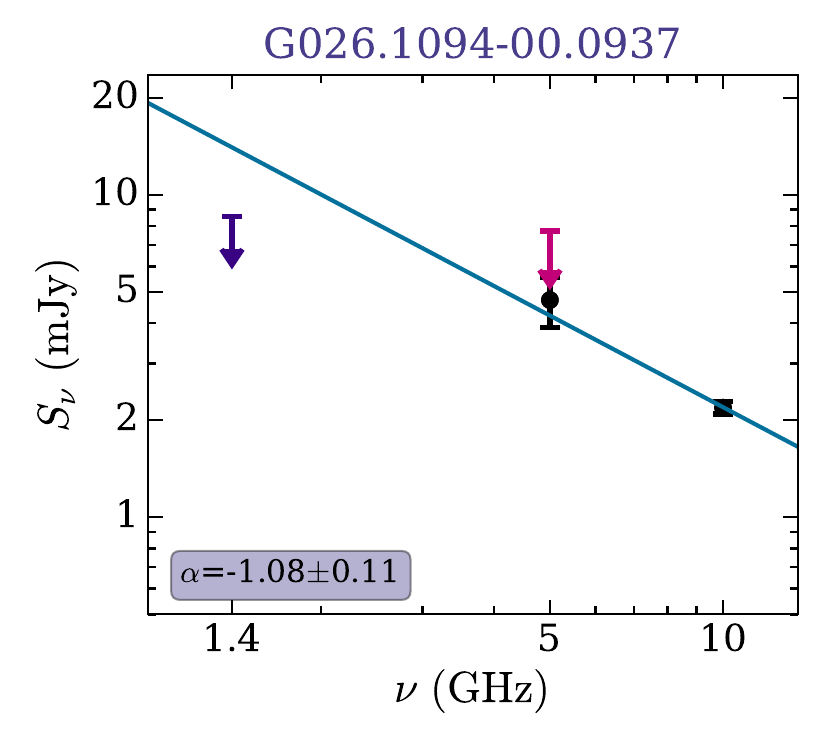} \hfill
\includegraphics[width=0.26\textwidth]{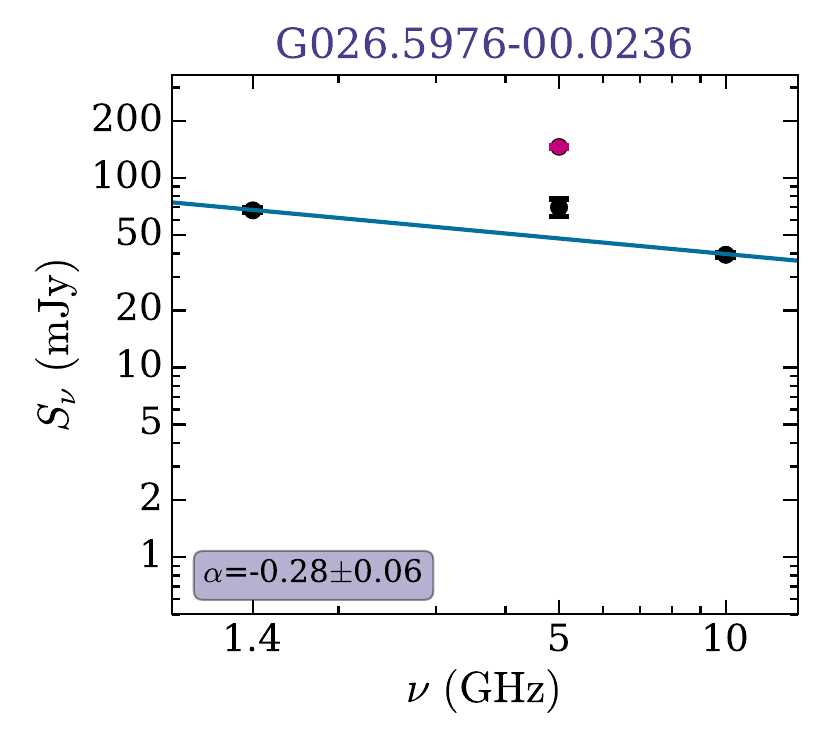} \hfill
\includegraphics[width=0.26\textwidth]{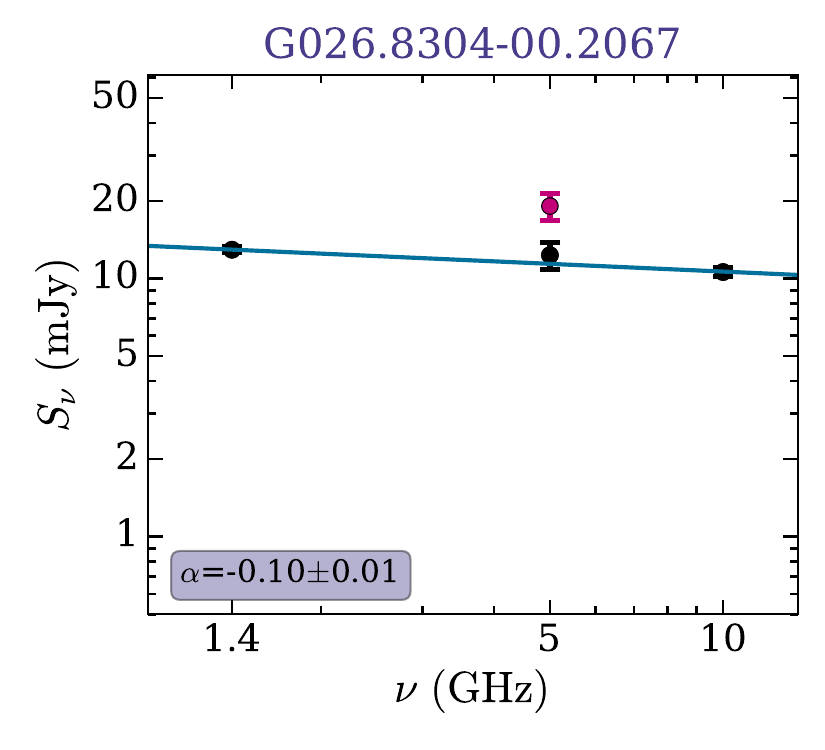}
\\[2mm]
\includegraphics[width=0.26\textwidth]{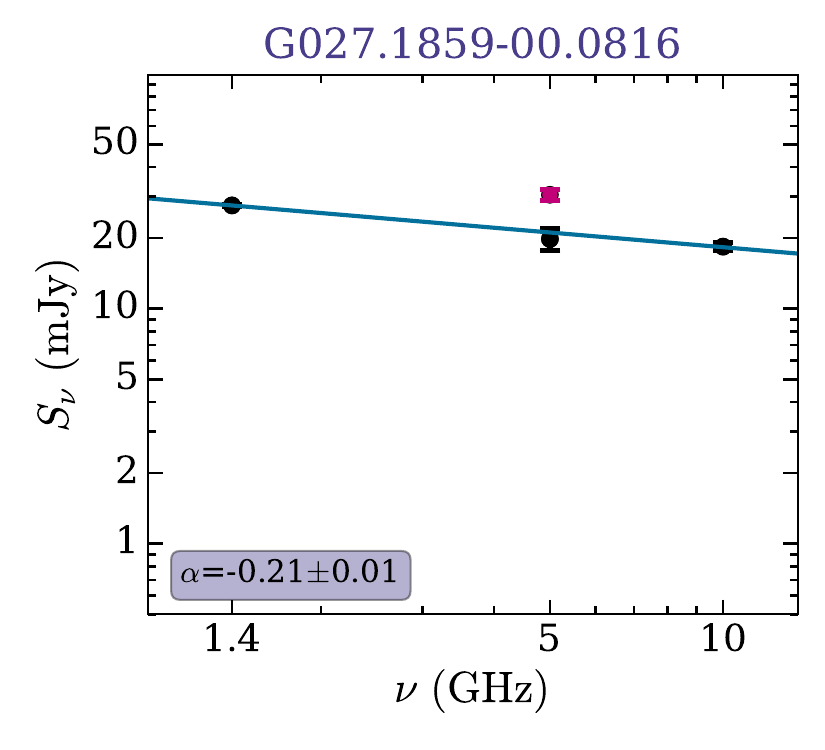} \hfill
\includegraphics[width=0.26\textwidth]{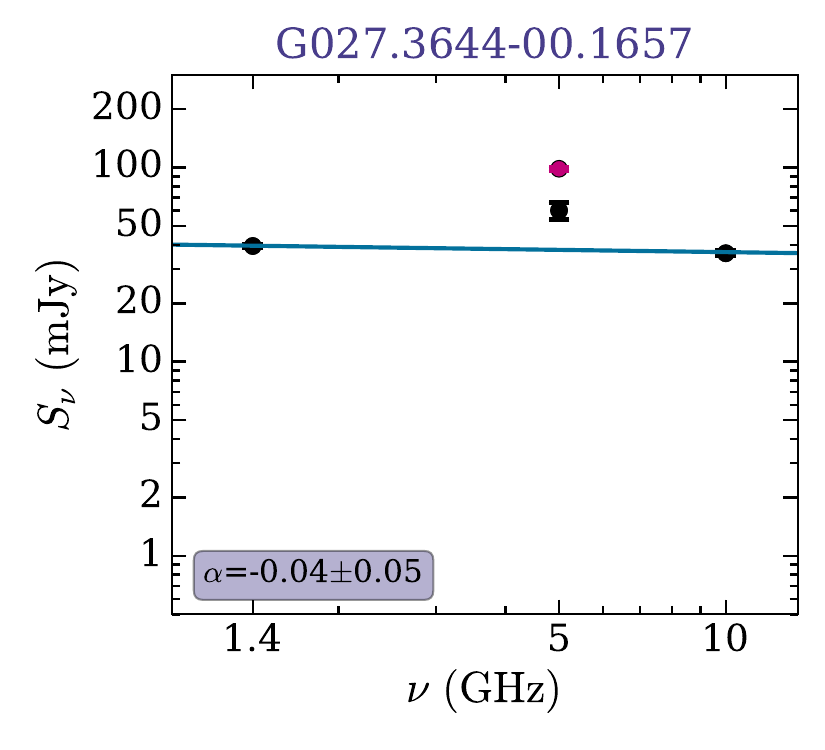} \hfill
\includegraphics[width=0.26\textwidth]{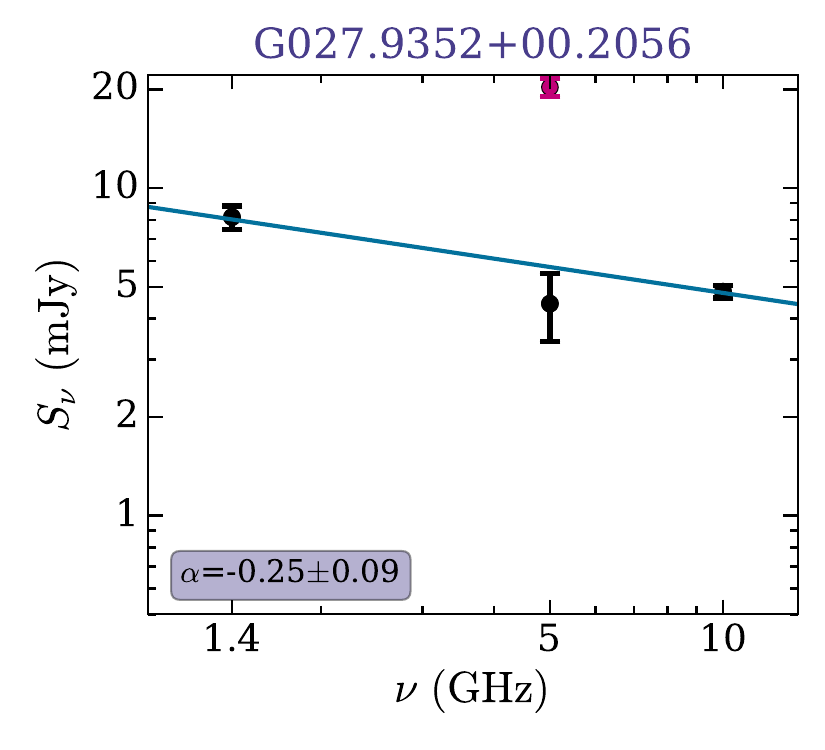}
\\[2mm]
\includegraphics[width=0.26\textwidth]{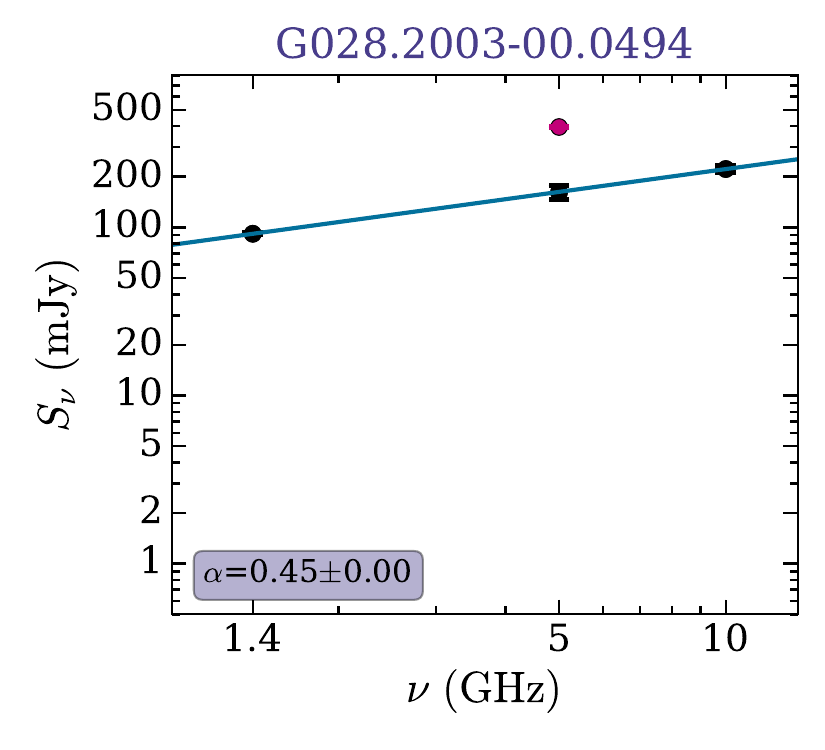} \hfill
\includegraphics[width=0.26\textwidth]{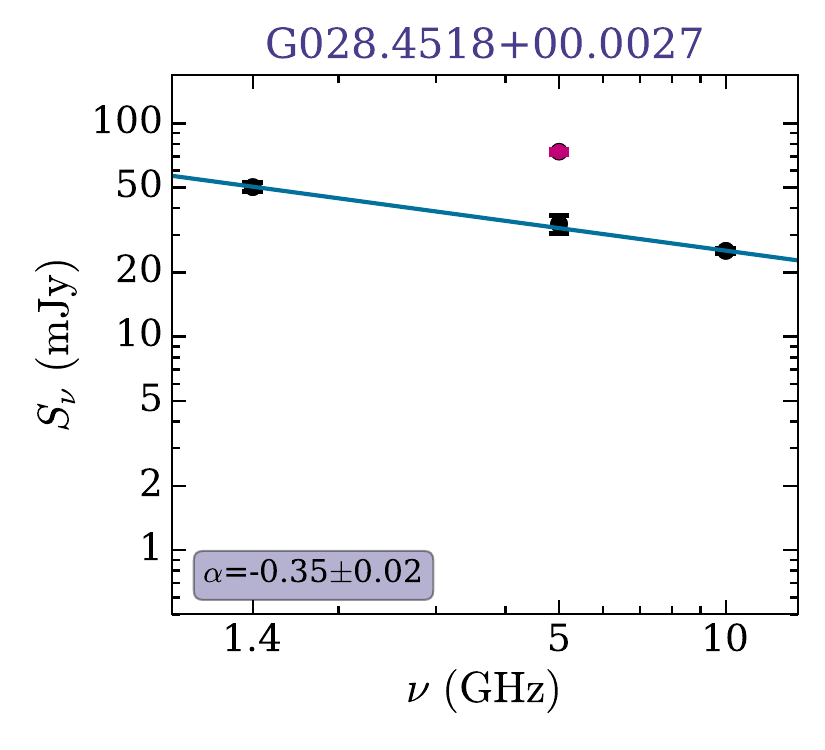} \hfill
\includegraphics[width=0.26\textwidth]{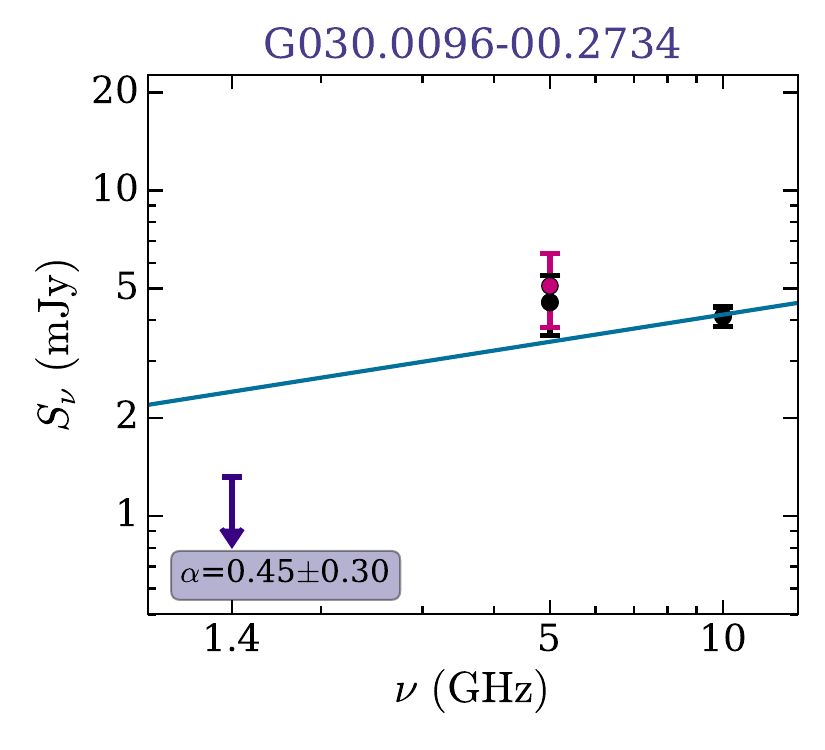}
\\[2mm]
\includegraphics[width=0.26\textwidth]{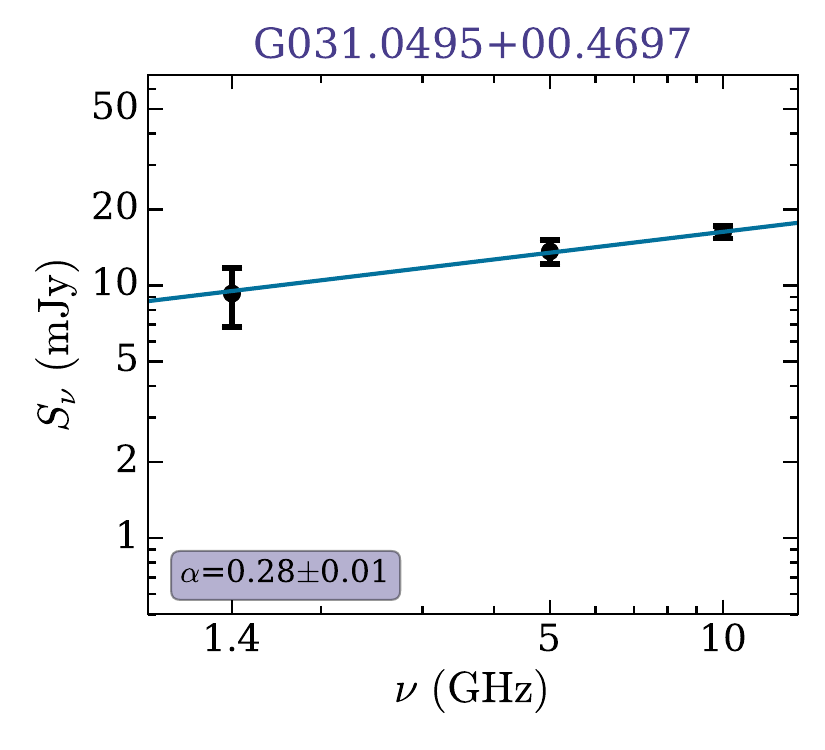} \hfill
\includegraphics[width=0.26\textwidth]{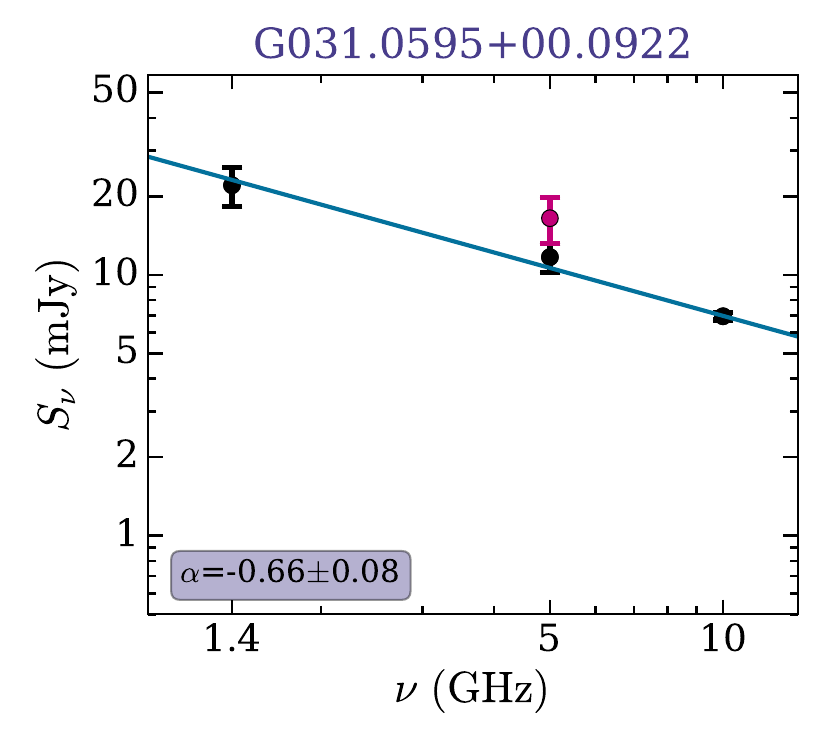} \hfill
\includegraphics[width=0.26\textwidth]{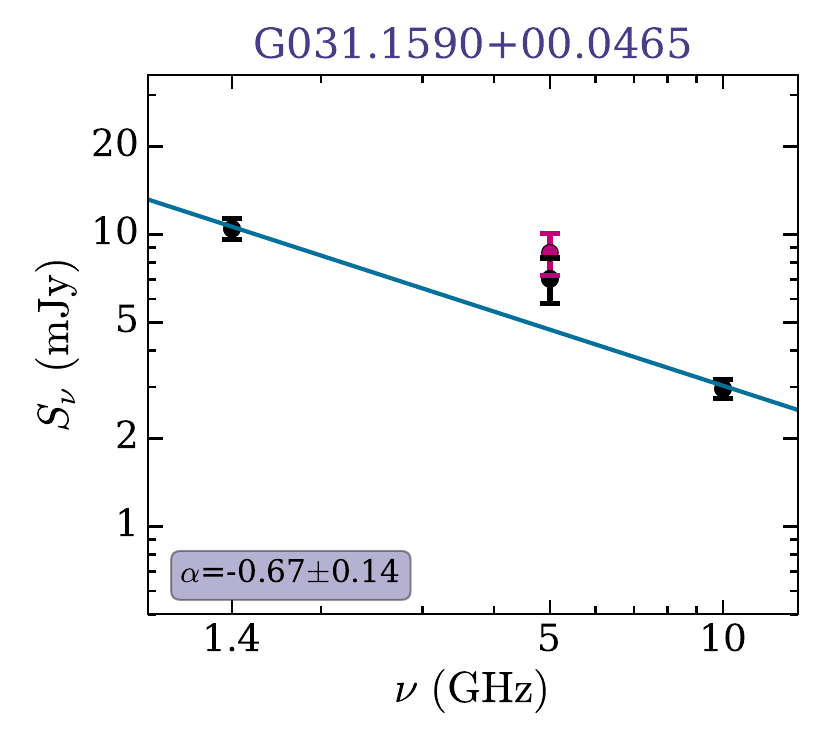}
\\[2mm]
\includegraphics[width=0.26\textwidth]{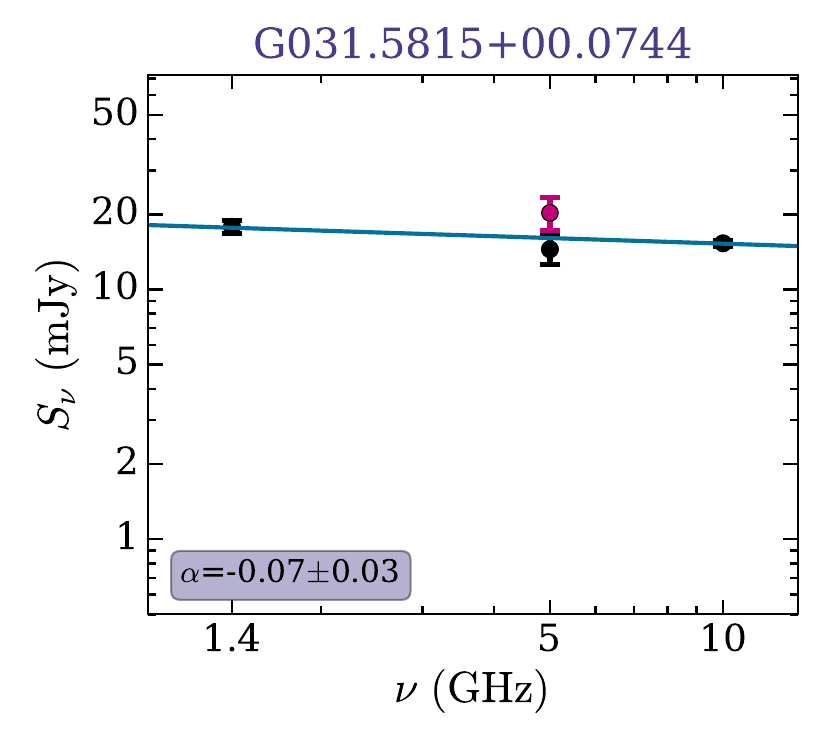} \hfill
\includegraphics[width=0.26\textwidth]{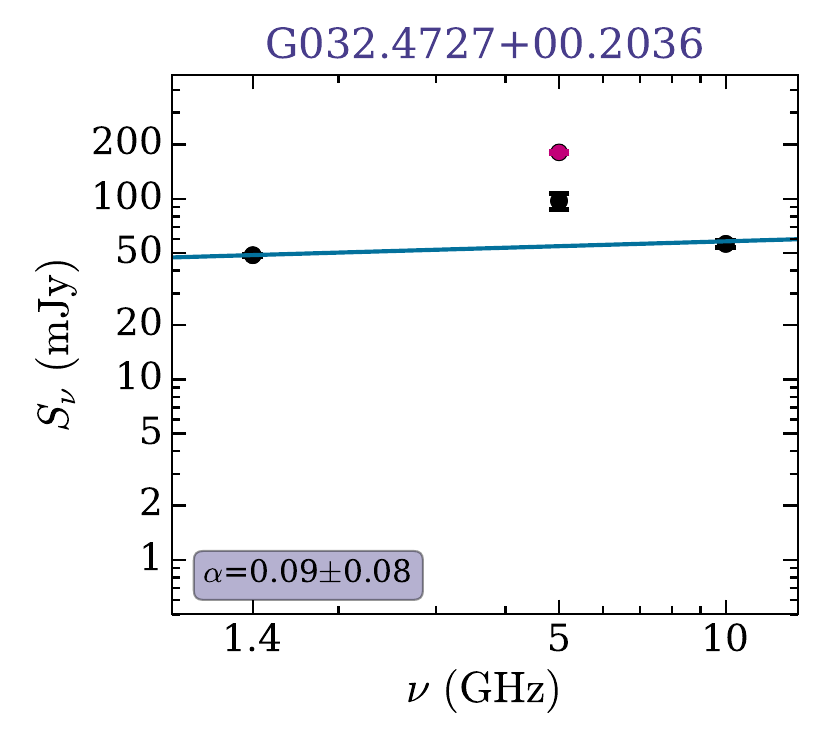} \hfill
\includegraphics[width=0.26\textwidth]{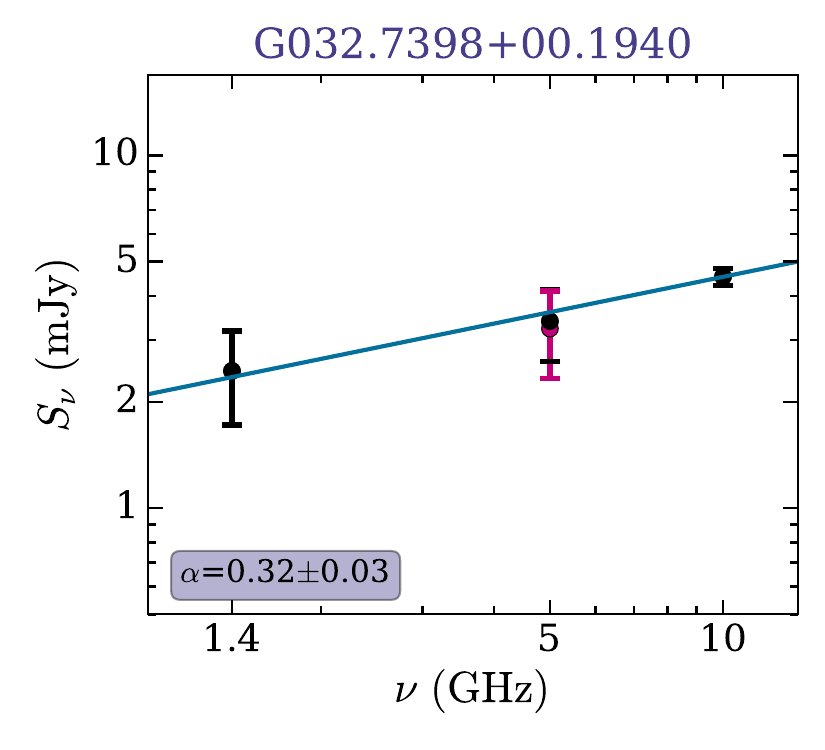}
\\[2mm]
\caption[Radio SEDs of JVLA targets, continued]{Radio spectral energy distributions, continued.}
\end{figure*}
\clearpage

\begin{figure*}
\ContinuedFloat
\centering
\includegraphics[width=0.26\textwidth]{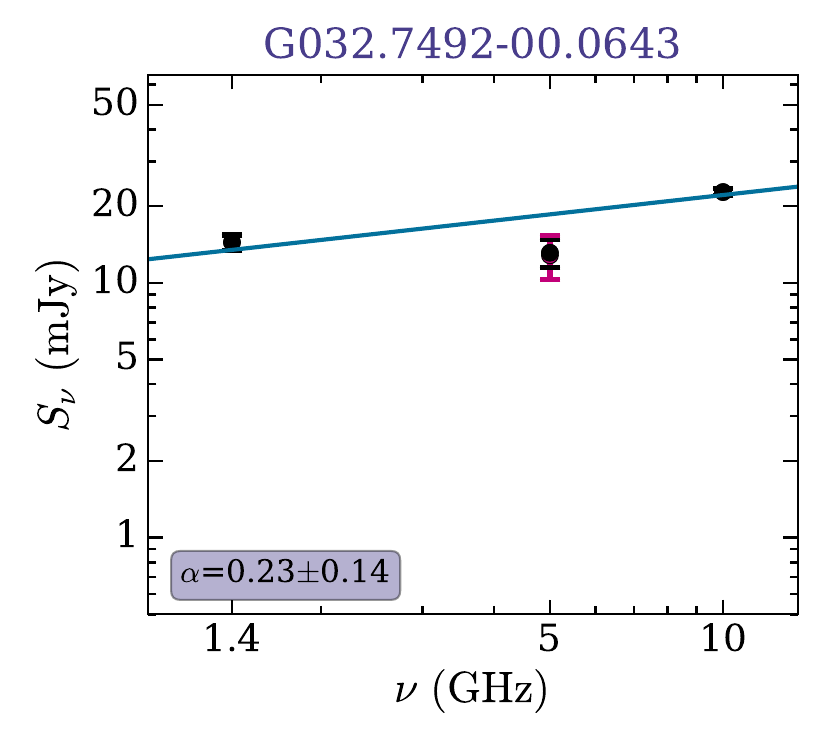} \hfill
\includegraphics[width=0.26\textwidth]{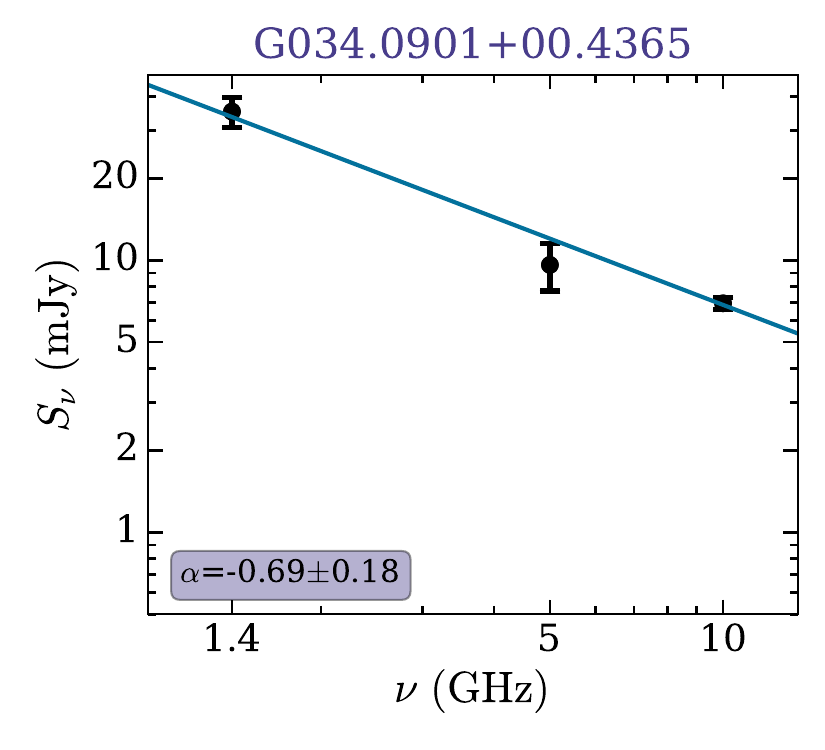} \hfill
\includegraphics[width=0.26\textwidth]{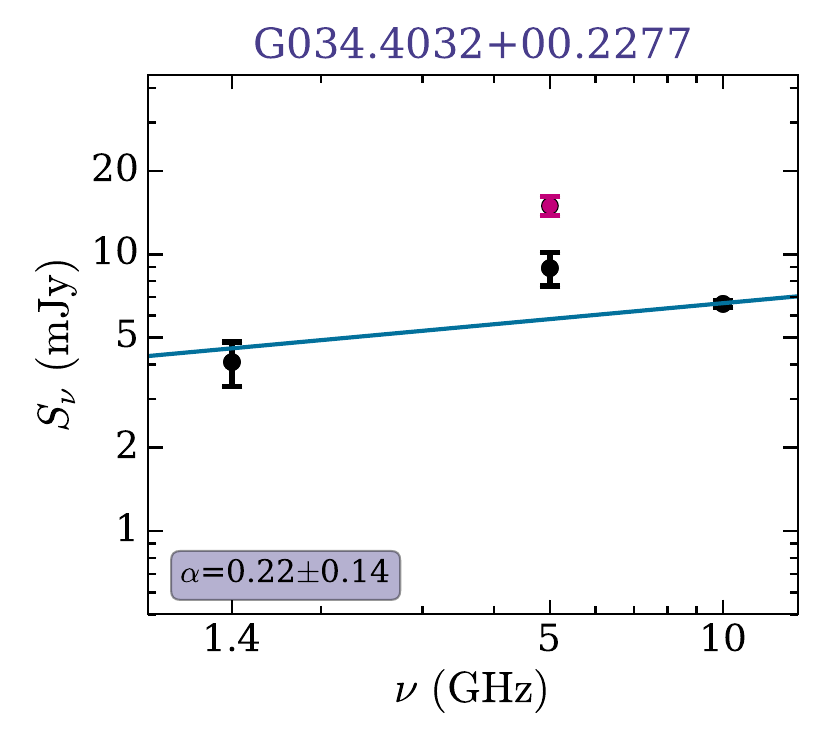}
\\[2mm]
\includegraphics[width=0.26\textwidth]{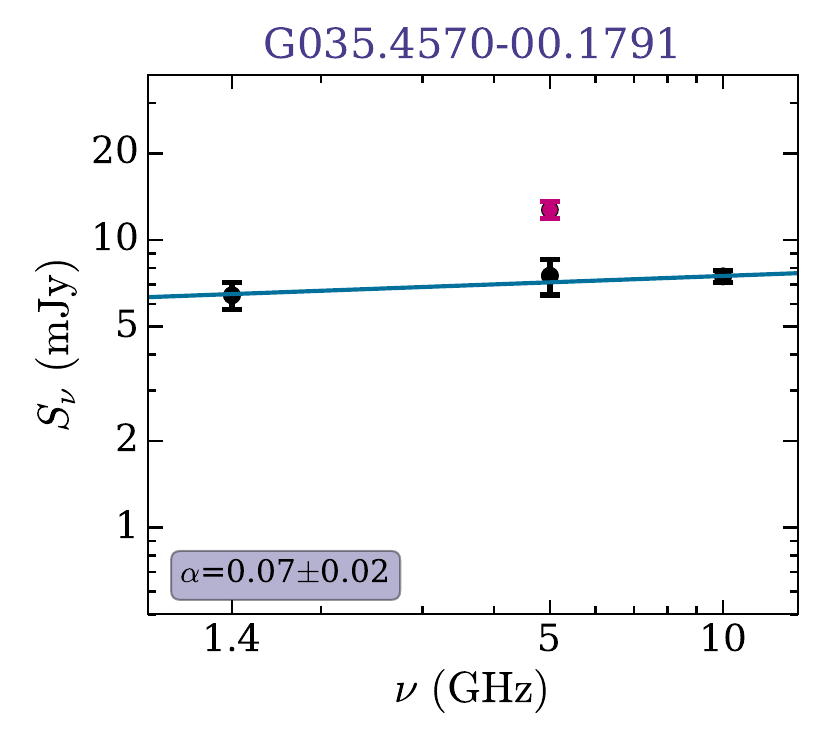} \hfill
\includegraphics[width=0.26\textwidth]{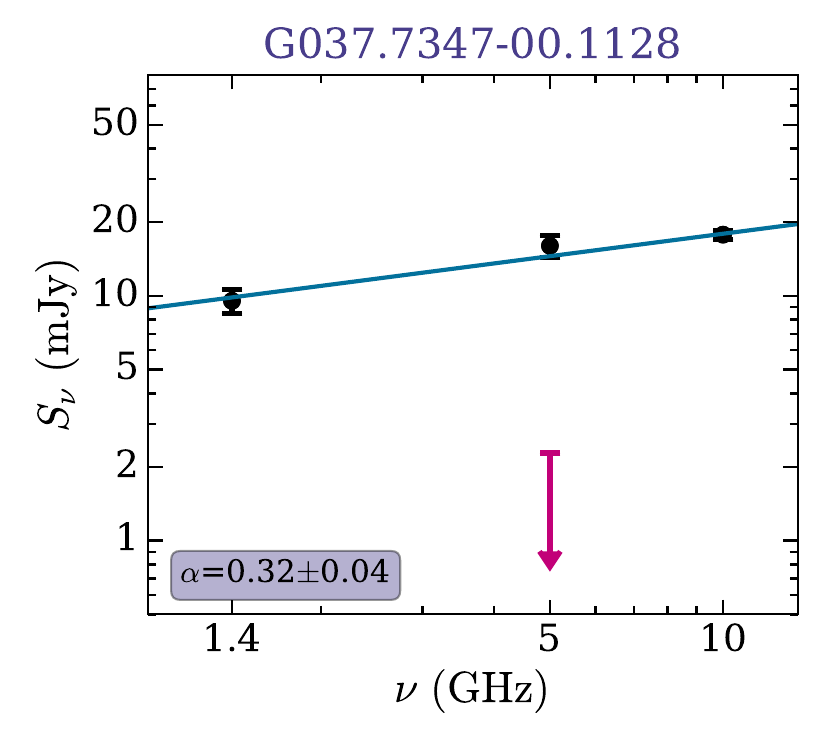} \hfill
\includegraphics[width=0.26\textwidth]{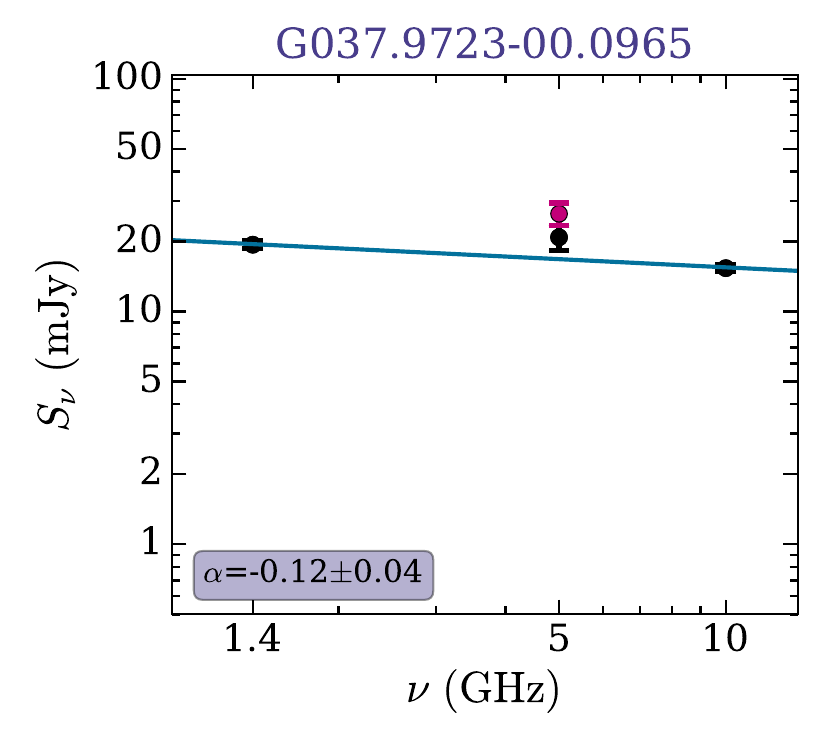}
\\[2mm]
\includegraphics[width=0.26\textwidth]{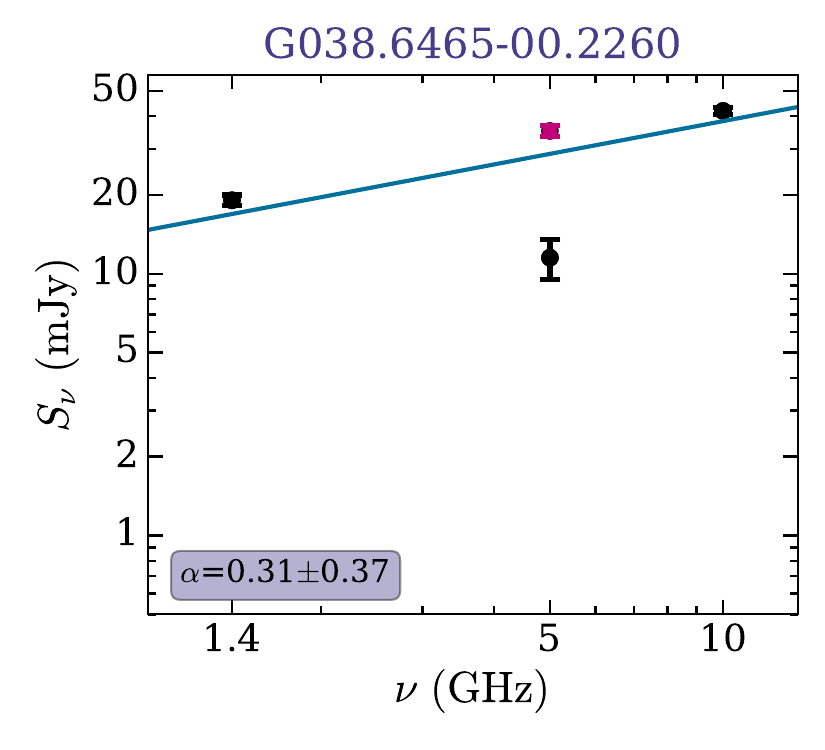} \hfill
\includegraphics[width=0.26\textwidth]{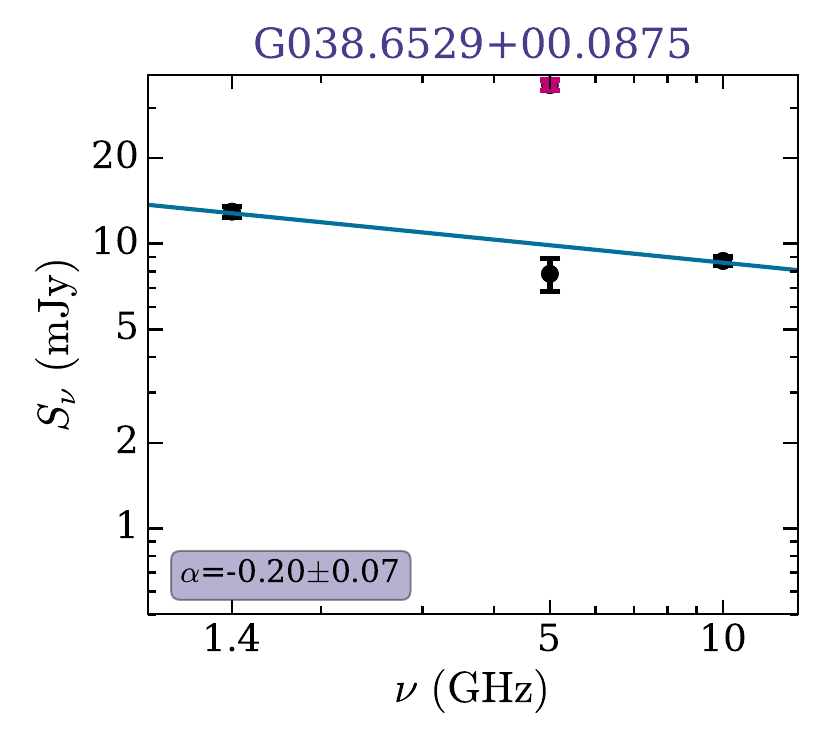} \hfill
\includegraphics[width=0.26\textwidth]{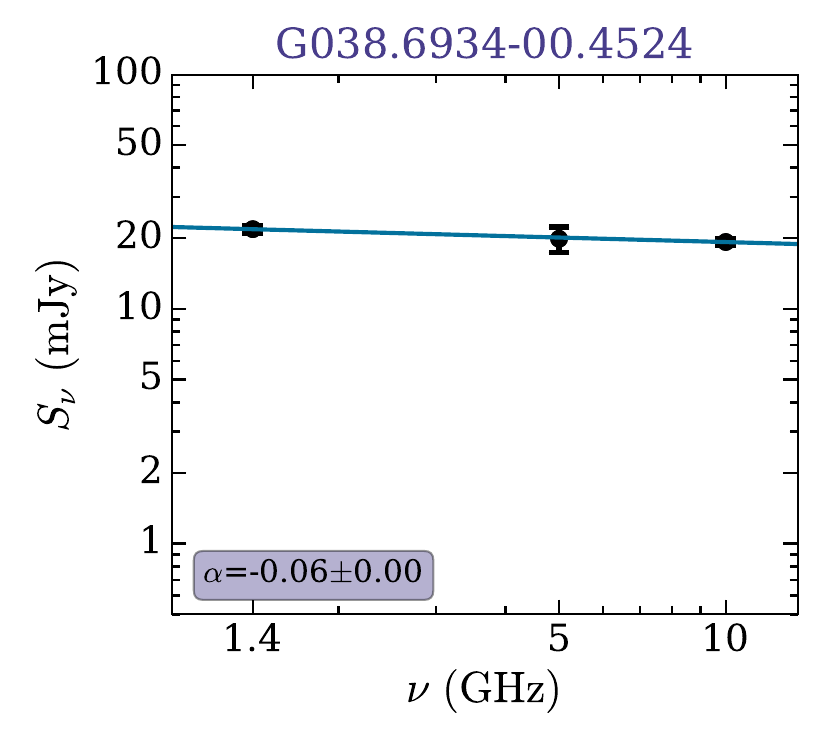}
\\[2mm]
\includegraphics[width=0.26\textwidth]{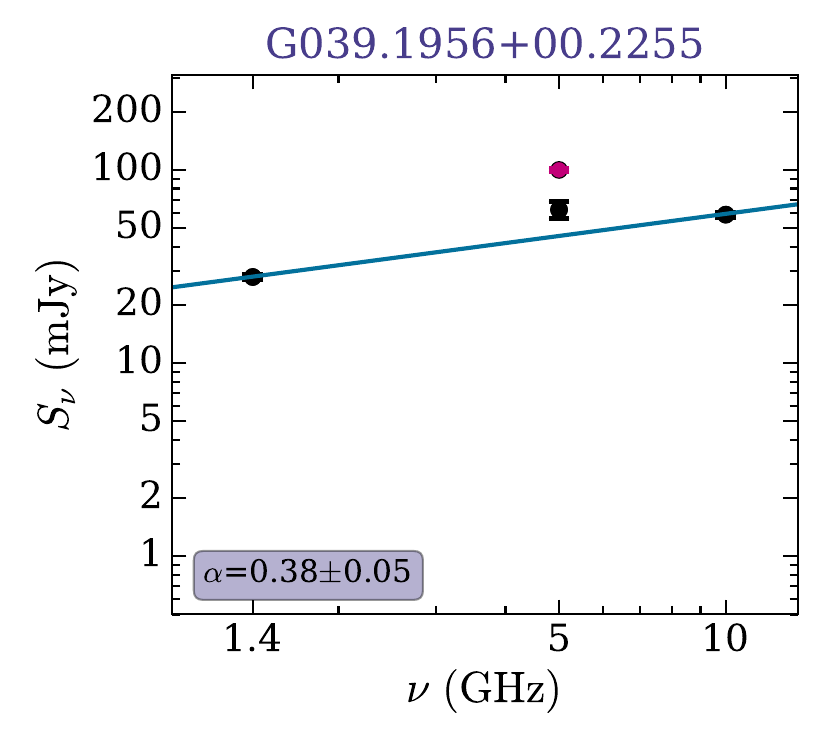} \hfill
\includegraphics[width=0.26\textwidth]{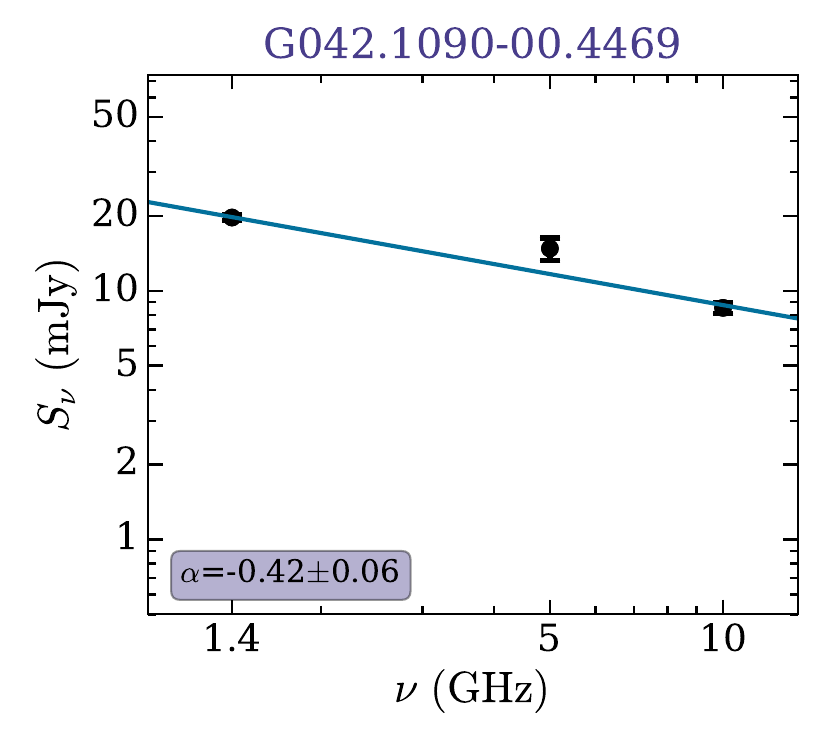} \hfill
\includegraphics[width=0.26\textwidth]{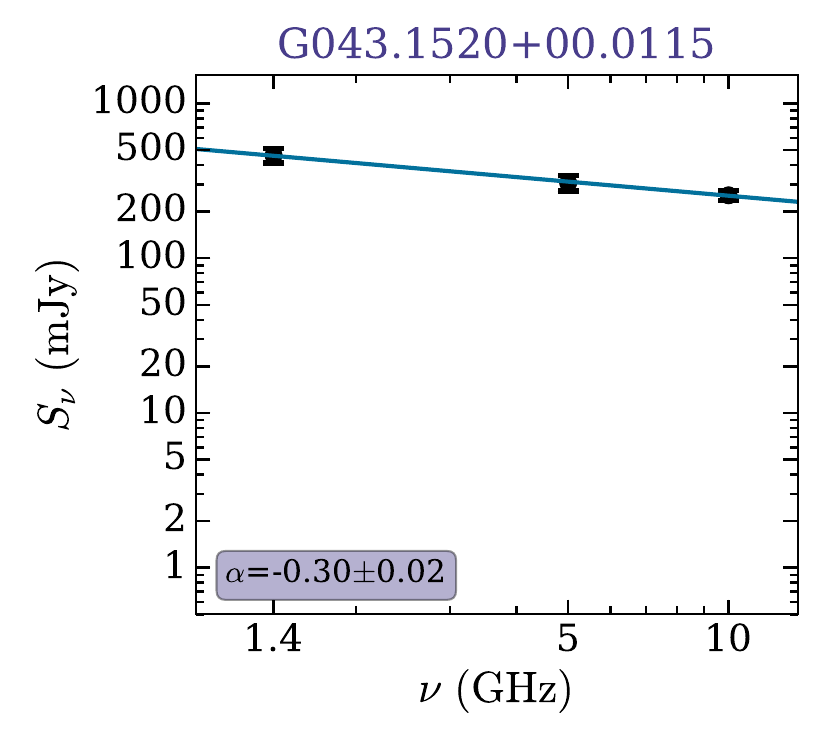}
\\[2mm]
\includegraphics[width=0.26\textwidth]{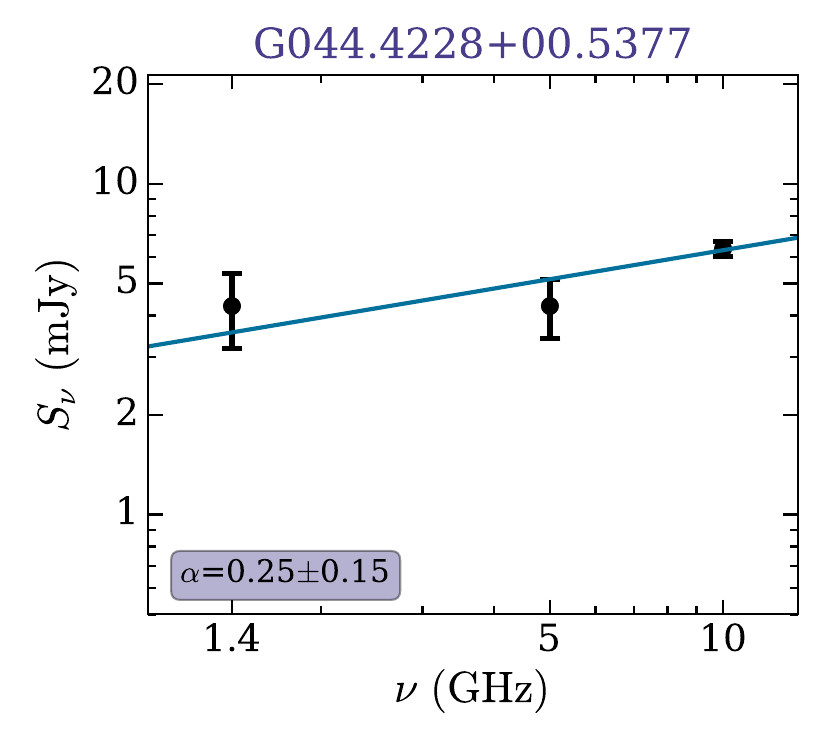} \hspace{6em}%
\includegraphics[width=0.26\textwidth]{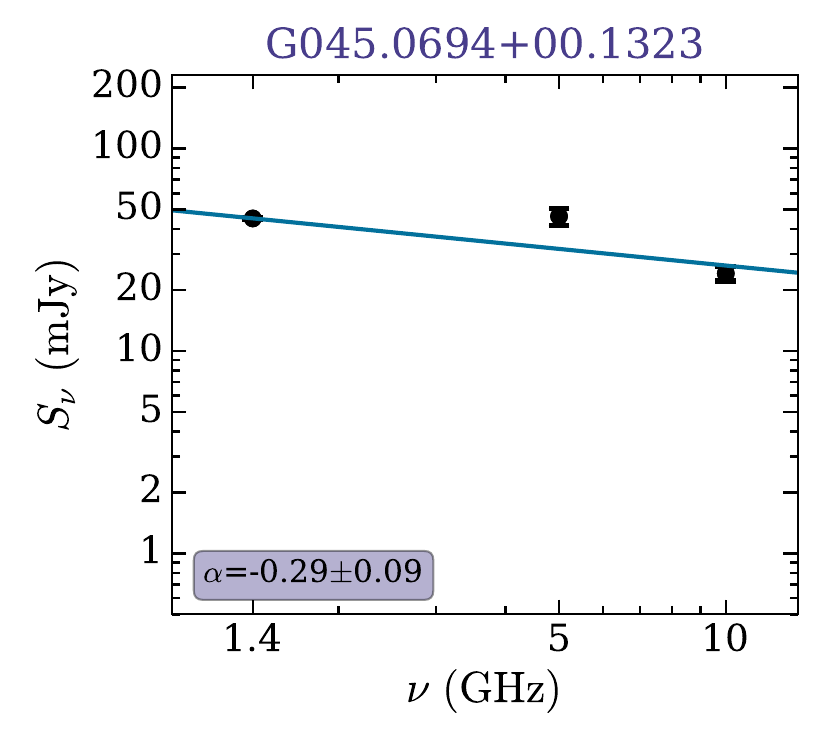}
\\[2mm]
\caption[Radio SEDs of JVLA targets, continued]{Radio spectral energy distributions, continued.}
\end{figure*}
\clearpage

\onecolumn
\begin{landscape}
\phantomsection
\refstepcounter{section}\label{app:appendixD}
\addcontentsline{toc}{section}{\thesection\ JVLA follow-up -- photometric results}
{\bfseries APPENDIX D: JVLA FOLLOW-UP -- PHOTOMETRIC RESULTS}\par\vspace{4mm}
\setlength\LTcapwidth{\linewidth}
\begin{longtable}{lll S[table-format=1.2] S[table-format=1.2] S[table-format=1.2(2)] S[table-format=1.2(2)] S[table-format=3.1(3.1)] S[table-format=1.2(2)] S[table-format=3.2(2.2)] S[table-format=2.2(1.2)]}
\caption[Photometric results for the higher-resolution data targets]{Photometric results for the higher-resolution data targets at 10~GHz. New components of CORNISH sources that were previously merged into single unresolved sources have been measured as individual sources, where appropriate, and assigned a letter for distinction of the name. The measured source coordinates, clean beam axes, deconvolved source axes and position angles, source diameters (geometric mean of the deconvolved axes), integrated and peak fluxes are presented, in this order. $^{\dagger}$Residual rms of the Gaussian fit exceeds 15 per cent of the peak flux density. $^{\ddagger}$Flux densities corrected for primary-beam attenuation (W49A field). For the unresolved G027.9352+00.2056, the convolved sizes are given as upper limits.}
\label{table:fluxes_evla} \\
\toprule
  \multicolumn{1}{l}{Name} &
  \multicolumn{1}{c}{$\alpha~\rm (J2000)$} &
  \multicolumn{1}{c}{$\delta~\rm (J2000)$} &
  \multicolumn{1}{c}{$\theta_{\rm _{beam}^{Maj}}$} &
  \multicolumn{1}{c}{$\theta_{\rm _{beam}^{min}}$} &
  \multicolumn{1}{c}{$\theta_{\rm M}$} &
  \multicolumn{1}{c}{$\theta_{\rm m}$} &
  \multicolumn{1}{c}{$\theta_{\rm PA}$} &
  \multicolumn{1}{c}{$\theta_{\rm source}$} &
  \multicolumn{1}{c}{$F_{\rm int}$} &
  \multicolumn{1}{c}{$F_{\rm peak}$} \\

  \multicolumn{1}{l}{} &
  \multicolumn{1}{c}{(hh:mm:ss)} &
  \multicolumn{1}{c}{(dd:mm:ss)} &
  \multicolumn{1}{c}{(\arcsec)} &
  \multicolumn{1}{c}{(\arcsec)} &
  \multicolumn{1}{c}{(\arcsec)} &
  \multicolumn{1}{c}{(\arcsec)} &
  \multicolumn{1}{c}{($\degree$)} &
  \multicolumn{1}{c}{(\arcsec)} &
  \multicolumn{1}{c}{(mJy)} &
  \multicolumn{1}{c}{($\rm mJy~beam^{-1}$)} \\

\hline
\endfirsthead

\multicolumn{11}{c}%
{{\bfseries \tablename\ \thetable{} -- continued from previous page}} \\
\toprule
  \multicolumn{1}{l}{Name} &
  \multicolumn{1}{c}{$\alpha~\rm (J2000)$} &
  \multicolumn{1}{c}{$\delta~\rm (J2000)$} &
  \multicolumn{1}{c}{$\theta_{\rm _{beam}^{Maj}}$} &
  \multicolumn{1}{c}{$\theta_{\rm _{beam}^{min}}$} &
  \multicolumn{1}{c}{$\theta_{\rm M}$} &
  \multicolumn{1}{c}{$\theta_{\rm m}$} &
  \multicolumn{1}{c}{$\theta_{\rm PA}$} &
  \multicolumn{1}{c}{$\theta_{\rm source}$} &
  \multicolumn{1}{c}{$F_{\rm int}$} &
  \multicolumn{1}{c}{$F_{\rm peak}$} \\

  \multicolumn{1}{l}{} &
  \multicolumn{1}{c}{(hh:mm:ss)} &
  \multicolumn{1}{c}{(dd:mm:ss)} &
  \multicolumn{1}{c}{(\arcsec)} &
  \multicolumn{1}{c}{(\arcsec)} &
  \multicolumn{1}{c}{(\arcsec)} &
  \multicolumn{1}{c}{(\arcsec)} &
  \multicolumn{1}{c}{(\arcsec)} &
  \multicolumn{1}{c}{(\arcsec)} &
  \multicolumn{1}{c}{(mJy)} &
  \multicolumn{1}{c}{($\rm mJy~beam^{-1}$)} \\

\hline
\endhead

\hline \multicolumn{11}{r}{{Continued on next page}} \\ \hline
\endfoot

\endlastfoot

G016.1448+00.0088       &  18:20:04.58 & $-$14:53:25.43 &             0.29 &             0.17 &    0.70 \pm 0.03 &    0.43 \pm 0.02 &     53.6 \pm 3.5 &    0.55 \pm 0.02 &   12.82 \pm 0.45 &    1.64 \pm 0.05 \\
G017.0299$-$00.0696     &  18:22:05.08 & $-$14:08:48.90 &             0.28 &             0.17 &    0.61 \pm 0.04 &    0.45 \pm 0.03 &    164.7 \pm 8.4 &    0.52 \pm 0.02 &    2.03 \pm 0.11 &    0.30 \pm 0.01 \\
G017.5549+00.1654       &  18:22:14.79 & $-$13:34:22.10 &             0.28 &             0.17 &    1.58 \pm 0.06 &    1.42 \pm 0.06 &       156 \pm 20 &    1.50 \pm 0.04 &    5.98 \pm 0.23 &    0.12 \pm 0.00 \\
G018.4433$-$00.0056     &  18:24:34.65 & $-$12:52:05.58 &             0.28 &             0.17 &    2.34 \pm 0.06 &    2.21 \pm 0.06 &       121 \pm 26 &    2.27 \pm 0.04 &     97.1 \pm 2.4 &    0.88 \pm 0.02 \\
G018.6654+00.0294a      &  18:24:52.60 & $-$12:39:20.51 &             0.27 &             0.17 &    0.55 \pm 0.04 &    0.29 \pm 0.03 &     16.2 \pm 5.3 &    0.40 \pm 0.02 &     3.21 \pm 0.2 &    0.69 \pm 0.04 \\
G018.6654+00.0294b      &  18:24:52.57 & $-$12:39:20.55 &             0.27 &             0.17 &    0.23 \pm 0.03 &    0.19 \pm 0.03 &       119 \pm 41 &    0.21 \pm 0.02 &    0.75 \pm 0.04 &    0.37 \pm 0.02 \\
G018.7106+00.0002       &  18:25:04.15 & $-$12:37:45.82 &             0.28 &             0.17 &    0.89 \pm 0.03 &    0.78 \pm 0.02 &       114 \pm 12 &    0.83 \pm 0.02 &    101.2 \pm 2.5 &    6.43 \pm 0.15 \\
G018.7612+00.2630       &  18:24:12.88 & $-$12:27:41.73 &             0.28 &             0.17 &    0.87 \pm 0.02 &    0.76 \pm 0.02 &     12.8 \pm 9.6 &    0.81 \pm 0.01 &   44.71 \pm 0.97 &    2.93 \pm 0.06 \\
G018.8250$-$00.4675     &  18:26:59.04 & $-$12:44:46.54 &             0.27 &             0.17 &    1.68 \pm 0.08 &    1.39 \pm 0.07 &       105 \pm 11 &    1.53 \pm 0.05 &    7.85 \pm 0.38 &    0.15 \pm 0.01 \\
G019.0767$-$00.2882     &  18:26:48.87 & $-$12:26:23.79 &             0.28 &             0.17 &    1.21 \pm 0.05 &    1.12 \pm 0.04 &        84 \pm 25 &    1.16 \pm 0.03 &         91 \pm 3 &    3.03 \pm 0.10 \\
G019.7549$-$00.1282     &  18:27:31.60 & $-$11:45:55.16 &             0.26 &             0.17 &    0.45 \pm 0.02 &    0.40 \pm 0.01 &       156 \pm 16 &    0.42 \pm 0.01 &     37.2 \pm 1.1 &    7.28 \pm 0.19 \\
G020.0797$-$00.1337     &  18:28:09.79 & $-$11:28:48.53 &             0.27 &             0.17 &    0.96 \pm 0.07 &    0.92 \pm 0.07 &        79 \pm 84 &    0.94 \pm 0.05 &   14.07 \pm 0.65 &    0.69 \pm 0.03 \\
G020.7619$-$00.0646     &  18:29:12.35 & $-$10:50:38.47 &             0.26 &             0.17 &    1.98 \pm 0.09 &    1.29 \pm 0.06 &       82.6 \pm 4 &    1.59 \pm 0.05 &    5.89 \pm 0.25 &    0.10 \pm 0.00 \\
G020.9636$-$00.0744a    &  18:29:37.31 & $-$10:40:10.76 &             0.26 &             0.18 &    3.15 \pm 0.09 &    2.15 \pm 0.06 &       98.1 \pm 3 &    2.60 \pm 0.05 &   19.21 \pm 0.55 &    0.13 \pm 0.00 \\
G020.9636$-$00.0744b    &  18:29:37.32 & $-$10:40:12.56 &             0.26 &             0.18 &    0.65 \pm 0.06 &    0.56 \pm 0.05 &       116 \pm 30 &    0.60 \pm 0.04 &    2.59 \pm 0.18 &    0.28 \pm 0.02 \\
G021.3571$-$00.1766     &  18:30:43.92 & $-$10:22:06.86 &             0.25 &             0.17 &    0.71 \pm 0.07 &    0.45 \pm 0.05 &    139.2 \pm 9.1 &    0.57 \pm 0.04 &   10.83 \pm 0.92 &    1.28 \pm 0.10 \\
G023.4835+00.0964       &  18:33:44.05 & $-$08:21:23.43 &             0.24 &             0.17 &    0.82 \pm 0.05 &    0.56 \pm 0.04 &    171.2 \pm 7.2 &    0.68 \pm 0.03 &      3.2 \pm 0.2 &    0.27 \pm 0.02 \\
G024.1839+00.1199       &  18:34:57.19 & $-$07:43:26.16 &             0.24 &             0.17 &    0.85 \pm 0.04 &    0.65 \pm 0.03 &     91.8 \pm 7.6 &    0.74 \pm 0.03 &     4.32 \pm 0.2 &    0.30 \pm 0.01 \\
G025.3970+00.5614       &  18:35:37.44 & $-$06:26:38.34 &             0.24 &             0.17 &    0.78 \pm 0.04 &    0.67 \pm 0.03 &       155 \pm 18 &    0.72 \pm 0.03 &    119.3 \pm 5.5 &    8.76 \pm 0.38 \\
G025.7157+00.0487       &  18:38:02.78 & $-$06:23:47.42 &             0.24 &             0.17 &    1.25 \pm 0.06 &    0.87 \pm 0.04 &    153.8 \pm 5.7 &    1.04 \pm 0.04 &   14.51 \pm 0.72 &    0.54 \pm 0.03 \\
G025.8011$-$00.1568     &  18:38:56.37 & $-$06:24:53.42 &             0.24 &             0.18 &    0.98 \pm 0.10 &    0.92 \pm 0.09 &       125 \pm 73 &    0.95 \pm 0.07 &     30.8 \pm 2.7 &    1.39 \pm 0.12 \\
G026.0083+00.1369       &  18:38:16.28 & $-$06:05:45.46 &             0.24 &             0.17 &    0.31 \pm 0.01 &    0.28 \pm 0.01 &        17 \pm 30 &    0.30 \pm 0.01 &    4.91 \pm 0.14 &    1.55 \pm 0.04 \\
G026.0916$-$00.0565     &  18:39:06.98 & $-$06:06:38.26 &             0.24 &             0.17 &    1.18 \pm 0.09 &    0.86 \pm 0.07 &       158 \pm 10 &    1.01 \pm 0.05 &     6.85 \pm 0.5 &    0.27 \pm 0.02 \\
G026.1094$-$00.0937     &  18:39:16.92 & $-$06:06:43.39 &             0.34 &             0.17 &    0.33 \pm 0.03 &    0.22 \pm 0.02 &       115 \pm 10 &    0.27 \pm 0.02 &    2.19 \pm 0.09 &    0.92 \pm 0.03 \\
G026.5976$-$00.0236     &  18:39:55.89 & $-$05:38:45.32 &             0.33 &             0.17 &    1.44 \pm 0.04 &    0.79 \pm 0.02 &    149.9 \pm 1.8 &    1.07 \pm 0.02 &     39.3 \pm 1.1 &    1.91 \pm 0.05 \\
G026.8304$-$00.2067     &  18:41:00.84 & $-$05:31:22.37 &             0.33 &             0.17 &    1.36 \pm 0.05 &    1.18 \pm 0.04 &       143 \pm 13 &    1.27 \pm 0.03 &   10.59 \pm 0.39 &    0.37 \pm 0.01 \\
G027.1859$-$00.0816     &  18:41:13.26 & $-$05:08:58.10 &             0.33 &             0.17 &    1.54 \pm 0.07 &    1.39 \pm 0.06 &        84 \pm 22 &    1.46 \pm 0.04 &   18.35 \pm 0.71 &    0.47 \pm 0.02 \\
G027.3644$-$00.1657     &  18:41:50.96 & $-$05:01:45.12 &             0.32 &             0.17 &    0.89 \pm 0.03 &    0.81 \pm 0.03 &       150 \pm 22 &    0.85 \pm 0.02 &     36.2 \pm 1.1 &    2.61 \pm 0.07 \\
G027.9352+00.2056       &  18:41:34.35 & $-$04:21:06.78 &             0.34 &             0.17 & {$<$0.33} & {$<$0.23} & {--} & {$<$0.28} &    4.83 \pm 0.21 &    3.77 \pm 0.10 \\
G028.2003$-$00.0494     &  18:42:58.11 & $-$04:13:57.59 &             0.34 &             0.17 &    0.93 \pm 0.04 &    0.57 \pm 0.03 &    148.9 \pm 3.6 &    0.73 \pm 0.02 &    222.3 \pm 9.3 &   22.18 \pm 0.84 \\
G028.4518+00.0027       &  18:43:14.60 & $-$03:59:06.59 &             0.34 &             0.17 &    1.83 \pm 0.06 &    1.63 \pm 0.05 &        72 \pm 12 &    1.73 \pm 0.04 &   25.23 \pm 0.73 &    0.49 \pm 0.01 \\
G030.0096$-$00.2734     &  18:47:04.72 & $-$02:43:31.23 &             0.33 &             0.17 &    0.26 \pm 0.04 &    0.17 \pm 0.07 &        15 \pm 21 &    0.21 \pm 0.04 &    4.11 \pm 0.28 &    2.09 \pm 0.10 \\
G030.5887$-$00.0428a    &  18:47:18.90 & $-$02:06:16.72 &             0.32 &             0.17 &    0.22 \pm 0.10 &    0.17 \pm 0.13 &       107 \pm 75 &    0.19 \pm 0.09 &     25.6 \pm 4.8 &   14.80 \pm 1.90 \\
G030.5887$-$00.0428b$^{\dagger}$    &  18:47:18.88 & $-$02:06:17.18 &             0.32 &             0.17 &    1.09 \pm 0.19 &    0.33 \pm 0.07 &    163.6 \pm 5.2 &    0.60 \pm 0.08 &        97 \pm 16 &   12.10 \pm 1.70 \\
G031.0495+00.4697       &  18:46:19.86 & $-$01:27:39.54 &             0.24 &             0.18 &    1.43 \pm 0.08 &    0.85 \pm 0.05 &     15.4 \pm 4.3 &    1.10 \pm 0.04 &   16.27 \pm 0.91 &    0.55 \pm 0.03 \\
G031.0595+00.0922       &  18:47:41.61 & $-$01:37:27.35 &             0.24 &             0.18 &    0.76 \pm 0.03 &    0.56 \pm 0.02 &     17.3 \pm 4.9 &    0.65 \pm 0.02 &    6.94 \pm 0.23 &    0.62 \pm 0.02 \\
G031.1590+00.0465       &  18:48:02.27 & $-$01:33:23.65 &             0.24 &             0.18 &    0.79 \pm 0.06 &    0.56 \pm 0.05 &       117 \pm 10 &    0.67 \pm 0.04 &    2.96 \pm 0.22 &    0.26 \pm 0.02 \\
G031.2435$-$00.1103     &  18:48:45.04 & $-$01:33:10.49 &             0.24 &             0.18 &    1.84 \pm 0.07 &    1.81 \pm 0.07 &       62 \pm 112 &    1.83 \pm 0.05 &       383 \pm 13 &    4.87 \pm 0.17 \\
G031.5815+00.0744       &  18:48:42.56 & $-$01:10:04.43 &             0.24 &             0.18 &    2.67 \pm 0.07 &    1.54 \pm 0.04 &    111.8 \pm 1.9 &    2.03 \pm 0.04 &     15.3 \pm 0.4 &    0.16 \pm 0.00 \\
G032.4727+00.2036       &  18:49:52.55 &   $-$00:18:56.63 &             0.25 &             0.18 &    1.07 \pm 0.05 &    0.58 \pm 0.03 &       92.5 \pm 3 &    0.78 \pm 0.03 &     56.3 \pm 2.6 &    3.68 \pm 0.16 \\
G032.7398+00.1940       &  18:50:23.83 &   $-$00:04:57.30 &             0.24 &             0.17 &    0.44 \pm 0.03 &    0.41 \pm 0.03 &       150 \pm 85 &    0.42 \pm 0.02 &    4.54 \pm 0.25 &    0.84 \pm 0.04 \\
G032.7492$-$00.0643     &  18:51:20.02 &  $-$00:11:30.89 &             0.24 &             0.18 &    2.09 \pm 0.07 &    2.04 \pm 0.07 &        20 \pm 70 &    2.07 \pm 0.05 &   22.72 \pm 0.75 &    0.23 \pm 0.01 \\
G032.7982+00.1937       &  18:50:30.27 & $-$00:01:49.76 &             0.24 &             0.18 &    2.21 \pm 0.21 &    0.80 \pm 0.08 &     23.2 \pm 3.2 &    1.33 \pm 0.14 &     16.8 \pm 1.6 &    0.39 \pm 0.04 \\
G033.8100$-$00.1864a    &  18:53:42.23 &   +00:41:47.32 &             0.24 &             0.18 &    0.68 \pm 0.05 &    0.26 \pm 0.03 &      179.4 \pm 3 &    0.42 \pm 0.03 &     39.3 \pm 2.9 &    7.28 \pm 0.45 \\
G033.8100$-$00.1864b    &  18:53:42.27 &   +00:41:45.71 &             0.24 &             0.18 &    0.70 \pm 0.05 &    0.42 \pm 0.03 &    134.4 \pm 5.8 &    0.54 \pm 0.03 &    9.02 \pm 0.62 &    1.15 \pm 0.07 \\
G034.0901+00.4365$^{\dagger}$       &  18:51:59.74 &   +01:13:47.63 &             0.24 &             0.18 &    2.34 \pm 0.12 &    1.36 \pm 0.07 &     92.9 \pm 3.6 &    1.79 \pm 0.06 &    6.95 \pm 0.35 &    0.09 \pm 0.00 \\
G034.4032+00.2277       &  18:53:18.67 &   +01:24:47.64 &             0.22 &             0.18 &    0.45 \pm 0.02 &    0.35 \pm 0.01 &     53.2 \pm 6.4 &    0.40 \pm 0.01 &    6.62 \pm 0.19 &    1.32 \pm 0.03 \\
G035.4570$-$00.1791     &  18:56:41.03 &   +02:09:55.33 &             0.22 &             0.18 &    0.96 \pm 0.05 &    0.83 \pm 0.04 &       126 \pm 14 &    0.89 \pm 0.03 &    7.48 \pm 0.37 &    0.36 \pm 0.02 \\
G036.4057+00.0226       &  18:57:41.98 &   +03:06:05.18 &             0.22 &             0.18 &    1.25 \pm 0.04 &    1.17 \pm 0.04 &       119 \pm 20 &    1.21 \pm 0.03 &   25.14 \pm 0.75 &    0.65 \pm 0.02 \\
G037.7347$-$00.1128     &  19:00:36.99 &   +04:13:18.56 &             0.22 &             0.18 &    0.85 \pm 0.04 &    0.84 \pm 0.04 &      116 \pm 161 &    0.85 \pm 0.03 &   17.79 \pm 0.73 &    0.91 \pm 0.04 \\
G037.8209+00.4125       &  18:58:54.03 &   +04:32:18.89 &             0.22 &             0.18 &    1.57 \pm 0.06 &    1.23 \pm 0.05 &     17.2 \pm 6.1 &    1.39 \pm 0.04 &   17.37 \pm 0.63 &    0.35 \pm 0.01 \\
G037.9723$-$00.0965     &  19:00:59.67 &   +04:26:26.59 &             0.22 &             0.18 &    1.63 \pm 0.06 &    1.22 \pm 0.04 &    116.8 \pm 4.8 &    1.41 \pm 0.03 &   15.39 \pm 0.53 &    0.30 \pm 0.01 \\
G038.6465$-$00.2260$^{\dagger}$     &  19:02:41.75 &   +04:58:49.87 &             0.22 &             0.18 &    3.84 \pm 0.12 &    2.76 \pm 0.08 &    124.2 \pm 3.7 &    3.26 \pm 0.07 &     41.9 \pm 1.3 &    0.15 \pm 0.00 \\
G038.6529+00.0875       &  19:01:35.25 &   +05:07:47.30 &             0.22 &             0.18 &    0.24 \pm 0.01 &    0.20 \pm 0.01 &     9.1 \pm 13.9 &    0.22 \pm 0.01 &     8.7 \pm 0.29 &    3.84 \pm 0.09 \\
G038.6934$-$00.4524     &  19:03:35.35 &   +04:55:06.22 &             0.21 &             0.17 &    1.40 \pm 0.05 &    1.26 \pm 0.05 &       124 \pm 15 &    1.33 \pm 0.03 &   19.27 \pm 0.69 &    0.37 \pm 0.01 \\
G039.1956+00.2255       &  19:02:05.58 &   +05:40:31.63 &             0.21 &             0.17 &    0.84 \pm 0.03 &    0.82 \pm 0.03 &        29 \pm 50 &    0.83 \pm 0.02 &     58.7 \pm 1.7 &    2.82 \pm 0.08 \\
G042.1090$-$00.4469     &  19:09:53.56 &   +07:57:15.18 &              0.2 &             0.17 &    0.74 \pm 0.04 &    0.50 \pm 0.03 &    114.1 \pm 5.7 &    0.61 \pm 0.02 &    8.54 \pm 0.44 &    0.71 \pm 0.03 \\
G043.1489+00.0130$^{\ddagger}$       &  19:10:11.05 &   +09:05:20.15 &             0.47 &             0.43 &    0.47 \pm 0.01 &    0.26 \pm 0.01 &      120 \pm 1.5 &    0.35 \pm 0.01 & 99.54 \pm 1.10 & 59.79 \pm 0.45 \\
G043.1520+00.0115$^{\ddagger}$       &  19:10:11.72 &   +09:05:27.79 &             0.47 &             0.43 &    2.16 \pm 0.17 &    1.11 \pm 0.09 &     80.4 \pm 4.4 &    1.55 \pm 0.09 &     335 \pm 26 & 26.02 \pm 1.84 \\
G043.1706$-$00.0003     &  19:10:16.36 &   +09:06:06.98 &             0.47 &             0.43 &    0.95 \pm 0.05 &    0.68 \pm 0.04 &     47.8 \pm 6.5 &    0.80 \pm 0.03 &       385 \pm 17 &   91.70 \pm 3.40 \\
G043.1720+00.0080$^{\ddagger}$       &  19:10:14.75 &   +09:06:25.66 &             0.47 &             0.43 &    1.16 \pm 0.04 &    1.08 \pm 0.04 &       174 \pm 24 &    1.12 \pm 0.03 &   65.4 \pm 2.0 &  9.11 \pm 0.24 \\
G043.7954$-$00.1274a    &  19:11:53.99 &   +09:35:50.30 &              0.2 &             0.17 &    0.31 \pm 0.01 &    0.20 \pm 0.01 &    103.3 \pm 2.8 &    0.25 \pm 0.01 &     54.6 \pm 1.3 &   18.53 \pm 0.33 \\
G043.7954$-$00.1274b    &  19:11:54.06 &   +09:35:49.77 &              0.2 &             0.17 &    0.46 \pm 0.06 &    0.40 \pm 0.05 &       178 \pm 40 &    0.43 \pm 0.04 &    6.91 \pm 0.81 &    1.06 \pm 0.11 \\
G043.9675+00.9939$^{\dagger}$       &  19:08:11.01 &   +10:16:03.12 &             0.24 &             0.15 &    0.76 \pm 0.06 &    0.46 \pm 0.04 &      1.5 \pm 6.5 &    0.59 \pm 0.03 &     30.7 \pm 2.2 &    2.84 \pm 0.18 \\
G044.3103+00.0410a      &  19:12:15.73 &   +10:07:53.54 &             0.24 &             0.15 &    0.80 \pm 0.06 &    0.70 \pm 0.06 &       156 \pm 26 &    0.75 \pm 0.04 &    3.66 \pm 0.25 &    0.22 \pm 0.01 \\
G044.3103+00.0410b      &  19:12:15.78 &   +10:07:52.58 &             0.24 &             0.15 &    0.17 \pm 0.02 &    0.03 \pm 0.02 &     31.7 \pm 4.2 &    0.07 \pm 0.02 &    0.70 \pm 0.03 &    0.56 \pm 0.02 \\
G044.4228+00.5377       &  19:10:40.99 &   +10:27:40.69 &             0.22 &             0.18 &    0.90 \pm 0.05 &    0.73 \pm 0.04 &       160 \pm 11 &    0.81 \pm 0.03 &    6.36 \pm 0.32 &    0.36 \pm 0.02 \\
G045.0694+00.1323       &  19:13:21.81 &   +10:50:48.44 &             0.22 &             0.18 &    1.33 \pm 0.11 &    0.67 \pm 0.06 &     67.6 \pm 4.4 &    0.94 \pm 0.06 &       24.1 \pm 2 &    1.01 \pm 0.08 \\
G048.9901$-$00.2988     &  19:22:26.13 &   +14:06:39.72 &             0.21 &             0.18 &    0.57 \pm 0.02 &    0.16 \pm 0.01 &     61.9 \pm 1.1 &    0.30 \pm 0.01 &    7.45 \pm 0.29 &    1.92 \pm 0.06 \\
G050.3152+00.6762       &  19:21:27.52 &   +15:44:20.78 &             0.21 &             0.18 &    1.13 \pm 0.04 &    0.98 \pm 0.04 &        46 \pm 11 &    1.05 \pm 0.03 &       50.3 \pm 2 &    1.68 \pm 0.06 \\
G053.9589+00.0320       &  19:31:05.23 &   +18:38:16.91 &              0.2 &             0.18 &    1.79 \pm 0.06 &    1.12 \pm 0.04 &    122.1 \pm 2.9 &    1.42 \pm 0.03 &     31.3 \pm 1.1 &    0.56 \pm 0.02 \\
G060.8842$-$00.1286     &  19:46:20.14 &   +24:35:29.31 &             0.23 &             0.14 &    0.75 \pm 0.04 &    0.69 \pm 0.04 &       144 \pm 28 &    0.72 \pm 0.03 &     26.5 \pm 1.1 &    1.57 \pm 0.06 \\
G061.7207+00.8630       &  19:44:23.64 &   +25:48:43.35 &             0.23 &             0.14 &    2.25 \pm 0.06 &    2.23 \pm 0.06 &      127 \pm 122 &    2.24 \pm 0.04 &    112.7 \pm 2.8 &    0.74 \pm 0.02 \\
\hline
\end{longtable}
\end{landscape}

\onecolumn
\begin{landscape}
\setlength\LTcapwidth{\linewidth}
\begin{longtable}{lll S[table-format=1.2] S[table-format=1.2] S[table-format=1.2(2)] S[table-format=1.2(2)] S[table-format=3.1(2.1)] S[table-format=1.2(2)] S[table-format=4.2(3.2)] S[table-format=2.2(1.2)]}
\caption[Photometric results for the higher-resolution data -- neighbour set]{Photometric results for CORNISH UCHII found within the same fields as the targets of the JVLA follow-up (referred to as the neighbour set in the text). All columns and symbols are defined as for the previous table.}
\label{table:fluxes_evla_n} \\
\toprule
  \multicolumn{1}{l}{Name} &
  \multicolumn{1}{c}{$\alpha~\rm (J2000)$} &
  \multicolumn{1}{c}{$\delta~\rm (J2000)$} &
  \multicolumn{1}{c}{$\theta_{\rm _{beam}^{Maj}}$} &
  \multicolumn{1}{c}{$\theta_{\rm _{beam}^{min}}$} &
  \multicolumn{1}{c}{$\theta_{\rm M}$} &
  \multicolumn{1}{c}{$\theta_{\rm m}$} &
  \multicolumn{1}{c}{$\theta_{\rm PA}$} &
  \multicolumn{1}{c}{$\theta_{\rm source}$} &
  \multicolumn{1}{c}{$F_{\rm int}$} &
  \multicolumn{1}{c}{$F_{\rm peak}$} \\

  \multicolumn{1}{l}{} &
  \multicolumn{1}{c}{(hh:mm:ss)} &
  \multicolumn{1}{c}{(dd:mm:ss)} &
  \multicolumn{1}{c}{($''$)} &
  \multicolumn{1}{c}{($''$)} &
  \multicolumn{1}{c}{($''$)} &
  \multicolumn{1}{c}{($''$)} &
  \multicolumn{1}{c}{($\degree$)} &
  \multicolumn{1}{c}{($''$)} &
  \multicolumn{1}{c}{(mJy)} &
  \multicolumn{1}{c}{($\rm mJy~beam^{-1}$)} \\

\hline
\endfirsthead

\multicolumn{11}{c}%
{{\bfseries \tablename\ \thetable{} -- continued from previous page}} \\
\hline
  \multicolumn{1}{l}{Name} &
  \multicolumn{1}{c}{$\alpha~\rm (J2000)$} &
  \multicolumn{1}{c}{$\delta~\rm (J2000)$} &
  \multicolumn{1}{c}{$\theta_{\rm _{beam}^{Maj}}$} &
  \multicolumn{1}{c}{$\theta_{\rm _{beam}^{min}}$} &
  \multicolumn{1}{c}{$\theta_{\rm M}$} &
  \multicolumn{1}{c}{$\theta_{\rm m}$} &
  \multicolumn{1}{c}{$\theta_{\rm PA}$} &
  \multicolumn{1}{c}{$\theta_{\rm source}$} &
  \multicolumn{1}{c}{$F_{\rm int}$} &
  \multicolumn{1}{c}{$F_{\rm peak}$} \\

  \multicolumn{1}{l}{} &
  \multicolumn{1}{c}{(hh:mm:ss)} &
  \multicolumn{1}{c}{(dd:mm:ss)} &
  \multicolumn{1}{c}{($''$)} &
  \multicolumn{1}{c}{($''$)} &
  \multicolumn{1}{c}{($''$)} &
  \multicolumn{1}{c}{($''$)} &
  \multicolumn{1}{c}{($\degree$)} &
  \multicolumn{1}{c}{($''$)} &
  \multicolumn{1}{c}{(mJy)} &
  \multicolumn{1}{c}{($\rm mJy~beam^{-1}$)} \\

\hline
\endhead

\hline \multicolumn{11}{r}{{Continued on next page}} \\ \hline
\endfoot

\hline
\endlastfoot

G020.0809$-$00.1362$^{\dagger}$ &    18:28:10.38 & $-$11:28:48.45 &             0.27 &             0.17 &    4.11 \pm 0.29 &    1.40 \pm 0.10 &    113.9 \pm 2.2 &    2.40 \pm 0.12 &       597 \pm 41 &    4.69 \pm 0.32 \\
G025.3983+00.5617   &    18:35:37.55 & $-$06:26:33.51 &             0.24 &             0.17 &    2.68 \pm 0.08 &    2.43 \pm 0.07 &        42 \pm 13 &    2.55 \pm 0.05 &     96.6 \pm 2.8 &    0.61 \pm 0.02 \\
G031.1596+00.0448   &    18:48:02.71 & $-$01:33:24.73 &             0.24 &             0.18 &    0.99 \pm 0.04 &    0.89 \pm 0.04 &       175 \pm 24 &    0.94 \pm 0.03 &   24.19 \pm 0.95 &    1.14 \pm 0.04 \\
G032.7966+00.1909a  &    18:50:31.00 &   $-$00:01:56.90 &             0.24 &             0.18 &    1.07 \pm 0.06 &    0.96 \pm 0.05 &       173 \pm 53 &    1.02 \pm 0.04 &       267 \pm 14 &   10.65 \pm 0.52 \\
G032.7966+00.1909b$^{\dagger}$  &    18:50:30.80 &   $-$00:01:53.81 &             0.24 &             0.18 &    3.96 \pm 0.41 &    1.48 \pm 0.15 &     15.3 \pm 3.7 &    2.42 \pm 0.18 &       301 \pm 31 &    2.18 \pm 0.22 \\
G032.7966+00.1909c$^{\dagger}$  &    18:50:30.73 &   $-$00:01:57.91 &             0.24 &             0.18 &    1.34 \pm 0.11 &    0.75 \pm 0.07 &       43 \pm 5.6 &    1.00 \pm 0.06 &       145 \pm 12 &    5.88 \pm 0.47 \\
G032.7966+00.1909d  &    18:50:30.68 &   $-$00:02:00.39 &             0.24 &             0.18 &    2.02 \pm 0.07 &    1.32 \pm 0.04 &       50.2 \pm 3 &    1.63 \pm 0.04 &      1121 \pm 36 &   17.64 \pm 0.56 \\
G036.4062+00.0221a  &    18:57:42.14 &   +03:06:06.22 &             0.22 &             0.18 &    0.34 \pm 0.01 &    0.32 \pm 0.01 &       167 \pm 34 &    0.33 \pm 0.01 &   10.71 \pm 0.21 &    2.88 \pm 0.05 \\
G036.4062+00.0221b  &    18:57:42.23 &   +03:06:05.25 &             0.22 &             0.18 &    0.26 \pm 0.04 &    0.19 \pm 0.03 &        19 \pm 48 &    0.22 \pm 0.03 &    0.82 \pm 0.10 &    0.37 \pm 0.03 \\
G043.1651$-$00.0283$^{\ddagger}$ &    19:10:21.78 &   +09:05:02.64 &             0.47 &             0.43 &    3.79 \pm 0.13 &    1.31 \pm 0.05 &      170.9 \pm 1 &    2.23 \pm 0.05 &    1706 \pm 57 & 65.38 \pm 2.06 \\
G043.1652+00.0129$^{\ddagger}$   &    19:10:12.89 &   +09:06:11.77 &             0.47 &             0.43 &    0.67 \pm 0.05 &    0.47 \pm 0.05 &       132 \pm 12 &    0.56 \pm 0.04 &     256 \pm 16 & 99.05 \pm 4.60 \\
G043.1657+00.0116$^{\ddagger}$   &    19:10:13.22 &   +09:06:11.19 &             0.23 &             0.14 &    0.83 \pm 0.06 &    0.75 \pm 0.06 &        91 \pm 33 &    0.79 \pm 0.04 &     339 \pm 22 & 15.91 \pm 0.98 \\
G043.1665+00.0106$^{\dagger\ddagger}$   &    19:10:13.55 &   +09:06:12.43 &             0.47 &             0.43 &    4.89 \pm 0.30 &    2.93 \pm 0.18 &     95.7 \pm 4.5 &    3.79 \pm 0.16 &   3856 \pm 238 & 53.78 \pm 3.24 \\
G043.1674+00.0128$^{\ddagger}$   &    19:10:13.16 &   +09:06:18.61 &             0.47 &             0.43 &    0.85 \pm 0.07 &    0.78 \pm 0.07 &        84 \pm 61 &    0.81 \pm 0.05 &     327 \pm 22 & 76.74 \pm 4.33 \\
G043.1684+00.0087$^{\ddagger}$   &    19:10:14.16 &   +09:06:15.82 &             0.47 &             0.43 &    2.03 \pm 0.17 &    1.67 \pm 0.14 &       130 \pm 19 &    1.84 \pm 0.11 &     324 \pm 26 & 18.29 \pm 1.37 \\
G043.1699+00.0115$^{\ddagger}$   &    19:10:13.76 &   +09:06:24.91 &             0.47 &             0.43 &    1.64 \pm 0.13 &    1.16 \pm 0.10 &    156.5 \pm 9.4 &    1.38 \pm 0.08 &   77.8 \pm 6.2 &  7.48 \pm 0.54 \\
G043.1701+00.0078$^{\ddagger}$   &    19:10:14.54 &   +09:06:19.48 &             0.47 &             0.43 &    5.37 \pm 0.26 &    1.39 \pm 0.07 &      153.1 \pm 1 &    2.73 \pm 0.09 &     344 \pm 16 &  8.84 \pm 0.41 \\
G043.1716+00.0001   &    19:10:16.41 &   +09:06:10.26 &             0.23 &             0.14 &    2.92 \pm 0.18 &    1.06 \pm 0.07 &     57.9 \pm 2.1 &    1.76 \pm 0.08 &     67.8 \pm 4.1 &    0.66 \pm 0.04 \\
G045.0712+00.1321   &    19:13:22.08 &   +10:50:53.25 &             0.22 &             0.18 &    0.78 \pm 0.03 &    0.57 \pm 0.03 &    106.6 \pm 5.6 &    0.66 \pm 0.02 &       331 \pm 14 &   26.60 \pm 1.00 \\

\hline

\end{longtable}
\bsp
\label{lastpage}
\end{landscape}

\end{document}